\documentclass[12pt]{report}

\usepackage[online]{suthesis-2e}
\usepackage{array}
\usepackage{graphicx}
\usepackage{verbatim}
\usepackage{color}
\usepackage{amsmath}
\usepackage{longtable}
\usepackage[square,sort&compress,numbers]{natbib}
\usepackage{rotating}
\usepackage[table,usenames,dvipsnames]{xcolor}
\usepackage{multirow}
\usepackage{booktabs}
\usepackage{enumerate}
\usepackage[hidelinks]{hyperref}

\newcommand{\ket}[1]{\vert{#1}\rangle}
\newcommand{\bra}[1]{\langle{#1}\vert}

\begin{document}

    \title{Logic Synthesis for Fault-Tolerant Quantum Computers}
    \author{Nathan Cody Jones}
    \dept{Electrical Engineering}
    \principaladviser{Yoshihisa Yamamoto}
    \firstreader{Jelena Vuckovic}
    \secondreader{William D. Oliver,\\MIT Lincoln Laboratory}

    \beforepreface
    \prefacesection{Abstract}

        Efficient constructions for quantum logic are essential since quantum computation is experimentally challenging.  This thesis develops quantum logic synthesis as a paradigm for reducing the resource overhead in fault-tolerant quantum computing.  The model for error correction considered here is the surface code.  After developing the theory behind general logic synthesis, the resource costs of magic-state distillation for the $T = \exp(i \pi (I-Z)/8)$ gate are quantitatively analyzed.  The resource costs for a relatively new protocol distilling multi-qubit Fourier states are calculated for the first time.  Four different constructions of the fault-tolerant Toffoli gate, including two which incorporate error detection, are analyzed and compared.  The techniques of logic synthesis reduce the cost of fault-tolerant quantum computation by one to two orders of magnitude, depending on which benchmark is used.

        Using resource analysis for $T$~gates and Toffoli gates, several proposals for constructing arbitrary quantum gates are compared, including ``Clifford+$T$'' sequences, $V$-basis sequences, phase kickback, and programmable ancilla rotations.  The application of arbitrary gates to quantum algorithms for simulating chemistry is discussed as well.  Finally, the thesis examines the techniques which lead to efficient constructions of quantum logic, and these observations point to even broader applications of logic synthesis.

    \prefacesection{Acknowledgements}
        As you might expect, there is a long list of people to thank for contributing to a doctoral education.  I owe a large debt of gratitude to my advisor, Yoshi Yamamoto.  From the beginning, he knew that the best way to guide me was a hands-off approach.  I was allowed to make my own mistakes (and fix them), knowing that I could count on his gentle counsel when it was needed.

        There are so many members of the Yamamoto group that deserve thanks.  Among the people that I worked with and learned from are: Darin Sleiter, Leo Yu, Shruti Puri, Chandra Natarajan, Na Young Kim, Kaoru Sanaka, Wolfgang Nitsche, Zhe Wang, Georgios Roumpos, Kai Wen, Dave Press, and Susan Clark.  Kristiaan De Greve and Peter McMahon deserve special mention for their parts in (mostly) high-minded conversations, whether at Ike's or the Rose and Crown, and for being true friends.  Thaddeus Ladd only overlapped with me for three months in the group, but I learned a tremendous amount from him.  He is largely responsible for developing my interest in quantum computer architecture, and I continue to learn from his astonishing breadth of knowledge (except in Starcraft, perhaps the only subject I know better).  I must also thank Yurika Peterman and Rieko Sasaki, who work tirelessly in the background to keep the group functioning.

        There are many academic collaborators from whom I learned a great deal.  Jungsang Kim, a Yamamoto group alumnus, helped me focus on the engineering aspects of quantum architecture.  In Japan, I found kindred spirits to my architecture research in Rod Van Meter, Simon Devitt, and Bill Munro.  Austin Fowler taught me a great deal about the surface code, and maybe someday I will know half as much as he does on the matter.  At Harvard, I learned far more about quantum chemistry than I ever would have expected from James Whitfield, Man-Hong Yung, and Al\'{a}n Aspuru-Guzik.  I was privileged to work on simulation and dynamical decoupling with Bryan Fong, Jim Harrington, and Thaddeus at HRL Laboratories, and I look forward to joining them as a colleague.

        I want to thank the National Science Foundation, which supported my graduate studies through the Graduate Research Fellowships Program.

        Finally, I want to thank my brother, Jason, and my loving parents, who made me into the man I am.

    \afterpreface


\chapter{Introduction to Quantum Computing}
\label{Ch01}
Quantum computing is a research field that promises to solve problems that normal computers cannot, using quantum physics.  The notion of a powerful computer built on the esoteric rules of quantum mechanics sounds like an idea from science fiction, and the field has generated considerable interest in technical communities and the general public.  However, quantum computing is built on very sound principles.  There is ample theoretical analysis that shows the concept is viable, under the right conditions~\cite{Shor1996,Aharonov1997,Preskill1998}.  Furthermore, intense experimental work is steadily improving the reliability of quantum hardware~\cite{Ladd2010,Mariantoni2011,Monroe2013}.  Quantum computing is not science fiction, and the best evidence for this assertion is that the future of the field is not novel discoveries in physics, but rather steady advances in engineering.

The topic of this thesis is quantum logic synthesis.  This is just one component of designing a quantum computer, but I will argue several times that it is a very important component.  Logic synthesis is concerned with arranging the instructions in a quantum computer to minimize resource costs, such as number of quantum bits (``memory'') and gates (``calculations'').  Based on current understanding of experimental hardware, error correction will be the most costly feature of a quantum computer, and logic synthesis will play a crucial role in managing these costs.

As an introduction to the subject, this chapter gives a high-level overview of quantum computing.  I start with applications, just to show why the field has attracted the attention of so many.  The next section gives a basic primer on quantum bits, gates, and measurement.  The last section covers the greatest challenge for quantum information, noise and errors.  For a more comprehensive introduction, I refer the reader to Ref.~\cite{Nielsen2000}.

\section{Applications of Quantum Computing}
Quite a few applications for quantum computers have been identified.  A website maintained by Stephen Jordan has the most comprehensive list that I have seen (``Quantum Algorithm Zoo,''~\cite{Jordan_zoo}), which currently counts 50 different algorithms. The performance advantage for each over a classical computer varies, from polynomial to exponential to unknown.  Some applications are very general, while others address narrowly defined problems.  This section discusses just a handful of these algorithms, the ones which I think will have the greatest impact.

Shor's integer-factoring algorithm~\cite{Shor1999} is one of the oldest and most widely known applications of quantum computing.  Due to the connection to RSA cryptography~\cite{RSA1978}, the integer factoring problem was already a problem of considerable interest to computer science.  The heart of the matter is that multiplication of integers is computationally efficient (polynomial-bounded time and space complexity), but no efficient method in classical computing is known to decompose an integer into its prime factors.  For some time, the digital security firm RSA Security (founded by the creators of the protocol) held an open challenge to factor numbers typical of the RSA protocol~\cite{RSA_challenge}.  In this case, the number to be factored is $N = pq$, where $p$ and $q$ are prime numbers, typically both very large in size (\emph{e.g.} around 1000 bits).  Shor's algorithm demonstrates that quantum computers can factor such a number in polynomial-bounded time and space.  Nevertheless, the computation is rather complex when error correction is included, requiring perhaps millions of qubits and billions of gates~\cite{Isailovic2008,Jones2012_PRX,Fowler2012_Architecture}.

Simulating quantum physics is another problem ideal for quantum computers~\cite{Lloyd1996,Aspuru2005}.  In this application, the state of a quantum system is encoded into quantum bits, and the time evolution of this state is reproduced in simulated time using quantum gates.  Many useful quantities can be calculated, such as energy eigenvalues and chemical reaction rates~\cite{Kassal2011}. Simulation algorithms will be examined in more detail in Chapter~\ref{Ch08}.  Multiple forms of encoding are possible, and the choice has consequences for the way logic is synthesized.  More recently, a closely related linear systems algorithm has been proposed~\cite{Harrow2009}, which may have promising applications like solving partial differential equations for electromagnetic scattering~\cite{Clader2013}.

In these and other cases, the quantum computer solves a particular computational task better than a conventional computer.  Even in doing so, the quantum computer requires substantial classical computing support for pre- and post-processing, as well as managing the considerable task of quantum error correction~\cite{Devitt2010,Levy2011}.  For these reasons, quantum computers are appropriately viewed as ``co-processors'' that perform specialized tasks in a classical-quantum hybrid computing environment.

\section{Quantum States, Operations, and Measurement}
The information states in a quantum computer are normalized, complex-valued vectors.  The elements of each such vector correspond to the projection into a basis.  The most common basis will be the ``computational basis'', which is spanned by binary values.  For example, a single quantum bit, or qubit, is a superposition of the states $\ket{0}$ and $\ket{1}$ (the ``ket'' notation is a convention for identifying states).  Quantum states must be normalized, so an arbitrary qubit state can be specified by
\begin{equation}
\ket{\psi} = \alpha \ket{0} + \beta \ket{1},
\end{equation}
subject to the constraint that
\begin{equation}
\left|\alpha\right|^2 + \left|\beta\right|^2 = 1.
\end{equation}
Normalization ensures that, for measurement processes, the sum of probabilities for all outcomes sums to one.  Quantum states can consist of multiple qubits.  Furthermore, ``global phase,'' which is a scalar $e^{i\phi}$ coefficient to any state, is meaningless in quantum mechanics.  In general, the possible $n$-qubit states span a basis of dimension $2^n$, with the additional degrees of freedom of allowing complex amplitudes.  For simplicity, states consisting of multiple qubits use abbreviated state notation, such as $\ket{1} \otimes \ket{0} \otimes \ket{1} \equiv \ket{101}$.

Some notational shorthand will be used throughout this thesis.  The Pauli spin operators, which are used to define operations on a qubit, will be denoted as $X \equiv \sigma^x$, $Y \equiv \sigma^y$, and $Z \equiv \sigma^z$.  Similarly, $I$ is the identity operator, with dimensionality appropriate for its context.  As an example, one might write the projector onto the $\ket{0}$ state as $\ket{0}\bra{0} = \frac{1}{2}(I + Z)$, where $I$, like $Z$, has dimension two.

The state of a quantum system is modified using gates.  Each gate $U$ is unitary operator, meaning $U^{\dag}U = I$, where ``$\dag$'' denotes conjugate transpose (Hermitian adjoint).  One example is $X = \ket{0}\bra{1} + \ket{1}\bra{0}$.  When applied to a state, each righthand ``bra'' such as $\bra{0}$ combines with a ket to form an inner product, which is a scalar-valued quantity.  For example, $\langle 0 | 0 \rangle = 1$, because the bra and ket vectors are parallel.  Likewise, $\langle 1 | 0 \rangle = 0$, because the two unit vectors are orthogonal.   As a result, the action of $X$ on $\ket{0}$ is $X\ket{0} = \left(\ket{0}\bra{1} + \ket{1}\bra{0}\right)\ket{0} = \ket{0}(\langle 1 | 0 \rangle) + \ket{1}(\langle 0 | 0 \rangle) = \ket{1}$.

\begin{figure}
  \centering
  \includegraphics[width=7cm]{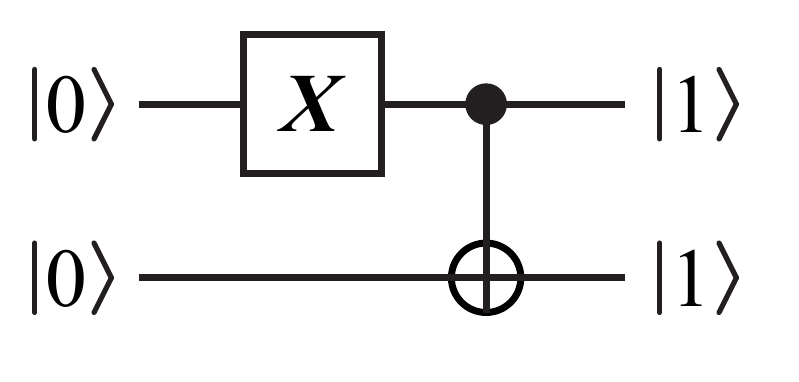}\\
  \caption[Circuit: Example of a circuit diagram]{Example of a quantum circuit diagram.  The $X$ gate flips the top qubit to $\ket{1}$, then the CNOT will flip the state of the bottom qubit.}
  \label{Ch1_circuit_example}
\end{figure}

A sequence of gates is often illustrated with a ``circuit diagram,'' as shown in Fig.~\ref{Ch1_circuit_example}.  Operations read left to right in time, so first an $X$ gate flips the state of the top qubit, then a controlled-NOT (CNOT) gate will apply $X$ to the bottom qubit if the first is $\ket{1}$, otherwise it does nothing.  The CNOT is a two-qubit gate, meaning that it couples the state of two qubits.  The output state is $\ket{11}$.

In addition to gates, quantum computation also relies on measurement to reveal the underlying state.  However, the superposition nature of quantum states means that there is no single basis in which to measure states.  This thesis will only consider strong projective measurements, where the measurement process projects the quantum system into one of several orthogonal states.  This can be described using projectors, such as $P_a = \ket{a}\bra{a}$.  A complete measurement basis is defined by the set $\{P_i\}$ such that $\sum_i P_i = I$.  For some state $\ket{\psi}$, the probability of measuring outcome $a$ and projecting the system into $\ket{a}$ is given by $\bra{\psi} P_a \ket{\psi}$.  The state after measurement result $a$ is determined by
\begin{equation}
\ket{a} = e^{i\phi}\frac{P_a\ket{\psi}}{\sqrt{\bra{\psi} P_a \ket{\psi}}},
\end{equation}
which is known as the projection postulate of quantum mechanics (as noted above, the complex phase here has no effect).  Informally, the system becomes consistent with the observed measurement.  Frequently, measurement bases are those of Pauli operators, which play an important role in error correction~\cite{Gottesman1997,Nielsen2000}.  For example, the computational basis is $P_0 = \frac{1}{2}(I + Z)$ and $P_1 = \frac{1}{2}(I - Z)$.  This common measurement operation will be denoted $M_Z$.

In addition to gates and measurement, one must be able to initialize to a well-defined quantum state.  The reason for saving this process for last is that initialization and measurement are dual operations.  They are both non-unitary with respect to the logical space of computable states.  Furthermore, the measurement process can be used to perform initialization, using the projection postulate of quantum mechanics.  Other methods, such as cooling the system to a ground state, are also used in practice.

\section{Noise and Decoherence}
Quantum operations are not perfect.  Gates will probabilistically introduce errors, and even idle qubits will experience ``decoherence,'' disturbance from the original state due to interactions with the environment.  In all hardware platforms considered so far, the error rates are so high as to make error correction mandatory for reliable computing.  Error correction will be the subject of later chapters, so I will just briefly review quantum errors here.

In general, an error is any change in the state of a quantum system that is not perfectly known by the system controller.  Errors are modeled with a quantum distribution known as a density matrix.  Previously, I introduced vectors like $\ket{0}$ that are ``pure'' states, having no error and being perfectly defined.  A density matrix is a probability-weighted distribution of pure states, which gives the likelihood of the system being in each of those states.  For example, $\rho = 0.9 \ket{0}\bra{0} + 0.1 \ket{1}\bra{1}$ is ``mixed'' state for a system with 90\% probability of being $\ket{0}$.  The normalization for a density matrix is having a trace (sum of diagonal entries) of one: $\mathrm{tr}(\rho) = 1$.

Density matrices are useful for modeling quantum noise.  Imagine one starts in the pure state $\ket{+} = (1/\sqrt{2})(\ket{0} + \ket{1})$, and this state experiences dephasing noise, which is a positive probability of the phase being flipped by operator $Z$.  Consider the dephasing channel $D(\rho) = 0.9 \rho + 0.1 Z \rho Z$.  The initial density matrix is
\begin{equation}
\ket{+}\bra{+} = \frac{1}{2}\left[\begin{array}{cc} 1 & 1 \\ 1 & 1 \end{array}\right].
\end{equation}
After dephasing, the state is
\begin{equation}
D(\ket{+}) = \frac{1}{2}\left[\begin{array}{cc} 1 & 0.8 \\ 0.8 & 1 \end{array}\right].
\end{equation}
If the qubit continues to dephase through $t$ such time intervals, the state is
\begin{equation}
D^{(t)}(\ket{+}) = \frac{1}{2}\left[\begin{array}{cc} 1 & 0.8^t \\ 0.8^t & 1 \end{array}\right],
\end{equation}
where it becomes clear that dephasing is damping the off-diagonal terms of the density matrix.  In the limit $t \rightarrow \infty$, dephasing turns the quantum state into an evenly mixed distribution of states $\ket{0}$ and $\ket{1}$.  Because phase is critically important to most quantum algorithms, this type of error will corrupt data.

A dephasing event can be modeled with the probabilistic application of operator $Z$.  Another common error channel is the depolarizing channel,
\begin{equation}
E(\rho) = (1-\epsilon) \rho + (\epsilon/3)(X \rho X + Y \rho Y + Z \rho Z)
\end{equation}
which applies one of the Pauli errors with probability $\epsilon/3$ each, for total probability of error $\epsilon$.  Viewing error events this way avoids the need to explicitly write density matrices, allowing one to analyze error correction without knowing the underlying state.  At an abstract level, error correction will act as a filter to catch these errors by performing measurements which reveal what error (if any) has occurred to the system.  This technique will be used often in Chapters~\ref{Ch05}--\ref{Ch07}. 


\chapter{Architecture of a Quantum Computer}
\label{Ch02}
Quantum computing as an engineering discipline is still in its infancy.  Although the physics is well understood, developing devices which compute with quantum mechanics is technologically daunting.  While experiments to date manipulate only a handful of quantum bits~\cite{Ladd2010}, this chapter considers what effort is required to build a large-scale quantum computer.  One must consider the faulty quantum hardware, with errors caused by both the environment and deliberate control operations; when error correction is invoked, classical processing is required; constructing arbitrary gate sequences from a limited fault-tolerant set requires special treatment, and so on.  Quantum computer architecture, the subject of this chapter, is a framework to address the complete challenge of designing a quantum computer.

This chapter provides an overview of the steps for designing a quantum computer, and it is based on Ref.~\cite{Jones2012_PRX}.  The chapter concludes with resource estimates for large-scale quantum computation.  Although they were the best estimates when Ref.~\cite{Jones2012_PRX} was published, the logic being used was not optimized.  The daunting numbers, such as $\sim 100$ million qubits, serve as a pretext for why resource-reduction techniques through logic synthesis, the subject of this thesis, are so important.

\section{Layered Architecture Overview}
Many researchers have presented and examined components of large-scale quantum computing.  This chapter considers how these components may be combined in an efficient design, and later chapters introduce methods to improve the quantum computer.  This engineering pursuit is quantum computer architecture, which is developed here in layers.  An \emph{architecture} decomposes complex system behaviors into a manageable set of operations.  A \emph{layered architecture} does this through layers of abstraction where each embodies a critical set of related functions.  Each ascending layer brings the system closer to an ideal quantum computing environment by suppressing errors and hiding implementation details not needed elsewhere.  This section reviews the field of quantum computer architecture, then discusses the layered architecture of Ref.~\cite{Jones2012_PRX}.

\subsection{Prior Work on Quantum Computer Architecture}
Many different quantum computing technologies are under experimental investigation~\cite{Ladd2010}, but for each a scalable system architecture remains an open research problem.  Since DiVincenzo introduced his fundamental criteria for a viable quantum computing technology~\cite{DiVincenzo2000} and Steane emphasized the difficulty of designing systems capable of running quantum error correction (QEC) adequately~\cite{Steane1998,Steane2002,Steane2007}, several groups of researchers have examined the architectural needs of large-scale systems~\cite{Spiller2005,VanMeter2006_Architecture}.  As an example, small-scale interconnects have been proposed for many technologies, but the problems of organizing subsystems using these techniques into a complete architecture for a large-scale system have been addressed by only a few researchers.  In particular, the issue of heterogeneity in system architecture has received relatively little attention.

The most important subroutine in fault-tolerant quantum computers considered thus far is the preparation of ancilla states for fault-tolerant circuits, because very many ancillas are required~\cite{Isailovic2008,Jones2012_PRX,Fowler2012_Architecture}.  Taylor~\mbox{\emph{et al.}} proposed a design with alternating ``ancilla blocks'' and ``data blocks'' in the device layout~\cite{Taylor2005}.  Steane introduced the idea of ``factories'' for creating ancillas~\cite{Steane1998}, as examined later in this chapter.  Isailovic~\mbox{\emph{et al.}}~\cite{Isailovic2008} studied this problem for ion trap architectures and found that, for typical quantum circuits, approximately 90\% of the quantum computer must be devoted to such factories in order to calculate ``at the speed of data,'' or where ancilla-production is not the rate-limiting process.  The results in this chapter are in close agreement with this estimate.  Metodi~\mbox{\emph{et al.}} also considered production of ancillas in ion trap designs, focusing instead on a 3-qubit ancilla state used for the Toffoli gate~\cite{Metodi2005}, which is an alternative pathway to a universal fault-tolerant set of gates.

Some researchers have studied the difficulty of moving data in a quantum processor.  Kielpinski \emph{et al.}~proposed a scalable ion trap technology utilizing separate memory and computing areas~\cite{Kielpinski2002}.  Because quantum error correction requires rapid cycling across all physical qubits in the system, this approach is best used as a unit cell replicated across a larger system.   Other researchers have proposed homogeneous systems built around this basic concept.  One common structure is a recursive H tree, which works well with a small number of layers of a Calderbank-Shor-Steane (CSS) code, targeted explicitly at ion trap systems~\cite{Copsey2003,Svore2006}.   Oskin~\mbox{\emph{et al.}}~\cite{Oskin2003}, building on the Kane solid-state NMR technology~\cite{Kane1998}, proposed a loose lattice of sites, explicitly considering the issues of classical control and movement of quantum data in scalable systems, but without a specific plan for QEC.  In the case of quantum computing with superconducting circuits, the quantum von Neumann architecture specifically considers dedicated hardware for quantum memories, zeroing registers, and a quantum bus~\cite{Mariantoni2011}.

Long-range coupling and communication is a significant challenge for quantum computers.  Cirac~\emph{et al.} proposed the use of photonic qubits to distribute entanglement between distant atoms~\cite{Cirac1997}, and other researchers have investigated the prospects for optically-mediated nonlocal gates~\cite{vanEnk1999,Steane2000,Duan2001,VanMeter2007,Duan2010}.  Such photonic channels could be utilized to realize a modular, scalable distributed quantum computer~\cite{Kim2009}.  Conversely, Metodi~\mbox{\emph{et al.}} consider how to use local gates and quantum teleportation to move logical qubits throughout their ion-trap QLA architecture~\cite{Metodi2005}.  Fowler~\mbox{\emph{et al.}}~\cite{Fowler2007} investigated a Josephson junction flux qubit architecture considering the extreme difficulties of routing both the quantum couplers and large numbers of classical control lines, producing a structure with support for CSS codes and logical qubits organized in a line.  Whitney~\mbox{\emph{et al.}}~\cite{Whitney2007,Whitney2009} have investigated automated layout and optimization of circuit designs specifically for ion trap architectures, and Isailovic~\mbox{\emph{et al.}}~\cite{Isailovic2006,Isailovic2008} have studied interconnection and data throughput issues in similar ion trap systems, with an emphasis on preparing ancillas for teleportation gates~\cite{Gottesman1999}.

Other work has studied quantum computer architectures with only nearest-neighbor coupling between qubits in an array~\cite{Levy2001,Fowler2004_Linear,Aliferis2007_BaconShor,Levy2009,Levy2011,Jones2012_PRX,Fowler2012_Architecture}, which is appealing from a hardware design perspective.  With the recent advances in the operation of the topological codes and their desirable characteristics such as having a high practical threshold and requiring only nearest-neighbor interactions, research effort has shifted toward architectures capable of building and maintaining large two- and three-dimensional cluster states~\cite{Weinstein2005,Stock2009,Devitt2010,Devitt2011_HighPerformance}. These systems rely on topological error correction models~\cite{Kitaev2002,Raussendorf2007_PRL,Raussendorf2007_NJP,Landahl2011}, whose higher tolerance to error often comes at the cost of a larger footprint in the hardware, relative to, for example, implementations based on the Steane code~\cite{Oskin2002}.  The surface code~\cite{Raussendorf2007_PRL,Raussendorf2007_NJP,Fowler2009_PRA,Fowler2010_Error,Fowler2012_Architecture}, which is studied throughout this thesis, belongs to the topological family of codes.

Recent attention has been directed at distributed models of quantum computing.  Devitt~\mbox{\emph{et al.}} studied how to distribute a photonic cluster-state quantum computing network over different geographic regions~\cite{Devitt2011_Perpetual}.  The abstract framework of a quantum multicomputer recognizes that large-scale systems demand heterogeneous interconnects~\cite{VanMeter2006_Distributed}.  For most quantum computing technologies, it may be challenging to build monolithic systems that contain, couple, and control billions of physical qubits.  Van~Meter~\mbox{\emph{et al.}}~\cite{VanMeter2009} extended this architectural framework with a design based on nanophotonic coupling of electron spin quantum dots that explicitly uses multiple levels of interconnect with varying coupling fidelities (resulting in varying purification requirements), as well as the ability to operate with a very low yield of functional devices.  Although that proposed system has many attractive features, concerns about the difficulty of fabricating adequately high quality optical components and the desire to reduce the surface code lattice cycle time led to the system design proposed in Ref.~\cite{Jones2012_PRX}.

\subsection{Layered Framework}
A good architecture must have a simple structure while also efficiently managing the complex array of resources in a quantum computer.  Layered architectures are a conventional approach to solving such engineering problems in many fields of information technology.  For example, Ref.~\cite{Svore2006} presents a layered architecture for quantum computer design software.  The architecture developed in Ref.~\cite{Jones2012_PRX} describes the physical design of the quantum computer, which consists of five layers, where each layer has a prescribed set of duties to accomplish.  The interface between two layers is defined by the services a lower layer provides to the one above it.  To execute an operation, a layer must issue commands to the layer below and process the results.  Designing a system this way ensures that related operations are grouped together and that the system organization is hierarchical.  Such an approach allows quantum engineers to focus on individual challenges, while also seeing how a process fits into the overall design.  The architecture is organized in layers to deliberately create a modular design for the quantum computer.

\begin{figure}
  \centering
  \includegraphics[width=10cm]{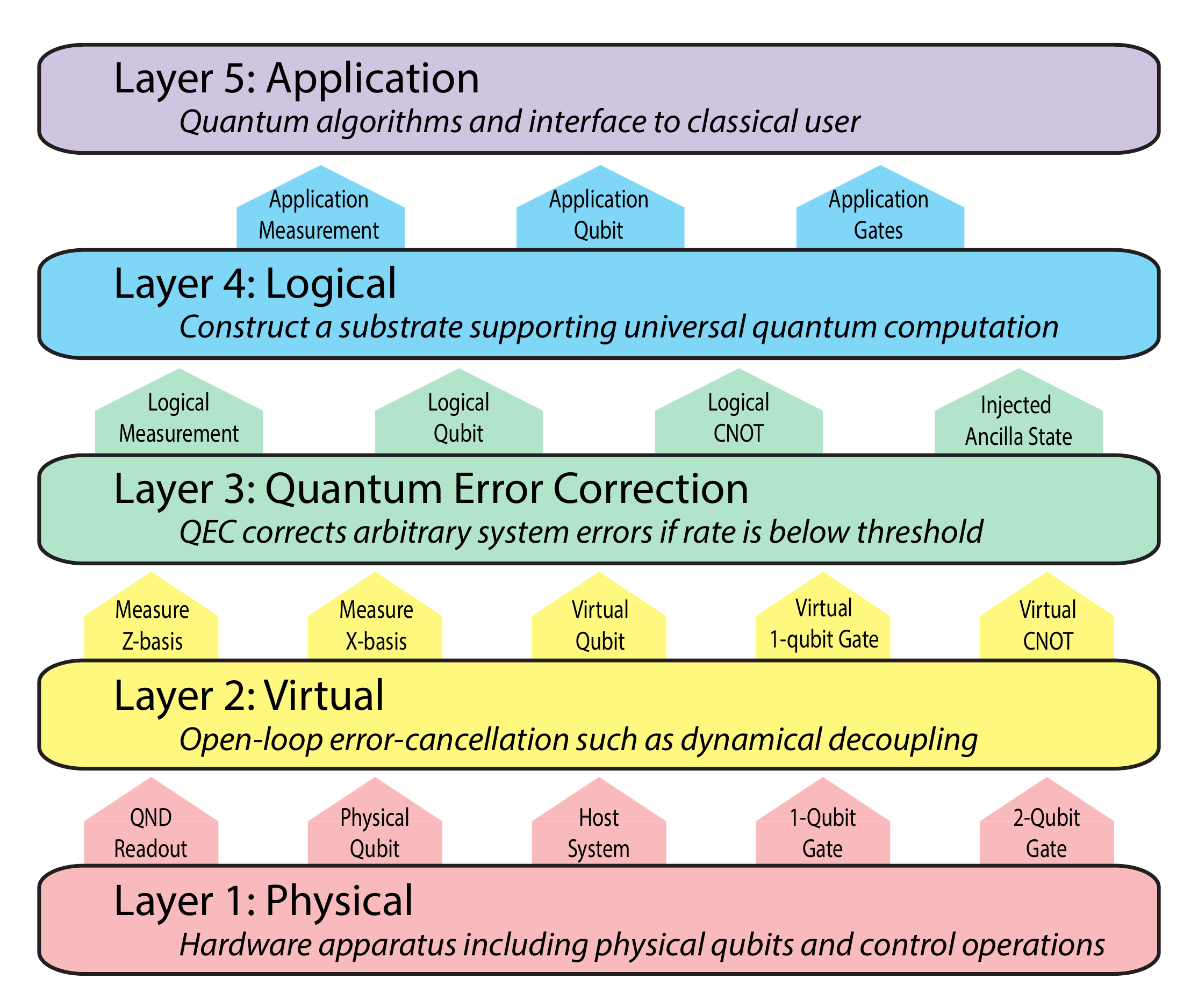}\\
  \caption[Layered architecture control stack]{Layered control stack which forms the framework of a quantum computer architecture.  Vertical arrows indicate services provided to a higher layer.  Originally published in Ref.~\cite{Jones2012_PRX}.}
  \label{LayeredControlStack}
\end{figure}

The layered framework can be understood by a control stack composed of the five layers in the architecture.  Figure~\ref{LayeredControlStack} shows an example of the control stack for the specific quantum dot architecture considered in this chapter~\cite{Jones2012_PRX}, but the particular interfaces between layers will vary according to the physical hardware, quantum error correction scheme, \emph{etc.} that one chooses to implement.  At the top of the control stack is the Application layer, where a quantum algorithm is implemented and results are provided to the user. The bottom Physical layer hosts the raw physical processes supporting the quantum computer. The layers between (Virtual, Quantum Error Correction, and Logical) are essential for shaping the faulty quantum processes in the Physical layer into a system of reliable fault-tolerant~\cite{Preskill1998,Preskill1998_FTQC,Nielsen2000,Steane2002,Steane2007,Devitt2013_Idiot} qubits and quantum gates at the Application layer.

\subsection{Interaction between Layers}
\begin{figure}
  \centering
  \includegraphics[width=\textwidth]{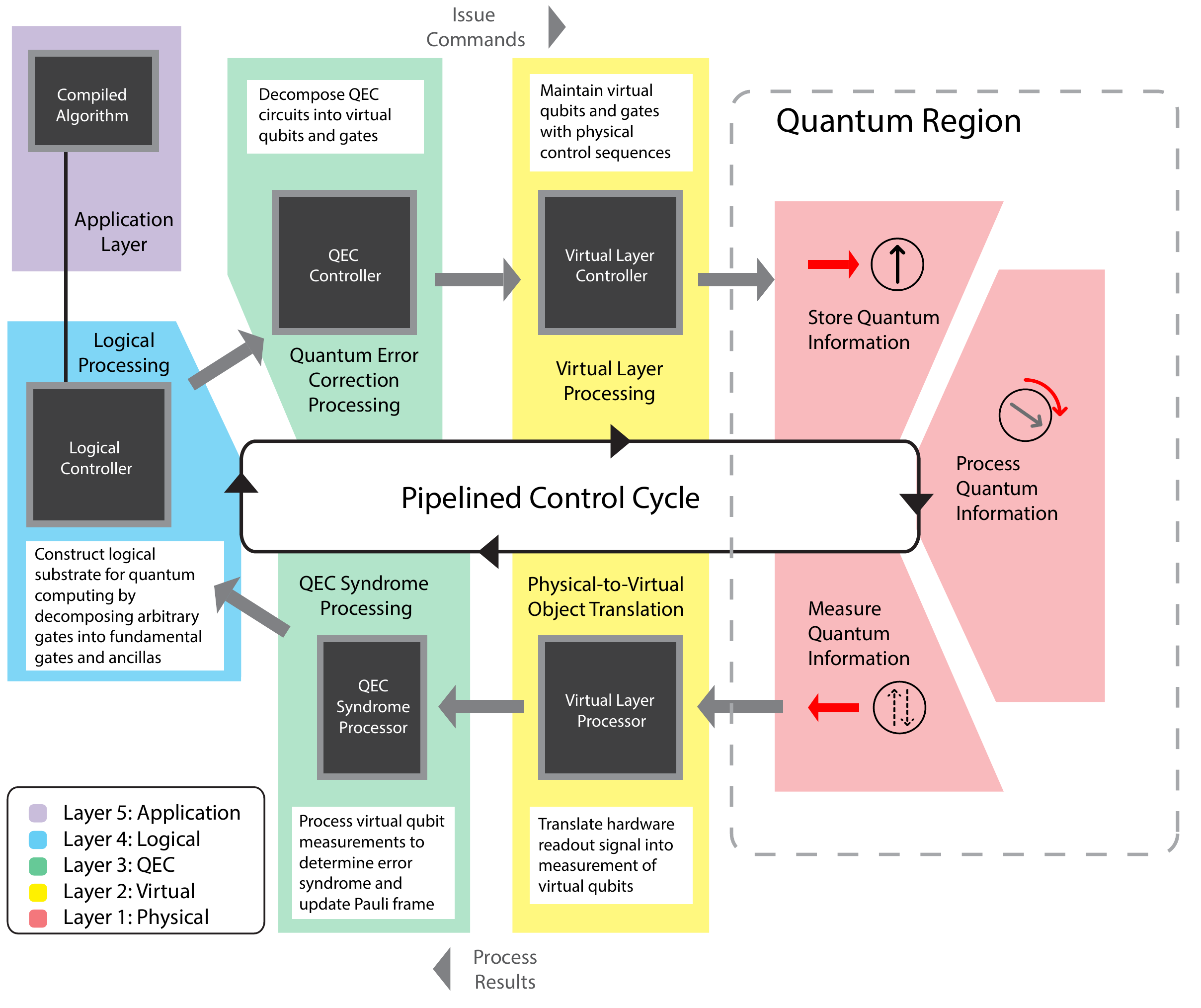}\\
  \caption[Layered architecture control cycle]{Primary control cycle of the layered architecture quantum computer.  Whereas the control stack in Fig.~\ref{LayeredControlStack} dictates the interfaces between layers, the control cycle determines the timing and sequencing of operations.  The dashed box encircling the Physical layer indicates that all quantum processes happen exclusively here, and the layers above process and organize the operations of the Physical layer.  The Application layer is external to the loop since it functions without any dependence on the specific quantum computer design.  Originally published in Ref.~\cite{Jones2012_PRX}.}
  \label{LayeredControlCycle_Abstract}
\end{figure}

Two layers meet at an interface, which defines how they exchange instructions or the results of those instructions.  Many different commands are being executed and processed simultaneously, so one must also consider how the layers interact dynamically. For the quantum computer to function efficiently, each layer must issue instructions to layers below in a tightly defined sequence. However, a robust system must also be able to handle errors caused by faulty devices. To satisfy both criteria, a control loop must handle operations at all layers simultaneously while also processing syndrome measurements to correct errors that occur.  A prototype for this control loop is shown in Fig.~\ref{LayeredControlCycle_Abstract}.

The primary control cycle defines the dynamic behavior of the quantum computer in this architecture since all operations must interact with this loop. The principal purpose of the control cycle is to successfully implement quantum error correction.  The quantum computer must operate fast enough to correct errors; still, some control operations necessarily incur delays, so this cycle does not simply issue a single command and wait for the result before proceeding --- pipelining is essential~\cite{Shen2005,Isailovic2008}.  A related issue is that operations in different layers occur on drastically different timescales, as discussed later in Section~\ref{Sec_Large_Scale}.  Figure~\ref{LayeredControlCycle_Abstract} also describes the control structure needed for the quantum computer. Processors at each layer track the current operation and issue commands to lower layers.  Layers~1 to 4 interact in the loop, whereas the Application layer interfaces only with the Logical layer, making the algorithm independent of the hardware.

\section{Quantum Hardware and Control}
\begin{figure}
  \centering
  \includegraphics[width=\textwidth]{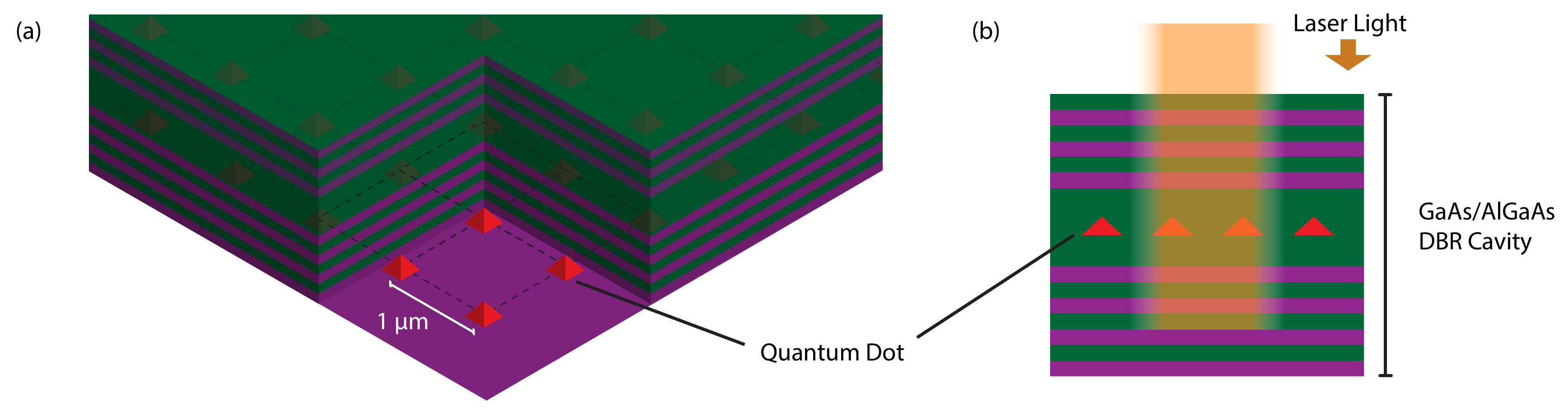}\\
  \caption[Perspective view of QuDOS hardware]{Quantum dots in a planar optical microcavity form the basis of the \mbox{QuDOS} hardware platform.  (a)~The quantum dots are arranged 1~$\mu$m apart in a two-dimensional square array.  The quantum dots trap single electrons, whose spins will be used for quantum information processing.  (b)~Side view. The electron spins are manipulated with laser pulses sent into the optical cavity from above, and two neighboring quantum dots can be coupled by a laser optical field which overlaps them.  The purple and green layers are AlGaAs and GaAs, grown by molecular beam epitaxy.  The alternating layers form a distributed Bragg reflector (DBR) optical cavity which is planar, confining light in the vertical direction and extending across the entire system in horizontal directions.  Originally published in Ref.~\cite{Jones2012_PRX}.}
  \label{Cavity_Perspective}
\end{figure}

The essential requirements for the Physical layer are embodied by the DiVincenzo criteria~\cite{DiVincenzo2000}.  The layered framework for quantum computing was developed in tandem with a specific hardware platform, known as \mbox{QuDOS} (\textbf{qu}antum \textbf{d}ots with \textbf{o}ptically-controlled \textbf{s}pins).  The \mbox{QuDOS} platform uses electron spins within quantum dots for qubits.  The quantum dots are arranged in a two-dimensional array; Figure~\ref{Cavity_Perspective} shows a cut-away rendering of the quantum dot array inside an optical microcavity, which facilitates control of the electron spins with laser pulses.  Reference~\cite{Jones2012_PRX} argued that the \mbox{QuDOS} design is a promising candidate for large-scale quantum computing, and I use it here as a model for generating concrete resource estimates.

The physical qubit used by \mbox{QuDOS} is the spin of an electron bound within an InGaAs self-assembled quantum dot (QD) surrounded by GaAs substrate~\cite{Bjork1994,Imamoglu1999,Bonadeo2000,Guest2002,Hours2005,Yamamoto2009}. These QDs can be optically excited to trion states (a bound electron and exciton), which emit light of wavelength $\sim 900$ nm when they decay.  A transverse magnetic field splits the spin levels into two metastable ground states~\cite{Bayer2002}, which form the computational basis states for a qubit.  The energy separation of the spin states is important for two reasons related to controlling the electron spin~\cite{Press2008}.  First, the energy splitting facilitates control with optical pulses.  Second, there is continuous phase rotation between spin states $\ket{\uparrow}$ and $\ket{\downarrow}$ around the $\hat{Z}$-axis on the qubit Bloch sphere, which in conjunction with timed optical pulses provides complete unitary control of the electron spin vector.

The electron spin is bound within a quantum dot.  These quantum dots are embedded in an optical microcavity, which will facilitate quantum gate operations via laser pulses.  To accommodate the two-dimensional array of the surface code detailed in Layer 3, this microcavity must be planar in design, so the cavity is constructed from two distributed Bragg reflector (DBR) mirrors stacked vertically with a $\lambda/2$ cavity layer in between, as shown in Fig.~\ref{Cavity_Perspective}.  This cavity is grown by molecular beam epitaxy (MBE). The QDs are embedded at the center of this cavity to maximize interaction with antinodes of the cavity field modes.  Using MBE, high-quality (\(Q > 10^{5}\)) microcavities can be grown with alternating layers of GaAs/AlAs~\cite{Reitz2007}.  The nuclei in the quantum dot and surrounding substrate have nonzero spin, which is an important source of noise that must be suppressed through control techniques like dynamical decoupling~\cite{Hahn1950,Carr1954,Haeberlen1968,Viola1999,Viola2003,Khodjasteh2005,Uhrig2007,Ng2009,Biercuk2009,Jones2012_PRX}.

Control in \mbox{QuDOS} uses laser pulses which selectively target quantum dots; see Ref.~\cite{Jones2012_PRX} for details.  The 1-qubit operations are developed using a transverse magnetic field and ultrafast laser pulses~\cite{Press2008,Yamamoto2009}.  The construction of a practical, scalable 2-qubit gate in \mbox{QuDOS} remains the most challenging element of the hardware, and various methods are currently under development.  A fast, all-optically controlled 2-qubit gate would certainly be attractive, and early proposals~\cite{Imamoglu1999} identified the importance of employing the nonlinearities of cavity QED.  Reference~\cite{Imamoglu1999} suggests the application of two lasers for both single-qubit and 2-qubit control; more recent developments have indicated that both single-qubit gates~\cite{Economou2006,Clark2007,Press2008} and 2-qubit gates~\cite{Ladd2011} can be accomplished using only a single optical pulse.

\begin{figure}
  \centering
  \includegraphics[width=8cm]{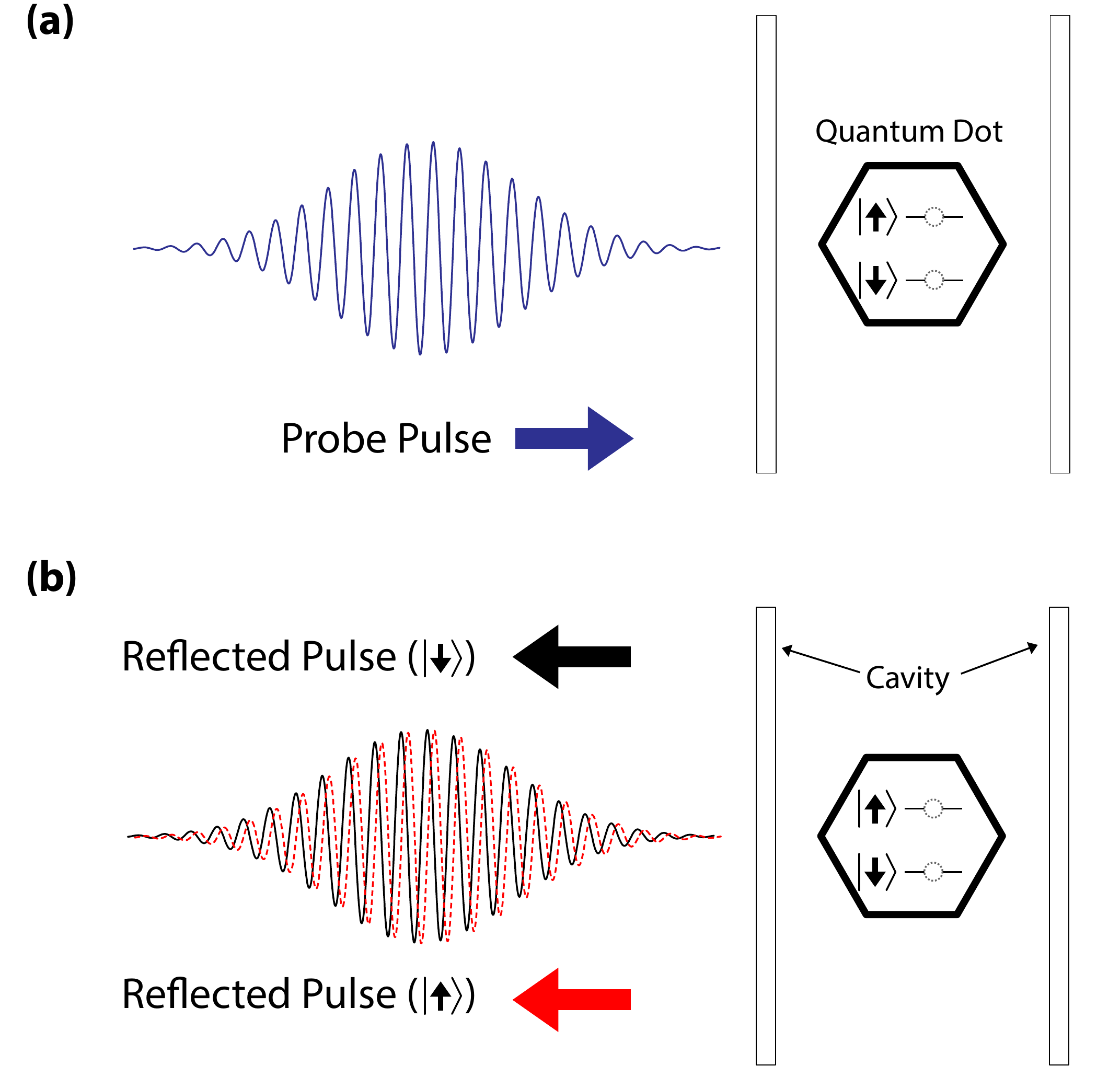}\\
  \caption[Measurement proposal for QuDOS]{A dispersive quantum non-demolition (QND) readout scheme for \mbox{QuDOS}.  (a)~A probe pulse is sent into a microcavity containing a charged quantum dot.  (b)~The cavity-enhanced dispersive interaction between the pulse and the electron spin creates a state-dependent phase shift in the light which leaves the cavity.  Measurement of the phase shift can perform projective measurement on the electron spin.  Originally published in Ref.~\cite{Jones2012_PRX}.}
  \label{QND}
\end{figure}

\mbox{QuDOS} requires a measurement scheme that is still under experimental development.  The proposed mechanism (shown in Fig.~\ref{QND}) is based on Faraday/Kerr rotation.  The underlying physical principle is as follows: an off-resonant probe pulse impinges on a quantum dot, and it receives a different phase shift depending on whether the quantum dot electron is in the spin-up or spin-down state (these are separated in energy by the external magnetic field).  Sensitive photodetectors combined with homodyne detection measure the phase shift to enact a projective QND measurement on the electron spin. Several results in recent years have demonstrated the promise of this mechanism for measurement: multi-shot experiments by Berezovsky \emph{et al.}~\cite{Berez2006} and Atat\"{u}re \emph{et al.}~\cite{Atature2007} have measured spin-dependent phase shifts in charged quantum dots, and Fushman \emph{et al.}~\cite{Fushman2008} observed a large phase shift induced by a neutral quantum dot in a photonic crystal cavity.  Most recently, Young \emph{et al.} observed a significantly enhanced phase shift from a quantum dot embedded in a micropillar cavity~\cite{Young2011}.

\section{Error Correction and Fault Tolerance}
\label{Sec_Error_Correction_Overview}
Error correction is essential to quantum computation, given current understanding of hardware technology.  Some of the best experimental results achieve an error-per-operation of about $10^{-4}$ or $10^{-5}$~(Refs.~\cite{Ladd2010,Monroe2013}, and references therein), but even these impressive feats are not close to the $10^{-12}$ to $10^{-15}$ error rates needed for large-scale quantum algorithms, as will be demonstrated below.  The gap can be bridged with fault-tolerant quantum computing~\cite{Shor1996,Preskill1998,Preskill1998_FTQC,Nielsen2000}, so long as error rates in the hardware are below a threshold value which is specific to the code being used~\cite{Aharonov1997}.

The threshold theorem has garnered significant attention in the community, but it is sometimes mistakenly presumed that the threshold itself is the target performance.  A functioning quantum error correction system must operate below threshold, and a \emph{practical} system must operate well below threshold.  Later chapters show that the resources required for error correction become manageable when the hardware error rate is about an order of magnitude below the threshold of the chosen code.  The code used throughout this thesis is the surface code~\cite{Kitaev2002,Raussendorf2007_PRL,Raussendorf2007_NJP,Fowler2009_PRA}, which is distinguished by requiring only nearest-neighbor operations in two dimensions and by having a high threshold around 1\% error per physical gate~\cite{Wang2010,Fowler2010_Error,Fowler2012_Architecture}.

This section discusses some of the features of error correction that are salient to quantum computer architecture.  First, I briefly outline the advantages of the surface code.  Second, I discuss the use of Pauli frames, which is a simple but effective technique for reducing the number of gates implemented.  Finally, I give an overview of magic-state distillation, which is a powerful technique in fault tolerance and the subject more intense investigation in Chapter~\ref{Ch05}.

\subsection{Surface Code Error Correction}
As just mentioned, the primary justifications for the surface code are that it requires only a two-dimensional geometry of nearest-neighbor gates in the hardware, yet still has one of the highest threshold error rates of any code considered thus far~\cite{Raussendorf2007_NJP,Fowler2012_Architecture}.  There is also evidence that the surface code might have lower overhead than other codes, when the demands of fault tolerance are considered~\cite{Fowler2012_Bridge}.

In this thesis, I base nearly all of my analysis on a hypothetical quantum computer that uses surface code error correction.  A complete explanation of the code and its properties is a subject of active research, so I defer to the literature~\cite{Kitaev2002,Raussendorf2007_PRL,Raussendorf2007_NJP,Fowler2009_PRA,Devitt2009_Architecture,Wang2010,Fowler2010_Error,Devitt2011_HighPerformance,Fowler2012_Practical,Fowler2012_Timing,Fowler2012_Hadamard,Fowler2012_NonCyclic,Fowler2012_Bridge,Fowler2012_Architecture,Fowler2013_BlockCodes}. Some of the features will be examined throughout the thesis.  Chapter~\ref{Ch03} shows how to depict the dynamic implementation of operations in the surface code, as well as calculating a power-law approximation to how resources scale with increasing levels of error correction.  Still, I only touch on the aspects immediately relevant to my analysis, and otherwise I assume the reader is familiar with the mechanics of the surface code.

\subsection{Pauli Frames}
\label{Pauli_frames}
A Pauli frame~\cite{Knill2005,DiVincenzo2007} is a simple and efficient classical computing technique to track the result of applying a series of Pauli gates ($X$, $Y$, or $Z$) to single qubits.  The Gottesman-Knill Theorem implies that tracking Pauli gates can be done efficiently on a classical computer~\cite{Simon2006}.  Many quantum error correction codes, such as the surface code, project the encoded state into a perturbed codeword with erroneous single-qubit Pauli gates applied (relative to states within the code subspace).  The syndrome reveals what these Pauli errors are, up to undetectable stabilizers and logical operators, and error correction is achieved by applying those same Pauli gates to the appropriate qubits (since Pauli gates are Hermitian and unitary).  However, quantum gates are faulty, and applying additional gates may introduce more errors into the system.

Rather than applying every correction operation, one can keep track of what Pauli correction operation \emph{would be applied}, and continue with the computation.  This is possible because the operations needed for error correction are in the Clifford group.  When a measurement in a Pauli $X$, $Y$, or $Z$ basis is finally made on a qubit, the result is modified based on the corresponding Pauli gate which should have been applied earlier.  This stored Pauli gate is called the Pauli frame~\cite{Knill2005,DiVincenzo2007}, since instead of applying a Pauli gate, the quantum computer \emph{changes the reference frame for the qubit}, which can be understood by remapping the axes on the Bloch sphere, rather than moving the Bloch vector.

I want to emphasize that the Pauli frame is a \emph{classical object} stored in the digital circuitry that handles error correction. Pauli frames are nonetheless very important to the functioning of a surface code quantum computer.  Layer~3 in the control stack (Fig.~\ref{LayeredControlStack}) uses a Pauli frame with an entry for each qubit in the error-correcting code.  As errors occur, the syndrome processing step identifies a most-likely pattern of Pauli errors.  Instead of applying the recovery step directly, the Pauli frame is updated in classical memory.  The Pauli gates form a closed group under multiplication (and global phase of the quantum state is unimportant), so a Pauli frame only tracks one of four values ($X$, $Y$, $Z$, or $I$) for each qubit in the hardware.

The Pauli frame is maintained as follows.  Denote the Pauli frame at time $t$ as $F_t$:
\begin{equation}
F_t = \bigotimes_j P_t(j),
\end{equation}
where $P_t(j) = \{I,X,Y,Z\}$ is an element from the Pauli group corresponding to qubit $j$ at time $t$.  Any Pauli gate in the quantum circuit is multiplied into the Pauli frame and \emph{is not implemented} in hardware, so $F_{t+1} = \left(\bigotimes_j U_{\{\footnotesize{I,X,Y,Z}\}}\right) F_t$ for all Pauli gates $U_{\{\footnotesize{I,X,Y,Z}\}}$ in the circuit at time $t$.  Other gates $U_{\mathrm{C}}$ in the Clifford group \emph{are implemented} in hardware, but they also transform the Pauli frame according to
\begin{equation}
\label{Pauli_Frame_Clifford}
F_{t+1} = U_{\mathrm{C}} F_t U_{\mathrm{C}}^{\dag}.
\end{equation}
When using Pauli frames, the flow of the computation proceeds in the same manner as if Pauli gates were being implemented, with the only change being how the final measurement of that qubit is interpreted.  The set of Clifford gates is sufficient for implementing surface code error correction, though one also needs to implement non-Clifford logical operations for universal quantum computing.

Quantum algorithms need to apply gates $U_{\mathrm{NC}}$ outside the Clifford group.  When using a Pauli frame, the gate that is actually implemented, $U_{\mathrm{NC}}'$, is given by:
\begin{equation}
U_{\mathrm{NC}}' = F_t U_{\mathrm{NC}} F_t^{\dag}.
\end{equation}
Note the distinction between this expression and Eqn.~(\ref{Pauli_Frame_Clifford}).  In Eqn.~(\ref{Pauli_Frame_Clifford}), the Pauli frame is changed by application of Clifford-group gate, but here an unchanging Pauli frame modifies the gate that is applied.

\subsection{Magic-State Distillation}
\label{Sec_MS_Distill_QuDOS}
In the layered framework, the Logical layer takes the fault-tolerant resources from Layer~3 and creates a logical substrate for universal quantum computing.  This task requires additional processing of error-corrected gates and qubits to produce any arbitrary gate required in the Application layer~\cite{Jones2012_PRX}.  Quantum error correction provides only a limited set of gates, such as the Clifford group (or only a subset thereof, as in the surface code~\cite{Fowler2009_PRA}).  Although circuits from this set can be simulated efficiently on a classical computer by the Gottesman-Knill Theorem~\cite{Nielsen2000}, the Clifford group forms the backbone of quantum logic.

The Logical layer constructs arbitrary gates from circuits of fundamental gates and ancillas injected into the error-correcting code~\cite{Fowler2009_PRA,Jones2012_PRX}.  For example, surface code architectures inject and purify the ancillas $\ket{Y} = \frac{1}{\sqrt{2}}\left(\ket{0} + i\ket{1}\right)$ and $\ket{A} = \frac{1}{\sqrt{2}}\left(\ket{0} + e^{i\pi/4}\ket{1}\right)$; then the surface code consumes these ancillas in quantum circuits to produce $S = \exp(i \pi(I - Z)/4)$ and $T = \exp(i \pi(I - Z)/8)$ gates, respectively~\cite{Nielsen2000,Raussendorf2007_PRL,Raussendorf2007_NJP,Fowler2009_PRA}.  Because the ancillas are faulty, they must be purified through a process known as magic-state distillation~\cite{Bravyi2005,Raussendorf2007_NJP,Fowler2009_PRA,VanMeter2009,Jones2012_PRX,Meier2012,Bravyi2012,Jones2013_Multilevel}.

\begin{figure}
  \centering
  \includegraphics[width=\textwidth]{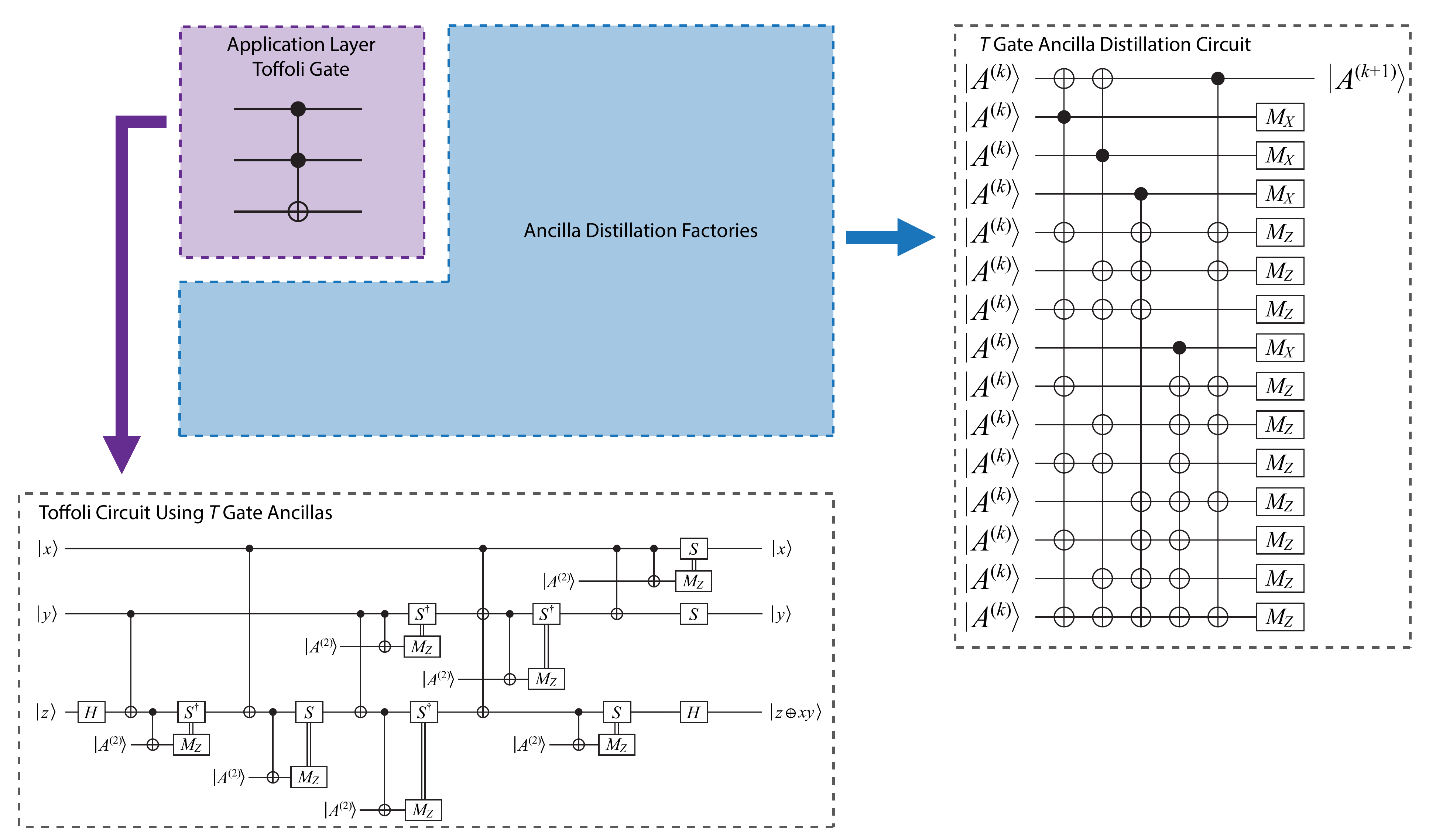}\\
  \caption[Toffoli construction using ancilla factory]{A Toffoli gate ($\ket{x,y,z} \rightarrow \ket{x,y,z \oplus xy}$) at the Application layer is constructed with assistance from the Logical layer, using the decomposition in Ref.~\cite{Nielsen2000}.  There are only three application qubits, but substantially more logical qubits are needed for distillation circuits in Layer 4.  The $|A^{(2)}\rangle$ ancillas are the result of two levels of distillation ($|A^{(0)}\rangle$ is an injected state) on the ancilla required for $T$~gates.  Note that each time an ancilla is used with measurement, the Pauli frame may need to be updated.  The ancilla-based circuit for $S$~gates is not shown here, for clarity.  Modified from version published in Ref.~\cite{Jones2012_PRX}.}
  \label{Toffoli_decomposition}
\end{figure}

Magic-state distillation will be examined at length in Chapter~\ref{Ch05}.  For now, I only want to explain the simple method that was used in Ref.~\cite{Jones2012_PRX}.  Consider the process of distilling the ancilla state $\ket{A}$ that is used to construct the $T$~gate~\cite{Bravyi2005,Raussendorf2007_NJP,Fowler2009_PRA,Jones2012_PRX}.  Figure~\ref{Toffoli_decomposition} provides an illustration of why this process is important by showing the fault-tolerant construction of a Toffoli gate at the Application layer using distilled ancillas at the Logical layer.  Two separate analyses contend that ancilla distillation circuits constitute over 90\% of the computing effort for a single Toffoli gate~\cite{Isailovic2008,Jones2012_PRX}.  Viewed another way, for every qubit used by the algorithm, approximately 10 qubits are working in the background to generate the necessary distilled ancillas.

The circuit in Fig.~\ref{Toffoli_decomposition} shows one level of $\ket{A}$ distillation, but a lengthy computation like Shor's algorithm will typically require two levels, where the outputs of the first round are distilled again.  Moreover, since perhaps trillions of distilled $\ket{A}$ ancillas will be needed for the entire algorithm, \mbox{QuDOS} uses a ``distillation factory''~\cite{Steane1998,VanMeter2009}, which is a dedicated region of the computer that continually produces these states as fast as possible.  Speed is important, because ancilla distillation can be the rate-limiting step in quantum circuits~\cite{Isailovic2008}.  Figure~\ref{Toffoli_decomposition} shows how to construct a Toffoli gate, but the $T$~gates can be used to approximate any other gate as well (see Ref.~\cite{Nielsen2000}; more details in Chapter~\ref{Ch07}).

Each $\ket{A}$ distillation circuit will require 15 lower-level $\ket{A}$ states, but they are not all used at the same time.  For simplicity, set the ``clock cycle time'' for each gate equal to the time to implement a logical CNOT, so that with initialization and measurement, the distillation circuit requires 6 cycles.  By only using $\ket{A}$ ancillas when they are needed, the circuit can be compacted to require at most 12 logical qubits at any time.  The computing effort can be characterized by a ``circuit volume,'' which is the product of logical memory space (\emph{i.e.} area of the computer) and time.  The circuit volume of $\ket{A}$ distillation is $V\left(\ket{A^{(1)}}\right) = (12 \textrm{ logical qubits}) \times (6 \textrm{ clock cycles}) = 72$.  A two-level distillation will require 16 distillation circuits, or a circuit volume of $V\left(\ket{A^{(2)}}\right) = 1152$.  An efficient distillation factory with area $A_{\mathrm{factory}}$ will produce on average $A_{\mathrm{factory}}/V\left(\ket{A^{(2)}}\right)$ distilled ancillas per clock cycle.  Table~\ref{distillation_factory} summarizes of these results.

\begin{table}
  \centering
  \begin{tabular}{l l l l}
    \toprule
    \textbf{Parameter} & \textbf{Symbol}                & \textbf{Value} \\ \midrule
    Circuit depth      &                                & 6 clock cycles \\
    Circuit area       & $A_{\text{distill}}$           & 12 logical qubits\\
    Circuit volume     & $V\left(\ket{A^{(1)}}\right)$  & 72 qubits$\times$cycles \\
    Factory rate (level $n$) & $R_{\mathrm{factory}}\left(\ket{A^{(n)}}\right)$ & $A_{\mathrm{factory}}/V\left(\ket{A^{(n)}}\right)$ \newline ancillas/cycle \\ \bottomrule
  \end{tabular}
  \caption[Magic-State Distillation Resources in QuDOS]{Resource analysis for a distillation factory.  These factories are crucial to quantum computers which require ancillas for universal gates.  Magic-state distillation uses Clifford gates and measurement, so the circuit can be deformed to reduce depth and increase area, or vice versa, while keeping volume approximately constant.}
  \label{distillation_factory}
\end{table}

As a research effort, magic-state distillation has exploded in the last year.  Chapter~\ref{Ch05} will cover these matters in more detail, but many new results were produced in the short time since Ref.~\cite{Jones2012_PRX} was published.  Fowler and Devitt developed a highly efficient implementation of distillation in the surface code, along with good estimates of resources~\cite{Fowler2012_Bridge}.  Several new schemes for distilling $\ket{A}$ states were also developed~\cite{Meier2012,Bravyi2012,Jones2013_Multilevel}.  Section~\ref{Sec_Multilevel_Distillation} will examine my proposal for ``multilevel distillation,'' which is asymptotically very efficient but perhaps too complicated to be useful in practice.  These developments, and those in other chapters, will dramatically lower the cost of fault-tolerant quantum computing.  As I mentioned at the outset to this chapter, one of the purposes of the calculations given here is to provide contrast for the new methods developed later.

\section{Quantum Algorithms}
The Application layer is where quantum algorithms are executed.  The efforts of Layers~1 through 4 have produced a computing substrate that supplies any arbitrary gate needed.  The Application layer is therefore not concerned with the implementation details of the quantum computer---it is an ideal quantum programming environment.  This section deals with estimating the resources required for a target application.  This analysis can indicate the feasibility of a proposed quantum computer design, which is a worthwhile consideration when evaluating the long-term prospects of a quantum computing research program.

A quantum engineer could start here in Layer~5 with a specific application in mind and work down the layers to determine the system design necessary to achieve desired functionality.  I take this approach for \mbox{QuDOS} by examining two interesting quantum algorithms: Shor's factoring algorithm and simulation of quantum chemistry.  A rigorous system design is beyond the scope of the present work, but this section considers the computing resources required for each application in sufficient detail that one may gauge the engineering effort necessary to design a quantum computer based on \mbox{QuDOS} technology.

\subsection{Elements of the Application Layer}
The Application layer is composed of \emph{application} qubits and gates that act on the qubits.  Application qubits are logical qubits used explicitly by a quantum algorithm.  As discussed in Section~\ref{Sec_MS_Distill_QuDOS}, many logical qubits are also used to distill ancilla states necessary to produce a universal set of gates, but these distillation logical qubits are not visible to the algorithm in Layer 5.  When an analysis of a quantum algorithm quotes a number of qubits without reference to fault-tolerant error correction, often this means the number of application qubits~\cite{Beauregard2003,Aspuru2005,Zalka2006,Takahashi2006}.  Similarly, Application-layer gates are equivalent in most respects to logical gates; the distinction is made according to what resources are visible to the algorithm or deliberately hidden in the machinery of the Logical layer, which affords some discretion to the computer designer.

A quantum algorithm could request any arbitrary gate in Layer 5, but not all quantum gates are equal in terms of resource costs.  As shown in Section~\ref{Sec_MS_Distill_QuDOS}, distilling $\ket{A}$ ancillas for $T$ gates is a very expensive process.  For example, Fig.~\ref{Toffoli_decomposition} shows how Layers~4 and~5 coordinate to produce an Application-layer Toffoli gate, illustrating the extent to which ancilla distillation consumes resources in the computer.  When ancilla preparation is included, $T$~gates can account for over 90\% of the circuit complexity in a fault-tolerant quantum algorithm~\cite{Isailovic2008,Jones2012_PRX}.

When analyzing algorithms, it is convenient to count resources in terms of Toffoli gates.  This is a natural choice, because the level of ancilla distillation, number of virtual qubits, \emph{etc.} depend on the choice of hardware, error correction, and many other design-specific parameters; by comparison, number of Toffoli gates is machine-independent since this quantity depends only on the algorithm (much like the number of application qubits mentioned above).  To determine error correction or hardware resources for a given algorithm, one can take the Layer~5 resource estimates and work down through Layers~4 to 1, which is an example of modularity in this architecture framework.  As shown in Ref.~\cite{Jones2012_PRX}, an Application-layer Toffoli gate in \mbox{QuDOS} has an execution time of 930 $\mu$s (31 logical gate cycles including the $S$~gate circuits).

\subsection{Shor's Integer-Factoring Algorithm}
\label{Shor}
Perhaps the most well-known application of quantum computers is Shor's algorithm, which decomposes an integer into its prime factors~\cite{Shor1999}.  Solving the factoring problem efficiently would compromise the RSA cryptosystem~\cite{RSA1978}.  Because of the prominence of Shor's algorithm in the field of large-scale, fault-tolerant quantum computing, I estimate the resources required to factor a number of size typical for RSA.

\begin{figure}
  \centering
  \includegraphics[width=12cm]{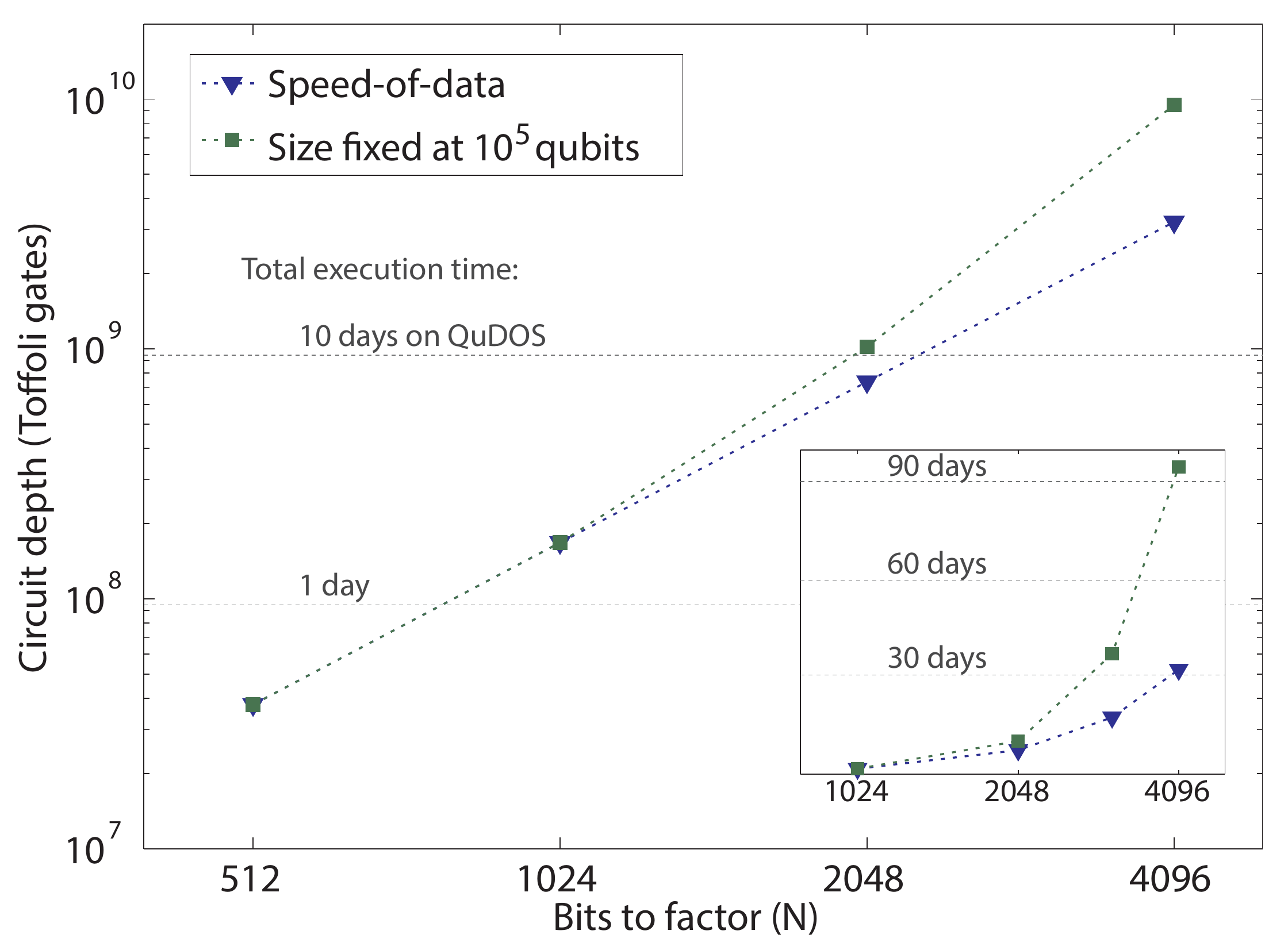}\\
  \caption[Execution time for Shor's algorithm]{Execution time for Shor's algorithm, using the same circuit implementation as Ref.~\cite{VanMeter2009}.  The vertical axis shows circuit depth, in terms of Toffoli gates, and the plot is labeled with estimated runtime on the \mbox{QuDOS} architecture.  The blue trace is a quantum computer whose size in logical qubits scales as necessary to compute at the speed of data (no delays).  The green trace is a machine limited to $10^5$ logical qubits, which experiences rapidly increasing delays as problem size increases beyond 2048 bits.  The problem is that insufficient resources are available to distill ancillas for $T$~gates, which are used to produce universal logic.  The inset shows the same data on a linear vertical scale, illustrating when the quantum computer experiences delays for lack of enough qubits.  Originally published in Ref.~\cite{Jones2012_PRX}.}
  \label{Shor_Toffoli}
\end{figure}

A common key length for RSA public-key cryptography is 1024 bits.  Factoring a number this large is not trivial, even on a quantum computer, as the following analysis shows.  Figure~\ref{Shor_Toffoli} shows the expected run time on \mbox{QuDOS} for one iteration of Shor's algorithm versus key length in bits for two different quantum computers: one where system size increases with the problem size, and one where the system size is limited to $10^5$ logical qubits (including application qubits).  For the fixed-size quantum computer, the runtime begins to grow faster than the minimal circuit depth when factoring numbers 2048 bits and higher.  Fixing the machine size highlights the importance of the ancilla distillation factories.  For this instance of Shor's algorithm, about 90\% of the machine should be devoted to distillation; if insufficient resources are devoted to distillation, performance of the factoring algorithm plummets.  For example, the 4096-bit factorization devotes $\sim 75\%$ of the machine to distillation, but about $3\times$ as many factories would be needed to achieve maximum execution speed in the lower trace in Fig.~\ref{Shor_Toffoli}.  I should also mention here that Shor's algorithm is probabilistic, so a few iterations may be required~\cite{Shor1999}.

\subsection{Simulation of Quantum Chemistry}
Quantum computers were inspired by the problem that simulating quantum systems on a classical computer is fundamentally difficult.  Feynman postulated that one quantum system could simulate another much more efficiently than a classical processor, and he proposed a quantum processor to perform this task~\cite{Feynman1982}.  Quantum simulation is one of the few known quantum algorithms that solves a useful problem believed to be intractable on classical computers, so I estimate the resource requirements for quantum simulation in QuDOS, and more details are available in Ref.~\cite{Jones2012_PRX}.

This section specifically considers fault-tolerant quantum simulation.  Other methods of simulation are under investigation~\cite{Buluta2009,Barreiro2011,Biamonte2010}, but they lie outside the scope of this work.  The particular example selected here is simulating the Schr\"{o}dinger equation for time-independent Hamiltonians in first-quantized form, where each Hamiltonian represents the electron/nuclear configuration in a molecule~\cite{Zalka1998,Kassal2008,Kassal2011,Jones2012_NJP}.  An application of such a simulation is to determine ground- and excited-state energy levels in a molecule.  This analysis focuses on first-quantized instead of second-quantized form for better resource scaling at large problem sizes~\cite{Kassal2011}.  Digital quantum simulation will also be examined in Chapter~\ref{Ch08}.

\begin{figure}
  \centering
  \includegraphics[width=12cm]{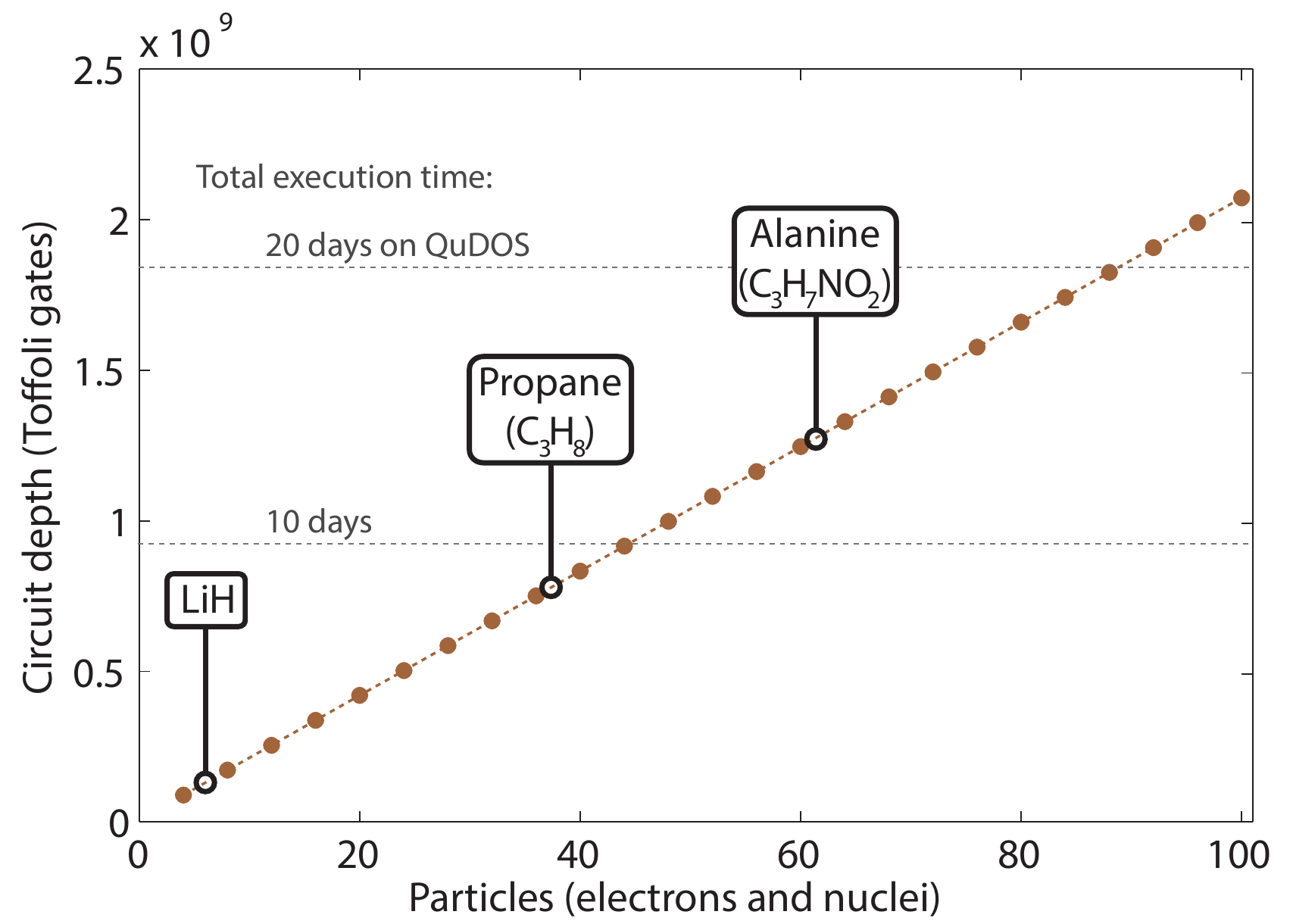}\\
  \caption[Execution time for quantum chemistry simulation]{Execution time for simulation of a molecular Hamiltonian in first-quantized form, as a function of problem size.  The horizontal axis is number of particles being simulated, and the plot is labeled with some interesting examples from chemistry.  The vertical axis is circuit depth in Toffoli gates, and the plot is labeled with estimated runtime on \mbox{QuDOS}.  Each simulation uses 12-bit spatial precision in the wavefunction and $2^{10}$ time steps for 10-bit precision in readout, or at most $\sim 3$ significant figures.  The linear scaling in algorithm runtime versus problem size is due to two-body potential energy calculations, which constitute the majority of the quantum circuit.  The number of potential energy calculations increases quadratically with problem size, but through parallel computation they require linear execution time~\cite{Jones2012_PRX,Jones2012_NJP}.  Originally published in Ref.~\cite{Jones2012_PRX}.}
  \label{Simulation_depth}
\end{figure}

Figure~\ref{Simulation_depth} shows the time necessary to execute the simulation algorithm for determining an energy eigenstate on the \mbox{QuDOS} computer as a function of the size of the simulation problem, expressed in number of electrons and nuclei.  First-quantized form stores the position-basis information for an electron wavefunction in a quantum register, and the complete Hamiltonian is a function of one- and two-body interactions between these registers, so this method does not depend on the particular molecular structure or arrangement; hence, the method is very general.  Note that the calculation time scales linearly in problem size, as opposed to the exponential scaling seen in exact classical methods.  The precision of the simulation scales with the number of time steps simulated~\cite{Aspuru2005}, and this example uses $2^{10}$ time steps for a maximum precision of about 3 significant figures.

\section{Quantum Computing and the Need for Logic Synthesis}
\label{Sec_Large_Scale}
The factoring algorithm and quantum simulation represent interesting applications of large-scale quantum computing, and for each the computing resources required of a layered architecture based on \mbox{QuDOS} are listed in Table~\ref{Computing_Resources}.  The algorithms are comparable in total resource costs, as reflected by the fact that these two example problems require similar degrees of error correction.  The simulation algorithm is more compact than Shor's, requiring fewer logical qubits for distillation, which is a consequence of this algorithm performing fewer arithmetic operations in parallel.  However, Shor's algorithm has a shorter execution time owing to its use of parallel computation.  Both algorithms can be accelerated through parallelism if the quantum computer has more logical qubits available~\cite{VanMeter2005,Jones2012_NJP}.

\begin{table}
  \centering
  \begin{tabular}{l l l l}
    \toprule
    & & \textbf{Shor's}& \textbf{Molecular}\\
    \multicolumn{2}{c}{\textbf{Computing Resource}} & \textbf{Algorithm} & \textbf{Simulation} \\
    & & (1024-bit) & (alanine) \\
    \midrule
    \multirow{2}{*}{Layer 5} & Application qubits          & 6144     & 6650 \\
                             & Circuit depth (Toffoli)     & $1.68 \times 10^8$  & $1.27\times10^9$ \\ 
    \multirow{2}{*}{Layer 4} & Log. distillation qubits    & 66564    & 15860 \\
                             & Logical clock cycles        & $5.21\times10^9$   & $3.94\times10^{10}$ \\ 
    \multirow{2}{*}{Layer 3} & Code distance               & 31       & 31 \\
                             & Error per lattice cycle     & $2.58\times10^{-20}$ & $2.58\times10^{-20}$ \\ 
    \multirow{2}{*}{Layer 2} & Virtual qubits              & $4.54\times10^8$     & $1.40\times10^8$ \\
                             & Error per virtual gate      & $1.00\times10^{-3}$  & $1.00\times10^{-3}$ \\ 
    \multirow{2}{*}{Layer 1} & Quantum dots                & $4.54\times10^8$ & $1.40\times10^8$ \\
                             & (area on chip)              & (4.54 $\mathrm{cm}^2$) & (1.40 $\mathrm{cm}^2$) \\
    \midrule
    \multicolumn{2}{c}{\textbf{Execution time (est.)}} & \textbf{1.81 days} & \textbf{13.7 days} \\ \bottomrule
  \end{tabular}
  \caption[QuDOS Resources with Naive Logic Designs]{Summary of the computing resources in a layered architecture based on the \mbox{QuDOS} platform, for Shor's algorithm factoring a 1024-bit number (same implementation as Ref.~\cite{VanMeter2009}) and the ground state simulation of the molecule alanine ($\mathrm{C}_3\mathrm{H}_7\mathrm{N}\mathrm{O}_2$) using first-quantized representation.}
  \label{Computing_Resources}
\end{table}

\begin{figure}
  \centering
  \includegraphics[width=\textwidth]{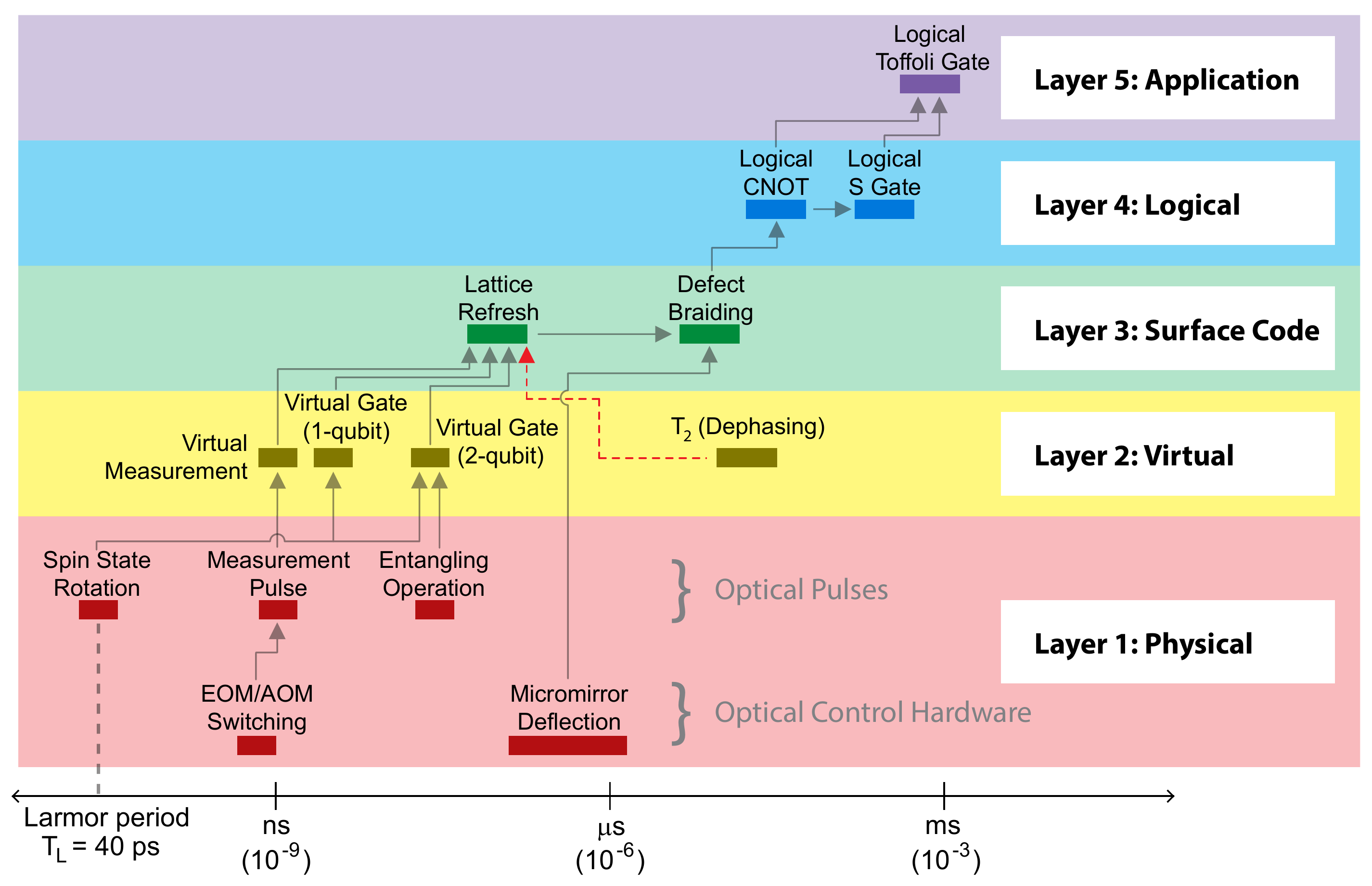}\\
  \caption[Relatives timescales of operations in QuDOS]{Relative timescales for critical operations in \mbox{QuDOS} within each layer.  Each bar indicates the approximate timescale of an operation, and the width indicates that some operation times may vary with improvements in technology.  The arrows indicate dependence of higher operations on lower layers. The red arrow signifies that the surface code lattice refresh must be 2--3 orders of magnitude faster than the dephasing time in order for error correction to function.  The Application layer is represented here with a Toffoli gate, which is a common building block of quantum algorithms.  Complete algorithm runtimes can vary significantly, depending on both the capabilities of the quantum computer and the specific way each algorithm is compiled, such as to what extent calculations are performed in parallel.  Originally published in Ref.~\cite{Jones2012_PRX}.}
  \label{LayeredRelativeTimescales}
\end{figure}

Precise timing and sequencing of operations are crucial to making an architecture efficient. In the layered framework presented by Ref.~\cite{Jones2012_PRX}, an upper layer in the architecture depends on processes in the layer beneath, so that logical gate time is dictated by QEC operations, and so forth. This system of dependence of operation times is depicted for \mbox{QuDOS} in Fig.~\ref{LayeredRelativeTimescales}.  The horizontal axis is a logarithmic scale in the time to execute an operation at a particular layer, while the arrows indicate fundamental dependence of one operation on other operations in lower layers.

Examining Fig.~\ref{LayeredRelativeTimescales}, the operation timescales increase as one moves to higher layers.  This is because a higher layer must often issue multiple commands to layers below.  A crucial point shown in Fig.~\ref{LayeredRelativeTimescales} is that the time to implement a logical quantum gate is four orders of magnitude greater than the duration of each individual physical gate, such as a laser pulse.  For large-scale quantum computing, the speed of error-corrected operations is the crucial figure of merit, and the substantial overhead for fault tolerance shown in Fig.~\ref{LayeredRelativeTimescales} indicates that improved methods are needed.

The findings in Table~\ref{Computing_Resources} and Fig.~\ref{LayeredRelativeTimescales} were, more or less, the key results of Ref.~\cite{Jones2012_PRX}.  In an earnest attempt to design the architecture of a quantum computer, it was revealed that a few error correction processes accounted for a substantial portion of the resource overhead.  These include magic-state distillation, Toffoli gates, and approximations to arbitrary gates.  These tasks all involve the synthesis of fault-tolerant quantum logic, and it soon became apparent to other researchers and myself that significant improvements are possible by optimizing the logic constructions.  Quantum logic synthesis is the subject of my thesis, and the following chapters develop methodology and novel techniques for this new field of research.  The processes listed above are considered explicitly in Chapters~\ref{Ch05}--\ref{Ch07}.  The methods in those chapters will improve on the resource costs given here by about a factor of 500.


\chapter{Preliminaries for Quantum Logic}
\label{Ch03}
Quantum logic is the result of composition.  Every \emph{quantum program} is a sequence of instructions, each being one of three types: preparing quantum states (qubits), applying unitary operations (gates), and performing projective measurement.  In addition to quantum logic, classical logic is often included when gates are conditioned on the result of an earlier measurement.  Because the order of operations is important, quantum programs can be quite complicated.  This chapter examines how quantum programs are specified, how programs are represented in diagrams, and how the resource costs are calculated for a program in the surface code.

Informative diagrams are essential for quantum circuit designers to see the action of a sequence of operations.  Having easy-to-understand pictorial diagrams helps to: design programs, adapt using previous results, identify mistakes, and communicate results.  This chapter discusses two types of quantum logic diagrams.  The first is the familiar quantum circuit, which was introduced in Chapter~\ref{Ch01}.  The second type is a surface code topology diagram, which is a three-dimensional rendering of how quantum logic is implemented using surface code error correction.  Surface codes are preferred in this work for reasons outlined in Chapter~\ref{Ch02}.  What is particularly useful about this diagram is that it provides both visual and quantitative assessment of actual resource costs at the hardware level; the disadvantage is that such diagrams are difficult to interpret alone.  Circuit diagrams and surface code diagrams will play complementary roles in this thesis.

Analyzing resource costs is essential to quantum logic synthesis.  The objective is to compose logic in a way that minimizes costs while ensuring reliable execution of the quantum program.  This chapter concludes by explaining how to quantitatively estimate resource costs in the surface code.  I also introduce the concept of the Trivial Upper Bound (TUB), which for any program is the resource costs for using a naive, ``worst case'' compilation.  The TUB represents the cost of a program that surely works but is probably not optimal, and TUBs will be used as benchmarks to demonstrate the efficacy of logic synthesis.

\section{Quantum Programs}
\label{quantum_programs_section}
A quantum program is any sequence of operations on a quantum state.  As mentioned in the introduction, there are three types of operations: initialization of quantum states, unitary gates, and measurement.  A program is defined by this sequence of operations and any input or output states that are fixed externally.  A program might not have an input state; if this is the case, the program initializes all of its quantum data.  A program also might not have an output state, returning only classical information from internal measurements.  A key concept for logic synthesis is that two programs are \emph{logically equivalent} if they produce the same outputs from the same inputs, within some specified error tolerance.

In practice, most operations belong to finite sets.  Within an error correcting code, such as the surface code~\cite{Raussendorf2007_PRL,Raussendorf2007_NJP,Fowler2009_PRA}, the logical operations are constrained.  The Eastin-Knill theorem and related results indicate that it is impossible for all logical operations to be native to the code~\cite{Zeng2007,Eastin2009}.  Moreover, the available error-corrected operations are often discrete.  The error-corrected operations supported by the surface code are:
\begin{itemize}
\item Initialize $\ket{0}$ or $\ket{+}$ ($Z$-basis or $X$-basis, respectively);
\item Unitary $X$, $Y$, or $Z$ gate (via Pauli frame, see~\cite{Knill2005,DiVincenzo2007,Jones2012_PRX});
\item Unitary $H$ gate (Hadamard);
\item Unitary CNOT gate;
\item Measurement $M_X$ or $M_Z$ ($X$-basis or $Z$-basis, respectively).
\end{itemize}
These operations are not universal for quantum computing, but they will account for most of the operations in quantum programs.

The final operation in the surface code is the ability to initialize a single qubit in any arbitrary state, though it has error probability proportional to the hardware error rate.  The qubit is called an ``injected state,'' because it was teleported into the code using faulty methods~\cite{Raussendorf2007_PRL,Raussendorf2007_NJP,Fowler2009_PRA,Devitt2009_Architecture,Fowler2012_Architecture}.  The error probability is a sum of error rates in the hardware.  Reference~\cite{Fowler2013_BlockCodes} estimates an injection error that is 10 times the gate error probability $p_g$, so the injected state could have error on the order of 1\% for $p_g = 10^{-3}$.  These faulty states are essential for universal quantum computation, but fault tolerance requires that they be purified in some manner using the error corrected operations listed above.  The choice of program to ``clean up'' these noisy inputs will have a dramatic impact on resource costs, as will be considered in detail in Chapters~\ref{Ch05} and~\ref{Ch06}.

The simplest way to implement a program is to initialize all the states that one might need at the beginning, then apply all of the operations using unitary gates, then perform all of the measurements at the end.  However, the same output can often be achieved by performing some initialization and measurement in the middle of the program.  Doing so can lower resource costs in several ways.  For one thing, idle quantum states still require error correction at the hardware level, so if initialization can be delayed until the state is needed or if measurement can be performed as soon as possible, then the program should do so.  Moreover, sometimes a unitary gate can be replaced by a non-unitary sequence of operations that has lower resource cost.

\begin{figure}
  \centering
  \includegraphics[width=9cm]{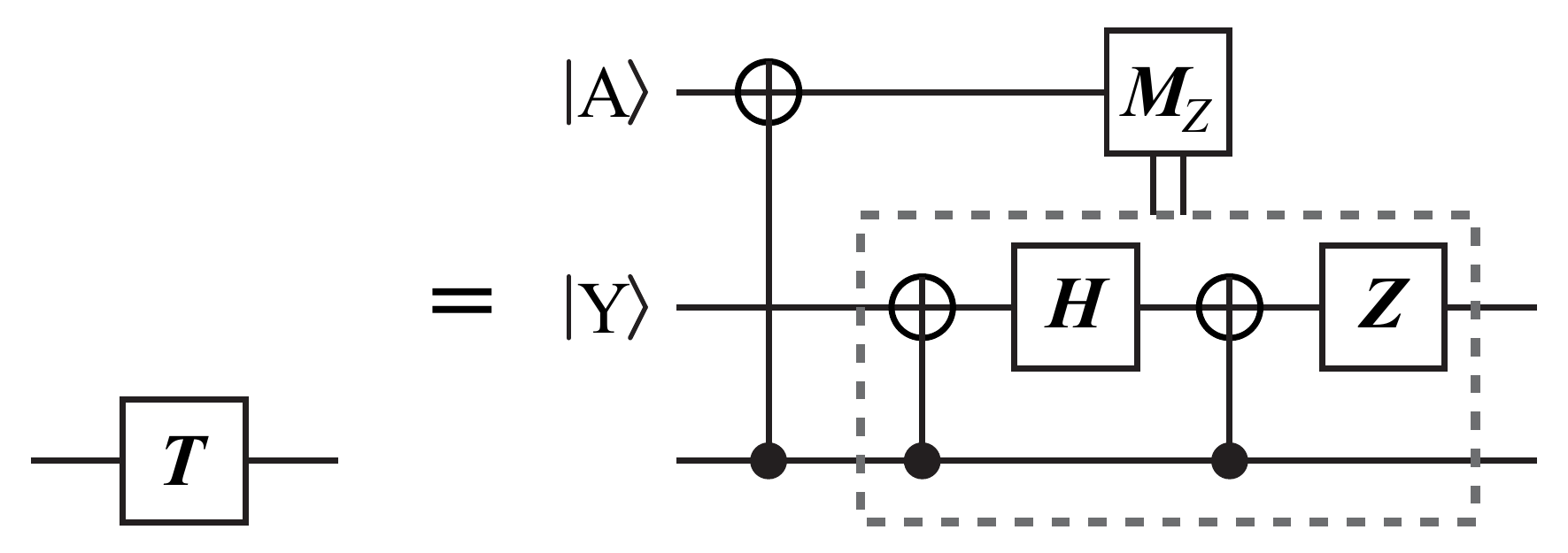}\\
  \caption[Circuit: $T$-gate teleportation as example.]{Circuit for implementing a non-native gate $T$ using specially prepared ancilla states $\ket{A} = (1/\sqrt{2})(\ket{0} + e^{i\pi/4}\ket{1})$ and $\ket{Y} = (1/\sqrt{2})(\ket{0} + i \ket{1})$.  This technique is sometimes referred to as a ``teleportation gate.''  The circuit enclosed in the dashed box is implemented only if the measurement result is $\ket{1}$.}
  \label{T_example}
\end{figure}

The technique of replacing unitary logic with non-unitary logic will be used frequently in later chapters.  It may seem counterintuitive to replace a single gate with multiple non-unitary operations, as the latter appears more complex.  However, some unitary gates are very expensive, so replacing them with a non-unitary sequence of operations can lead to a net reduction in resources.  Consider the circuit in Fig.~\ref{T_example} as an example.  On the left, one would like to implement the gate $T = \exp(i\pi(I-Z)/8)$, but this gate is not available (\emph{i.e.} it has infinite cost).  However, the logically equivalent program on the right uses an ancilla state (injected and purified), $H$, CNOT, and measurement.  The gates enclosed in the dashed box are conditionally applied based on the measurement outcome.  Neglecting for now the way in which the injected state is purified (Chapter~\ref{Ch05} covers this in detail), it is clear that all operations are available in the surface code, so this program has lower (finite) cost.

Considering the list of available error-corrected operations above, only trivial quantum programs can be implemented with unitary gates in the surface code.  This list is a subset of the Clifford group, and even programs that use the full Clifford group can be simulated classically using the Gottesman-Knill Theorem~\cite{Gottesman1998,Nielsen2000}.  Therefore, all useful programs in the surface code require purified injected states, and using these states requires non-unitary operations.  Hence all useful quantum programs in the surface code are non-unitary, at some level.  However, a quantum program can encapsulate the non-unitary details, so that the external world only sees the program perform a unitary mapping of an input state to an output state.  When some arbitrary program, such as a quantum algorithm, needs to be implemented in a fault-tolerant manner, the synthesis procedure will replace many unitary operations with logically equivalent, non-unitary programs so as to minimize resource costs.

Finally, quantum programs can call subprograms.  Using some inductive reasoning, any program is a valid composite operation because it is composed of valid operations.  Hence, programs can be structured in a hierarchical fashion.  This is a common technique in classical programming, but it plays a special role in quantum computing.  Later chapters show that certain choices of subprograms can be easily verified, thereby lowering the costs of error correction substantially.  Logic synthesis will tend to produce hierarchical quantum programs.

\section{Quantum Logic Diagrams}
Quantum logic diagrams provide a visual aid for understanding properties of quantum logic.  Moreover, each type of diagram is useful for a different purpose.  This section covers two frequently used diagrams, quantum circuits and surface code depictions.  Quantum circuits are one of the oldest methods to represent quantum programs, and they are straightforward to interpret.  Time progress left to right, like a musical score, and each horizontal line is a qubit.  By contrast, surface code diagrams are challenging to interpret, but they explicitly account for the resource costs of implementing a program.  When used together, the diagrams explain both the action of a program and its costs, which are the main concerns of logic synthesis.

Quantum circuits were introduced in Chapter~\ref{Ch01}, so I will be brief.  In a quantum circuit diagram, each qubit is a horizontal line, and operations affecting a certain qubit touch the corresponding line.  The line begins where the qubit is initialized or at an input to the program, and it ends where the qubit is measured or at an output of the program.  In some contexts, multi-qubit states are grouped into one line, often borrowing the digital-logic notation of a slash ``/'' through the line to denote multiple bits.  Figure~\ref{T_example} is a quantum circuit, and Nielsen and Chuang provide a more detailed overview of quantum circuits (Ref.~\cite{Nielsen2000}, Ch.~1).  In the List of Figures, I denote circuit diagrams by the prefix ``Circuit.''

The second type of logic diagram, the surface code diagram, is a three-dimensional geometric depiction.  Two dimensions are space, and one dimension is time.  In most cases, I will set the viewing angle such that time flows left to right, making the spatial dimensions vertical and out-of-page.  The diagram represents how the surface code implements encoded gates with many physical gates and qubits.  By using surface code diagrams, I implicitly assume that the quantum computer implements surface code error correction at the lowest level.  This is justified by arguments in Chapter~\ref{Ch02}, which in essence reduce to the following: surface codes are the best error correction scheme published so far when hardware gates are constrained to a nearest-neighbor, two-dimensional geometry~\cite{Raussendorf2007_PRL,Raussendorf2007_NJP,Fowler2009_PRA,Devitt2009_Architecture,Fowler2012_Architecture}.  In the List of Figures, I denote surface code diagrams by the prefix ``Surface Code.''

Surface code diagrams are useful for two reasons.  First, this type of diagram accurately represents the total resource costs of quantum logic, because there is a direct correspondence between the features of the diagram and the operations at the hardware level, in both space and time.  Such information is not readily available in circuit diagrams, where the costs associated with two different gates may differ by orders of magnitude.  Second, surface code diagrams provide a visual way to modify or optimize logic while maintaining the error-correction capacity of the surface code.  In this work, I make use only on the first purpose, though optimization within the surface code is actively being studied elsewhere~\cite{Fowler2012_Bridge,Fowler2013_BlockCodes,Paetznick2013_Compaction}.

For all their utility, surface code diagrams have a notable downside.  Owing to the way that quantum logic in the surface code depends on topology~\cite{Raussendorf2007_PRL,Raussendorf2007_NJP}, it is virtually impossible to determine the underlying logic being shown, as will become apparent in the examples which follow.  For this reason, a surface code diagram should always be paired with a quantum circuit diagram, because the two are complementary.  The quantum circuit shows what the logic does, while the surface code diagram shows how the logic is implemented and what the resource costs are.  This complementarity will be used frequently to demonstrate logic synthesis in later chapters.

\begin{figure}
  \centering
  \includegraphics[width=10cm]{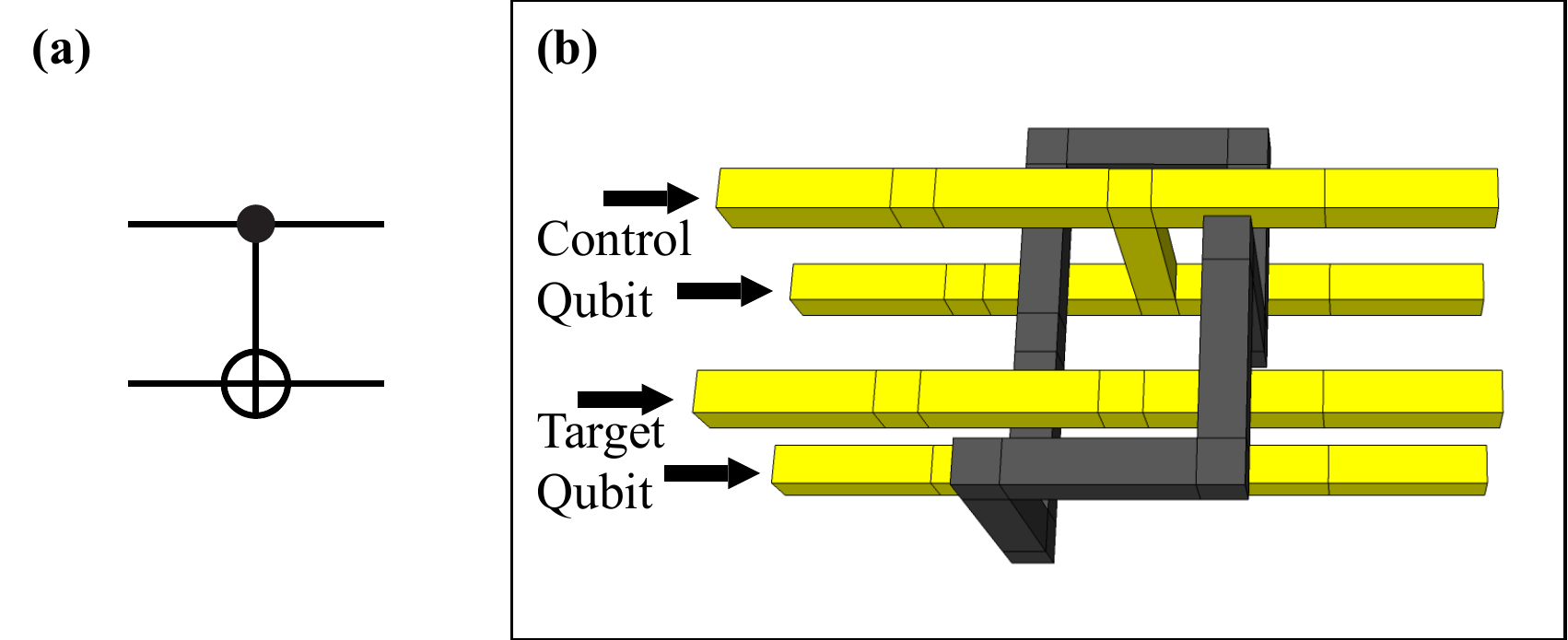}\\
  \caption[Circuit and Surface Code: CNOT gate]{Implementation of CNOT in the surface code. (a)~Circuit diagram for CNOT.  (b)~Perspective rendering of CNOT implemented in the surface code.  Each horizontal pair of yellow pipes corresponds to the qubit on the left in the same vertical position.  Each logical qubit is a pair of yellow defects, arranged along the out-of-page dimension.}
  \label{CNOT_combined}
\end{figure}

\begin{figure}
  \centering
  \includegraphics[width=10cm]{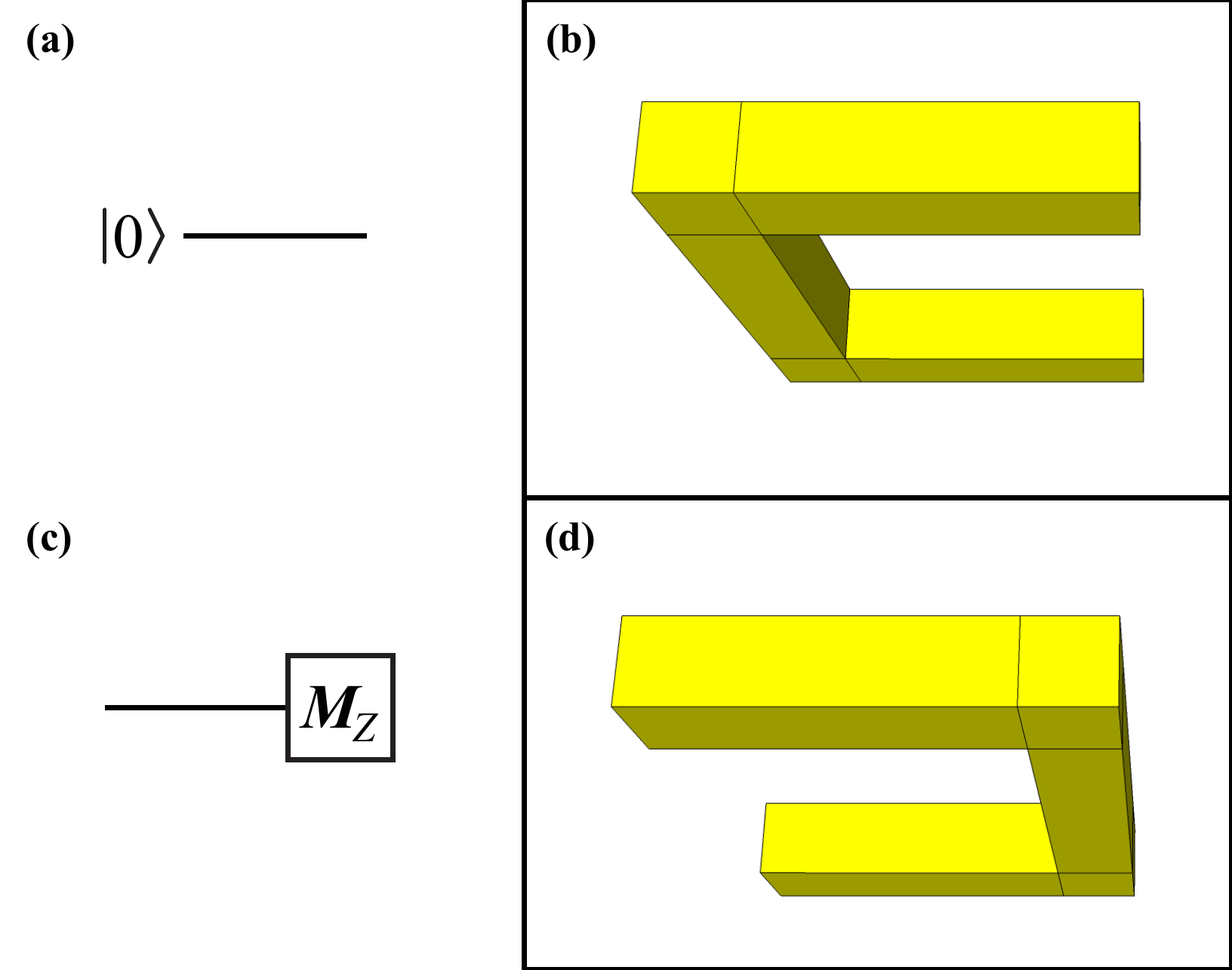}\\
  \caption[Circuit and Surface Code: Initialization and measurement]{Initialization and measurement operations in the surface code.  Primal defects are shown, and the equivalent operations for dual defects would initialize $\ket{+}$ or measure in $X$ basis. (a)~Circuit element for initializing $\ket{0}$.  (b)~Initializing $\ket{0}$ in the surface code.  (c)~Circuit element for $Z$-basis measurement.  (d)~$Z$-basis measurement in surface code.}
  \label{Init_measure_combined}
\end{figure}

\begin{figure}
  \centering
  \includegraphics[width=10cm]{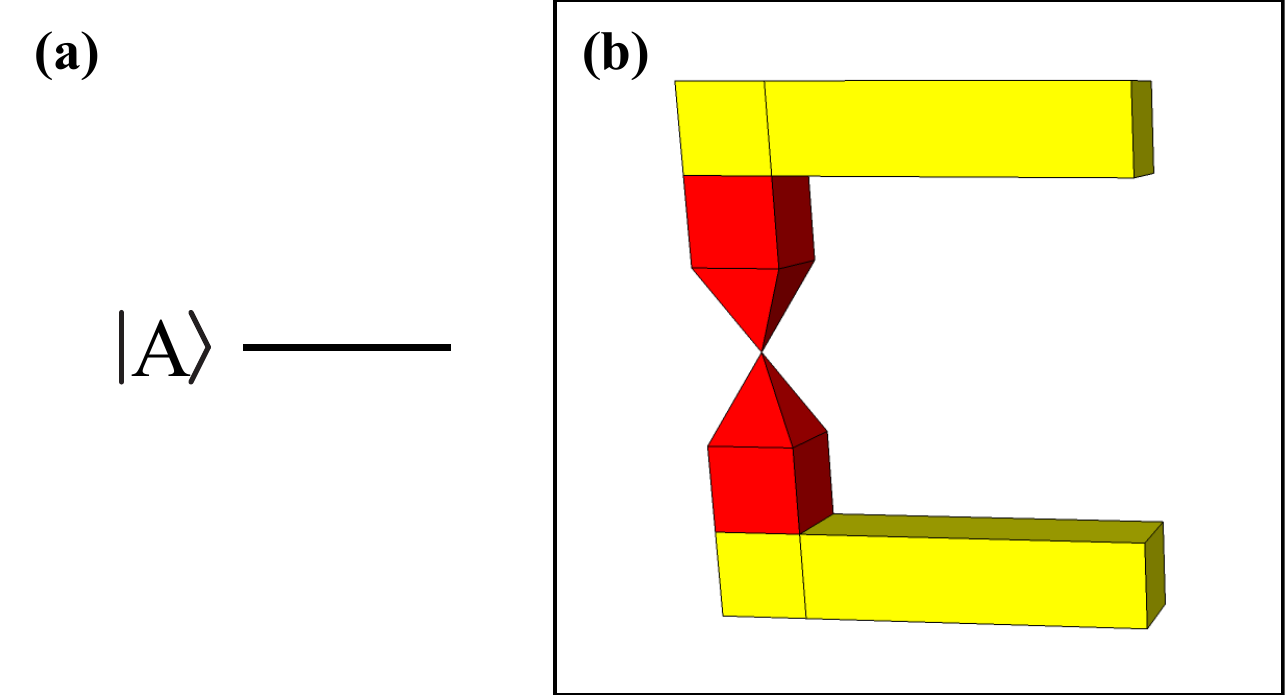}\\
  \caption[Circuit and Surface Code: State injection]{State injection in the surface code. (a)~Circuit diagram for initialization of the state $\ket{A} = (1/\sqrt{2})(\ket{0} + e^{i\pi/4}\ket{1})$.  This is an example of a commonly injected state, but in principle any single-qubit state can be injected.  (b)~Depiction of state injection in the surface code.  The viewing angle is from the side to provide better perspective.  The injection process uses two pyramids which are point defects expanding in circumference.  The pyramids are colored differently from other defects to stand out visually.}
  \label{state_injection_combined}
\end{figure}

An example of a surface code diagram is shown in Fig.~\ref{CNOT_combined}.  The left side is a simple circuit with a CNOT gate acting on two qubits, while the right shows how this might be implemented in the surface code.  CNOT gates in the surface are determined by the topology of defects in the code (shown here as yellow and black pipes) braid around each other.  Each defect is a hole of sorts in the surface code lattice, and Refs.~\cite{Fowler2009_PRA,Fowler2012_Architecture} give a good explanation of how this is implemented at the hardware level.  Some other common circuit primitives are initialization and measurement (Fig.~\ref{Init_measure_combined}), which at this level are mirror images in time, and state injection (Fig.~\ref{state_injection_combined}).  In Fig.~\ref{state_injection_combined}, the tip of the pyramids is a single physical qubit, whose state is converted into a surface code logical qubit contained in the defect.  As mentioned earlier, state injection is a critical process in surface code programs, and Refs.~\cite{Raussendorf2007_PRL,Raussendorf2007_NJP,Fowler2009_PRA,Devitt2009_Architecture,Fowler2012_Architecture} give a proper explanation.  The Hadamard gate is also important, but it is not shown in braiding diagrams because it requires some manipulation of the code properties; see Refs.~\cite{Fowler2012_Hadamard,Fowler2012_Architecture} for details.  In other codes, the Hadamard gate may be the ``hard'' operation~\cite{Paetznick2013_CCZ}.

\section{Resource Calculations}
\label{Sec_Resource_Methodology}
The objective of quantum logic synthesis is to minimize resource costs while executing a reliable quantum program.  There are many resources that require consideration for running a quantum computer, but this work will focus on only two: qubits and gates used for fault-tolerant computation.  Suppose that qubits are regularly spaced on a two-dimensional grid and that gates are regularly separated in time, or ``clocked.''  Using this model, one can account for resource costs by the three-dimensional volume (space and time) required to execute the program, which corresponds exactly to the volume required to implement the braiding topology in the surface code.  Volume is a useful measure for resource cost because it depends mostly on the underlying logic of the program and the error rates of the hardware, and less on the sequence of gates in the program.

The reliability of a quantum program is the probability that the output does not have an error.  A program is reliable if the output error probability is below some target value.  Errors are suppressed using techniques of quantum error correction, but these are costly in terms of resources~\cite{Preskill1998,Nielsen2000,Knill2005,Isailovic2008,Jones2012_PRX,Fowler2012_Architecture}.  The overhead associated with fault tolerance depends on error rates in the hardware and the chosen code.  Generally speaking, the cost scales as $O(\log^c (1/p_{\mathrm{out}}))$, where $p_{\mathrm{out}}$ is an upper bound on the logical error of the program and exponent $c > 1$ is a constant that depends on the logic synthesis method.  Instead of relying on asymptotic estimates, a more precise resource analysis described below will be used in later chapters to give quantitative resource costs.

\begin{figure}
  \centering
  \includegraphics[width=10cm]{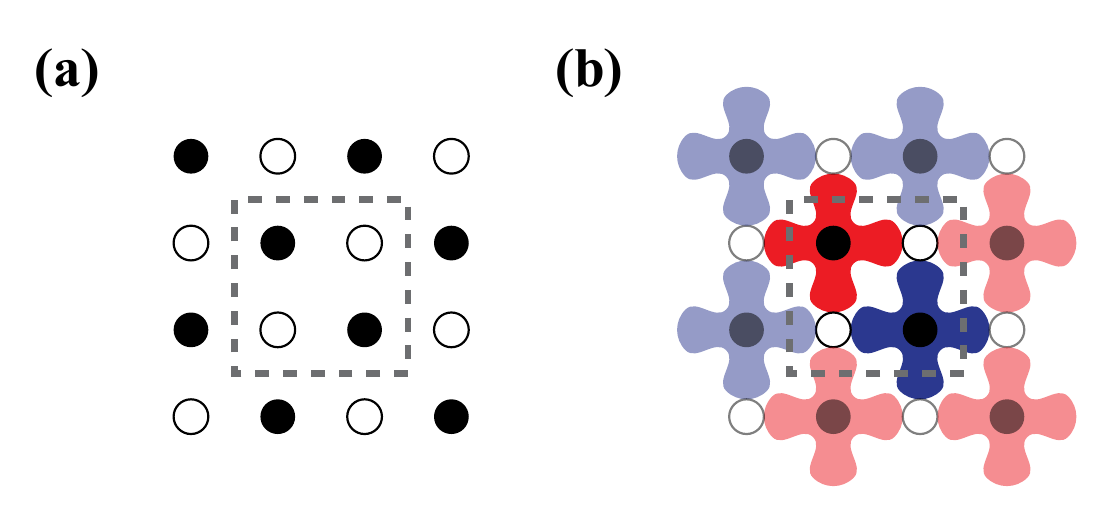}\\
  \caption[Unit cell in the surface code]{Diagram showing the unit cell of the surface code.  (a)~A 2D square array, where each circle represents a qubit.  The open and filled circles play different roles in the surface code.  A unit cell encompasses four qubits, two of each type.  (b)~A unit cell in the surface code includes two stabilizers.  The red and blue ``plus'' shapes are four-body, nearest-neighbor stabilizer measurements of $XXXX$ or $ZZZZ$ (see Refs.~\cite{Fowler2009_PRA,Fowler2012_Architecture} for details).  Neighboring stabilizers are shown for illustration.}
  \label{SC_unit_cell}
\end{figure}

Operations in the surface code are convenient to analyze at a high level of abstraction, where one only considers the arrangement of the braiding surfaces.  Surface code diagrams exist at this level, as the details of hardware operations are not shown.  Apart from visual clarity, this abstraction also gives the diagram a sense of scale invariance, because the same topology, hence same program, could be implemented in two instances of a surface code, where each has a different code distance.  The code distance, often denoted $d$, determines how far apart the braid surfaces must be separated in terms of qubits (space) or stabilizer measurements (time).  Because of this fundamental spacing, one can define a unit cell as two stabilizer measurements, one of each type ($X$ and $Z$), as shown in Fig.~\ref{SC_unit_cell}.  The surface code consists of these unit cells tiled across the 2D plane in space, and repeated in time.  Viewed this way, the surface code is a crystal, in the abstract sense, where the unit cell is repeated in three dimensions.  Logic is implemented with defects, or holes, in the repeated pattern~\cite{Raussendorf2007_PRL,Raussendorf2007_NJP,Fowler2009_PRA}, but the volume can still be accounted in terms of these unit cells, which is the methodology I use throughout this manuscript.

A relevant example of the resource overhead required for fault-tolerant quantum computing is the cost of making some cubic region of the surface code sufficiently reliable.  First, let me explain some rules for surface code logic.  Fowler, Devitt, and collaborators~\cite{Fowler2012_Bridge,Fowler2013_BlockCodes} develop a simple set of design rules for spacing defects.  For a given code distance $d$, the rules are:
\begin{enumerate}[(1)]
\item two defects (or other boundaries) of the same type must be separated by $d$;
\item any defect must have circumference greater than or equal to $d$, so square defects must have side length $\lceil d/4 \rceil$;
\item given (1) and (2), two defects of different types must be separated by $\lceil d/8 \rceil$.
\end{enumerate}
A simple strategy to follow these rules is to design braiding patterns using cubic regions of with side length $d + \lceil d/4 \rceil$.  The finite set of allowable braiding patterns are known as ``plumbing pieces,'' because visually they are pipes that connect together~\cite{Fowler2012_Bridge,Fowler2013_BlockCodes}.  A simple estimate for the probability of error in a plumbing piece with distance $d$ is derived in Ref.~\cite{Fowler2013_BlockCodes}:
\begin{equation}
P_L(p_g,d) \approx d (100 p_g)^{(d+1)/2},
\label{Ch3_Surface_Code_Error}
\end{equation}
where $p_g$ is the error per hardware gate and the factor 100 comes from numerical data fitting in Refs.~\cite{Fowler2012_Practical,Fowler2012_Architecture,Fowler2012_NonCyclic}.

\begin{figure}
  \centering
  \includegraphics[width=\textwidth]{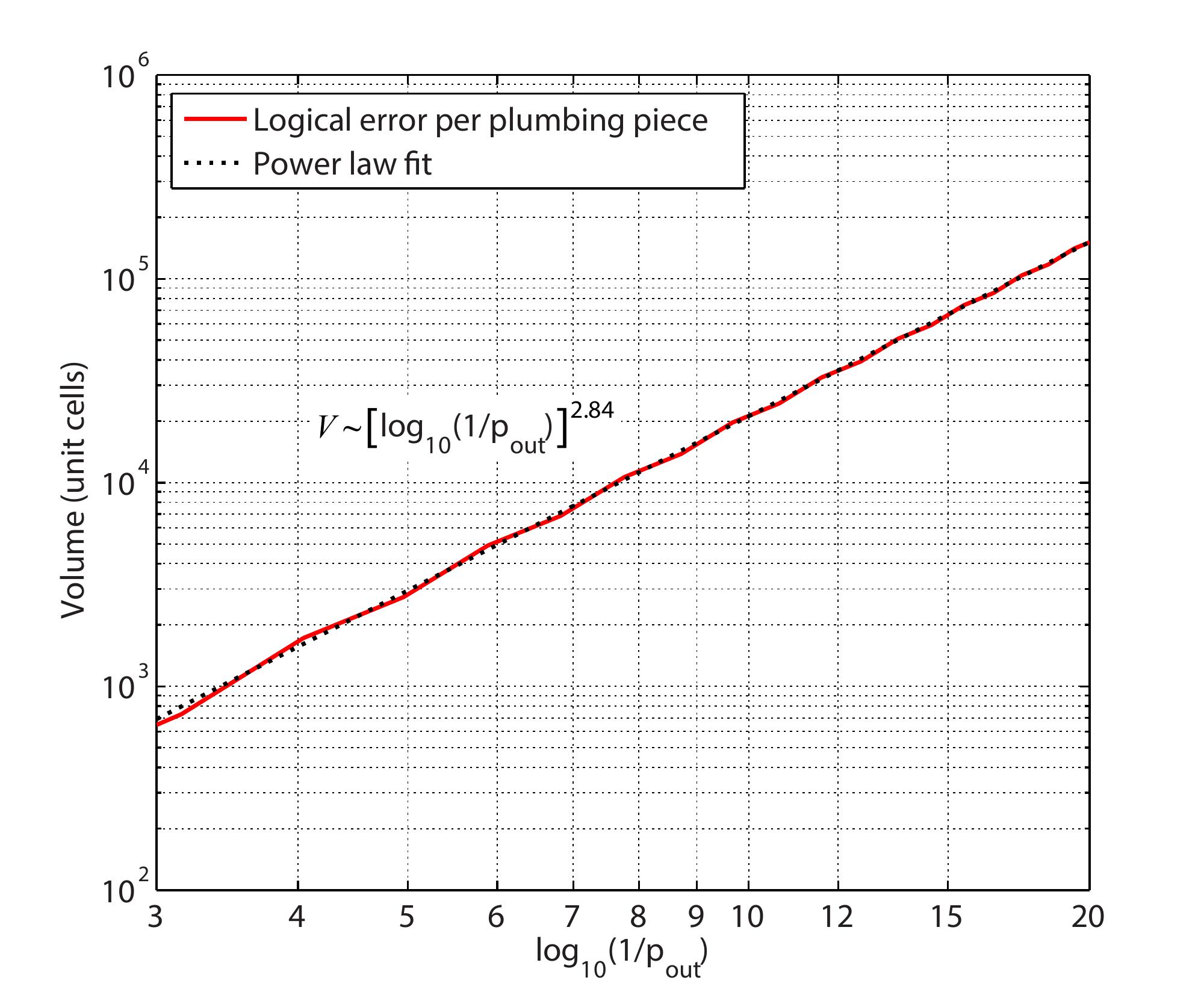}\\
  \caption[Resource costs for a plumbing piece in the surface code]{Resource costs for a plumbing piece in the surface code.  The hardware error rate is $p_g = 10^{-3}$.  Lower logical error rate is achieved by increasing code distance, which also increases the volume of the plumbing piece.  A power law fit shows that volume scales with an exponent of 2.84.}
  \label{Plumbing_piece_resources}
\end{figure}

The volume of a plumbing piece as a function of is logical error probability is plotted in Fig.~\ref{Plumbing_piece_resources}, where $p_g = 10^{-3}$.  This type of plot will be used many times throughout this manuscript to quantify the resource cost of making a quantum program sufficiently reliable.  The volume is measured in unit cells of the surface code, as discussed earlier.  The only notable feature of this plot is that the resource scaling obeys a power law (dashed line) very well: $V \approx 30.4\left(\log_{10}(1/p_{\mathrm{out}})\right)^{2.84}$ unit cells.  This is in close agreement to other findings that the ``scaling exponent'' should be 3~\cite{Fowler2012_Bridge}.  The exponent is less than 3 here only because of the coefficient $d$ in Eqn.~(\ref{Ch3_Surface_Code_Error}), whose presence skews the estimated error rate up more at lower values of $d$.  Indeed, the fitted exponent will approach 3 as $d \rightarrow \infty$, but the plot in Fig.~\ref{Plumbing_piece_resources} only shows the range relevant to practical quantum computing.  This is a good time to remark that power law fits should only be used for estimating quantities like resources, not revealing some deep meaning about quantum information.

On the subject of surface code scaling trends, I would also like to note that the error bound in Eqn.~(\ref{Ch3_Surface_Code_Error}) may overstate error probability at low values of $d$, because there is numerical evidence that the surface code actually performs better than the asymptotic fits for low code distance~\cite{Fowler2012_Practical,Fowler2012_Architecture}.  Hence a more accurate error rate (as a function of $d$) may come closer to the expected scaling coefficient of 3.  The reason for this behavior is that the edges and corners of the surface code become more important at low distance, and these stabilizers have lower weight (two or three, instead of four), which reduces the possible sources of error at the physical level.

A primary concern of this thesis is optimizing quantum logic, and any statement of improvement requires some point of reference.  For comparison purposes, it is useful to define the worst-case resource cost for implementing a quantum program.  Given a quantum program composed of some fundamental operations, the \emph{Trivial Upper Bound} (TUB) is the resource cost associated with the simplest logic design.  For example, one could make the probability of error in each fundamental operation so low that, when summed together, the total probability is small and the entire program is guaranteed to be reliable.  This approach is usually not optimal, but it is a starting point that is useful for comparison.  For a particular program, the difference between TUB and optimized logic shows how important logic synthesis can be.


\chapter{Quantum Logic Synthesis}
\label{Ch04}
The purpose of logic synthesis is to execute a quantum program in a way that minimizes resource costs.  The previous chapter introduced quantum programs to encapsulate quantum logic, diagrams to depict quantum logic, and ways to estimate resources.  These are the tools required for logic synthesis.  This chapter gives an overview of the common synthesis techniques, while the subsequent chapters provide detailed examples with resource analysis.

\section{Generalized Teleportation Gates}
A crucial development for fault-tolerant quantum logic was the teleportation gate~\cite{Shor1996,Preskill1998_FTQC,Gottesman1999,Zhou2000}.  Instead of teleporting a quantum state from one position to another, this procedure implements a logical gate using a sequence of operations fueled by a special quantum state.  In effect, the quantum state changes through teleportation, even though it may not change its physical location.  The novelty of this proposal is that a gate can be encoded into a quantum state, so long as one knows how to ``read'' this information.

Let me introduce the notion of a ``quantum look-up table'' (QLUT).  Take any $N$-dimensional unitary operator $U$ and represent it in the spectral decomposition using eigenvalues $\{\lambda_j\}$ and eigenvectors $\{\ket{u_j}\}$:
\begin{equation}
U = \sum_{j=1}^N \lambda_j \ket{u_j}\bra{u_j}.
\label{Spectral_Decomposition}
\end{equation}
Let $\ket{\psi} = \frac{1}{N} \sum_{j=1}^N \ket{u_j}$ be a uniform superposition over the eigenvectors.  The QLUT for $U$ is
\begin{equation}
U\ket{\psi} = \frac{1}{N} \sum_{j=1}^N \lambda_j \ket{u_j}.
\label{QLUT_eqn}
\end{equation}
There is a clear similarity between the RHS of Eqns.~(\ref{Spectral_Decomposition}) and~(\ref{QLUT_eqn}).  The reason I call this a ``look-up table'' is that that the QLUT is a state that encodes the action of $U$.  For any eigenvector of $U$, the QLUT has the associated eigenvalue stored in its state.  In many contexts, these are also called ``magic states,'' for precisely the same reason.  For example, the magic state for $T = \exp(i \pi(I - Z)/8)$ is $T\ket{+}$, where $\ket{+} = (1/\sqrt{2})(\ket{0} + \ket{1})$ is the uniform superposition over the eigenvectors of $T$.

\begin{figure}
  \centering
  \includegraphics[width=\textwidth]{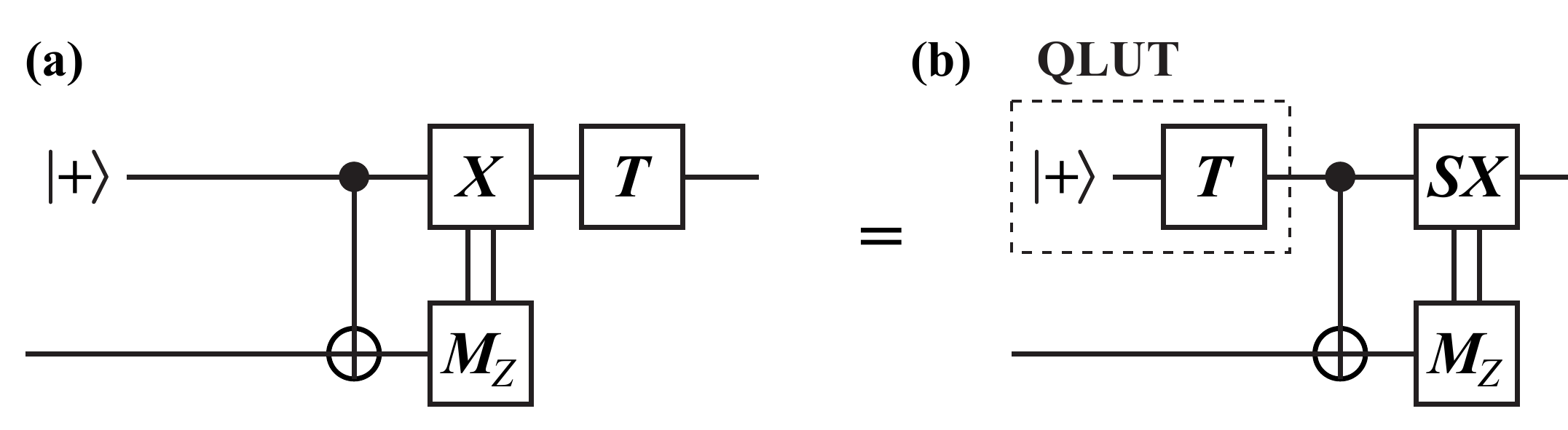}\\
  \caption[Circuit: Constructing a QLUT for $T$~gate]{Circuit technique for creating the $T$-gate QLUT.  (a)~Generic teleportation, followed by $T$~gate.  (b)~The $T$~gate is moved (using commutation rules) to just before $\ket{+}$ to form a QLUT.  Here, the commutation affects the conditional $X$ operation, as $TX = (TXT^{\dag})T$, and $TXT^{\dag} = e^{-i \pi/4} SX$.  The global phase $e^{-i \pi/4}$ is dropped.}
  \label{QLUT_circuit}
\end{figure}

One way to compile a quantum program into a QLUT is to begin with a teleportation circuit that takes $\ket{\psi}$ as an input.  Specifically, the circuit teleports an arbitrary qubit onto the ancilla $\ket{\psi}$, then implements $U$.  This process is depicted in Fig.~\ref{QLUT_circuit}(a) for $U = T$ and $\ket{\psi} = \ket{+}$.  The QLUT may be formed by using commutation rules to move $U$ to before the teleportation circuit, as in Fig.~\ref{QLUT_circuit}(b) (\emph{cf.}~\cite{Nielsen2000}, p.487).  In general, this commutation step modifies the teleportation procedure, so it is crucial that the new circuit has an efficient fault-tolerant construction.  Developing other general procedures for designing teleportation gates is an area of future research.

\section{Off-Line Validation and Fault Tolerance}
The technique of compiling a quantum program into a QLUT can take much of the computational effort off of the data path.  The data path is the sequence of operations which come in direct contact with data qubits in an algorithm.   If there is a failure here, the data is corrupted.  By contrast, operations off of the data path (``off-line'') may be expendable; if an error is detected, the faulty states are removed without affecting the rest of the computation.  Reference~\cite{Isailovic2008} also discusses how off-line preparation of QLUTs enables fast computation.

A QLUT can be compiled in a faulty manner, then validated using a procedure that checks for error in the QLUT.  Using the quantum measurement postulate, successful validation projects the QLUT into a higher-fidelity state.  This is essentially a variant of post-selected quantum computation~\cite{Knill2005}.  Fault tolerance is achieved by bringing the QLUT to sufficient fidelity for interaction on the data path.

At first glance, the strategy of moving quantum programs into QLUTs would appear to just redistribute the effort of error correction from one place to another.  However, the ability to discard states which fail validation is quite valuable.  Validation only requires error \emph{detection} instead of correction, and the former is more efficient.  A \mbox{distance-$d$} code can correct $\lfloor \frac{d-1}{2}\rfloor$ errors, leading to an output error of order $O(p^{\lfloor (d+1)/2 \rfloor})$~\cite{Nielsen2000}.  By contrast, the same code can detect $d-1$ errors, leading to a validated output state with error $O(p^d)$.  Moreover, error detection is almost always less taxing on classical control hardware, which can be a concern in some contexts~\cite{Devitt2010}.  For these reasons, performing validation can lead to substantial reductions in the overhead for fault-tolerant computation.

The steps for off-line logic synthesis are: (1)~identify an important and frequently used quantum program; (2)~compile this program into a teleportation gate using a QLUT; (3)~develop an efficient procedure to validate the QLUT.  This design methodology will be demonstrated repeatedly in Chapters~\ref{Ch05}--\ref{Ch07}.  Chapter~\ref{Ch09} will examine common features of these techniques, which may be useful both for developing new methods and for understanding limitations of this approach. 


\chapter{Distillation Protocols}
\label{Ch05}
Distillation protocols are a special case of error detection where many noisy copies of a quantum state are ``distilled'' into fewer low-error copies of the same state.  A common theme for this chapter is that, in many circumstances, an important but difficult operation can be encoded into a well-characterized quantum state, such as a quantum look-up table (QLUT; see Chapter~\ref{Ch04}).  After injecting noisy copies of the desired state into the surface code, they are distilled before being used by computation.  Error detection is often employed with a quantum code that uses only operations that are themselves error-corrected by the surface code (see Section~\ref{quantum_programs_section} for a list).  However, at the end of the chapter, I discuss Fourier-state distillation, which distills a special class of multi-qubit quantum states.  This is a new protocol that relies on Toffoli gates, which are not native to the surface code.  The Toffoli gates require techniques developed in Chapter~\ref{Ch06}, and this example shows that distillation can be applied to produce useful multi-qubit states beyond just satisfying the minimum requirements of universal computation.

Distillation protocols hold an important place in fault-tolerant quantum computing.  For example, entanglement distillation demonstrated that arbitrarily long-range quantum entanglement was achievable in principle, using quantum repeaters~\cite{Bennett1996,Deutsch1996,Briegel1998}.  The advent of magic-state distillation made the prospect of large-scale quantum computing more plausible~\cite{Knill2005}.  There are alternative ways to achieve universal, fault-tolerant quantum computing~\cite{Shor1995,Steane1996_Interference,Calderbank1996,Steane1996_Theory,Gottesman1997,Preskill1998,Nielsen2000,Steane2002}, but the magic-state techniques developed by Knill~\cite{Knill2004,Knill2005} and Bravyi and Kitaev~\cite{Bravyi2005} are compatible with broader sets of codes, including the surface code.

When viewed as a quantum program, a distillation protocol takes many copies of the same state as inputs and returns fewer copies of the same state as outputs.  By assumption, the input states have independent errors, which is essential for the technique.  Moreover, it is often assumed that the errors are also identically distributed, but this is not necessary.  Because the inputs and outputs are of the same form, distillation protocols can be executed recursively.  Recursive distillation is needed when just one round does not purify the desired state to sufficiently low error probability.  The different rounds have different requirements for error correction, and hence different resource costs.

Resource costs for magic-state distillation can dominate the total resources required for quantum computing~\cite{Isailovic2008,Jones2012_PRX,Fowler2012_Bridge,Fowler2012_Architecture,Fowler2013_BlockCodes}.  A recursive distillation protocol used to make a single gate, such as $T = \exp(i\pi(I-Z)/8)$, requires very many fundamental gates in the surface code.  Fowler and Devitt estimate that a single $T$~gate requires 46 times the surface code volume as a single CNOT~\cite{Fowler2012_Bridge}.  Resource costs will be a central concern for this chapter, as the distillation protocols examined here will be the first concrete demonstrations of the techniques of logic synthesis.

\section{Magic-State Distillation}
\label{Sec_MS_Distillation}
Magic-state distillation purifies a quantum look-up table (QLUT) for a gate that is otherwise unavailable within the chosen code.  For example, the surface code is usually implemented with two distinct types of magic state distillation.  The gates $S = \exp(i \pi (I-Z)/4)$ and $T = \exp(i \pi (I-Z)/8)$ are required for universal computation, and they may be produced using magic states $\ket{Y} = (1/\sqrt{2})(\ket{0} + i \ket{1})$ and $\ket{A} = (1/\sqrt{2})(\ket{0} + e^{i\pi/4}\ket{1})$, respectively~\cite{Raussendorf2007_NJP,Fowler2009_PRA,Jones2012_PRX,Fowler2012_Architecture}.  This section focuses on distilling $\ket{A}$ because this process is more costly than distilling $\ket{Y}$; however, $S$~gates are a necessary part of $\ket{A}$ distillation, as discussed later.

\begin{figure}
  \centering
  \includegraphics[width=10cm]{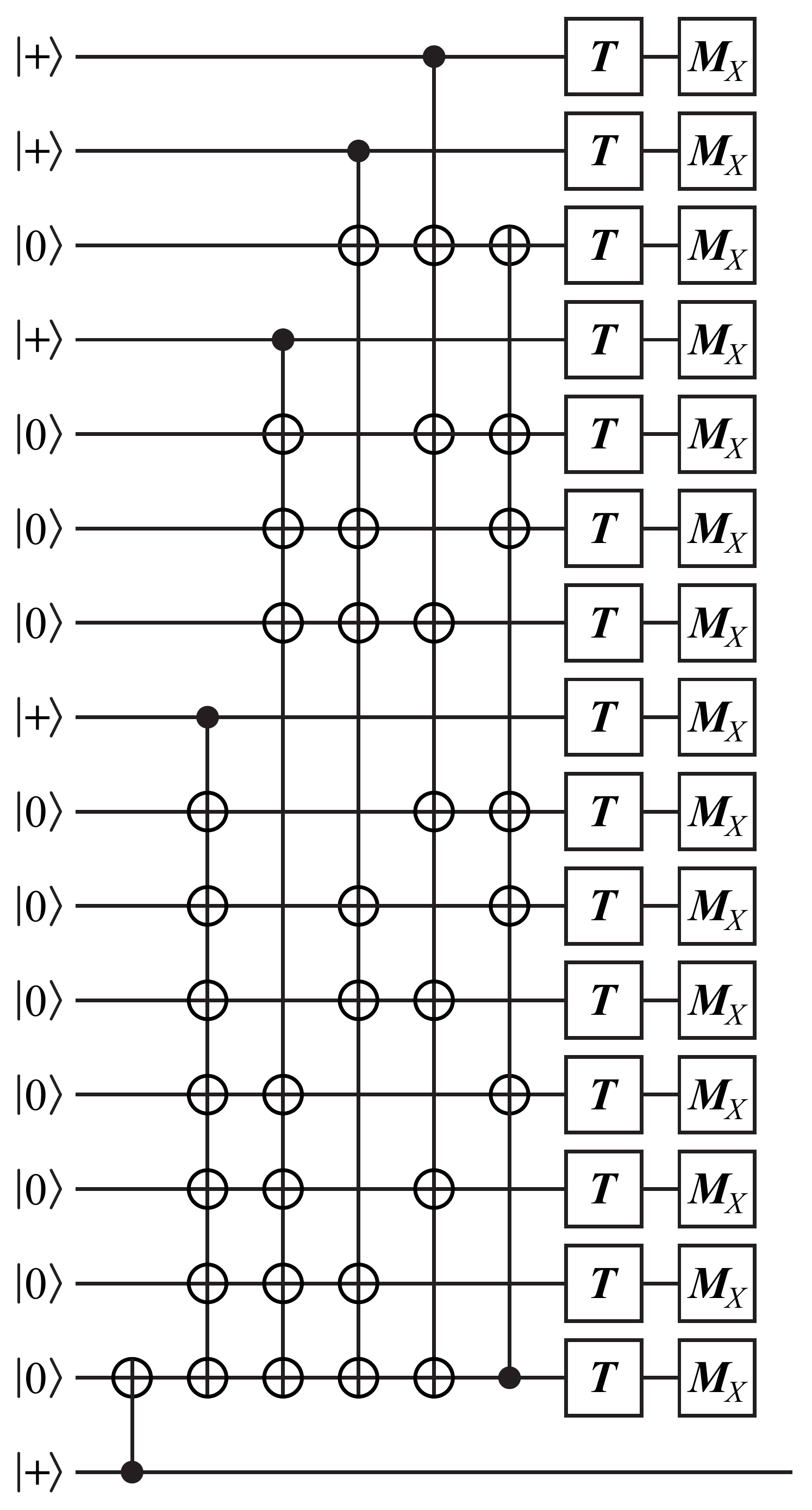}\\
  \caption[Circuit: $\ket{A}$ distillation]{Circuit for distilling the magic state $\ket{A}$.  Each $T$~gate is produced using $\ket{A}$, its QLUT, with the circuit in Fig.~\ref{T_teleportation}.  The bottom qubit is the output state.}
  \label{A_distillation_circuit}
\end{figure}

\begin{figure}
  \centering
  \includegraphics[width=7cm]{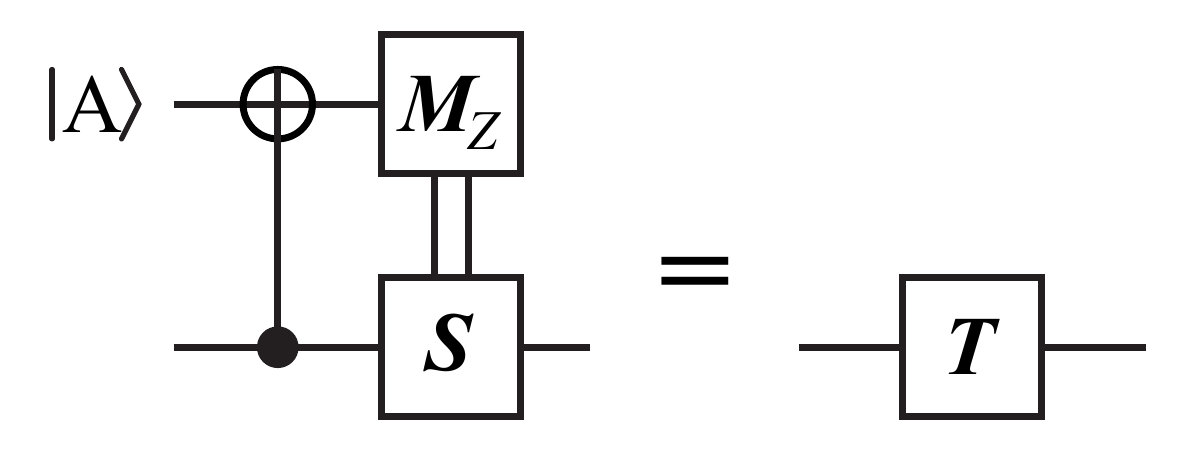}\\
  \caption[Circuit: $T$-gate teleportation using $\ket{A}$]{Circuit for teleporting a $T$~gate using the QLUT $\ket{A}$.  The error in the $T$ gate depends on the error the $\ket{A}$ magic state.}
  \label{T_teleportation}
\end{figure}

There are many proposals for distilling $\ket{A}$ states~\cite{Knill2004,Bravyi2005,Meier2012,Bravyi2012,Jones2013_Multilevel}, but I focus on the 15-to-1 Bravyi-Kitaev (or ``BK'') protocol, named for the authors of Ref.~\cite{Bravyi2005}.  The label ``15-to-1'' refers to the ratio of input states to output states, which is an important consideration for efficiency.  A circuit diagram for the BK protocol is shown in Fig.~\ref{A_distillation_circuit}.  Each $T$~gate is produced using a copy of $\ket{A}$, as shown in Fig.~\ref{T_teleportation}, so the BK distillation protocol takes 15 copies of $\ket{A}$ as inputs.  When each of the input states has independent error $\epsilon \ll 1$, the distilled output state has error $35 \epsilon^3$ to lowest non-vanishing order.

The BK protocol has an important advantage over many other competing protocols: it distills only one output state.  Other protocols~\cite{Meier2012,Bravyi2012,Jones2013_Multilevel} that distill two or more $\ket{A}$ states within the same code block inevitably lead to correlated errors at the output.  This poses a problem when one round of magic-state distillation is insufficient, so the output of the first round must be purified again.  In such a scenario, states with correlated errors must fan out to different second-round distillation blocks, \emph{etc}.  By not having this issue, the BK protocol is much simpler to analyze.  Still, recent analysis suggests there may be advantages to the multiple-output distillation methods~\cite{Meier2012,Bravyi2012,Jones2013_Multilevel}, if the routing considerations can be addressed.

\subsection{Bravyi-Kitaev Distillation in Surface Code}
\label{BK_surface_code_section}
Several works have analyzed the BK protocol assuming perfect Clifford gates~\cite{Bravyi2005,Reichardt2005,Jones2012_PRX} and the costs associated with making Clifford gates fault tolerant~\cite{Raussendorf2007_PRL,Raussendorf2007_NJP,Fowler2012_Bridge,Fowler2012_Architecture,Fowler2013_BlockCodes}.  Using the implementation from Ref.~\cite{Fowler2012_Bridge} of the BK protocol in the surface code, one can estimate the resources required to implement sufficient error correction for this distillation routine.  Moreover, the cost of a $T$~gate at any level of fidelity can be calculated by accounting for the costs of multi-round distillation, as explained below.

Following the methodology developed in Refs.~\cite{Fowler2012_Bridge,Fowler2013_BlockCodes}, one constructs programs in the surface code using regular-sized ``plumbing pieces'' (see also Chapter~\ref{Ch03}).  Each piece occupies a cube in the surface code with side length $d + \lceil d/4 \rceil$.  The probability of logical error in a single plumbing piece can be bounded from above by
\begin{equation}
P_L(p_g,d) \approx d (100 p_g)^{(d+1)/2},
\label{Surface_Code_Error}
\end{equation}
as derived in Ref.~\cite{Fowler2013_BlockCodes}. Subscript $L$ denotes logical error, $p_g$ is error-per-gate at the hardware level, $d$ is the distance of this implementation of the surface code, and the power law scaling is a fit to numerical simulations of surface code error correction~\cite{Wang2010,Fowler2010_Error,Fowler2012_Practical,Fowler2012_Timing}.  The error at the output of BK distillation is therefore bounded by the sum of probabilities for distillation error from input states and for error in the distillation circuit: $p_{\mathrm{out}} \le V \cdot P_L(p_g,d) + 35 {p_{\mathrm{in}}}^3$, where $V$ is the number of plumbing pieces and $p_{\mathrm{in}}$ is the error of the input $T$~gates.  The volume, in unit cells of the surface code, is the product of number of plumbing pieces and the volume of a single plumbing piece, which is $(d + \lceil d/4 \rceil)^3$ unit cells.

\begin{figure}
  \centering
  \includegraphics[width=12cm]{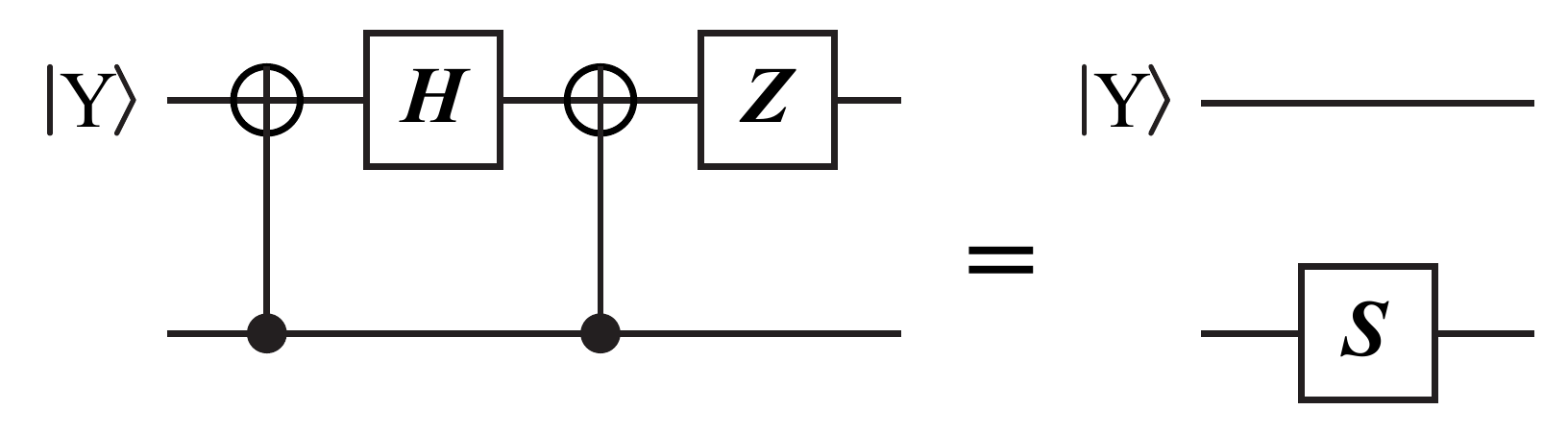}\\
  \caption[Circuit: $S$-gate using $\ket{Y}$]{Circuit for producing a $S$~gate using the QLUT $\ket{Y}$.  The circuit does not destroy the ancilla qubit, and the $Z$ gate is recorded in the Pauli frame.}
  \label{S_kickback}
\end{figure}

\begin{figure}
  \centering
  \includegraphics[width=\textwidth]{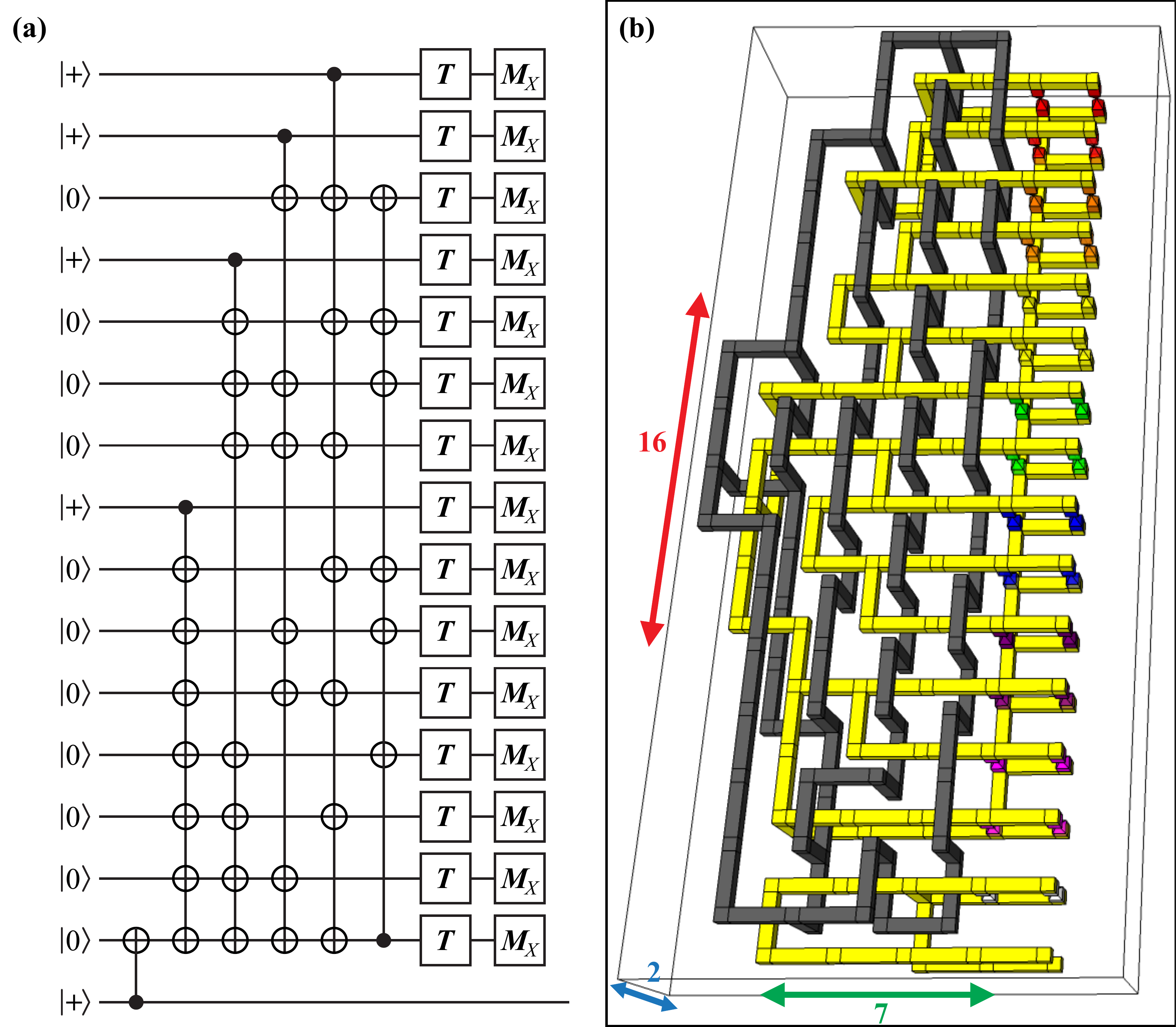}\\
  \caption[Circuit and Surface Code: Bravyi-Kitaev distillation]{Bravyi-Kitaev 15-to-1 distillation implemented in the surface code. (a)~Circuit diagram for BK distillation from Fig.~\ref{A_distillation_circuit}, repeated here for convenience.  (b)~Surface code braiding pattern, derived in Ref.~\cite{Fowler2012_Bridge}.  The horizontal primal defects (yellow pipes) correspond to qubits on the left.  Dual defects (black pipes) implement logical CNOT gates.  The colored pyramids on the right are state injection, corresponding to the $T$ gate followed by $X$-basis measurement in the circuit.  The bottom pair of primal defects is the output qubit.  The volume is $7 \times 16 \times 2 = 224$ plumbing pieces.  Extra volume is allotted for the conditional $S$ gates in the circuit diagram, as explained in the text.}
  \label{A30_combined}
\end{figure}

Fowler and Devitt constructed a version of the BK protocol in the surface code with $V = 192$ plumbing pieces~\cite{Fowler2012_Bridge}; however, this work considers the volume to be slightly larger.  An important issue for distilling $\ket{A}$ states in the surface code is that $T$~gates are implemented using the teleportation circuit in Fig.~\ref{T_teleportation}, which may also require an $S$-gate correction.  $S$ is not a native operation in the surface code, but it is still relatively inexpensive since it can be catalyzed by $\ket{Y}$ without destroying the magic state~\cite{Aliferis2007_Thesis,Jones2012_PRX}, as shown in Fig.~\ref{S_kickback}.  Moreover, the additional overhead is small because each $S$ gate need only have a fidelity on the same order as the $T$ gate input error $p_{\mathrm{in}}$, and hence lower code distance can be used for $S$ gates.  I estimate that the operations for these $S$ gates can be implemented in a depth of two plumbing pieces (there was already one allocated in the volume estimate above), making the total volume now $V = 7 \times 16 \times 2 = 224$ plumbing pieces for the Bravyi-Kitaev distillation protocol, as shown in Fig.~\ref{A30_combined}.  A bounding box serves as a guide to how the volume is estimated.

\subsection{Resource Analysis for Bravyi-Kitaev Distillation}
\label{Sec_BK_resources}
Determining the best combination of Bravyi-Kitaev distillation protocols at different code distances is a resource optimization problem.  In the first round of distillation, increasing code distance will increase volume and lower output error, until the probability of surface code error is negligible in comparison to the $35{p_{\mathrm{in}}}^3$ error from faulty input states.  To move beyond this limit, one must use two rounds of distillation.  Since the inputs to the second round require distillation, the total volume will be 15 times the volume in first round, plus the volume of the second round.  Furthermore, there is positive probability that any distillation circuit will fail, requiring repetition; I account for this by multiplying volume by $1/(1-p_{\mathrm{fail}})$, which gives the mean volume including repeated distillation.  For BK distillation, $p_{\mathrm{fail}} \approx 15p_{\mathrm{in}}$.  For a specified number of rounds, let $p_0$ denote the injected state error probability, $p_1$ the probability of error after one round of distillation, \emph{etc}.  The approximate volume and output error after round $r$ are given by the recursion relations:
\begin{equation}
p_r = 35{p_{r-1}}^3 + 224 P_L(p_g,d_r)
\label{output_error_recursion}
\end{equation}
\begin{equation}
V_r = 224 \sum_{s=1}^r \frac{15^{r-s}}{1-15p_s} (d_s + \lceil d_s/4 \rceil)^3
\label{volume_recursion}
\end{equation}
The factor 224 is the estimated size of the surface code program, in plumbing pieces.  The factor $15^{r-s}$ is the number of copies of round-$s$ distillation needed to feed into one instance of round-$r$ distillation.

\begin{figure}
  \centering
  \includegraphics[width=\textwidth]{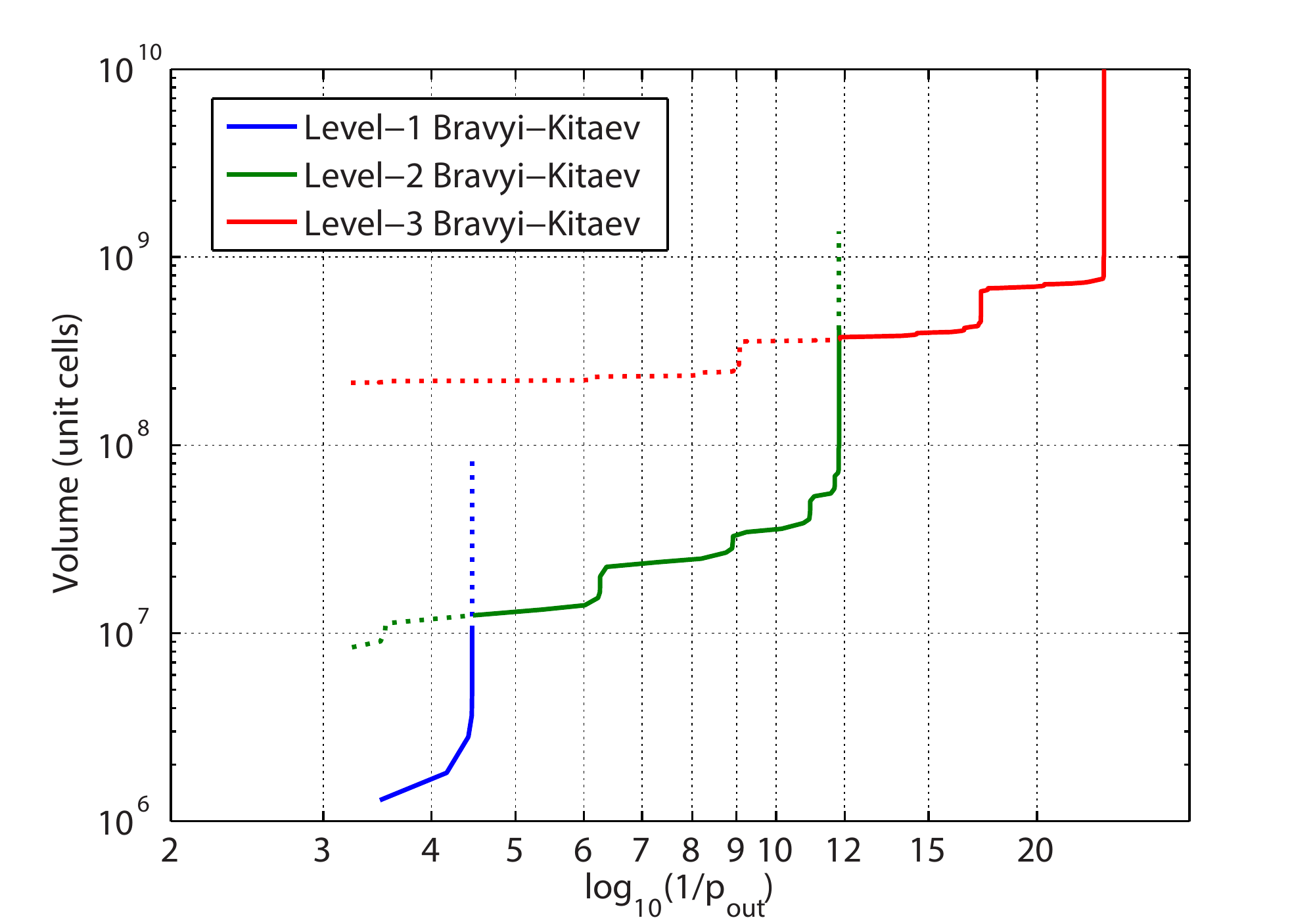}\\
  \caption[Resources for efficient Bravyi-Kitaev distillation]{Resource costs for Bravyi-Kitaev distillation protocols.  Color denotes the number of rounds of distillation.  For each of one, two or three rounds, the efficient frontier is plotted.  Dashed lines show protocols dominated by a protocol having different number of rounds, so the efficient frontier considering any number of rounds is the union of the solid lines.}
  \label{BK_protocols_resources}
\end{figure}

Using the formulas in Eqns.~(\ref{output_error_recursion}) and~(\ref{volume_recursion}), I calculated all possible combinations of BK distillation volume (in unit cells) and output error rate, as explained below.  The number of rounds ranged from one to three, the distance in each round ranged from $5 \le (d_1,d_2,d_3) \le 55$, and I calculated results for values of the hardware gate error such that $\log_{10}(1/p_g) = \{3,3.5,4,4.5,5\}$.  By conventional assumption, the injected $\ket{A}$ states used by the first round have $p_{\mathrm{in}} = 10 p_g$~\cite{Fowler2013_BlockCodes}, where the factor 10 accounts for the number of faulty hardware operations during injection and before error correction.  To narrow focus to useful results, I only make note of protocols on the ``efficient frontier,'' which consists of those protocols (each having a unique combination of parameters $d_1$,$d_2$,$d_3$) that are not dominated by any other protocol.  In terms of performance, one protocol dominates another if the first has both lower volume and lower output error rate; there is no reason to use a dominated protocol.  The results for $p_g = 10^{-3}$ are shown in Fig.~\ref{BK_protocols_resources}.  The results for other values of $p_g$ show effectively the same behavior, so they are not plotted.  In addition to this plot, Table~\ref{BK_resources_table} gives the estimated resource costs when using BK distillation for different input error rates $\log_{10}(1/p_g) = \{3,3.5,4,4.5,5\}$ and output error rates $\log_{10}(1/p_{\mathrm{out}}) = \{3,...,15\}$.  This can be compared with tables in Refs.~\cite{Meier2012,Bravyi2012,Jones2013_Multilevel,Fowler2013_BlockCodes}, most of which only consider cost in number of input magic states.  The difference between my results and those in Ref.~\cite{Fowler2013_BlockCodes} is due mostly to my definition of a unit cell, which contains four qubits when ancillas are used for stabilizer measurement.  Additionally, I estimate a slightly larger volume for BK distillation in the surface code (224 vs. 192 plumbing pieces).

\begin{table}
  \centering
  \begin{tabular}{c m{2cm} m{2cm} m{2cm} m{2cm} m{2cm}}
    \toprule
    \multirow{3}{*}{$\log_{10} \left(p_{\mathrm{out}}\right)$} & \multicolumn{5}{c}{Volume (unit cells)}\\
    & \multicolumn{5}{c}{$(d_1,d_2,d_3)$} \\
    & $p_g = 10^{-3}$ &  $p_g = 10^{-3.5}$ & $p_g = 10^{-4}$ & $p_g = 10^{-4.5}$ & $p_g = 10^{-5}$\\ \midrule
     \multirow{2}{*}{-3} & $1.1 \times 10^{6}$ \cellcolor{Cerulean} & $3.5 \times 10^{5}$ \cellcolor{Cerulean} &  &  & \\
     & (13) \cellcolor{Cerulean} & (9) \cellcolor{Cerulean} &  &  &  \\
     \multirow{2}{*}{-4} & $1.5 \times 10^{6}$ \cellcolor{Cerulean} & $3.5 \times 10^{5}$ \cellcolor{Cerulean} & $1.4 \times 10^{5}$ \cellcolor{Cerulean} & $6.6 \times 10^{4}$ \cellcolor{Cerulean} & \\
     & (15) \cellcolor{Cerulean} & (9) \cellcolor{Cerulean} & (7) \cellcolor{Cerulean} & (5) \cellcolor{Cerulean} &  \\
     \multirow{2}{*}{-5} & $1.1 \times 10^{7}$ \cellcolor{YellowGreen} & $5.5 \times 10^{5}$ \cellcolor{Cerulean} & $3.4 \times 10^{5}$ \cellcolor{Cerulean} & $1.4 \times 10^{5}$ \cellcolor{Cerulean} & $6.6 \times 10^{4}$ \cellcolor{Cerulean}\\
     & (11,17) \cellcolor{YellowGreen} & (11) \cellcolor{Cerulean} & (9) \cellcolor{Cerulean} & (7) \cellcolor{Cerulean} & (5) \cellcolor{Cerulean} \\
     \multirow{2}{*}{-6} & $1.2 \times 10^{7}$ \cellcolor{YellowGreen} & $3.2 \times 10^{6}$ \cellcolor{YellowGreen} & $3.4 \times 10^{5}$ \cellcolor{Cerulean} & $1.4 \times 10^{5}$ \cellcolor{Cerulean} & $6.6 \times 10^{4}$ \cellcolor{Cerulean}\\
     & (11,19) \cellcolor{YellowGreen} & (7,13) \cellcolor{YellowGreen} & (9) \cellcolor{Cerulean} & (7) \cellcolor{Cerulean} & (5) \cellcolor{Cerulean} \\
     \multirow{2}{*}{-7} & $2.0 \times 10^{7}$ \cellcolor{YellowGreen} & $3.5 \times 10^{6}$ \cellcolor{YellowGreen} & $5.3 \times 10^{5}$ \cellcolor{Cerulean} & $3.3 \times 10^{5}$ \cellcolor{Cerulean} & $1.4 \times 10^{5}$ \cellcolor{Cerulean}\\
     & (13,21) \cellcolor{YellowGreen} & (7,15) \cellcolor{YellowGreen} & (11) \cellcolor{Cerulean} & (9) \cellcolor{Cerulean} & (7) \cellcolor{Cerulean} \\
     \multirow{2}{*}{-8} & $2.1 \times 10^{7}$ \cellcolor{YellowGreen} & $6.5 \times 10^{6}$ \cellcolor{YellowGreen} & $2.7 \times 10^{6}$ \cellcolor{YellowGreen} & $3.3 \times 10^{5}$ \cellcolor{Cerulean} & $1.4 \times 10^{5}$ \cellcolor{Cerulean}\\
     & (13,23) \cellcolor{YellowGreen} & (9,15) \cellcolor{YellowGreen} & (7,11) \cellcolor{YellowGreen} & (9) \cellcolor{Cerulean} & (7) \cellcolor{Cerulean} \\
     \multirow{2}{*}{-9} & $2.4 \times 10^{7}$ \cellcolor{YellowGreen} & $7.3 \times 10^{6}$ \cellcolor{YellowGreen} & $3.1 \times 10^{6}$ \cellcolor{YellowGreen} & $1.3 \times 10^{6}$ \cellcolor{YellowGreen} & $3.3 \times 10^{5}$ \cellcolor{Cerulean}\\
     & (13,27) \cellcolor{YellowGreen} & (9,17) \cellcolor{YellowGreen} & (7,13) \cellcolor{YellowGreen} & (5,9) \cellcolor{YellowGreen} & (9) \cellcolor{Cerulean} \\
     \multirow{2}{*}{-10} & $3.1 \times 10^{7}$ \cellcolor{YellowGreen} & $7.9 \times 10^{6}$ \cellcolor{YellowGreen} & $3.1 \times 10^{6}$ \cellcolor{YellowGreen} & $1.5 \times 10^{6}$ \cellcolor{YellowGreen} & $3.3 \times 10^{5}$ \cellcolor{Cerulean}\\
     & (15,27) \cellcolor{YellowGreen} & (9,19) \cellcolor{YellowGreen} & (7,13) \cellcolor{YellowGreen} & (5,11) \cellcolor{YellowGreen} & (9) \cellcolor{Cerulean} \\
     \multirow{2}{*}{-11} & $3.5 \times 10^{7}$ \cellcolor{YellowGreen} & $7.9 \times 10^{6}$ \cellcolor{YellowGreen} & $3.4 \times 10^{6}$ \cellcolor{YellowGreen} & $1.5 \times 10^{6}$ \cellcolor{YellowGreen} & $1.3 \times 10^{6}$ \cellcolor{YellowGreen}\\
     & (15,31) \cellcolor{YellowGreen} & (9,19) \cellcolor{YellowGreen} & (7,15) \cellcolor{YellowGreen} & (5,11) \cellcolor{YellowGreen} & (5,9) \cellcolor{YellowGreen} \\
     \multirow{2}{*}{-12} & $3.1 \times 10^{8}$ \cellcolor{Lavender} & $1.2 \times 10^{7}$ \cellcolor{YellowGreen} & $3.4 \times 10^{6}$ \cellcolor{YellowGreen} & $1.9 \times 10^{6}$ \cellcolor{YellowGreen} & $1.5 \times 10^{6}$ \cellcolor{YellowGreen}\\
     & (11,15,33) \cellcolor{Lavender} & (11,21) \cellcolor{YellowGreen} & (7,15) \cellcolor{YellowGreen} & (5,13) \cellcolor{YellowGreen} & (5,11) \cellcolor{YellowGreen} \\
     \multirow{2}{*}{-13} & $3.2 \times 10^{8}$ \cellcolor{Lavender} & $1.3 \times 10^{7}$ \cellcolor{YellowGreen} & $4.2 \times 10^{6}$ \cellcolor{YellowGreen} & $3.1 \times 10^{6}$ \cellcolor{YellowGreen} & $1.5 \times 10^{6}$ \cellcolor{YellowGreen}\\
     & (11,17,33) \cellcolor{Lavender} & (11,23) \cellcolor{YellowGreen} & (7,17) \cellcolor{YellowGreen} & (7,13) \cellcolor{YellowGreen} & (5,11) \cellcolor{YellowGreen} \\
     \multirow{2}{*}{-14} & $3.3 \times 10^{8}$ \cellcolor{Lavender} & $1.3 \times 10^{7}$ \cellcolor{YellowGreen} & $7.1 \times 10^{6}$ \cellcolor{YellowGreen} & $3.1 \times 10^{6}$ \cellcolor{YellowGreen} & $1.5 \times 10^{6}$ \cellcolor{YellowGreen}\\
     & (11,17,35) \cellcolor{Lavender} & (11,23) \cellcolor{YellowGreen} & (9,17) \cellcolor{YellowGreen} & (7,13) \cellcolor{YellowGreen} & (5,11) \cellcolor{YellowGreen} \\
     \multirow{2}{*}{-15} & $3.4 \times 10^{8}$ \cellcolor{Lavender} & $2.1 \times 10^{7}$ \cellcolor{YellowGreen} & $7.7 \times 10^{6}$ \cellcolor{YellowGreen} & $3.4 \times 10^{6}$ \cellcolor{YellowGreen} & $1.9 \times 10^{6}$ \cellcolor{YellowGreen}\\
     & (11,19,37) \cellcolor{Lavender} & (13,25) \cellcolor{YellowGreen} & (9,19) \cellcolor{YellowGreen} & (7,15) \cellcolor{YellowGreen} & (5,13) \cellcolor{YellowGreen} \\ \bottomrule \\
     & Key: & Level-1 BK\cellcolor{Cerulean} & Level-2 BK\cellcolor{YellowGreen} & Level-3 BK\cellcolor{Lavender} \\
  \end{tabular}
  \caption[Surface Code Resources for Bravyi-Kitaev Distillation]{Resources for Bravyi-Kitaev magic-state distillation as a function of gate error $p_g$ and output state error $p_{\mathrm{out}}$.  Volume is given in unit cells of the surface code.  Beneath volume, the code distance for each round is given.  The background color for each protocol provides a visual guide to the number of rounds.  Each reported protocol has the lowest resource cost for given $p_g$ while producing an output state with error below $p_{\mathrm{out}}$.  Injected magic states have error $p_{\mathrm{in}} = 10 p_g$.  Empty cells do not require distillation because $p_{\mathrm{in}}$ is at or below the target $p_{\mathrm{out}}$.}
  \label{BK_resources_table}
\end{table}

\begin{figure}
  \centering
  \includegraphics[width=\textwidth]{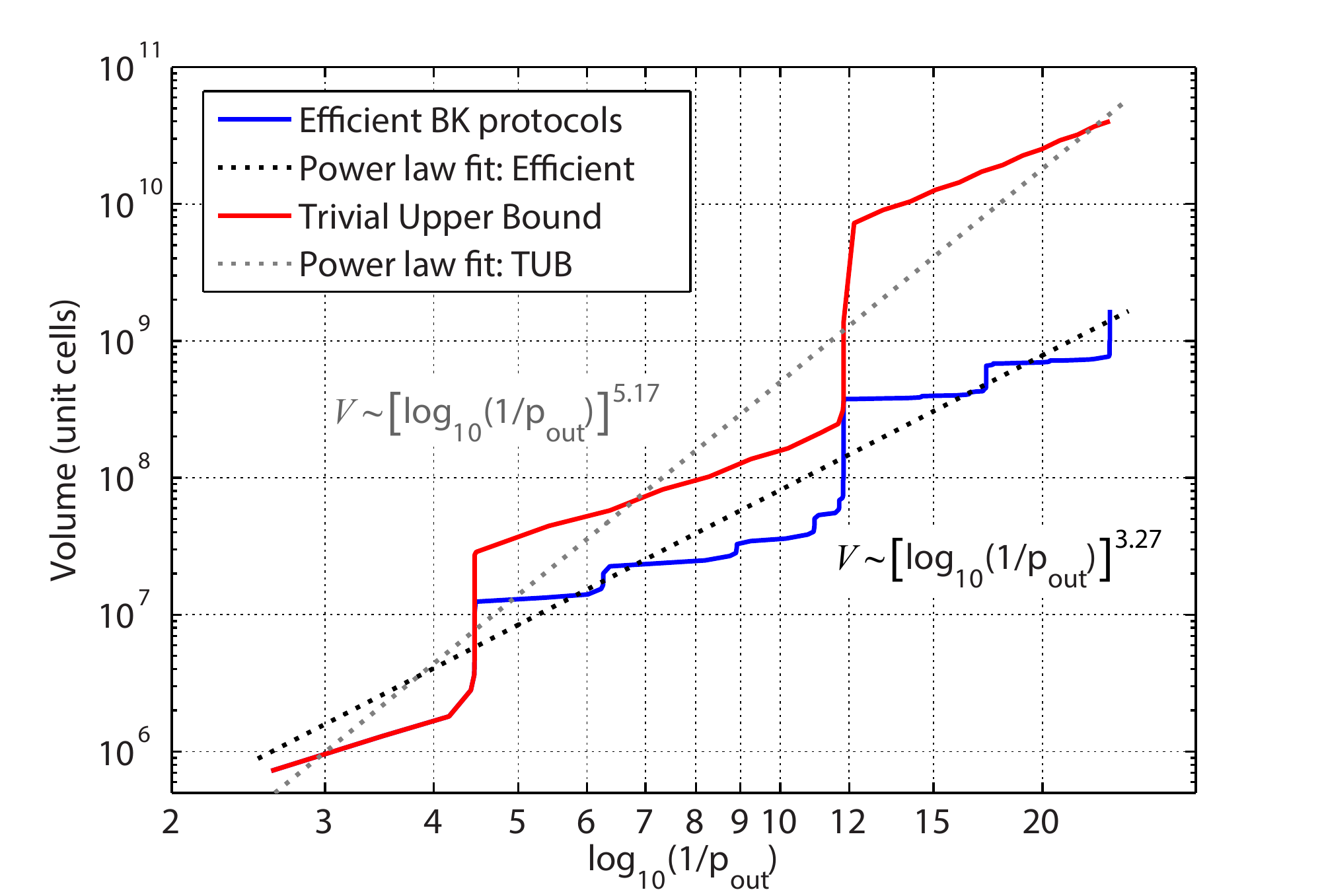}\\
  \caption[Scaling of efficient and naive programs for BK distillation]{Scaling of efficient and naive programs for BK distillation. The lower blue trace is the efficient frontier for multi-round BK distillation with code distance varying by round.  The black dashed line is a power law fit with exponent 3.27.  The red trace is the trivial upper bound, \emph{i.e.} the naive approach with all rounds of distillation using the same code distance; red overlaps blue for just one round of distillation.  The gray dashed line shows a power law fit with exponent 5.17.  For both solid traces, the abrupt jumps in volume occur where the distillation procedure must use an additional round.}
  \label{BK_resource_scaling}
\end{figure}

In the numerical optimization above, I allowed the distance in each round of distillation to be independent of the other rounds, as suggested in Ref.~\cite{Fowler2012_Architecture}.  The protocols on the efficient frontier approximately double the code distance from one round to the next.  This is an optimization that lowers the burden of error correction; with increasing strength of error correction, the probability of surface code failure becomes negligible compared to the distillation error term $35{p_{\mathrm{in}}}^3$.  To see what difference is made by this optimization, consider what happens when the distance is the same in all rounds.  The code distance will need to be large enough for total probability of logical error in the surface code to be well below the target error probability, which makes the entire distillation procedure very expensive.  Figure~\ref{BK_resource_scaling} shows the efficient frontier from Fig.~\ref{BK_protocols_resources} compared to this naive approach.

As mentioned in the introduction, I will use a Trivial Upper Bound (TUB) for each program to demonstrate the importance of logic synthesis.  In Fig.~\ref{BK_resource_scaling}, the naive approach will be the TUB for magic-state distillation.  The power law fits show how the volume required by each method scales as demands on output error are increased.  The optimized logic requires a surface code volume of approximately $(4.37 \times 10^4) \left[\log_{10}(1/p_{\mathrm{out}})\right]^{3.27}$ unit cells, while the TUB requires $(3.39 \times 10^3) \left[\log_{10}(1/p_{\mathrm{out}})\right]^{5.17}$.  As demonstrated in Section~\ref{Sec_Resource_Methodology}, error per plumbing piece in the surface code scales with exponent 2.84, and the asymptotic scaling exponent of BK distillation is $\gamma = \log_3(15) \approx 2.46$.  A simple guess would suppose that the naive protocol would have scaling coefficient $2.84 + 2.46 = 5.30$, which is rather close to the fitting parameter calculated above.

The optimized protocol has scaling exponent 3.27, which is surprisingly close to the lower bound of 2.84 for a surface code operation.  Compressing early rounds of distillation by lowering code distance works very well to improve performance over the TUB.  For example, at output error $10^{-12}$, the optimized protocols reduce resources by about a factor of 10.  The resource estimates in Ref.~\cite{Jones2012_PRX} and Chapter~\ref{Ch02} used the naive method; moreover, $T$-gate production was found to dominate those resource totals.  Using the optimized BK distillation protocols examined here would dramatically lower the estimated costs to execute those quantum algorithms.  This is the first example of how logic synthesis can reduce the overhead of fault tolerance by a sizable factor relative to naive constructions, but this chapter and those that follow will examine several more.

\subsection{Alternative Protocols}
Many other magic-state distillation protocols have been developed but are not analyzed here.  Notable examples are:
\begin{itemize}
\item the 7-to-1 protocol for distilling $\ket{Y}$ with output error $O(p^3)$~\cite{Raussendorf2007_PRL,Raussendorf2007_NJP,Fowler2009_PRA,Fowler2012_Bridge};
\item a 2-to-1, $O(p^2)$ protocol for $\ket{Y}$~\cite{Aliferis2007_Thesis};
\item the Meier-Eastin-Knill 10-to-2, $O(p^2)$ protocol for distilling $\ket{H} = \cos(\pi/8)\ket{0} + \sin(\pi/8)\ket{1}$ (which can be converted to a protocol for distilling $\ket{A}$)~\cite{Meier2012};
\item Bravyi-Haah triorthogonal codes distilling $(3k+8)$-to-$k$ with output error $O(p^2)$~\cite{Bravyi2012};
\item Landahl-Cesare generalized Reed-Muller codes distilling the family of states $(1/\sqrt{2})(\ket{0} + \exp(i \pi/2^n)\ket{1})$ for integers $n \ge 3$~\cite{Landahl2013};
\item a distillation protocol for Toffoli magic states~\cite{Aliferis2007_Thesis}
\item a block-code distillation protocol for distilling controlled-controlled-$Z$ (locally equivalent to Toffoli)~\cite{Jones2013_Multilevel,Paetznick2013_CCZ}.
\end{itemize}
In addition to analysis in the original proposals, there has been work to understand the resource costs associated with fault tolerance for some of these protocols.  Several researchers have considered the limits on errors in magic states for distillation to succeed~\cite{Reichardt2005,Campbell2010}. Reference~\cite{OConnor2013} considers the effectiveness of magic-state distillation using faulty gates.  An equivalent form of the triorthogonal codes called ``block codes,'' implemented with surface code error correction, was analyzed in Ref.~\cite{Fowler2013_BlockCodes}.  Finally, it is worth mentioning that a handful of codes, such as the Steane code and some topological color codes, use a fault-tolerant state injection method rather than magic-state distillation.  While this technique is appealing for its simplicity, the surface code still appears to have better performance for the reasons outlined in Section~\ref{Sec_Error_Correction_Overview}.

\section{Multilevel Distillation}
\label{Sec_Multilevel_Distillation}
I developed another distillation protocol for $\ket{H} = \cos(\pi/8)\ket{0} + \sin(\pi/8)\ket{1}$ states (or $\ket{A}$, through slight modification) to probe the limits of magic-state distillation~\cite{Jones2013_Multilevel}.  Let the efficiency of a distillation protocol be measured only in terms of input and output states with $\gamma = \log(k/n)/\log(d)$, where $n$ is the number of input states, $k$ is the number of output states, and the output error is $O(p^d)$.  In their work on triorthogonal codes, Bravyi and Haah conjectured that $\gamma \ge 1$~\cite{Bravyi2012}.  All previous distillation methods obey this limit, but there was no protocol for $\ket{A}$ states that approached $\gamma = 1$.  The multilevel protocols described below come arbitrarily close to $\gamma = 1$ in certain limits, which is interesting theoretically.  However, I ultimately conclude that these methods are probably not useful for quantum computing.  This is an instructive example for logic synthesis because it shows that narrowly improving one aspect of fault-tolerant computation may not be effective at lowering overall resource costs.  Still, the techniques developed below may be useful in other applications.

\subsection{Block Codes with Transversal Hadamard}
As a preliminary step, I define a family of CSS quantum codes that encode $k$ logical qubits, where $k$ is even, using $(k+4)$ physical qubits.  Furthermore, these codes possess a transversal Hadamard operation, so I call them collectively ``$H$~codes'' and denote $H_n$ as the code using $n = k+4$ physical qubits.  Any $H$~code may be defined as follows.  The stabilizer generators are $S^{(1)} = X_1 X_2 X_3 X_4$, $S^{(2)} = Z_1 Z_2 Z_3 Z_4$, $S^{(3)} = X_1 X_2 X_5 X_6 \ldots X_n$, $S^{(4)} = Z_1 Z_2 Z_5 Z_6 \ldots Z_n$, where subscripts index over physical qubits and tensor product between Pauli operators is implicit.  The logical Pauli operators (corresponding to logical qubits), denoted with an over bar and indexed by $i = 1 \ldots k$, are $\overline{X}_i = X_1 X_3 X_{i+4}$ and $\overline{Z}_i = Z_1 Z_3 Z_{i+4}$.  The Hadamard transform exchanges $X$ and $Z$ operators, so application of transversal Hadamard gates at the physical level enacts a transversal Hadamard operation at the logical level, which will be a useful property when I later concatenate these codes.  All $H$~codes have distance two, which means they can detect a single physical Pauli error.  The product of two logical Pauli operators of the same type for two distinct logical qubits has weight two (number of non-identity physical, single-qubit Pauli operators); the product of same-type Pauli operators on all logical qubits is also weight-two at the physical level.  The stabilizers come in matched $X$/$Z$ pairs, so there are no weight-one logical operators.

\begin{figure}
  \centering
  \includegraphics[width=12cm]{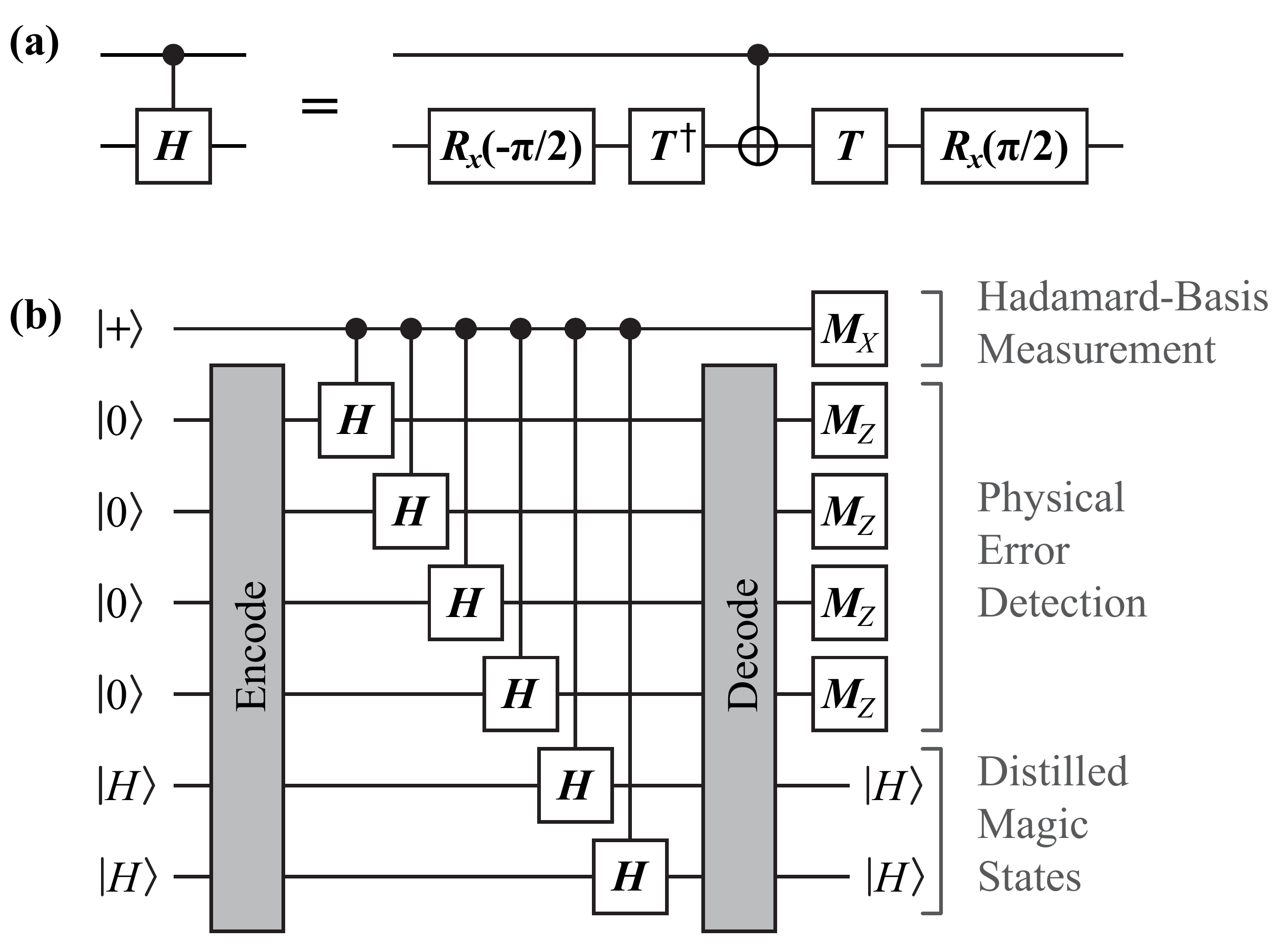}\\
  \caption[Circuit: $H$-code distillation]{Distillation of $\ket{H}$ magic states using an $H$~code.  \textbf{(a)}~Controlled-Hadamard gate is constructed using $T$~gates, each of which requires one $\ket{H}$ state (or $\ket{A}$).  \textbf{(b)}~Initial $\ket{H}$ states (left) are encoded with four additional qubits, initialized to $\ket{0}$ here.  The boxes ``Encode'' and ``Decode'' represent quantum circuits for encoding and decoding, which are not shown here.  Modified from version published in Ref.~\cite{Jones2013_Multilevel}.}
  \label{Had_distill_figure}
\end{figure}

The $(+1)$~eigenstate $\ket{H}$ of the Hadamard operator $H = (1/\sqrt{2})(X + Z)$ is a magic state for universal quantum computing~\cite{Knill2004,Bravyi2005,Reichardt2005,Meier2012,Bravyi2012}.  In particular, two of these magic states can be consumed to implement a controlled-$H$ operation~\cite{Knill2004,Meier2012}, enabling one to measure in the basis of $H$ (see Fig.~\ref{Had_distill_figure}(a)).  The distillation procedure is as follows: (a)~encode faulty $\ket{H}$ magic states in an $H$~code; (b)~measure in the basis of the transversal Hadamard gate by consuming $\ket{H}$ ancillas; (c)~reject the output states if either the measure-Hadamard or code-stabilizer circuits detect an error.  For example, when an $H_{(k+4)}$~code is used for distillation, $k$ $\ket{H}$ states are encoded as logical qubits using $(k+4)$ physical qubits.  Each transversal controlled-Hadamard gate consumes two $\ket{H}$ states~\cite{Meier2012}, and this gate is applied to all physical qubits, which results in the $(3k+8)$-to-$k$ input/output distillation efficiency of these codes.  A diagram of the quantum circuit for distillation using $H_6$ is shown in Fig.~\ref{Had_distill_figure}(b).

\subsection{Multilevel Protocols}
\label{Sec_Multilevel_Protocols}
Multilevel distillation uses concatenated codes with transversal Hadamard for distillation, in such a manner that the protocol uses two classes of input magic states, where the classes have different levels of infidelity and enter at different concatenation levels in the code.  The $\ket{H}$ ancillas consumed for transveral controlled-Hadamard measurement are of lower fidelity than the encoded logical $\ket{H}$ states being distilled.  When two quantum codes with transversal Hadamard are concatenated, the resulting code also has transversal Hadamard.  Under appropriate conditions, the distance of the concatenated code is the product of the distances for the individual codes: $d' = d_1 d_2$~\cite{Meier2012}.  Thus the concatenation of two $H$~codes yields a distance-4 code with transversal Hadamard, and $r$-level concatenation has distance $2^r$.

\begin{figure}
  \centering
  \includegraphics[width=12cm]{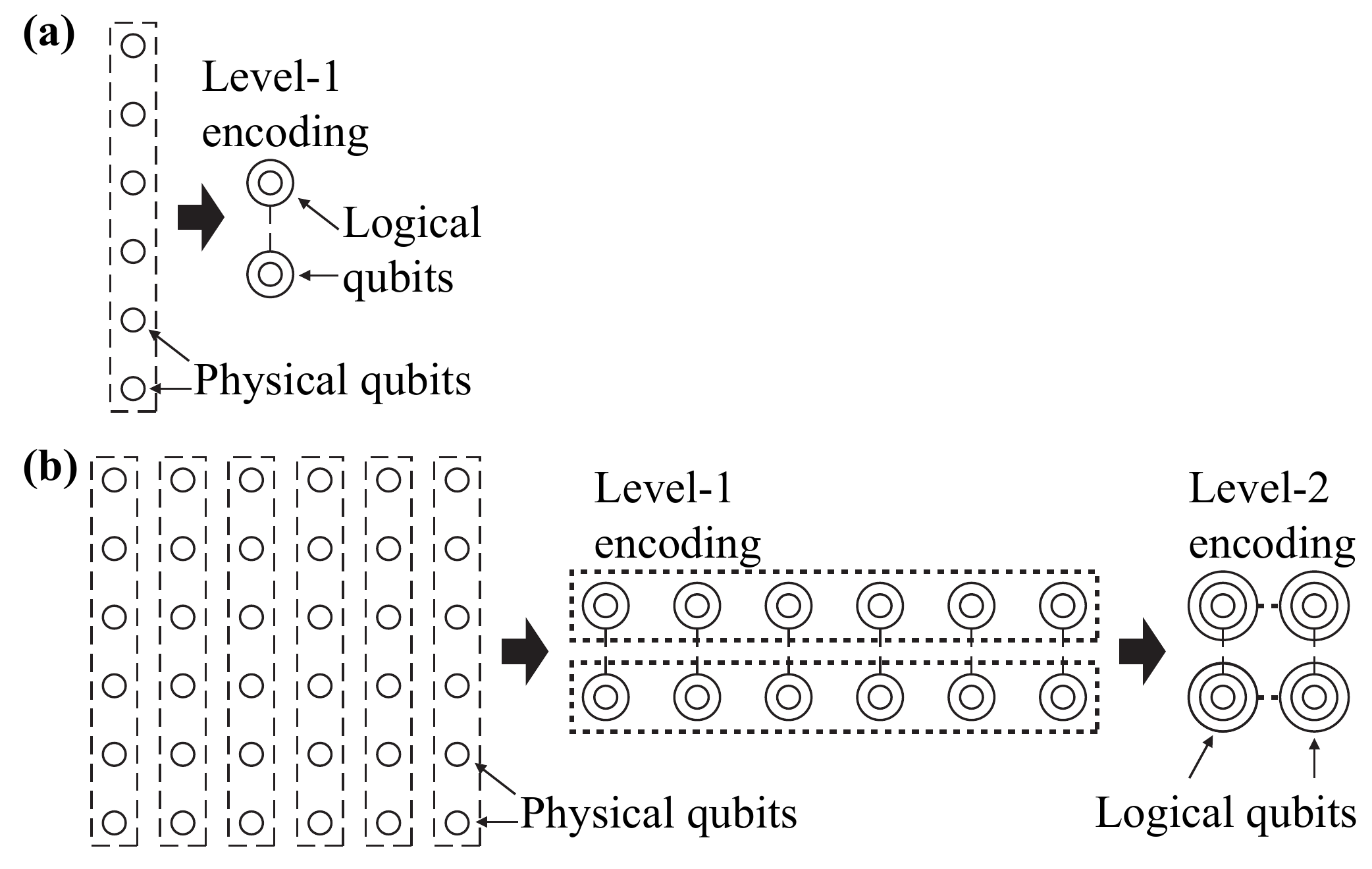}\\
  \caption[Illustration of $H$-code concatenation]{Concatenation of $H$~codes.  \textbf{(a)} Six physical qubits are coupled into an $H_6$~code with two logical qubits  \textbf{(b)} A $6 \times 6$ array of physical qubits are coupled into a concatenated two-level $H_6$~code.  Originally published in Ref.~\cite{Jones2013_Multilevel}.  {\copyright}2013 American Physical Society.}
  \label{H_code_concatenation}
\end{figure}

The concatenation conditions for $H$~codes are that, through all levels of concatenation, any pair of physical qubits have at most one encoding block (at any level) in common.  The reasons for this restriction are that logical errors in the same block are correlated and that the statement above regarding distance multiplying through concatenation assumes independence of errors.  Consequently, two logical qubits from the same encoding block can never be paired again in a different encoding block.  The required arrangement of qubits can be given a geometric interpretation.  Arrange all physical qubits at points on a Cartesian grid in the shape of a rectangular solid, with the number of dimensions given by the number of levels of concatenation.  A square, cube, or hypercube are possible examples at dimensionality two, three, or four.  Each dimension is associated with a level of concatenation, and there must be an even $n \ge 6$ qubits in each dimension to form an $H$~code.  Construct $H$~codes in the first dimension by forming an encoding block with each line of qubits in this direction, as in Fig.~\ref{H_code_concatenation}(a).  This will give rise to $k = n-4$ logical qubits along each line in this direction.  Repeat this procedure by grouping these first-level logical qubits in lines along the second dimension to form logical qubits in a two-level concatenated code, as in Fig.~\ref{H_code_concatenation}(b).  Continuing in this fashion through all dimensions ensures that any pair of qubits have at most one encoding block in common.

As with the $H$~codes, multilevel codes use a transversal logical Hadamard-basis measurement to detect whether any one encoded qubit has an error (an even number of encoded errors would not be detected).  If the logical $\ket{H}$ states have independent error probabilities $\epsilon_l$, then the distilled states will have infidelity $O({\epsilon_l}^2)$ with perfect distillation.  One must also consider whether the Hadamard-basis measurement has an error.  For a two-level code arranged as a square of side length $n$, the transversal controlled-Hadamard gates at the lowest physical level require $(2 n^2)$ $\ket{H}$ magic states, each of which has infidelity $\epsilon_p$.  However, this is a distance-4 code, so for independent input error rates, the probability of failing to detect errors at the physical level is $O({\epsilon_p}^4) + O(\epsilon_l{\epsilon_p}^2)$ (analysis is provided in Section~\ref{Sec_Multilevel_Error}).  The code can detect more errors in the magic states at the lower physical level, so these $\ket{H}$ states can be of lower fidelity than the magic states encoded as logical qubits and successfully perform distillation.  This is the essential distinction between multilevel distillation and all prior distillation protocols.  When multiple rounds of distillation are required~\cite{Jones2012_PRX,Fowler2012_Architecture}, low-fidelity magic states are less expensive to produce, so multilevel protocols achieve higher efficiency.

Multilevel distillation protocols are applied in rounds, beginning with a small protocol (such as an $H$~code) and progressing to concatenated multilevel codes.  Let us denote the output infidelity from a single round by the function $\epsilon_{\mathrm{out}} = E_t^{n_1 \times \ldots n_t}(\epsilon_l,\epsilon_p)$.  For each such function, $t$ is the dimensionality (number of levels of concatenation) and $n_1 \ldots n_t$ are the sizes of each dimension, which need not all be the same.  As before, $\epsilon_l$ and $\epsilon_p$ refer to the independent error probabilities on logical and physical magic states, respectively.  A typical progression of rounds using a source of $\ket{H}$ states with infidelity $\epsilon_0$ might be $\epsilon_1 = E_1^n(\epsilon_0,\epsilon_0)$, $\epsilon_2 = E_2^{n \times n}(\epsilon_1,\epsilon_0)$, \emph{etc}.

Multilevel protocols tend to be much larger in both qubits and gates than other protocols.  Because there can be many encoded qubits, the protocol is still very efficient, but the size of the overall circuit may be a concern for some quantum computing architectures.  At any number of levels, the distilled output states have correlated errors, so distilled magic-state qubits in multilevel distillation must never meet again in a subsequent distillation circuit (it is a requirement that errors are independent within the same encoding block, as in Refs.~\cite{Meier2012,Bravyi2012}).  Let us suppose that one performs two rounds of distillation, where the first round uses one-level distillers with $k$ encoded magic states and the second round uses two-level distillers with $k^2$ encoded states.  Because the inputs to each distiller in the second round must have independent errors, there must be $k^2$ independent distillation blocks in the first round.  Therefore, to distill $k^3$ output states through two rounds, the number of input states is
\begin{eqnarray}
N_{\mathrm{inputs}} & = & k^3 [\textrm{logical inputs}] + 2k^2(k+4) [\textrm{physical inputs}] \nonumber \\
& & + 2k(k+4)^2 [\textrm{physical inputs}] \nonumber \\
& = & 5k^3 + 24k^2 + 32k.
\label{inputs_two_rounds}
\end{eqnarray}

Consider a similar sequence through $r$ rounds with each distiller in round $q$ having $k^q$ encoded qubits.  The total number of \emph{logical} magic states is $k^r \times k^{r-1} \times \ldots k = k^{r(r+1)/2}$ to ensure that errors are independent between logical magic states in every round.  In the first round, the number of consumed magic states is $2(k+4)k^{r(r+1)/2 - 1}$; in any subsequent round $q \ge 2$, the number of consumed magic states is $2^{q-1} (k+4)^q k^{r(r+1)/2 - q}$ (recall that the Hadamard measurement is implemented $2^{q-2}$ times, meaning it is repeated for $q \ge 3$).  The total number of input magic states can thus be expressed as
\begin{equation}
N_{\mathrm{inputs}} = \left[1 + \frac{k+4}{k} + \sum_{q=1}^r 2^{q-1}\left(\frac{k+4}{k}\right)^q\right] k^{r(r+1)/2}.
\label{Eqn_Num_Input_States}
\end{equation}
For $r = 2$, this reproduces Eqn.~(\ref{inputs_two_rounds}).  What also becomes clear is that the total size of multilevel protocols becomes unwieldy as $r$ and $k$ increase.  For example, the case of $r=3$ and $k=10$ would require about $1.87 \times 10^7$ input magic states and a comparable number of gates to distill $10^6$ output magic states.  In general, the most efficient multilevel distillation protocols use large $k$ and multiple rounds, where efficiency is measured in the ratio of low-fidelity $\ket{H}$ input states consumed to yield a single high-fidelity $\ket{H}$ output.  Because of the complexity of such protocols, the greatest benefit from their application is seen in large-scale quantum computing, where a typical algorithm run may require $10^{12}$ magic states, each with error probability $10^{-12}$~\cite{Jones2012_PRX}.  It may be possible for alternative designs to circumvent these issues.  If the first round uses a different protocol without correlated errors across logical magic states, such as Bravyi-Kitaev 15-to-1 distillation, then having multiple distillation blocks is unnecessary in the second round using a two-level concatenated protocol, which would lead to smaller multi-round, multilevel protocols.  Indeed, the results below show that optimal protocols found by numerical search happen to take this approach.

The scaling exponent $\gamma$ of a distillation protocol characterizes its efficiency.  Specifically, $O(\log^{\gamma}(\epsilon_{\mathrm{in}}/\epsilon_{\mathrm{out}}))$ input states are required to distill one magic state of infidelity $\epsilon_{\mathrm{out}}$.  Scaling exponents for previous protocols are $\gamma \approx 2.46$ (``15-to-1''~\cite{Knill2004,Bravyi2005}), $\gamma \approx 2.32$ (``10-to-2''~\cite{Meier2012}), and $\gamma \approx 1.6$ (triorthogonal codes~\cite{Bravyi2012}).  Moreover, Bravyi and Haah conjecture that no magic-state distillation protocol has $\gamma < 1$~\cite{Bravyi2012}.  In this work, if each round of distillation uses one higher level of concatenation in the multilevel protocols, then the number of consumed inputs doubles.  In the limits of $k \rightarrow \infty$, $\epsilon \rightarrow 0$, multilevel protocols require $2^r + 1$ input states to each output state for $r$ rounds of distillation, where the $r^{\mathrm{th}}$ round is a level-$r$ distiller.  The final infidelity is $O((\epsilon_{\mathrm{in}})^{2^r})$, so the scaling exponent is $\gamma = \log(2^r + 1)/\log(2^r) \rightarrow 1$ as $r \rightarrow \infty$, which is the closest any protocol has come to reaching the conjectured bound.  I show later through numerical simulation that $\gamma \approx 1$ for error rates relevant to quantum computing.

\subsection{Error Analysis for Multilevel Distillation}
\label{Sec_Multilevel_Error}
For simplicity, make the conventional assumption that all quantum circuit operations are perfect, except for the initial $\ket{H}$ magic states being distilled.  This is a valid approximation if all operations are performed using fault-tolerant quantum error correction where the logical gate error is far below the final infidelity for distilled magic states~\cite{Raussendorf2007_NJP,Jones2012_PRX,Fowler2012_Architecture}; for a more explicit construction of fault-tolerant distillation circuits, see Ref.~\cite{Fowler2012_Bridge}.  Additionally, following the methodology in Refs.~\cite{Bravyi2005,Meier2012}, one can consider each magic state with infidelity $\epsilon$ as the mixed state $\rho = (1-\epsilon)\ket{H}\bra{H} + \epsilon\ket{-H}\bra{-H}$, where $\ket{-H}$ is the $(-1)$ eigenstate of the Hadamard operation.

Determining the infidelity at the output of distillation becomes simply a matter of counting the distinct ways that errors lead to the circuit incorrectly accepting faulty states, which is aided by the geometric picture from Section~\ref{Sec_Multilevel_Protocols}.  It is essential that error probabilities $\epsilon_l$ and $\epsilon_p$ for each input magic state are independent.  Then a one-level, $(3k+8)$-to-$k$ distiller using the $H_{(k+4)}$ code has output error rate on each $\ket{H}$ state as
\begin{equation}
E_1^{(k+4)}(\epsilon_l,\epsilon_p) = (k-1){\epsilon_l}^2 + (2k+2){\epsilon_p}^2 + (\ldots),
\label{Eqn_Multilevel_One}
\end{equation}
where higher order terms denoted $(\ldots)$ are omitted.  The numerical results justify the use of lowest-order approximations as higher-order terms are negligible in optimally efficient protocols.  The lowest-order error rates are both second order, because the Hadamard basis measurement and $H_{(k+4)}$ code can together detect a single error in any magic state.  The probability of the distiller detecting an error, in which case the output is discarded, is $k\epsilon_l + 2(k+4)\epsilon_p + (\ldots)$.  If $\epsilon_l = \epsilon_p = \epsilon$, then the output error rate of $(3k+1)\epsilon^2$ conditioned on success is the same as in Ref.~\cite{Bravyi2012}.  Using the two-level distiller constructed from concatenated $H_{(k+4)}$ codes, the output infidelity for each distilled $\ket{H}$ state is
\begin{eqnarray}
& E_2^{(k+4)\times(k+4)}(\epsilon_l,\epsilon_p) = & (k^2 - 1){\epsilon_l}^2 + 8(k^2 + 4k + 3){\epsilon_p}^4 \nonumber \\
& & + (k+4)^2\epsilon_l {\epsilon_p}^2 + (\ldots).
\label{Eqn_Multilevel_Two}
\end{eqnarray}
The probability of the two-level distiller detecting an error is
\begin{equation}
p_{\mathrm{error}} = k^2\epsilon_l + 2(k+4)^2\epsilon_p + 2k^2(k+4)^2 \epsilon_l\epsilon_p + (\ldots).
\end{equation}
Similar error suppression extends to higher multilevel protocols, as examined in more detail below.

The multilevel codes analyzed here use concatenated $H$~codes, though other codes could be concatenated.  When two $H$~codes are concatenated, the logical qubits of the first level of encoding are used as physical qubits for completely distinct codes at the second level.  Consider a two-level scheme: if the codes at first and second levels are $[[n_1,(n_1-4),2]]$ and $[[n_2,(n_2-4),2]]$, respectively, then the concatenated code is $[[n_1 n_2, (n_1-4)(n_2-4),4]]$, as shown in Fig.~\ref{H_code_concatenation}(b). This process can be extended to higher levels of concatenation.

Determining the potential errors and their likelihood in multilevel protocols requires careful analysis.  Let us enumerate the error configurations which are detected by the protocols; the error probability is given by summing the probability of all error configurations that are not detected and that lead to error(s) in the encoded $\ket{H}$ states.  As a first step, the analysis of multilevel codes is simplified by considering each input magic state to the quantum computer as having an independent probability of $Y$ error, as discussed in Refs.~\cite{Bravyi2005,Meier2012}.  Hence only one type of error stems from each magic state used in the protocol.

Identifying undetected error events in multilevel distillation, which lead to output error rate, is aided by the geometric picture introduced before.  Qubits which will form the code are arranged in a rectangular solid, then grouped in lines along each dimension for encoding.  There are two error-detecting steps which together implement distillation: the Hadamard-basis measurement and the error detection of the $H$~codes.  The Hadamard measurement registers an error for odd parity in the total of encoded state errors and physical-level errors in the first round of $T$~gates, and there is one of these for each qubit site in the code (see Fig.~\ref{Had_distill_figure}).

The second method for $H$~codes to detect errors is by measuring the code stabilizers.  The code stabilizers detect any configuration of errors which is not a logical operator in the concatenated code.  Because of the redundant structure using overlapping $H$~codes, only a very small fraction of error configurations evade detection.  Before moving on, note that at each qubit site, there are two faulty gates applied, and two errors on the same qubit will cancel (however, the first error will propagate to the Hadamard-basis measurement).  Conversely, a single error in one of the two gates will propagate to the stabilizer-measurement round, but only an error in the first gate will also propagate to the Hadamard measurement.  The stabilizer-measurement round will only ``see'' the odd/even parity of the number of errors at each qubit site.

One type of error event that occurs at concatenation levels three and higher requires special treatment.  If there is an error in an encoded magic state and errors on two physical states used for the same controlled-Hadamard gate at the physical level, then this combination of input errors is not detected by the distillation protocol, leading to logical output error.  This event leads to the $O(\epsilon_l{\epsilon_p}^2)$ error probability mentioned previously, which is not an issue for two-level protocols, but it must be addressed in levels three and higher.  The solution for $t$-level distillation, where $t \ge 3$, is to repeat the controlled-Hadamard measurement $2^{(t-2)}$ times, consuming $2^{(t-1)}$ magic states at the physical level.  After each transversal controlled-Hadamard, the code syndrome checks for detectable error patterns.  With this procedure, one encoded-state error would also require at least $2^{(t-1)}$ errors in physical-level magic states to go undetected, leading to probability of error that scales as $O(\epsilon_l{\epsilon_p}^{2^{t-1}})$.

Consider the pattern of errors after the two potentially faulty gates on each qubit in the $t$-dimensional Cartesian grid arrangement.  The many levels of error checking in the $H$ codes can detect a single error in any encoding block at any encoding level.  For this analysis, let us separate the $(k+4)$ qubits in a single $H$~code block into two groups: the first four qubits are ``preamble'' qubits, while the remaining $k$ qubits are index qubits.  The reason for this distinction is that the logical $\overline{Y}_i$ operators, which would also be undetected error configurations, have common physical-qubit operators in the preamble, with a degeneracy of two: $\overline{Y}_i = -Y_1 Y_3 Y_{i+4} = -Y_2 Y_4 Y_{i+4}$, because of the stabilizer $Y_1 Y_2 Y_3 Y_4$.  Conversely, the logical operators are distinguished by the $i^{\mathrm{th}}$ logical Pauli operator having a physical Pauli operator on the $i^{\mathrm{th}}$ index qubit (numbered $(i+4)$ when preamble is included).

The preamble/index distinction makes it easier to identify the most likely error patterns.  For any size $H$~code, there are two weight-2 errors in the preamble: $Y_1 Y_2$ and $Y_3 Y_4$.  Logically, these represent the product of $\overline{Y}$ operators on all encoded qubits.  In the index qubits, any pair of errors is logical: $Y_{i+4} Y_{j+4} = \overline{Y}_i \overline{Y}_j$.  However, a pair of errors split with one each in preamble and index is always detectable by the code stabilizers.  Thus, any single encoded qubit could have a logical error stemming from a pair of errors in two different configurations in the preamble or $(k-1)$ configurations in the index qubits.  There is also one weight-three error.  Each physical-state error configuration is multiplied by a degeneracy factor that is the number of ways an even number of errors occur before the CNOT in Fig.~\ref{Had_distill_figure}, thereby evading the Hadamard measurement.  Thus the probability of logical error is $2(k+1){\epsilon_p}^2 + 4{\epsilon_p}^3 + O({\epsilon_p}^4)$.  The Hadamard measurement fails to detect an even number of errors in the logical input states.  There are $(k-1)$ ways that a pair of encoded input errors could corrupt any given qubit and $(k-1)(k-2)(k-3)/6$ ways four errors could corrupt any given qubit (assuming $k \ge 4$).  This contributes error terms $(k-1){\epsilon_l}^2 + (1/6)(k-1)(k-2)(k-3){\epsilon_l}^4 + O({\epsilon}^6)$. Finally, it is possible for a single logical error and an odd number of physical errors before the CNOT in Fig.~\ref{Had_distill_figure}, potentially in conjunction with other physical errors after CNOT, to occur simultaneously in a way that evades both checks.  This contributes a term $(k+4)\epsilon_l{\epsilon_p}^2 + 8(k-1)\epsilon_l{\epsilon_p}^3 + O(\epsilon_l{\epsilon_p}^4)$.

The numerical analysis detailed below shows that efficient use of one-level $H$~codes has similar error rates for $\epsilon_l$ and $\epsilon_p$, and both are below 0.01, so the relevant terms in the error functions for one-level $H$~codes are $E_1^{(k+4)}(\epsilon_l,\epsilon_p) = (k-1){\epsilon_l}^2 + (2k+2){\epsilon_p}^2 + (\ldots)$, which reproduces Eqn.~(\ref{Eqn_Multilevel_One}).  As a result, the higher-order terms above can be neglected for this range of parameters so long as $k$ is not too large.  Simply put, if the higher terms become relevant (\emph{i.e.} $\epsilon_l$, $\epsilon_p$, or $k$ is sufficiently large in magnitude), then the distillation protocol is being used ineffectively, and it may in fact cause more errors than it corrects.  These findings are supported by the numerical search for optimal protocols, justifying this approximation.

When $H$~codes are concatenated, the analysis of undetected error patterns becomes more complicated.  In particular, logical errors from one layer of encoding must be ``matched'' with errors from other encoding blocks to go undetected at the next level.  Consider the case of the level-two-concatenated, square-array distiller, and focus on one of the encoded states.  As before, a pair of encoded-state input errors evades the Hadamard measurement, which contributes a term $(k^2 - 1){\epsilon_l}^2$.  The undetected errors resulting from consumed magic states are more complicated.  Within the upper encoding block, there are two ways a logical error could be caused by a pair of errors in the preamble, and $(k-1)$ possibilities for logical error from a pair of index errors.  However, each of the inputs to the second level are the logical qubits of distinct $H$~codes at the first level, which has additional error detection.  The most likely errors from the first level come in pairs, but these pairs are sent to different codes at the second level.  As a result, the error patterns from the first level must come in ``matched'' pairs that are also not detected at the second level.  For any particular error configuration going into a block at the second level, there are four preamble configurations and $(k-1)$ index configurations at the first level that could have caused it.  There are $(k+1)$ undetected error configurations at the second level, and the degeneracy factor of four physical errors is eight, so the consumed magic states contribute a term $8(k+1)(k+3){\epsilon_p}^4$.  Finally, the most likely way that physical and encoded errors can occur in conjunction is a logical error on the magic state in question and two physical errors on the same qubit anywhere, which has probability $(k+4)^2 \epsilon_l {\epsilon_p}^2$.  Combined, these error terms reproduce the results in Eqn.~(\ref{Eqn_Multilevel_Two}).  Higher-order terms can be neglected because they are found to be negligible in optimal protocols.  For example, the first optimal two-level protocol has parameters $k = 8$, $\epsilon_l = 3.5 \times 10^{-5}$, and $\epsilon_p = 9 \times 10^{-4}$, where both input types come from earlier rounds of distillation (Bravyi-Kitaev and Meier-Eastin-Knill, respectively).  More details of the numerical search are given below.

Continuing this approach, one can show the significant error terms at level $t \ge 3$ are given by
\begin{eqnarray}
& E_t^{(k+4)^t}(\epsilon_l,\epsilon_p) = & (k^t - 1){\epsilon_l}^2 + 2^{(2^t + t-3)}(k+1)(k+3)^{t-1} {\epsilon_p}^{(2^t)} \nonumber \\
& & + (k+4)^{t(2^{(t-2)})}\epsilon_l {\epsilon_p}^{(2^{(t-1)})} + (\ldots).
\end{eqnarray}
These terms incorporate degeneracy in error configurations and repeated Hadamard measurements.  The coefficients of the second and third terms on the RHS of Eqn.~(\ref{Eqn_Multilevel_One}) represent  physical error configurations and encoded+physical combinations, respectively, and these grow rapidly as a function of $r$.  Accordingly, the optimal-protocol search does not advocate the use of three-level protocols until the desired output error rate is below $10^{-25}$, which is beyond the needs of any quantum algorithm so far conceived.  No four-level protocols were found to be optimal for output error rates above $10^{-40}$, which under practical considerations means they are not likely to ever be used.

\subsection{Resource Analysis for Multilevel Distillation}
\begin{figure}
  \centering
  \includegraphics[width=\textwidth]{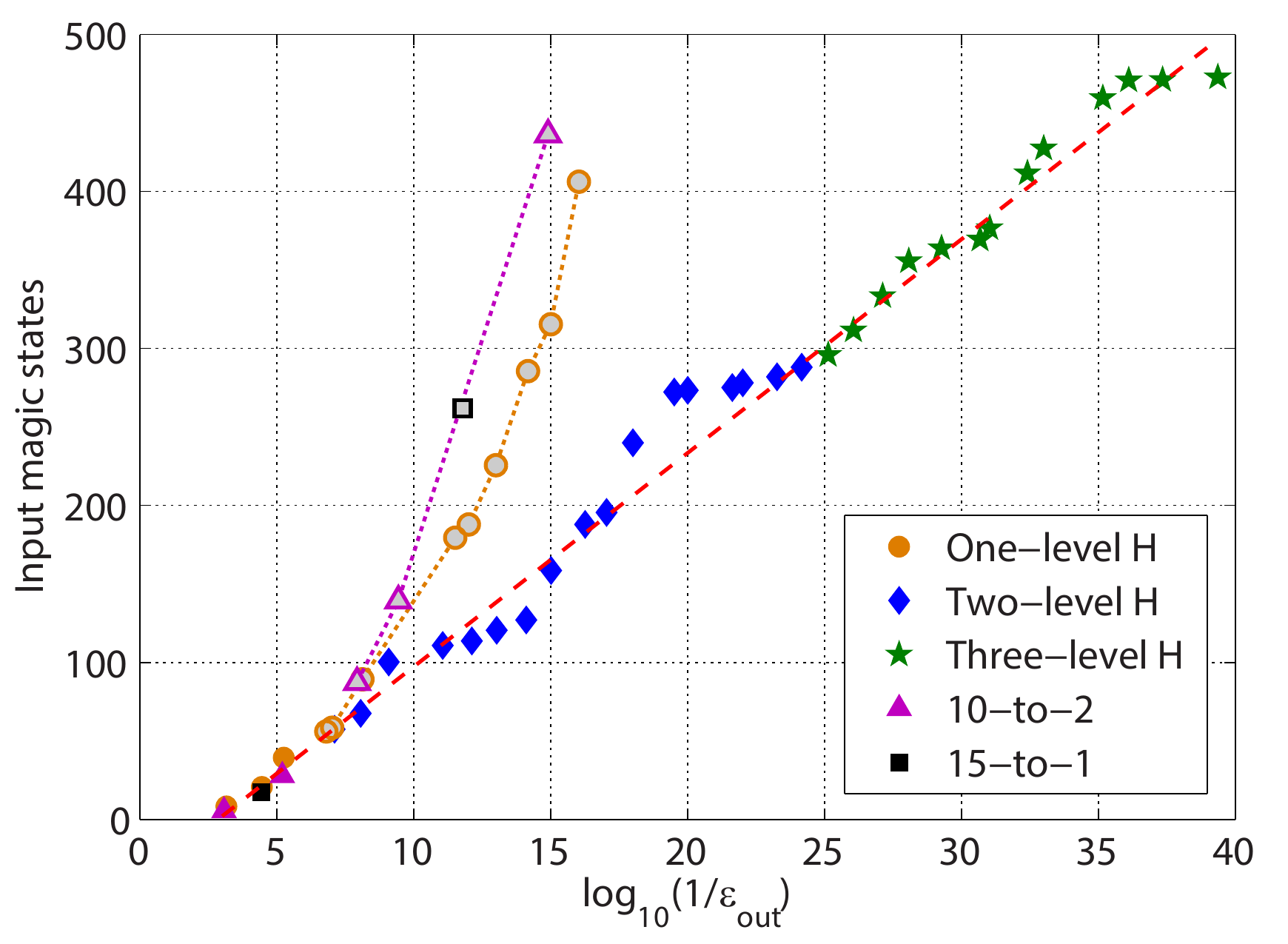}\\
  \caption[Multilevel distillation input states vs. output error]{Average number of input $\ket{H}$ states with $\epsilon_{\mathrm{in}} = 0.01$ consumed to produce a single output $\ket{H}$ state with fidelity $\epsilon_{\mathrm{out}}$.  Multiple-round distillation can use different protocols in each round, and the markers indicate just the last round of distillation.  The grey-shaded squares, triangles, and circles show, respectively, the best distillation possible with only \hbox{15-to-1}~\cite{Bravyi2005}, \hbox{10-to-2}~\cite{Meier2012}, and triorthogonal-code~\cite{Bravyi2012} protocols.  The dashed line is a linear fit $14\log_{10}(1/\epsilon_{\mathrm{out}}) - 40$.  Originally published in Ref.~\cite{Jones2013_Multilevel}.  {\copyright}2013 American Physical Society.}
  \label{Protocols_plot}
\end{figure}

Figure~\ref{Protocols_plot} shows the performance of optimal multi-round distillation protocols identified by numerical search, indicating the number of input states with $\epsilon_0 = 0.01$ required to reach a desired output infidelity $\epsilon_{\mathrm{out}}$.  The markers indicate the type of protocol in the last round of distillation, including Bravyi-Kitaev~\cite{Bravyi2005}, Meier-Eastin-Knill~\cite{Meier2012}, and multilevel $H$~codes.  The search attempts to identify the best distillation routines using any combination of known methods.  Note that the recent Bravyi-Haah protocols~\cite{Bravyi2012} have the same performance as one-level $H$~codes.  As expected, there is a trend of using higher-distance multilevel protocols in the last round as the output error rate $\epsilon_{\mathrm{out}}$ decreases (earlier rounds may use different protocols).  Where present, open markers indicate the best possible performance of previously studied protocols without the advent of multilevel distillation, and multilevel distillation is dominant for $\epsilon_{\mathrm{out}} \le 10^{-7}$, which is the regime pertinent to quantum computing.  Moreover, in this regime, input error rates are sufficiently small that only lowest-order terms in the $E(\cdot)$ output-error functions are significant.  The linear fit provides empirical evidence that the scaling exponent is $\gamma \approx 1$ in this regime, which demonstrates that multilevel protocols are close to the conjectured optimal performance in practice.

A single level of distillation using $H$~codes (or ``block codes'') in the surface code was analyzed by Fowler, Devitt, and myself~\cite{Fowler2013_BlockCodes}.  It should be noted that $H$~codes are essentially equivalent to the triorthogonal codes proposed by Bravyi and Haah~\cite{Bravyi2012}, so the resource costs are probably the same as well.  We found that some resource advantage over Bravyi-Kitaev distillation was possible with careful selection of the code distance.  The block code analyzed therein performs $(3k+8)$-to-$k$ distillation of the magic state $\ket{A} = (\ket{0} + e^{\pi i /4}\ket{1})/\sqrt{2}$, with output error $(3k+1)p^2$ to lowest order.  The number of surface code plumbing pieces was $96k + 216$ for $k$ output states; the average of $96 + 216/k$ can reach as low as half that of the equivalent results for the Bravyi-Kitaev protocol ($V = 192$ plumbing pieces was used in that work).  For the same output error rate, block codes may deliver a volume lower by a factor of 2 or 3.  However, this comes with significant caveats.  As discussed above, correlated errors are a nuisance.  The outputs of one block-code distillation must fan out to different protocols in the next round.  Magic states with correlated errors could not be used in the same error-detecting Toffoli gate, as discussed in Chapter~\ref{Ch06}.  Because it is not simple to handle these complications, block codes are not included in the resource analysis of this work.

Multilevel protocols, which are the concatenation of multiple block codes, have not (to my knowledge) been analyzed in the surface code.  As mentioned above, circuit diagrams for these protocols are difficult to construct, and a manual construction of the surface code braiding topology may not be practical.  Perhaps the development of automated synthesis tools for the surface code could solve this task~\cite{Paetznick2013_Compaction}.  However, there is good reason to believe that multilevel protocols are not very useful anyway.  These protocols only show advantage using the simple metric of counting input states to output states when error rates are very very low.  Furthermore, this metric does not capture the overhead associated with long-range interactions needed for the massive surface code topology.  Multilevel protocols are fascinating in theory because they are the only known form of distillation that comes arbitrarily close to the conjectured bound $\gamma \ge 1$~\cite{Bravyi2012}.  For the more practical matter of developing resource-efficient quantum logic, the Toffoli constructions in Chapter~\ref{Ch06} appear much more promising.

\section{Fourier-State Distillation}
\label{Sec_FS_Distillation}
Distillation protocols need not purify single-qubit states.  Single-qubit magic states like $\ket{A}$ are conceptually simpler to use, but there are many other useful quantum states, such as QLUTs for multi-qubit operations.  One example is the three-qubit QLUT for Toffoli gates.  Aliferis provides a distillation protocol for this state~\cite{Aliferis2007_Thesis}, and Chapter~\ref{Ch06} is devoted to other error-detecting programs to produce this state.  This section considers an even larger multi-qubit register that I call a ``Fourier state'' for its straightforward connection to the quantum Fourier transform~\cite{Nielsen2000}.  Fourier states are useful for producing arbitrary phase rotations, which Chapter~\ref{Ch07} examines in detail.  Moreover, the distillation protocol for Fourier states is intriguing because it invokes multiple levels of logic synthesis, and I discuss the importance of hierarchical logic synthesis in Chapter~\ref{Ch09}.

To give some context, a Fourier state of size $n$ qubits is defined as
\begin{equation}
\ket{\gamma^{(k)}} = \frac{1}{\sqrt{N}}\sum_{y=0}^{N-1} e^{i 2 \pi ky/N}\ket{y},
\label{define_Fourier_state}
\end{equation}
where $N = 2^n$.  Note that this sign convention is opposite of that in Ref.~\cite{Jones2012_NJP}.  These states are eigenstates of the modular addition operator $U_{\oplus 1}\ket{x} = \ket{x + 1 \; (\mathrm{mod} \; N)}$:
\begin{equation}
U_{\oplus 1}\ket{\gamma^{(k)}} = e^{-i 2 \pi k/N}\ket{\gamma^{(k)}}.
\end{equation}
Using this property, a phase-rotation gate can be approximated using a Fourier state and an addition circuit, which is known as phase kickback~\cite{Cleve1998,Cleve2000,Kitaev2002,Jones2012_NJP}.  This method can produce any rotation around the $Z$ axis of the Bloch sphere in units of $\pi/2^{n-1}$ radians, so the precision required by a quantum algorithm determines $n$.  A notable feature of phase kickback is that the register $\ket{\gamma^{(k)}}$ is preserved, which means it can be used repeatedly.  Many implementations of addition circuits are known~\cite{Cuccaro2004,VanMeter2005,Draper2006,Takahashi2010}, but the fault-tolerant preparation of Fourier states has received less attention.  Kitaev \emph{et~al.} propose a scheme based on phase estimation~\cite{Kitaev2002}, but this protocol suffers from two notable disadvantages.  First, the resulting state $\ket{\gamma^{(k)}}$ has random odd $k$.  Second, the protocol requires a Fourier transform, which requires phase rotations; since the purpose of phase kickback is to produce phase rotations, implementing the Fourier transform to produce the Fourier state requires an approximate, iterative procedure.  This paper develops a fault-tolerant distillation protocol for producing the frequently used $\ket{\gamma^{(1)}}$ state with $O(n \log n)$ gates from a finite set.  I also give a related protocol for constructing any $\ket{\gamma^{(k)}}$ using $O(n^2)$ gates.

\begin{figure}
  \centering
  \includegraphics[width=\textwidth]{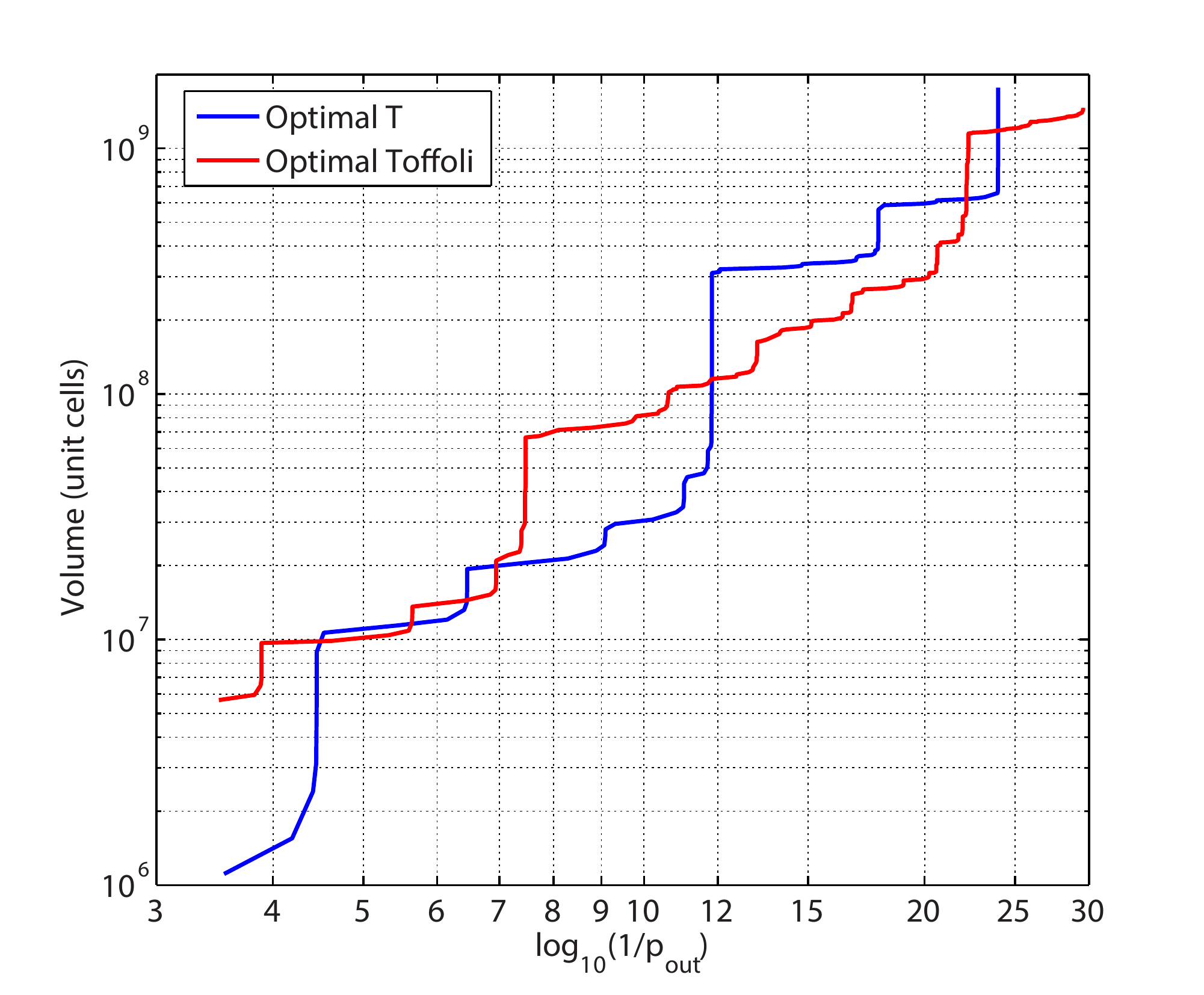}\\
  \caption[Resource comparison for optimal $T$ and Toffoli gates]{Resource costs for one $T$ gate \emph{vs.} one Toffoli gate, as a function of error probability.  The $T$ gate is produced using magic-state distillation, as analyzed in Section~\ref{Sec_MS_Distillation}.  The Toffoli gate resources are the best results from the methods analyzed in Chapter~\ref{Ch06}.}
  \label{T_vs_Toffoli}
\end{figure}

The analysis of resource costs in this thesis and in Refs.~\cite{Fowler2012_Bridge,Fowler2012_Architecture,Jones2013_Toffoli,Fowler2013_BlockCodes} indicates that, when using surface code error correction, producing a Toffoli gate with error around $10^{-8}$ -- $10^{-12}$ requires comparable physical resources to producing a single $T$~gate having the same error probability.  Figure~\ref{T_vs_Toffoli} shows the resource costs, in surface code volume, of one fault-tolerant $T$~gate \emph{vs}. one fault-tolerant Toffoli gate.  For each of the operations, Fig.~\ref{T_vs_Toffoli} shows the efficient frontier for the protocols considered in Section~\ref{Sec_MS_Distillation} and Chapter~\ref{Ch06}.  All of Chapter~\ref{Ch06} is devoted to analyzing Toffoli gates, but for now I use the final results (volume as function of $p_{\mathrm{out}}$) to analyze a distillation protocol for Fourier states.  What is striking is that a Toffoli gate is actually less expensive than a $T$~gate for some values of output error, which motivates the proposal in this section to construct arbitrary rotations from Toffoli gates.  Chapter~\ref{Ch06} will show how produce such efficient Toffoli gates, and Chapter~\ref{Ch07} will use the accumulated results to produce phase rotations.  In Section~\ref{Sec_Fundamental_Analysis}, I return to quantifying the resource costs of Fourier-state distillation.

Although Fig.~\ref{T_vs_Toffoli} suggests that using Toffoli gates could be more resource-efficient than $T$~gates, there are some remaining issues to resolve.  In particular, one must consider how to efficiently compile phase rotations, which appear in many quantum algorithms, using Toffoli gates.  The results of this section complete the phase-kickback protocol~\cite{Cleve1998,Kitaev2002,Jones2012_NJP} mentioned previously by showing that the discrete set of Clifford operations and Toffoli gates can efficiently approximate any phase rotation.  This approach competes with constructions that require fault-tolerant $T$~gates~\cite{Dawson2005,Fowler2011_ArbitraryGates,Amy2012,Kliuchnikov2012}, and I make such a comparison in Chapter~\ref{Ch07}.

The analysis of Fourier-state distillation is organized as follows.  Section~\ref{Sec_Fundamental_Distillation} presents a protocol for distilling the ``fundamental'' Fourier state $\ket{\gamma^{(1)}}$ from approximations produced using only Clifford gates.  Section~\ref{Sec_Fundamental_Analysis} analyzes the resource costs of the distillation protocol, which is summarized here.  The program for constructing an $n$-qubit $\ket{\gamma^{(1)}}$ state requires circuit width $2n + O(1)$ qubits, circuit depth $O(n)$ gates, and $O(n \log n)$ Toffoli gates in total.  Finally, Section~\ref{Sec_Arbitrary_Distillation} outlines a protocol for distilling $\ket{\gamma^{(k)}}$ with arbitrary $k$ using $O(n^2)$ Toffoli gates.

\subsection{Distilling the Fundamental Fourier State}
\label{Sec_Fundamental_Distillation}
The fundamental $\ket{\gamma^{(1)}}$ state of size $n$ qubits is required for phase-kickback constructions (see Chapter~\ref{Ch07}) for both single-qubit phase rotations and two-qubit, controlled phase rotations~\cite{Jones2012_NJP}.  This state is also useful for constructing a quantum Fourier transform (or its approximate version) through a special form of phase kickback called quantum-variable rotation~\cite{Jones2012_NJP}.  In this subsection, I show how to construct $\ket{\gamma^{(1)}}$ using a distillation protocol that uses quantum adders~\cite{Cuccaro2004,VanMeter2005,Draper2006,Takahashi2010}.  Later, I generalize the method to distill $\ket{\gamma^{(k)}}$ for arbitrary $k$, but the special case of $k = 1$ requires fewer quantum gates.

Any pure Fourier-basis state where $N$ in Eqn.~(\ref{define_Fourier_state}) is a power of 2 is separable into individual qubits.  Using a general $Z$-axis rotation $R_z(\phi) = \exp[i \pi \phi (I - Z)/2]$, a Fourier state can be decomposed as
\begin{equation}
\ket{\gamma^{(k)}} = \left[ R_z(k\pi/2^0)\ket{+} \right] \otimes \left[ R_z(k\pi/2^1)\ket{+} \right] \otimes \left[ R_z(k\pi/2^2)\ket{+} \right] \otimes (\ldots).
\label{Fourier_state_decomposition}
\end{equation}
The single-qubit state $\ket{+} = H \ket{0} = (1/\sqrt{2})(\ket{0} + \ket{1})$.  Using Eqn.~(\ref{Fourier_state_decomposition}), one can see that the quantum state
\begin{equation}
\ket{\tilde{\gamma}_0^{(1)}} = Z\ket{+} \otimes S\ket{+} \otimes I\ket{+} \otimes I\ket{+} \otimes (\ldots)
\label{initial_state}
\end{equation}
is an approximation of $\ket{\gamma^{(1)}}$ (denoted with tilde).  Moreover, $\ket{\tilde{\gamma}_0^{(1)}}$ can be produced using only Clifford gates.  The rotations of $\pi/4$, $\pi/8$, \emph{etc.} in Eqn.~(\ref{Fourier_state_decomposition}) that are omitted in Eqn.~(\ref{initial_state}) become exponentially close to identity (in gate fidelity) with increasing qubit index, so they may be approximated with identity gates.  This is the same justification behind neglecting small-angle rotations in the approximate quantum Fourier transform~\cite{Barenco1996}.  The fidelity between the approximate and ideal states is $\left|\left\langle \tilde{\gamma}_0^{(1)} \mid \gamma^{(1)} \right\rangle\right|^2 \ge 0.81$ for all values of $n$, as shown in Section~\ref{Fourier_analysis}.

Each approximate initial state can be expanded in the orthonormal Fourier-state basis as
\begin{equation}
\ket{\tilde{\gamma}_0^{(1)}} = \sum_{j = 0}^{N-1} a_j \ket{\gamma^{(j)}},
\label{Fourier_basis}
\end{equation}
where the dominant term among the complex coefficients is $a_1$, with magnitude $\left|a_1\right|^2 \approx 0.81$ from above.  I ignore complex phase because all of the analysis is performed in the Fourier basis, and the distillation protocol depends only on the magnitudes of the Fourier-basis coefficients.

Using two approximate $\ket{\tilde{\gamma}^{(1)}}$ states (dropping subscript for generality), the distillation protocol is very simple, requiring just two steps:
\begin{enumerate}
\item Add one register to the other.  Binary-encoded, mod-$2^n$ addition given by $$U_{\mathrm{add}}\ket{v}\ket{w} = \ket{v}\ket{w+v \; (\mathrm{mod} \; 2^n)}$$ has been studied extensively~\cite{Cuccaro2004,VanMeter2005,Draper2006,Takahashi2010}.  Notably, many addition circuits use the Toffoli gate as the non-Clifford operation.  In the Fourier basis, the addition circuit implements $$U_{\mathrm{add}}\ket{\gamma^{(k)}}\ket{\gamma^{(k')}} = \ket{\gamma^{(k-k')}}\ket{\gamma^{(k')}}.$$  As an aside, this is precisely phase kickback, where the Fourier index $k'$ of the second register determines the quantum-variable rotation applied to the first register (see Section~\ref{Sec_Phase_Kickback}).
\item Measure the first register in the Fourier basis, and postselect the cases where the result is $\ket{\gamma^{(0)}}$.  The resulting output has each of its Fourier-basis coefficients $\{a_k\}$ weighted by the probability that \emph{both} inputs to distillation were in the $\ket{\gamma^{(k)}}$ state.  If both inputs had sizable overlap with a particular state, then the fidelity conditioned on successful distillation is concentrated to a higher magnitude, and probability of being in unwanted basis states is suppressed.
\end{enumerate}

\begin{figure}
  \centering
  \includegraphics[width=14cm]{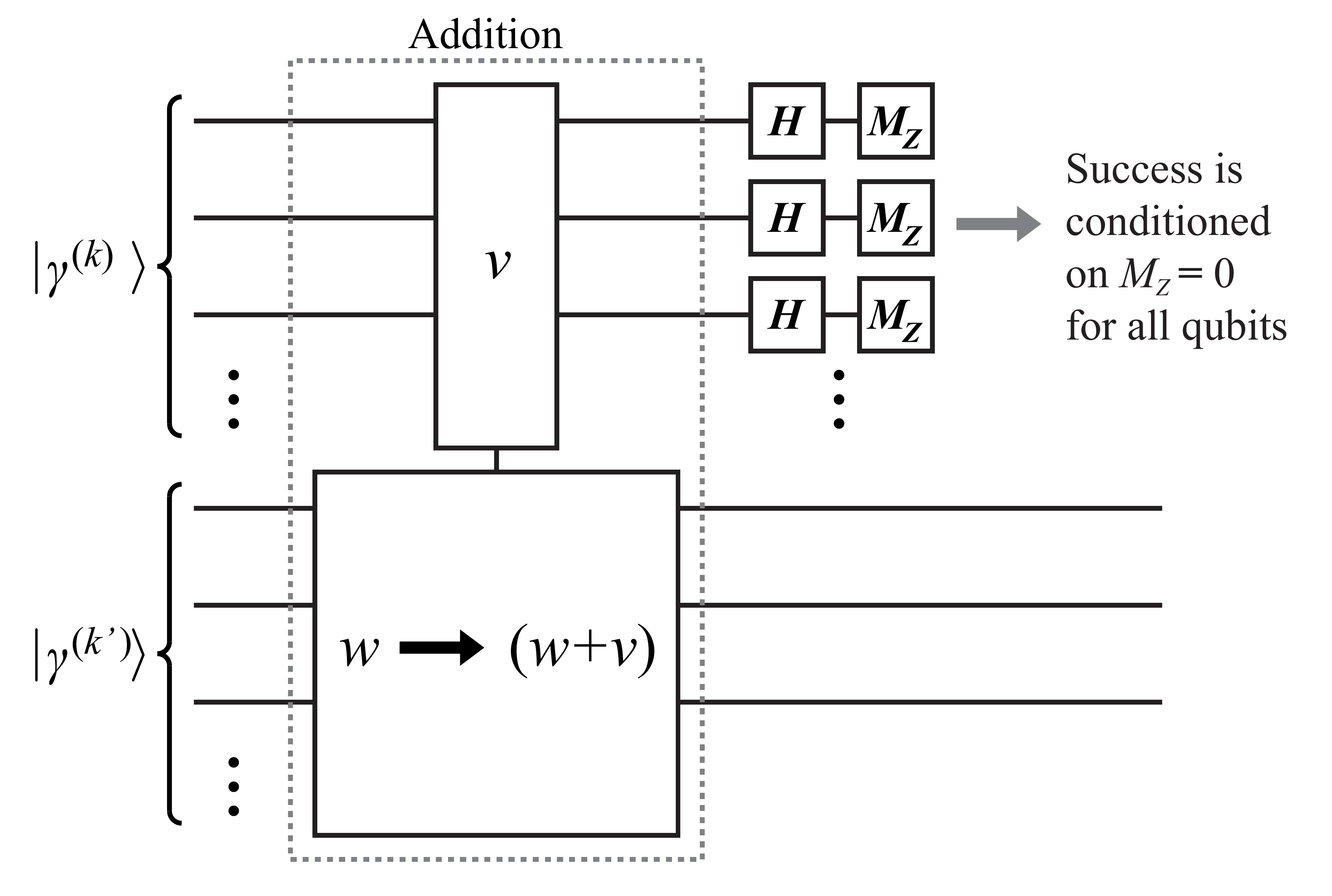}\\
  \caption[Circuit: Distillation of Fourier states]{Circuit for distilling Fourier states.  Each Fourier state is a multi-qubit register, and only three qubits are shown, with the rest indicated by vertical dots.  For clarity, this circuit shows the inputs as Fourier-basis states, but in actual distillation protocols, each input will be a superposition of Fourier-basis states.  The addition circuit shown in the dashed box would be decomposed into Clifford gates and Toffoli gates. The verification step is implemented with Hadamard $H$ and computational-basis measurement $M_Z$ on each qubit in the top register.}
  \label{distillation_circuit}
\end{figure}

The quantitative expressions for distillation success probability and projection of output state into the Fourier basis are simple to derive.  Let the two inputs to distillation have Fourier-basis coefficients $\{a_j\}$ and $\{a_j'\}$ as in Eqn.~(\ref{Fourier_basis}).  The probability of measuring $\ket{\gamma^{(0)}}$ (\emph{i.e.} distillation succeeds) is given by
\begin{equation}
P_{\mathrm{success}} = \sum_{y = 0}^{N-1}\left|a_y\right|^2\left|a_y'\right|^2.
\end{equation}
The output register of distillation will have Fourier coefficients $\{b_j\}$ with magnitudes
\begin{equation}
\left|b_j\right|^2 = \frac{\left|a_j\right|^2\left|a_j'\right|^2}{P_{\mathrm{success}}}.
\end{equation}
These expressions mirror those of entanglement distillation~\cite{Bennett1996,Deutsch1996}.  The fidelity of the output state is $F = \left|b_1\right|^2$, and the error probability in the distilled state is $\epsilon = 1-F$.

A general Fourier-basis measurement would pose a problem because it requires operations outside the Clifford group, but I show how to circumvent this issue with Clifford gates.  Since the quantum computer only supports computational-basis measurements, one would require a quantum Fourier transform (QFT) to map between the bases.  This is essentially the obstacle encountered by the Kitaev-Shen-Vyalyi protocol~\cite{Kitaev2002}, which addresses the matter with an iterative procedure of approximate QFTs.  However, the Fourier distillation protocol does not require a \emph{complete} Fourier-basis measurement; instead, one only needs to know if the first register is in state $\ket{\gamma^{(0)}}$.  This state happens to be the tensor product of $\ket{+}$ states, which are eigenstates in the $X$-basis.  Hence, one only needs the ability measure in the $X$ basis, as shown in Fig.~\ref{distillation_circuit}.  If each qubit is the $\ket{+}$ state, then the register was projected into $\ket{\gamma^{(0)}}$, and distillation succeeds.  Otherwise, reject the output and attempt again.  Since $H$ and $M_Z$ are in the Clifford group, they are considered inexpensive to produce relative to the preceding non-Clifford addition circuit.  In addition to preparing the specific state $\ket{\gamma^{(1)}}$ (as opposed to a random Fourier-basis state~\cite{Kitaev2002,Jones2012_NJP}), this measurement trick is how Fourier distillation improves on the method in Ref.~\cite{Kitaev2002}.

If the two input states are both $\ket{\tilde{\gamma}_0^{(1)}}$ from Eqn.~(\ref{initial_state}), then $\left|a_j\right|^2 = \left|a_j'\right|^2$ for all $j$.  In general, when the inputs satisfy $\left|a_j\right|^2 = \left|a_j'\right|^2$ for all $j$, I say the distillation is ``symmetric.''  The success probability is $P_{\mathrm{success}} = \sum_{y=0}^{N-1}\left|a_y\right|^4 \approx 0.67$, where the coefficients $\{a_j\}$ can be calculated using the method in Section~\ref{Fourier_analysis}.  The Fourier-basis weights of the output state are $\left|b_j\right|^2 = \left|a_j\right|^4/P_{\mathrm{success}}$.  The fidelity $F = \left|b_1\right|^2$ after one round of distillation is upper-bounded by $0.986$, so multiple rounds of distillation are needed to reach arbitrarily high fidelity.  This bounded fidelity also means that early rounds of distillation can use fewer than $n$ qubits to represent the intermediate Fourier states, as explained in Section~\ref{Sec_Fundamental_Analysis}; before going into those details, I  quantify the fidelity in each round.

Define an $n$-qubit, $r$-round distilled $\ket{\tilde{\gamma}_r^{(1)}}$ Fourier state as having ``sufficiently high fidelity'' if its fidelity with the pure Fourier state satisfies
\begin{equation}
1-\left|\langle \tilde{\gamma}_r^{(1)} \mid \gamma^{(1)} \rangle \right|^2 \le \sin^2\left(\pi/2^n\right).
\label{fidelity_threshold}
\end{equation}
Subscript $r$ denotes how many rounds of symmetric distillation have been successfully applied, which is why the initial state $\ket{\tilde{\gamma}_0^{(1)}}$ from Eqn.~(\ref{initial_state}) has subscript 0.  In phase kickback, the constraint in Eqn.~(\ref{fidelity_threshold}) represents the highest accuracy that is needed.  When the $\ket{\tilde{\gamma}_r^{(1)}}$ register is used for phase kickback, there are two sources of error that are considered here.  The first is that the register $\ket{\tilde{\gamma}_r^{(1)}}$ is not pure, meaning it has non-zero overlap with some other Fourier basis state.  The second error source is that any phase rotation is truncated to $n$ bits of precision.  As a result, the truncated angle error is at most $\pi/2^n$ radians, which results in an upper bound on the rotation-gate error probability of $\sin^2\left(\pi/2^n\right) \approx \left(\pi/2^n\right)^2$.  In phase kickback using an $n$-qubit $\ket{\tilde{\gamma}_r^{(k)}}$ state, the combination of the two errors means that any resulting rotations are accurate to $\pm \pi/2^{(n-1)}$ radians, or at least $(n-1)$ bits.  Ultimately, $n$ is chosen based on the gate-accuracy requirements of the quantum algorithm.  I arbitrarily choose to balance the error from a noisy $\ket{\tilde{\gamma}_r^{(1)}}$ with the worst-case truncation-of-angle error.  If $\ket{\tilde{\gamma}_r^{(1)}}$ is used for other applications, such as complex-instruction-set quantum computing~\cite{Landahl2013}, a different accuracy may be required.

\begin{figure}
  \centering
  \includegraphics[width=14cm]{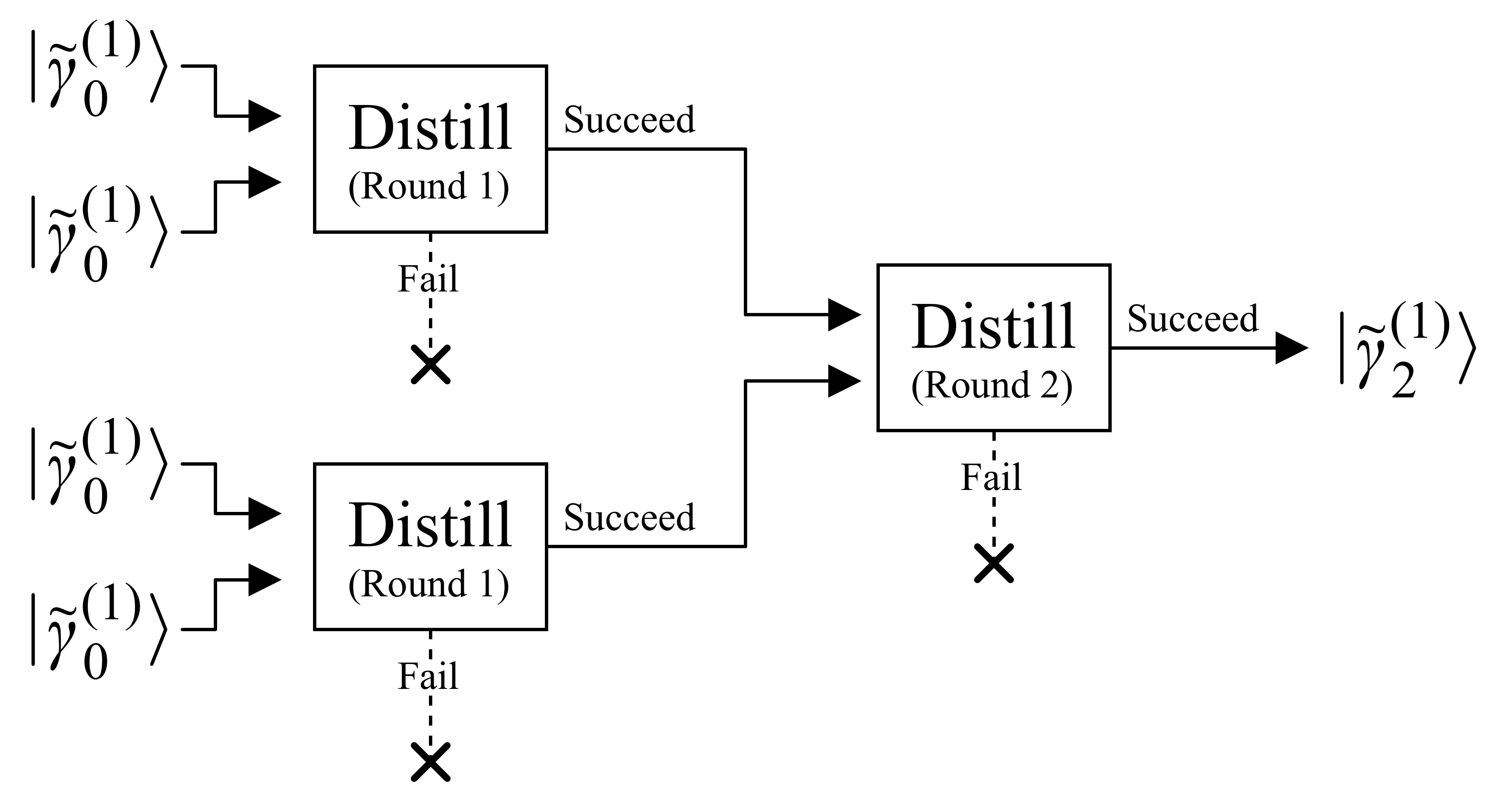}\\
  \caption[Symmetric Fourier distillation rounds]{Symmetric distillation of Fourier states, where two rounds are shown.  The approximate initial states are defined in Eqn.~(\ref{initial_state}).  Each distillation step is implemented using the quantum circuit in Fig.~\ref{distillation_circuit}.  Distillation is probabilistic, and when a distillation circuit fails, that circuit and preceding steps that feed into it must be repeated.  The probability of failure decreases super-exponentially in round number, so the overhead of distillation failure is bounded.}
  \label{distillation_tree}
\end{figure}

The first round of distillation will produce a Fourier state accurate to about 5 bits.  To construct an $n$-qubit Fourier state, one requires a distillation protocol consisting of multiple rounds of symmetric distillation.  The symmetric distillation subroutines are arranged in a binary tree, as shown in Fig.~\ref{distillation_tree}.  Multiple low-fidelity $\ket{\tilde{\gamma}_0^{(1)}}$ input states are distilled to arrive at a single output state, $\ket{\tilde{\gamma}_r^{(1)}}$.  Since subscript $r$ is the number of rounds of symmetric distillation, it is also the depth of this binary tree arrangement.  Choosing $r$ depends on the desired number of precision qubits $n$ in the Fourier state.  In the next section, I show that $r$ scales as $O(\log n)$.

\subsection{Fourier Analysis and Distillation Efficiency}
\label{Fourier_analysis}
The distillation protocol can be understood by viewing probability amplitudes of the input state in the computational basis as discrete samples in time of a function $f(t) = e^{i 2 \pi \nu(t)}$ such that $\ket{\tilde{\gamma}_0^{(1)}} = \sum_{y=0}^{N-1} f(y/N)\ket{y}$.  In this picture, the probability amplitudes of the quantum state in the Fourier basis are related to Fourier-series coefficients $\{c_j\}$ given by
\begin{equation}
f(t) = \sum_{j = -\infty}^{\infty} c_j e^{i 2 \pi j t}.
\end{equation}
The correspondence exists because a quantum Fourier transform maps between computational and Fourier bases.  The number of ``samples'' is $N = 2^n$, the number of computational basis states.  Discrete sampling causes aliasing according to the Shannon-Nyquist Theorem, so Fourier-basis probability amplitudes $\{a_j\}$ are related to the Fourier series of $f(t)$ by
\begin{equation}
a_j = \sum_{x = -\infty}^{\infty} c_{(Nx + j)}.
\label{coefficient_relationship}
\end{equation}
If $N$ is sufficiently large (\emph{e.g.} $n \ge 6$), then $a_j \approx c_j$ for $0 \le j < N/2$ or $a_{(N+j)} \approx c_j$ for $-N/2 \le j < 0$, because the coefficients $c_j$ decay in magnitude asymptotically as $\left|c_j\right|^2 \propto 1/j^2$, which means the error from neglecting aliased frequencies is suppressed exponentially in $n$.  This asymptotic upper bound follows from Parseval's theorem for any signal that is square-integrable over its period, a condition which corresponds to normalized quantum states.

\begin{figure}
  \centering
  \includegraphics[width=\textwidth]{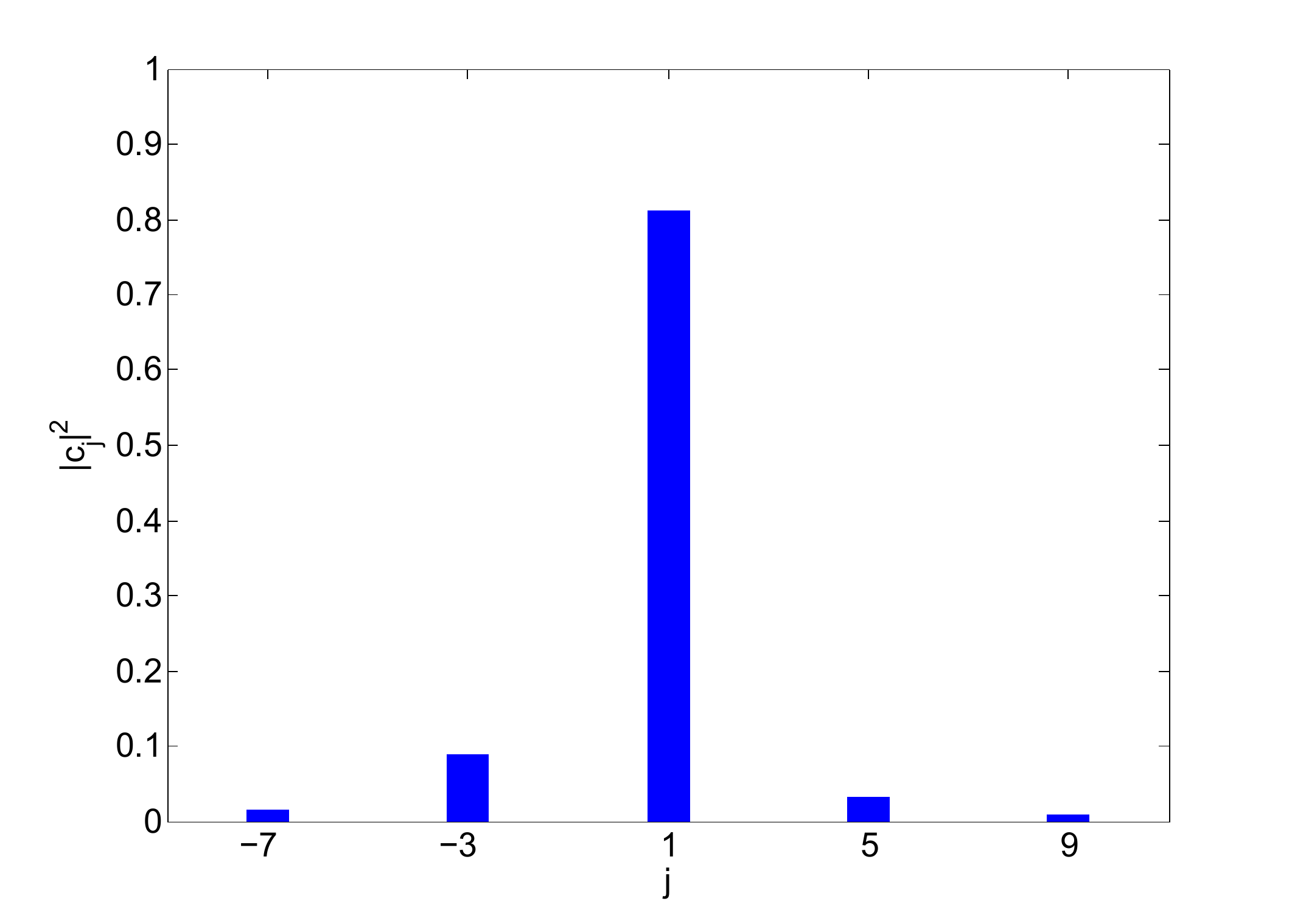}\\
  \caption[Spectrum of function $f(t)$ approximating Fourier-basis amplitudes]{Frequency spectrum of $f(t)$.  The height of a bar at $j$ corresponds to the squared magnitude $|c_j|^2$, where $\{c_j\}$ are Fourier-series coefficients of $f(t)$.  Eqn.~(\ref{coefficient_relationship}) relates these series coefficients to Fourier-basis amplitudes.}
  \label{spectrum}
\end{figure}

In each approximate $\ket{\tilde{\gamma}_0^{(1)}}$ register, the first qubit is the most significant bit in a binary encoding of equally-spaced time coordinates for samples of $f(t)$.  Using only $Z$ and $S$ rotations is effectively discretizing the phase of $f(t) = e^{i 2 \pi \nu(t)}$ to two bits of precision as a piecewise-constant function over four equally-sized intervals in the domain $t \in [0,1)$.  One can readily calculate the $j^{\mathrm{th}}$ Fourier series coefficient of this function:
\begin{eqnarray}
c_j & = & \int_{0}^{1} e^{i 2 \pi (\nu(t) - jt)} dt \nonumber \\
    & = & \sum_{m = 1}^4 \int_{(m-1)/4}^{m/4} e^{-i 2 \pi (jt - (m-1)/4)} dt \nonumber \\
    & = & \left(\frac{1-i}{2 \pi j}\right) \sum_{m=1}^4 e^{-i \pi m (j-1) /2},
\end{eqnarray}
which is valid everywhere except $j=0$, in which case $c_0 = 0$.  Sign convention follows Eqn.~(\ref{define_Fourier_state}).  The only nonzero terms occur for $j \equiv 1 \; (\mathrm{mod} \; 4)$, and they are $c_j = (2-2i)/(\pi j)$.  The squared magnitudes of the largest Fourier components for state $\ket{\tilde{\gamma}_0^{(1)}}$ are plotted in the spectrum in Fig.~\ref{spectrum}.  The expression for Fourier-series coefficients allows us to derive bounds on distillation performance and hence the necessary number of rounds of distillation.  For example, since there is no relative phase between these coefficients, initial Fourier states have $|a_1|^2 > |c_1|^2$, where $|c_1|^2 = 8/\pi^2 \approx 0.81$.  Likewise, the fidelity after one round of symmetric distillation is upper bounded by
\begin{equation}
\frac{|c_1|^4}{\sum_{j = -\infty}^{\infty} |c_j|^4} = \left(\sum_{x = 1,3,5,...} \frac{1}{x^4}\right)^{-1} \approx 0.986.
\end{equation}

I now derive a general method to estimate fidelity after each round of distillation.  The output state of the distillation protocol, conditioned on success, has modified Fourier components.  In symmetric distillation, the relative magnitude of each component to the fundamental harmonic is squared.  After normalization, the largest component at $j = 1$ is amplified, while the rest are suppressed.  The second-largest component is the ``sideband'' at $j = -3$ in Fig.~\ref{spectrum}.  Note that because of aliasing in the frequency spectrum, $c_{(-3)}$ maps to $a_{(N - 3)}$; I assume that Fourier-series term $c_{(N - 3)}$ is negligibly small.

Distillation proceeds until the sidebands are suppressed to a sufficiently low level.  The rate at which these sidebands are suppressed dictates how many rounds of distillation are required, which determines the total number of gates in the protocol.  This rate is limited by the ratio in magnitudes between the fundamental $j = 1$ harmonic and the second-largest sideband at $j = -3$.  This behavior is analogous to the rate of convergence in Markov-chain Monte Carlo, which depends on the magnitude of the second-largest eigenvalue of the state transition matrix (the largest eigenvalue of a stochastic matrix is 1).  Successful symmetric state distillation through $r$ rounds modifies each Fourier-basis amplitude from $a_j$ to $b_j^{(r)}$ according to
\begin{equation}
\left|b_j^{(r)}\right|^2 = \frac{1}{C}\left(\left|a_j\right|^2\right)^{2^{r}},
\end{equation}
where
\begin{equation}
C = \sum_{j=0}^{N-1} \left(\left|a_j\right|^2\right)^{2^{r}}
\end{equation}
is the normalization.  Since any sidebands to the fundamental harmonic ($j = 1$ in this case) will be suppressed super-exponentially in $r$, one need only focus on the magnitude of the largest sideband at $c_{(-3)}$, which will dominate the error in the output of distillation.  As a result, the error $\epsilon = 1 - \left|\left\langle \tilde{\gamma}_r^{(k)} \mid \gamma^{(k)} \right\rangle \right|^2$ in the distilled Fourier state is closely approximated by
\begin{equation}
\epsilon \approx \left(\left|c_{(-3)}\right|^2/\left|c_1\right|^2\right)^{2^{r}}.
\label{distillation_error}
\end{equation}
Consequently, the ratio $\left|c_{(-3)}\right|^2/\left|c_1\right|^2 = 1/9$ (exactly) dictates how fast error is suppressed through distillation.  Since the error tolerances require that $\epsilon \le \left(\pi/2^n\right)^2$, one can determine the number of rounds of distillation as
\begin{equation}
R = \left\lceil \log_2 \left(\frac{2n - 2 \log_2 \pi}{\log_2(|c_1|^2/|c_{(-3)}|^2)}\right) \right\rceil.
\label{num_rounds}
\end{equation}
This expression can be simplified to $R = \left\lceil \log_2(0.63n - 1.04) \right\rceil$ (approximately), which shows that $R$ scales as $O(\log n)$.  Moreover, Eqn.~(\ref{distillation_error}) shows that the error at the output of each successive round of distillation is squared.  Eqn.~(\ref{fidelity_threshold}) shows that the number of qubits needed to represent an approximate Fourier state is $O(\log \epsilon)$, so the smallest allowable size in qubits of intermediate distilled states will double after each round.  The next section uses this technique to save resources.

\subsection{Resource Analysis for Distilling the Fundamental Fourier State}
\label{Sec_Fundamental_Analysis}
This section shows that distilling the fundamental $n$-qubit Fourier state is efficient in the sense that it requires at most $O(n \log n)$ Toffoli gates and total gates, with circuit width $2n + O(1)$ qubits.  After each round of distillation, the number of bits of precision in the Fourier states doubles, so the protocol must also double the number of qubits going into the next round.  The procedure is: (a)~after one round of distillation, each Fourier state is accurate to $s$ qubits; (b)~append $s$ more qubits in the $\ket{+}$ state to each register; (c)~repeat distillation on the input states of size $2s$ qubits.  The additional error of making a larger approximate Fourier state by appending $s$ qubits in the $\ket{+}$ state is less than the error already present, so the fidelity is not reduced appreciably.  The extra qubits provide space for the output state of distillation (if it succeeds) to contain twice as many accurate qubits.

\begin{figure}
  \centering
  \includegraphics[width=\textwidth]{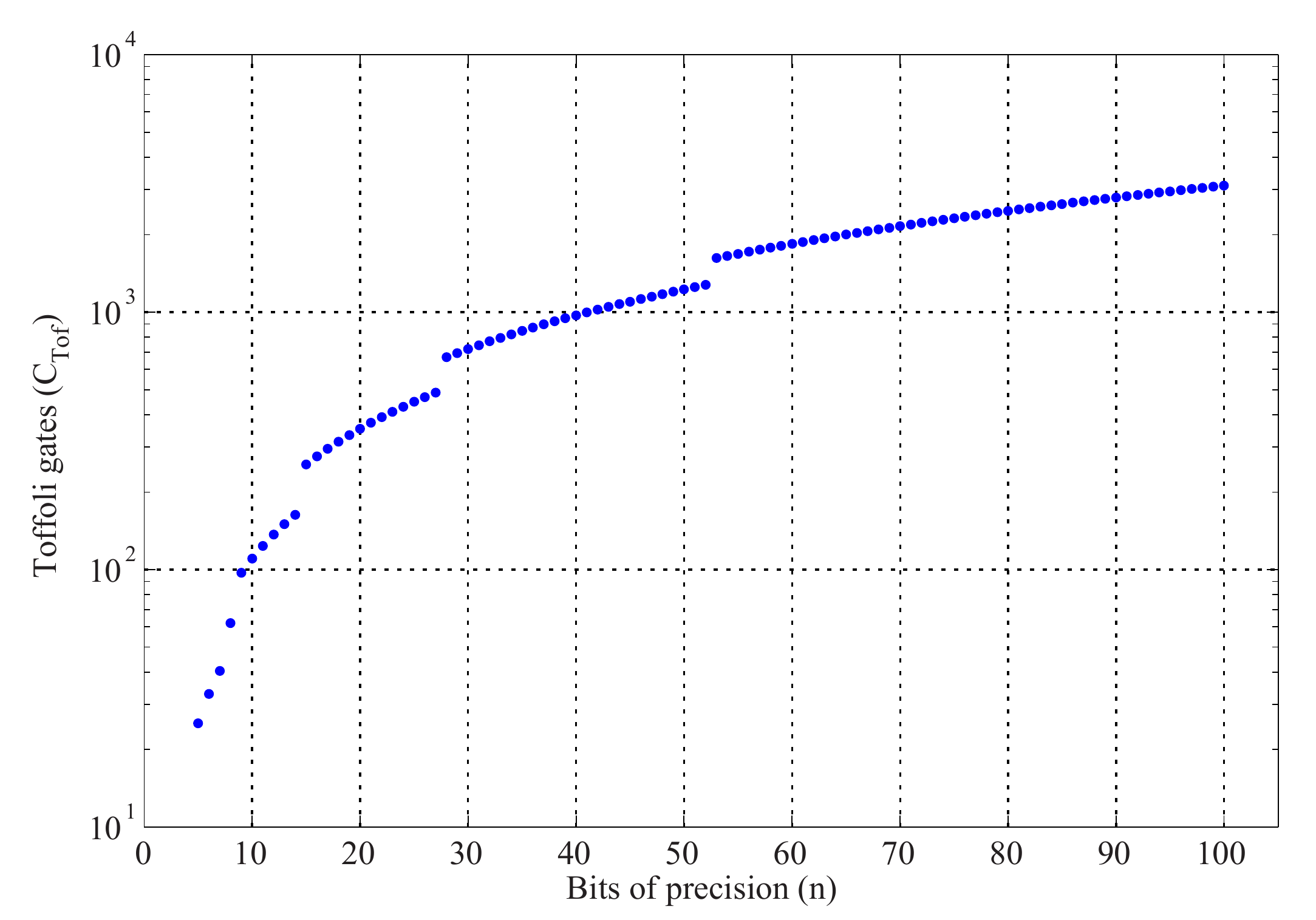}\\
  \caption[Plot of Toffoli gates consumed to produce fundamental Fourier state]{Expected number of Toffoli gates consumed in the distillation of an $n$-qubit Fourier state.  Uncertainty in number of Toffoli gates is a result of distillation success being probabilistic in each round.  No other non-Clifford gates are required, and the Toffoli gates dominate the cost of the adder circuits within distillation.}
  \label{Toffoli_cost}
\end{figure}

Each round of distillation on $s$-qubit registers uses addition circuits that each require $(2s-4)$ Toffoli gates for the CDKM adder~\cite{Cuccaro2004} (note that the carry-out Toffoli is unnecessary and removed).  If there are $R$ rounds of distillation, then the $r^{\mathrm{th}}$ round requires $2^{(R-r)}$ adder circuits.  If the protocol begins with $s$ qubits per Fourier state going into the first round and double the number of qubits in each subsequent round, then the total number of Toffoli gates in the entire distillation protocol is
\begin{equation}
C_{\mathrm{Tof}} = \sum_{r = 1}^R 2^{(R-r)} (2^{(r+1)}s-4) = 2^{(R+1)} R s - 2^{(R+2)} + 4.
\end{equation}
Since $R$ scales as $O(\log n)$, then $C_{\mathrm{Tof}}$ scales as $O(n \log n)$.  The protocol uses $s = 5$ since the output of the first round is accurate to about 5 bits of precision.  Eqn.~(\ref{num_rounds}) gives an exact expression for $R$.  Figure~\ref{Toffoli_cost} plots the number of Toffoli gates required for distillation with up to $n = 100$ bits of precision, which is the most precision one could imagine needing for a quantum algorithm.  If the rotations in some quantum algorithm require 100 bits of precision, this implies the error per gate is of order $2^{-200}$ (errors adding up incoherently), which is only necessary if there are around $2^{200}$ gates.  A computation of this size is far beyond any fathomable quantum hardware.  For example, 10 bits of precision is more than sufficient for 4096-bit Shor's algorithm~\cite{Fowler2004_Rotations}.  In this case the approximately 100 Toffoli gates needed to distill $\ket{\gamma^{(1)}}$ are negligible in comparison to the rest of the algorithm~\cite{Jones2012_PRX}.  I emphasize that even if multiple copies of a Fourier state are required, the distillation need only be performed once.  Fourier states of size $n$ qubits can be cloned using a single adder circuit requiring $(n-1)$ Toffoli gates~\cite{Jones2012_NJP}.

The last round of distillation uses $n + O(1)$ qubits for each input register, which may be less than $2^{(R-1)}s$.  The additive constant appears because one might need to distill to $(n+1)$ or $(n+2)$ qubits in the final output, compensating for errors introduced by truncating the size of Fourier states in earlier rounds.  Each round of symmetric distillation doubles the number of qubits per Fourier state, but the number of Fourier states is reduced by half.  Consequently, the circuit width of the protocol is at most $2n + O(1)$ qubits, because the final round uses two registers of size $n + O(1)$ qubits.

Compiling the Fourier distillation protocol in the surface code gives a more accurate representation of the true resource costs.  Each round requires an adder circuit, and I select the CDKM adder~\cite{Cuccaro2004}, which is appealing for its simple design and economical use of Toffoli gates.  Each Toffoli gate can be constructed using one of the methods in Chapter~\ref{Ch06}, and I use the calculations therein to estimate the volume of an adder circuit here.  Each $n$-bit adder consists of $2n-4$ Toffoli gates and $5n-8$ CNOT gates; one Toffoli and one CNOT are removed from the design in Ref.~\cite{Cuccaro2004} because, in mod-$2^n$ addition, a carry-out qubit is not needed.  If each Toffoli gate can be executed in the time of one plumbing piece, then the addition circuit has volume of $(2n-4) \times n \times 2 = 4n^2 - 8n$ plumbing pieces, plus the volume require to produce the Toffoli gates.  The CNOT gates act on otherwise idle qubits, so they do not increase the volume.

\begin{figure}
  \centering
  \includegraphics[width=\textwidth]{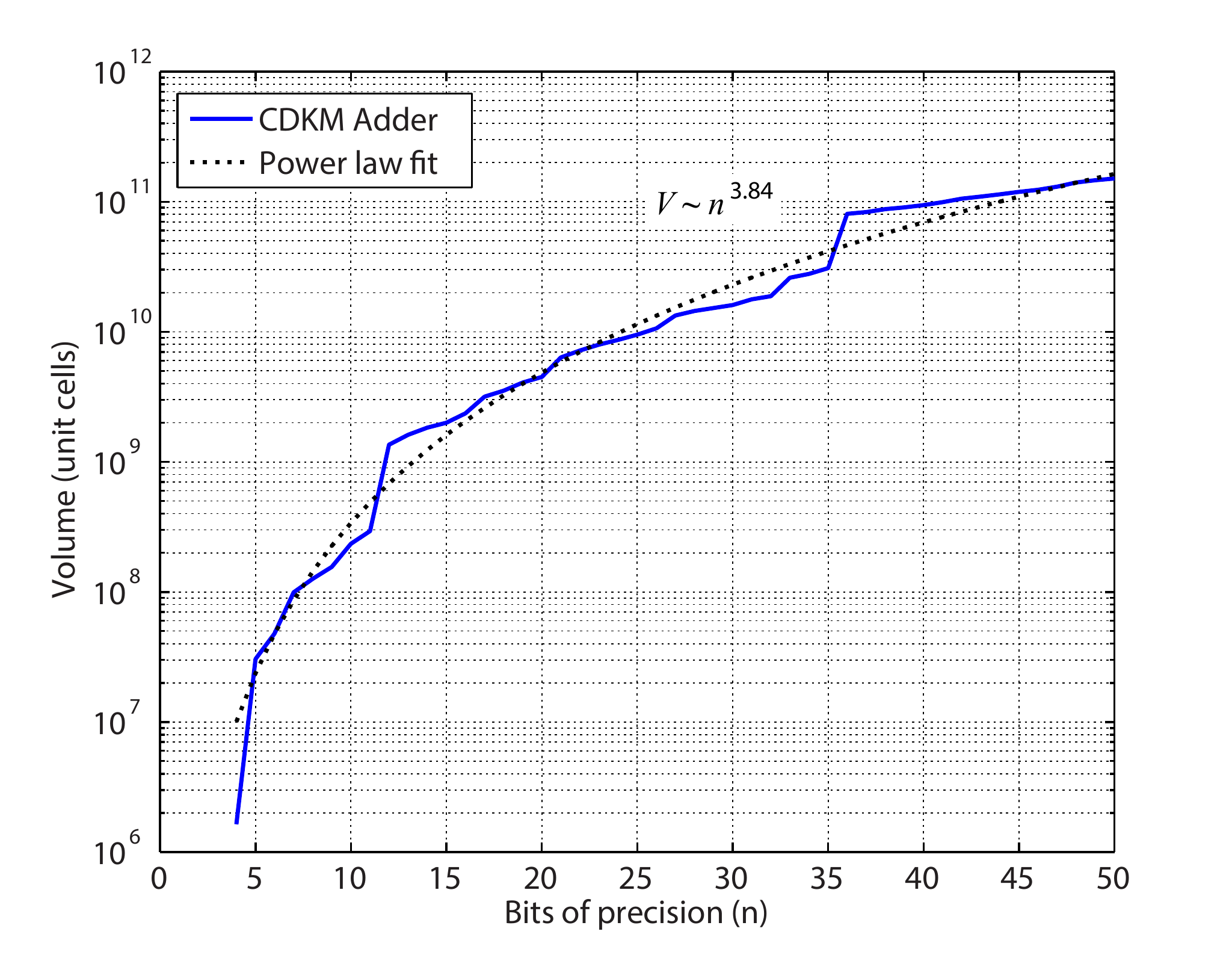}\\
  \caption[Resource costs for a CDKM quantum adder]{Resource cost for a CDKM quantum adder as a function of bits ($n$).  The hardware error rate is $p_g = 10^{-3}$.  A power-law fit gives the approximation $V_{\mathrm{add}}(n) \approx (4.9 \times 10^4) n^{3.84}$ unit cells.  Because these adders are designed for distilling Fourier states, the minimal-volume adder is found such that $p_{\mathrm{out}} < \sin^2(\pi/2^n)$.}
  \label{Adder_resources}
\end{figure}

\begin{figure}
  \centering
  \includegraphics[width=\textwidth]{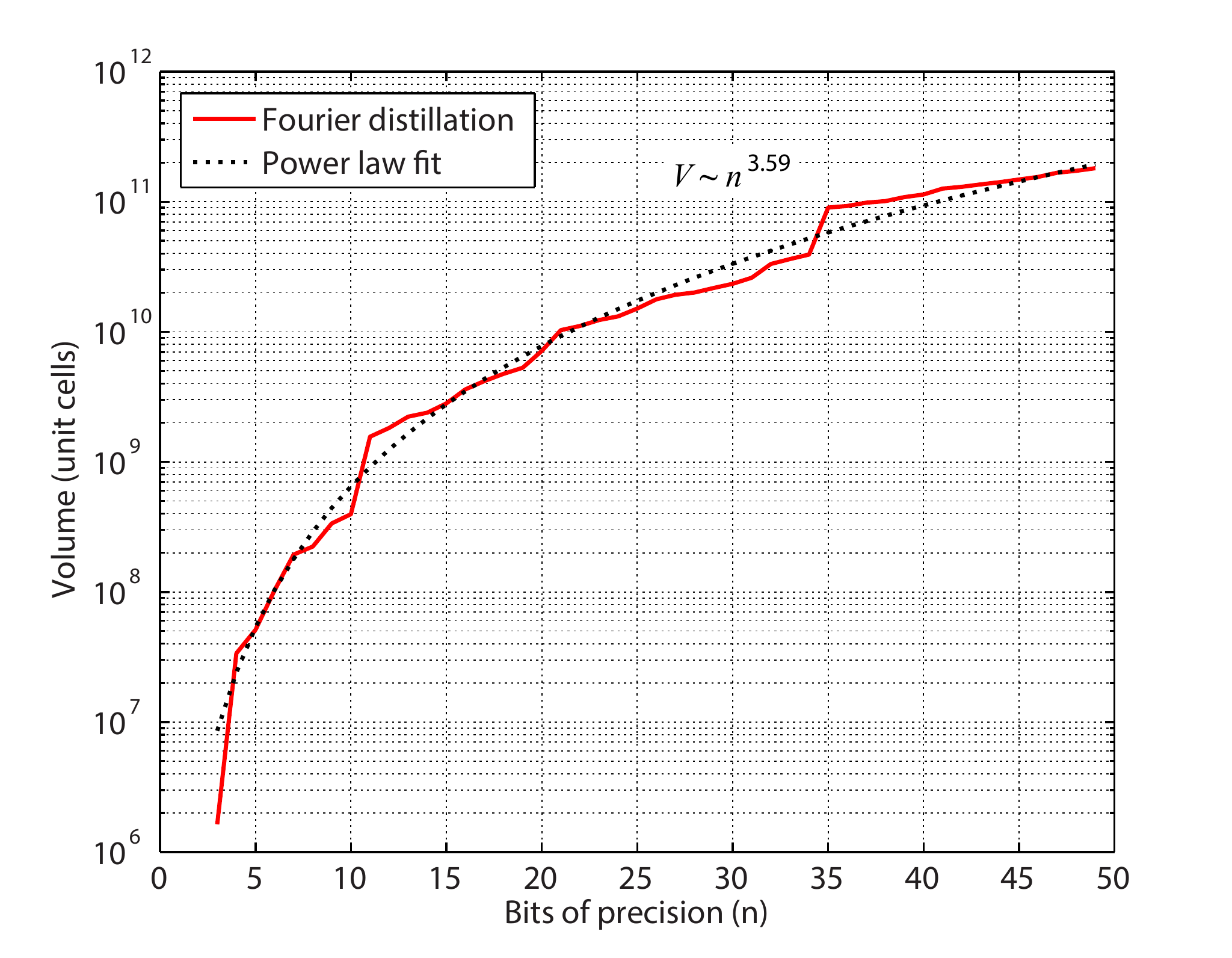}\\
  \caption[Resource costs for Fourier-state distillation]{Resource cost for distilling the fundamental Fourier state $\ket{\gamma^{(1)}}$ as a function of bits ($n$).  The hardware error rate is $p_g = 10^{-3}$.  A power-law fit gives the approximation $V_{\mathrm{Fourier}}(n) \approx (1.7 \times 10^5) n^{3.59}$ unit cells.  The output error of the entire distillation procedure is $p_{\mathrm{out}} < \sin^2(\pi/2^n)$.}
  \label{Fourier_distillation_resources}
\end{figure}

The volume of a CDKM adder in the surface code, as a function of number of bits, is plotted in Fig.~\ref{Adder_resources}.  The output error rate is chosen such that $p_{\mathrm{out}} < \sin^2(\pi/2^n)$.  By doing so, the $n$-bit adder has sufficient accuracy to distill an $n$-bit Fourier state.  Note that adders applied to other purposes may have different accuracy requirements.  A power-law fit gives a scaling coefficient of $V_{\mathrm{add}}(n) \propto n^{3.84}$.  Using these results, Fig.~\ref{Fourier_distillation_resources} calculates the volume for Fourier-state distillation to $n$ bits.  To ensure error tolerances are met, an $n$-bit Fourier state actually uses $(n+1)$-bit addition, with the extra output qubit being discarded through measurement in the $Z$-basis.  A power-law fit gives a scaling coefficient of $V_{\mathrm{Fourier}}(n) \propto n^{3.59}$.  The similarity of this plot to Fig.~\ref{Adder_resources} has a simple explanation.  The Fourier distillation volume is approximately given by the recurrence relation $V_{\mathrm{Fourier}}(n) \approx V_{\mathrm{add}}(n+1) + 2V_{\mathrm{Fourier}}(\lceil n/2 \rceil)$.  Because both $V_{\mathrm{add}}$ and $V_{\mathrm{Fourier}}$ have scaling coefficients much greater than one, the cost of the final adder dominates the total volume.

\subsection{Distilling Arbitrary Fourier States}
\label{Sec_Arbitrary_Distillation}
Finally, it is possible to distill any $\ket{\gamma^{(k)}}$ Fourier state, for any integer $k$.  I outline some possible methods but leave the detailed analysis for future work.  Arbitrary values of $k$ are need for QVR phase kickback~(see Chapter~\ref{Ch07}), which is useful in quantum simulation and some implementations of the linear-systems algorithm~\cite{Jones2012_NJP,Clader2013}.  The appendix of Ref.~\cite{Jones2012_NJP} gives a method for transforming any $n$-qubit $\ket{\gamma^{(k)}}$ with odd $k$ to any other $\ket{\gamma^{(k')}}$, using $O(n^2)$ gates.  Since it requires phase kickback with successively larger addition circuits, the number of Toffoli gates is
\begin{equation}
\sum_{s = 3}^{n-1}(s-2) = \frac{(n-3)(n-2)}{2}.
\end{equation}
That protocol is deterministic and does not require distillation, assuming one already has a Fourier state.  The protocol in Section~\ref{Sec_Fundamental_Distillation} could distill $\ket{\gamma^{(1)}}$, which could then be transformed into any $\ket{\gamma^{(k)}}$.

Approximations of any $\ket{\gamma^{(k)}}$ can also be distilled using a modified form of Fourier-state distillation.  To have reasonably good efficiency, the approximate initial states need to have substantial fidelity with the desired pure state, say $F > 0.5$.  There are at least two possible approaches.  One is to start $\ket{\gamma^{(0)}} = \ket{+}^{\otimes n}$, then apply QVR for each `1' bit in the binary representation of $k$, using a $\ket{\gamma^{(1)}}$ state truncated to $O(\log n)$ qubits.  Each QVR gate need only be accurate to error $O(2^{-\log n}) = O(1/n)$, as there are at most $n$ such operations, so the aggregate error is of order unity, meaning it is bounded it below 0.5.  These approximate states are then distilled, but they are all of size $n$ qubits, so the total number of Toffoli gates is $O(n^2)$.  Whether this method is more efficient than the deterministic construction is not yet clear.  A second way to prepare approximate states is to split the quantum register encoding a desired Fourier state into two registers of roughly equal size, and first prepare these approximately through distillation.  This method can be applied recursively until the input states are small, like the $s=5$ starting state above.  For the protocol in Section~\ref{Sec_Fundamental_Distillation}, many of the intermediate registers are $\ket{\gamma^{(0)}}$, which can be constructed with Clifford gates; this will not always hold for arbitrary $k$, so the number of Toffoli gates needed here is higher than the distillation of $\ket{\gamma^{(1)}}$ and may also be $O(n^2)$ for this method.  Since Fourier states used for QVR phase kickback can be reused, the number of \emph{different} values of $k$ used by an algorithm will dictate whether seeking optimized state-preparation protocols is a worthwhile endeavor.

Although I only use $Z$ and $S$ gates to initialize approximate Fourier states, one could also use smaller-angle magic states $R_z(\pi/2^x)\ket{+}$ for $x > 2$, which would increase success probability and decrease the number of rounds.  However, this approach would require distillation of those small-angle magic states or approximation of the small-angle rotations~\cite{Landahl2013}.  Adding a low-fidelity $T$~gate to produce a more accurate initial state might be advantageous, but dramatic improvements using smaller-angle rotations are not expected for typical quantum-computing parameters.  Conversely, Fourier states would readily enable the complex-instruction-set computing of Ref.~\cite{Landahl2013}, because each Fourier state is the tensor product of the desired small-angle magic states.  Fourier states can be cloned with just $(2n-4)$ Toffoli gates using QVR phase kickback~\cite{Jones2012_NJP}.  As such, phase kickback is a better way to produce these magic states than distilling them individually.


\chapter{Verification Protocols for Toffoli Gates}
\label{Ch06}
Verification protocols produce a logical quantum operation from noisy subprocesses, using some form of error detection.  The distillation protocols in Chapter~\ref{Ch05} create a low-error quantum state from many noisy copies of the same state, but a verification protocol is more general.  This chapter focuses exclusively on Toffoli gates that are built from $T$~gates (using results from Section~\ref{Sec_MS_Distillation}) and Clifford operations in the surface code.  Toffoli gates are ubiquitous in quantum computing, as they are useful for quantum arithmetic~\cite{Cuccaro2004,VanMeter2005,Draper2006} and conditional logic.  Since the resource costs due to Toffoli gates can strongly influence the performance of quantum algorithms~\cite{Isailovic2008,Jones2012_PRX}, optimizing this gate is a worthwhile pursuit.

This chapter introduces two verification protocols that can use faulty $T$~gates, because the protocols detect one or more errors.  As with the distillation protocols in Chapter~\ref{Ch05}, the output error probability of a verification protocol is suppressed if the input errors are independent and sufficiently small.  It is worth noting that distillation protocols for Toffoli gates are known~\cite{Aliferis2007_Thesis,Jones2013_Multilevel,Paetznick2013_CCZ}, but they do not appear to perform well in the surface code, because they require recursive protocols or large block codes.  By contrast, the verification protocols I examine lead to significant reduction in resource costs.  A commonly cited Toffoli gate without error detection and using seven $T$~gates will serve as the Trivial Upper Bound (TUB) to which the results in this chapter are compared.

The analysis at the end of this chapter shows the improvement to resource costs can be dramatic.  For output error probability $10^{-12}$, an optimized Toffoli beats the TUB by a factor of $10\times$, where both use optimized $T$~gates from Section~\ref{Sec_MS_Distillation}; the optimized Toffoli beats a naive Toffoli with naive $T$~gates (as in Chapter~\ref{Ch02} and the constructions in Ref.~\cite{Jones2012_PRX}) by a factor $500\times$.  However, since quantum algorithms are not composed entirely of Toffoli gates, the improvement in total performance for a computation will be less.  Still, given how frequently Toffoli gates are used and that, in aggregate, they tend to dominate resources, the improvement to most quantum algorithms will be substantial.

\begin{figure}
  \centering
  \includegraphics[width=8cm]{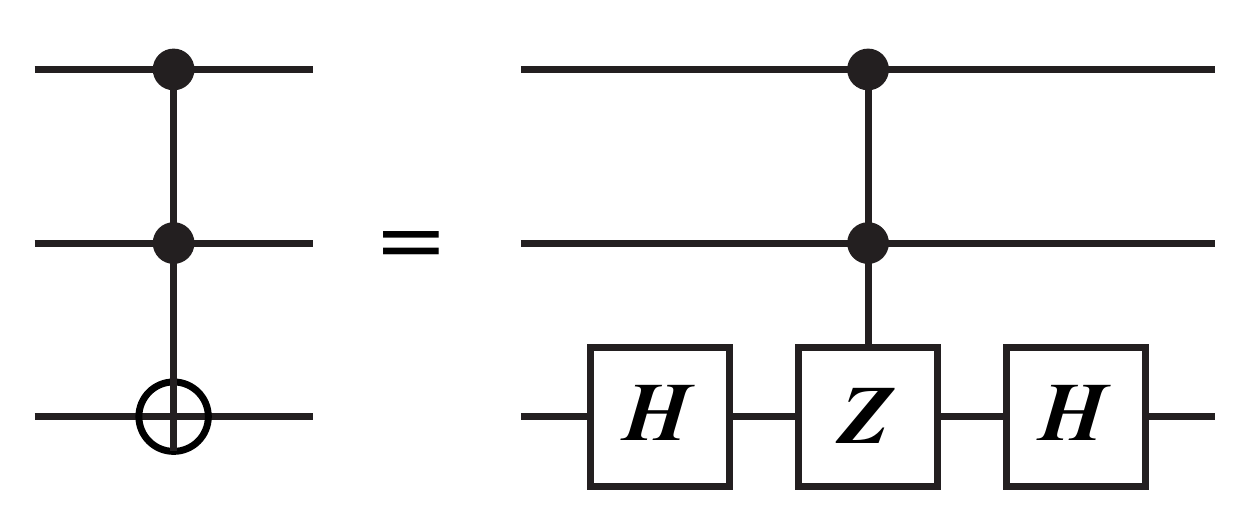}\\
  \caption[Circuit: Local equivalence of Toffoli and CCZ]{Circuit diagram showing Toffoli and CCZ are locally equivalent, with the target being specified with Hadamard gates.}
  \label{Toffoli_CCZ}
\end{figure}

As a final comment, all of the surface code programs in this chapter actually produce controlled-controlled-$Z$ (CCZ), a gate which imparts -1 phase to state $\ket{111}$, and +1 phase otherwise.  CCZ has the advantage of being symmetric in its inputs, and it is locally equivalent to Toffoli, as placing Hadamard gates on both sides of CCZ for a particular qubit produces a Toffoli gate targeting that qubit (see Fig.~\ref{Toffoli_CCZ}).  Using the CCZ constructions in this chapter, any orientation of Toffoli can be produced by proper insertion of Hadamard gates.

\section{Simple Toffoli from Seven $T$ Gates (7T)}
\label{Sec_7T}
Decompositions of the Toffoli gate into Clifford gates and a single non-Clifford gate, such as $T = \exp(i \pi(I-Z)/8)$, have been known going back at least as far as the work of Barenco~\emph{et al.}~\cite{Barenco1995}.  The Toffoli circuit on p.~181 of Ref.~\cite{Nielsen2000} is probably the most familiar decomposition; it uses seven $T$~gates, so I will refer to this as the ``7T'' construction.  Note that the inverse gate $T^{\dag}$ requires the same ancilla-based teleportation circuit as $T$, so these gates are equivalent in state-distillation cost and construction.  By dropping the Hadamard gates, one produces CCZ, and so 7T can implement either gate by inserting or removing Hadamard operations in the appropriate locations.  Because the 7T construction is both the simplest and most commonly used, it will be the source of the Trivial Upper Bound on resource costs for a Toffoli gate.

\begin{figure}
  \centering
  \includegraphics[width=\textwidth]{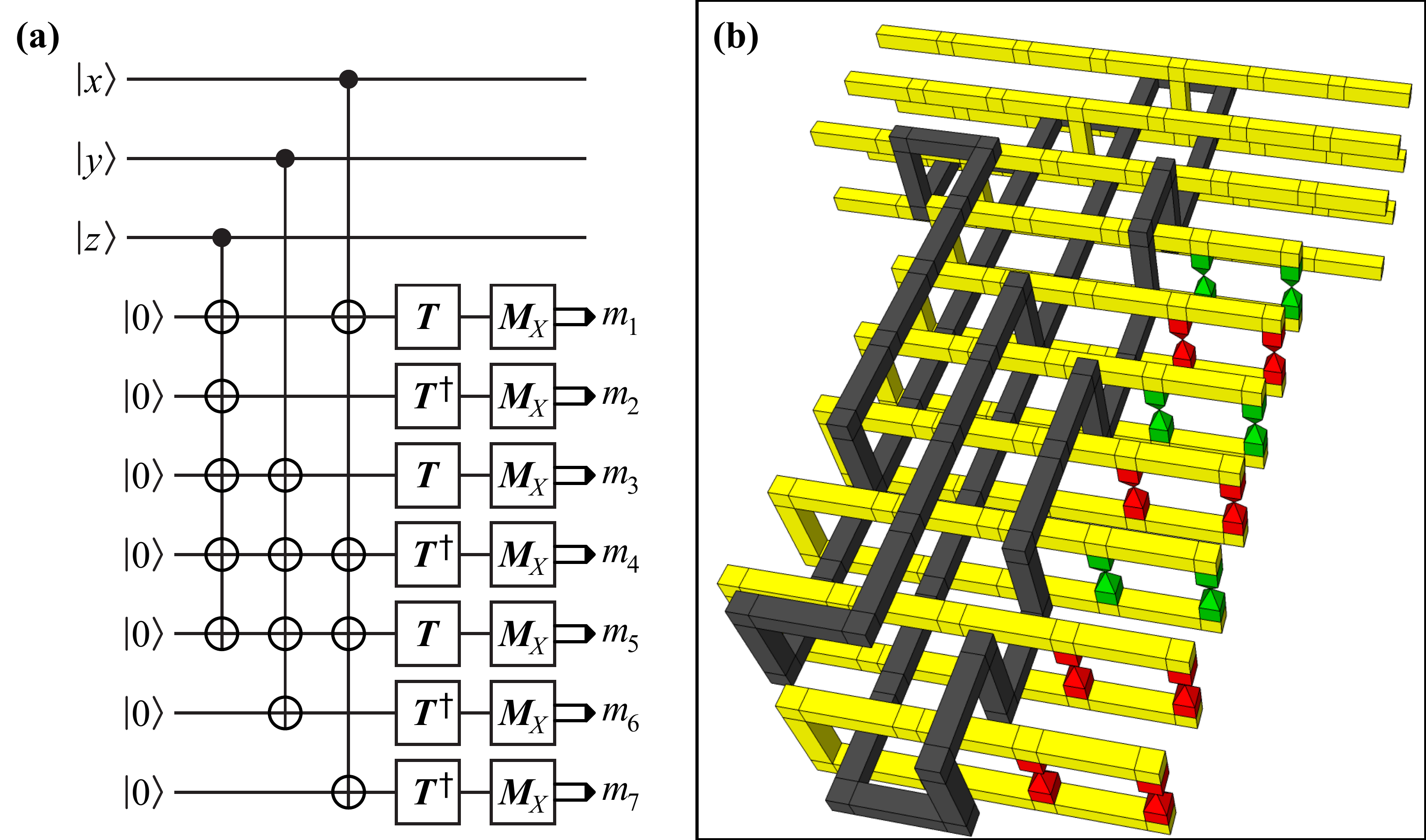}\\
  \caption[Circuit and Surface Code: 7T Toffoli]{Design for the 7T Toffoli. (a)~Circuit diagram for 7T that is amenable to surface code braiding.  (b)~Perspective rendering of 7T implemented in the surface code.  Each horizontal pair of yellow pipes corresponds to the qubit on the left in the same vertical position.  Green pyramids correspond to injection the $T$~gate (such as by the teleportation circuit of Fig.~\ref{T_teleportation}), whereas red pyramids correspond to $T^{\dag}$.  The bounding box is $7 \times 11 \times 2 = 154$ plumbing pieces.  As described in Section~\ref{BK_surface_code_section}, additional volume is allotted for the conditional $S$ gates.}
  \label{Seven_T_combined}
\end{figure}

The 7T design has no protection against errors.  If each of the $T$~gates have error probability $p$, the entire Toffoli will have error probability bounded by
\begin{equation}
p_{\mathrm{out,7T}} = 7p + V_{\mathrm{7T}} \cdot p_L(p_g,d),
\label{7T_error}
\end{equation}
where $p_L(p_g,d)$ is the probability of logical failure per ``plumbing piece'' as a function of gate error $p_g$ and code distance $d$, and $V_{\mathrm{7T}}$ is the volume of the circuit in plumbing pieces (see discussion in Section~\ref{Sec_Resource_Methodology}).  To analyze the 7T construction, I designed the circuit in Fig.~\ref{Seven_T_combined}(a).  The $T$~gates are placed on ancilla qubits, and every $T$~gate is followed by $X$-basis measurement.  This circuit is related to the result by Selinger where all of the $T$~gates are implemented simultaneously~\cite{Selinger2013_Tdepth}.  In my circuit, $X$-basis measurement replaces half of the CNOTs with teleportation.  Heuristically, this design methodology tends to produce compact surface code programs (see Refs.~\cite{Fowler2012_Bridge,Fowler2013_BlockCodes} for other examples).  The surface code braiding pattern for the 7T circuit is shown in Fig.~\ref{Seven_T_combined}(b).  The volume, including the input/output qubits, is $V_{\mathrm{7T}} = 154$ plumbing pieces.

\section{Simple Toffoli from Four $T$ Gates (4T)}
\label{Sec_4T}
\begin{figure}
  \centering
  \includegraphics[width=11cm]{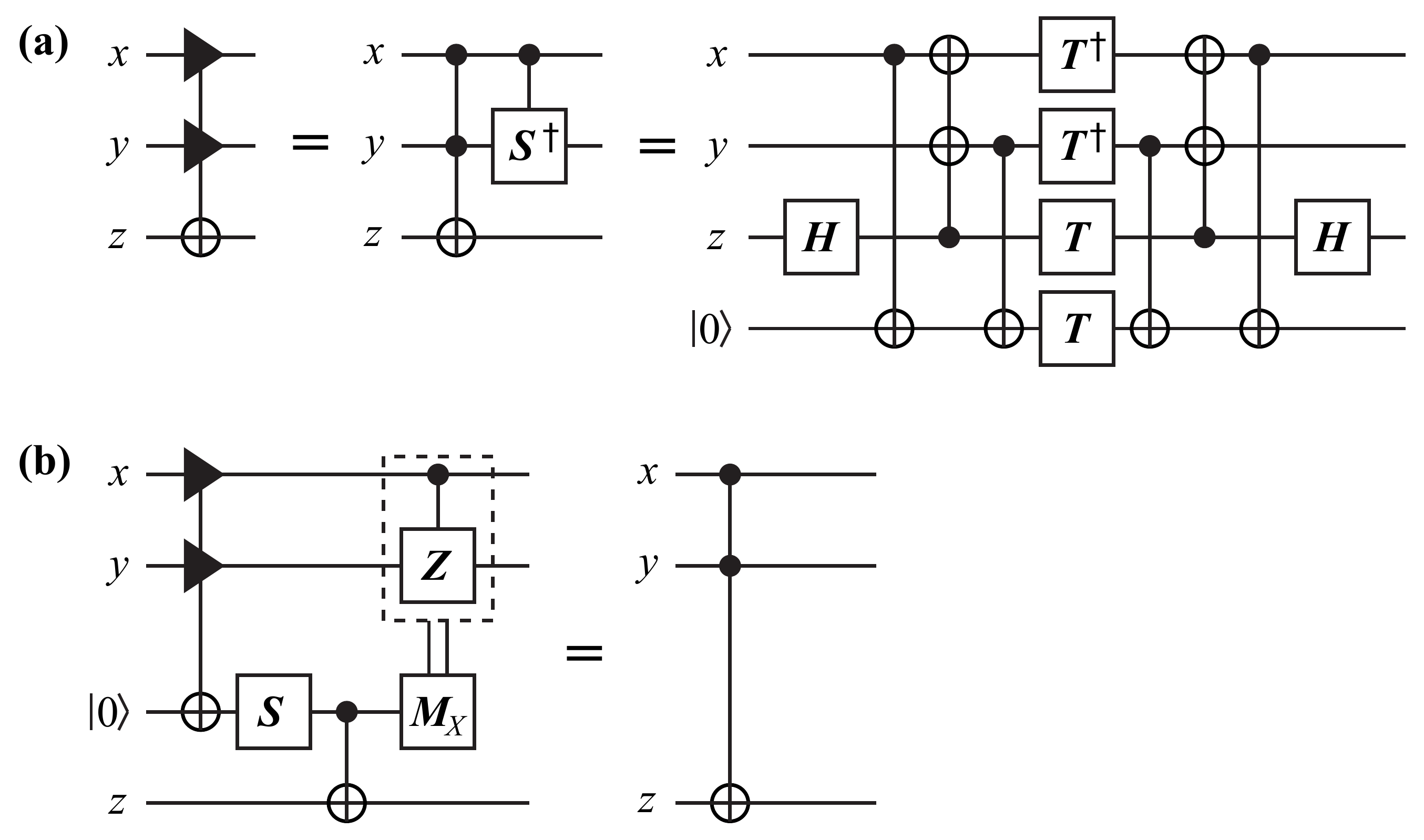}\\
  \caption[Circuit: Toffoli construction 4T using four $T$~gates]{A circuit construction for Toffoli using four $T$~gates (``4T'').  (a) The Toffoli$^{\star}$ circuit~\cite{Selinger2013_Tdepth} that is almost a Toffoli, with the difference being the controlled-$S^{\dag}$ operation.  (b) A circuit that combines Toffoli$^{\star}$ with a phase correction and teleportation to produce an exact Toffoli gate.  The measurement is in the $Z$ basis, and the double vertical lines indicate that the controlled-$Z$ correction is conditioned on the measurement result being $\ket{1}$.  Originally published in Ref.~\cite{Jones2013_Toffoli}.  {\copyright}2013 American Physical Society.}
  \label{four_T_circuit}
\end{figure}

One can improve on the 7T construction by replacing some $T$~gates with teleportation, reducing the required number to four $T$~gates.  Denote the Toffoli$^{\star}$ gate as the operation in Fig.~\ref{four_T_circuit}(a), which requires four $T$~gates and was introduced by Selinger~\cite{Selinger2013_Tdepth}.  Toffoli and Toffoli$^{\star}$ differ only by a controlled-$S^{\dag}$ gate between the control qubits $x$ and $y$.  Beginning with Toffoli$^{\star}$, one needs only an ancilla qubit, a phase gate $S$, and teleportation to implement the exact Toffoli gate, as shown in Fig.~\ref{four_T_circuit}(b).  First, apply the Toffoli$^{\star}$ using the same controls as the desired Toffoli but with an ancilla $\ket{0}$ as target.  Second, the erroneous controlled-$S^{\dag}$ is corrected by a simple $S$~gate applied to the ancilla.  Third, the CNOT and measurement teleport the doubly conditional NOT operation encoded in the ancilla to the target qubit of the desired Toffoli.  Finally, the measurement result determines whether a corrective gate of controlled-$Z$, which is in the Clifford group, is required to correct a $-1$ phase resulting from measurement back-action.  One can readily verify that only four $T$~gates are required in this procedure.  I will call the Toffoli design represented by the circuit in Fig.~\ref{four_T_circuit}(b) the ``4T'' construction.

As an aside, one can also use the 4T construction in Fig.~\ref{four_T_circuit}(b) to add control-qubit inputs to an existing controlled-$G$ gate, where $G$ is any unitary.  Replace the CNOT in Fig.~\ref{four_T_circuit}(b) with controlled-$G$ (targeting however many qubits $G$ acts on), and the result is controlled-controlled-$G$.  By iterating this procedure, one can add $n$ controls to controlled-$G$ using $4n$ $T$~gates.  The best prior result required $8n$ $T$~gates~\cite{Selinger2013_Tdepth}.

\begin{figure}
  \centering
  \includegraphics[width=\textwidth]{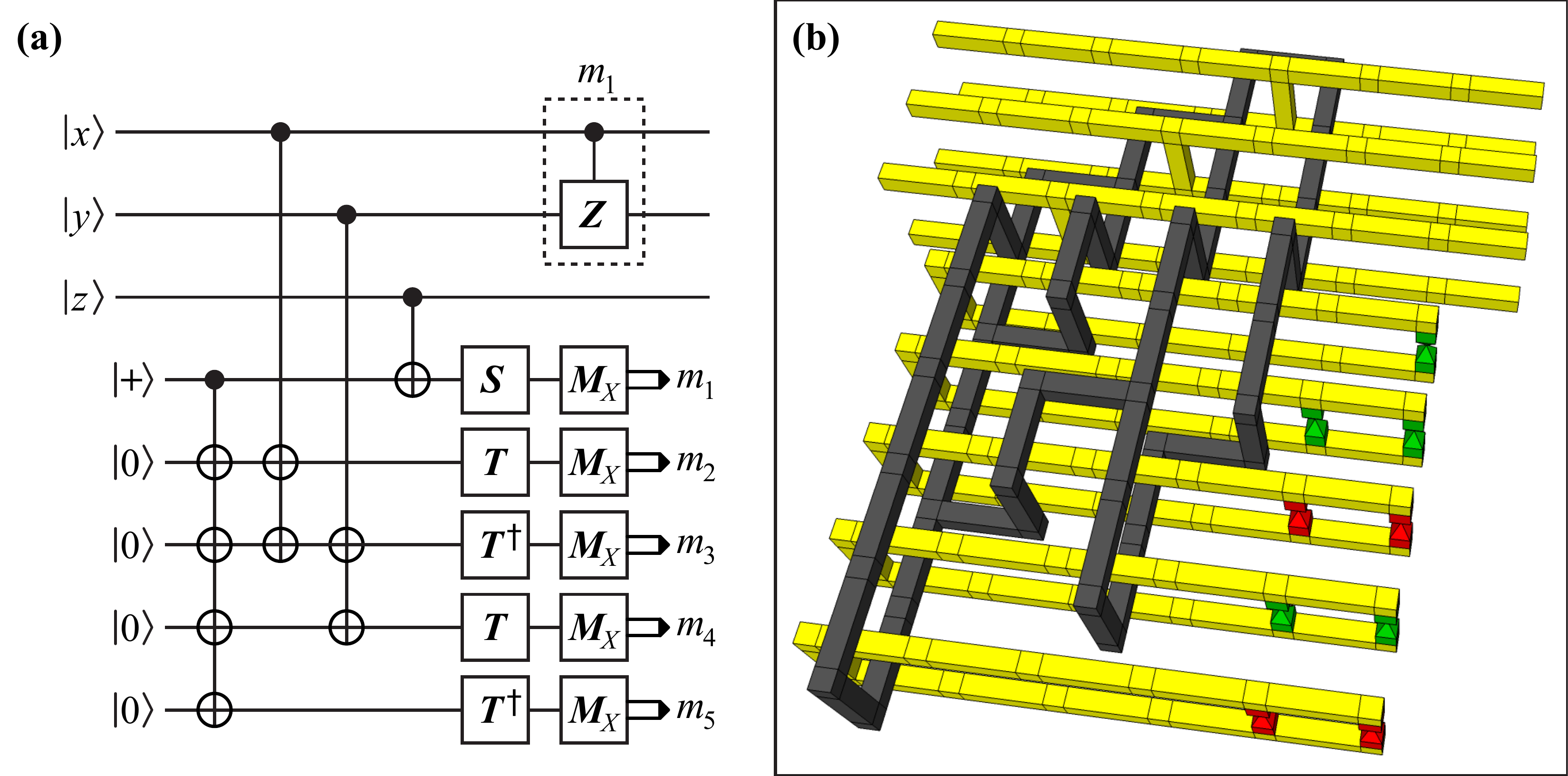}\\
  \caption[Circuit and Surface Code: 4T Toffoli]{Design for the 4T Toffoli. (a)~Circuit diagram for 4T that is amenable to surface code braiding.  (b)~Perspective rendering of 4T implemented in the surface code.  As in Fig.~\ref{Seven_T_combined}, the horizontal yellow defects correspond in vertical order to the qubits on the left.  The fourth qubit from the top ends with an injected $S$ gate (the lower four qubits end in conditional $S$~gates that depend on the injected $T$~gates).  The conditional controlled-$Z$ gate is not shown.  The bounding box is $7 \times 9 \times 2 = 126$ plumbing pieces.  As described in Section~\ref{BK_surface_code_section}, additional volume is allotted for the conditional $S$ gates.}
  \label{Four_T_combined}
\end{figure}

A circuit for 4T that is well suited to the surface code is shown in Fig.~\ref{Four_T_combined}(a).  As with 7T, the $T$~gates have been moved to ancilla qubits, implemented just prior to an $X$-basis measurement.  An implementation of 4T in the surface code, with direct correspondence to this circuit, is shown in Fig.~\ref{Four_T_combined}(b).  The volume is $V_{\mathrm{4T}} = 126$ plumbing pieces.  In analogy with Eqn.~(\ref{7T_error}), the error probability of the 4T construction can be be bounded by
\begin{equation}
p_{\mathrm{out,4T}} = 4p + V_{\mathrm{4T}} \cdot p_L(p_g,d).
\end{equation}
Although the volume of the 4T design is nearly the same as that of 7T, the use of only four $T$~gates will make a significant difference since the volume of each $T$~gate, including distillation, is at least 224 plumbing pieces, which is larger than either 4T or 7T design.

\section{Error-Detecting Toffoli Gate (D2)}
\label{Sec_D2}
As shown in Sections~\ref{Sec_7T} and~\ref{Sec_4T}, the cost of a Toffoli gate (or Toffoli$^{\star}$) is dominated by the cost of the underlying $T$ gates.  If one produces a Toffoli using the 4T construction with output error $p_{\mathrm{out}} = 10^{-12}$, then over 99\% of the total required volume is devoted to the four $T$ gates.  The cost of a $T$~gate depends on the magic-state distillation used (see Section~\ref{Sec_MS_Distillation}), with an approximate scaling where surface code volume is $O\left(\log^{3.27}(1/\epsilon)\right)$ for error probability $\epsilon$.  Since high-fidelity $T$~gates are so expensive, this section and the next investigate Toffoli constructions that detect errors occurring in the $T$~gates.  When $T$~gates can have higher error probability, they are less costly.

I designed a distance-two Toffoli gate that can detect an error in any one of eight $T$~gates~\cite{Jones2013_Toffoli}, so at least two errors must occur for verification to fail (Eastin independently derived an equivalent result~\cite{Eastin2013_Toffoli}).  Since this construction has distance two with respect to errors in $T$~gates, it is labeled ``D2.''  The error probability of the D2 Toffoli gate due to $T$-gate failure is $28p^2$ instead of $4p$, to lowest non-vanishing order.  The coefficient is given by $\binom{8}{2} = 28$, which comes from counting all configurations of distinct two-error events, each having probability $p^2$ since they are independent.  Even though eight $T$~gates are needed instead of four, each $T$~gate is permitted a higher error rate to achieve the same final error in the Toffoli circuit.  The result will be that, in total, the $T$~gates are less expensive to produce for the error-detecting construction.  To see why, consider two Toffoli constructions, where the first has no error detection and uses four $T$~gates having error $p_1$, and where the second has distance-two error detection and uses eight $T$~gates with error $p_2$.  Let the final Toffoli error probability be $10^{-12}$ for both designs.  The required input error rates are $4p_1 = 28{p_2}^2 = 10^{-12}$, so $p_1 = 2.5 \times 10^{-13}$ while $p_2 = 1.9 \times 10^{-7}$.  Because $p_2 > p_1$, the $T$~gates in the second case require fewer resources for distillation.

\begin{figure}
  \centering
  \includegraphics[width=8cm]{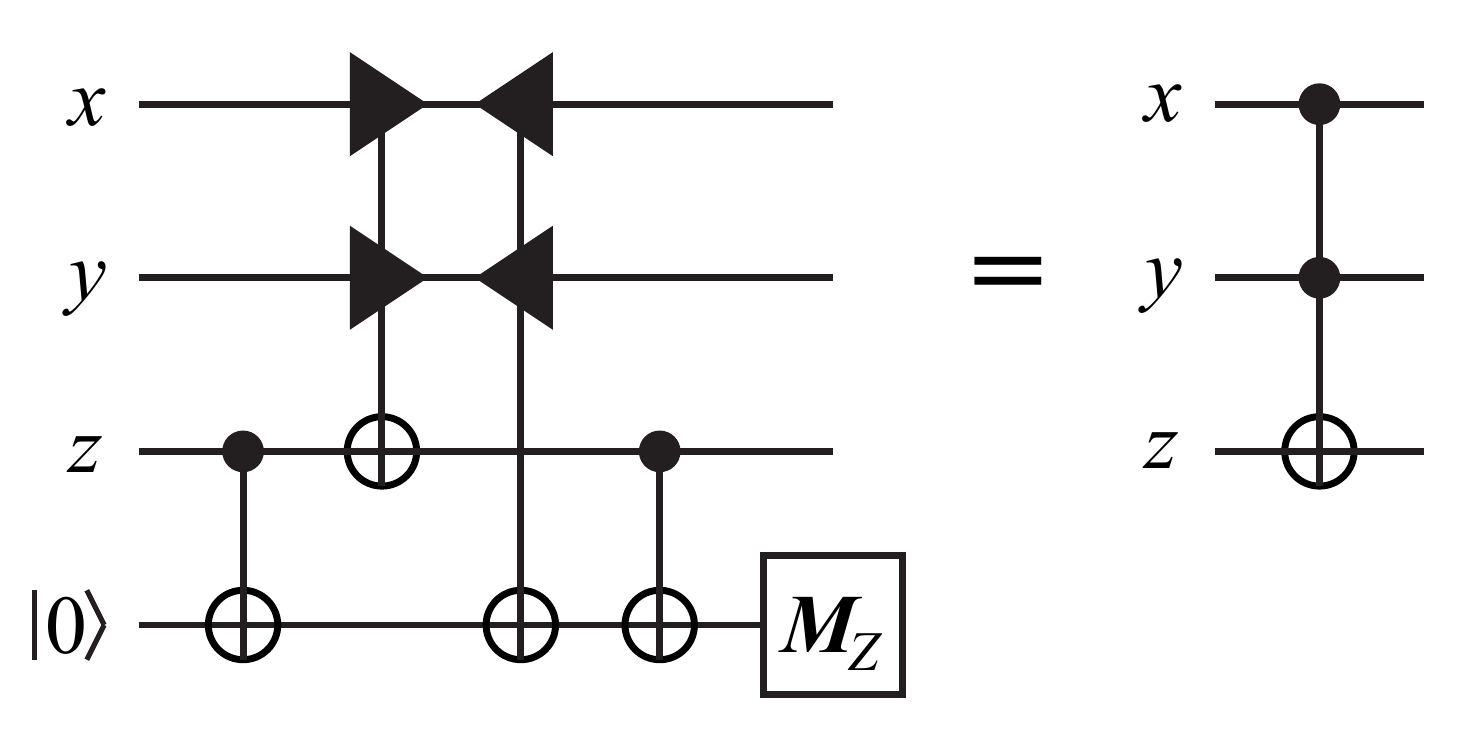}\\
  \caption[Circuit: Error-detecting Toffoli using Toffoli$^{\star}$]{Error-detecting Toffoli gate (D2) constructed using Toffoli$^{\star}$ and its inverse (denoted by reversed triangles).  The measurement is in the $Z$ basis, and obtaining result $\ket{1}$ indicates an error was detected, so the qubits should be discarded.  Originally published in Ref.~\cite{Jones2013_Toffoli}.  {\copyright}2013 American Physical Society.}
  \label{error_detect_circuit_simple}
\end{figure}

\begin{figure}
  \centering
  \includegraphics[width=12cm]{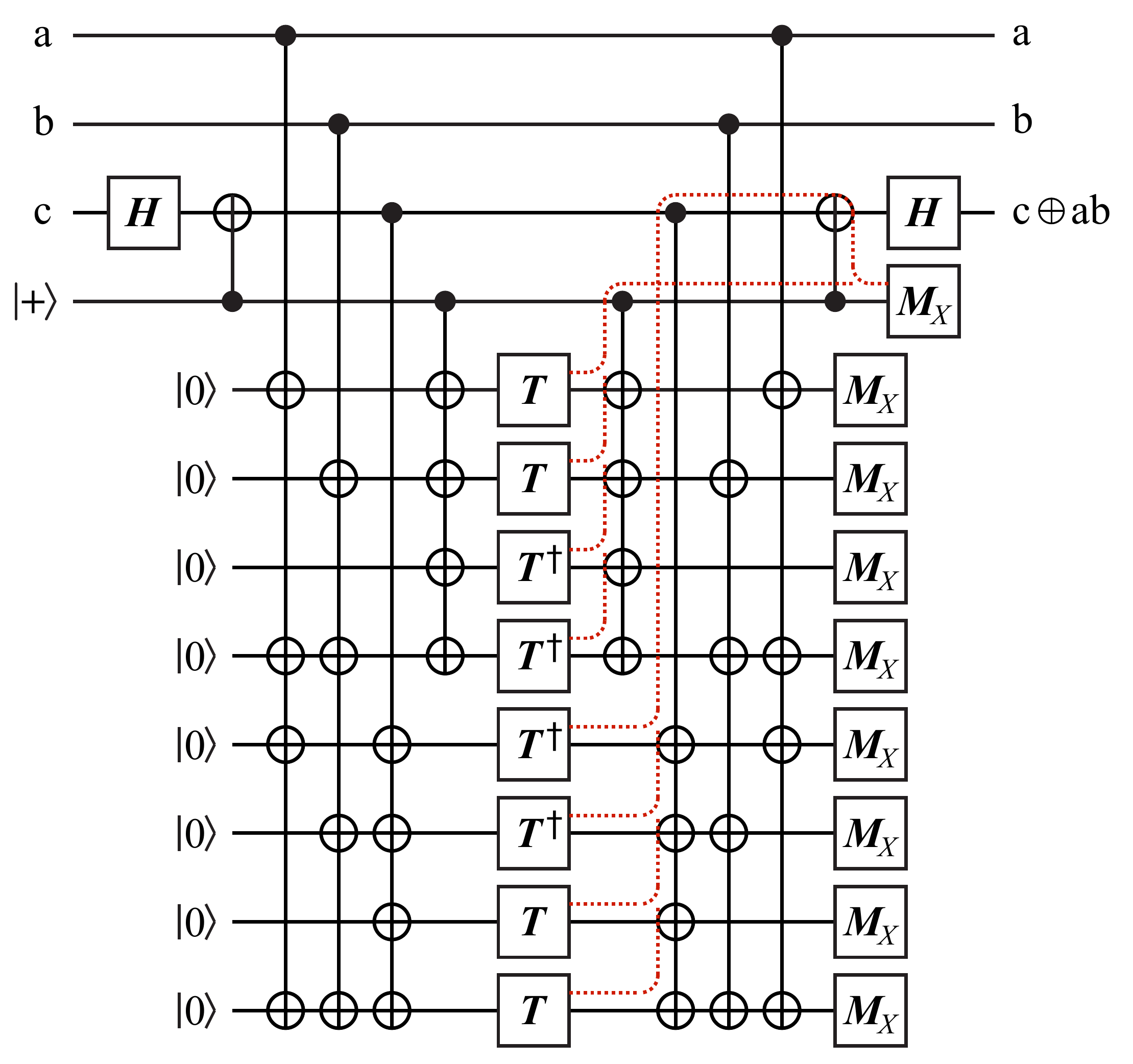}\\
  \caption[Circuit: Error-detecting Toffoli using $T$ gates]{Circuit for D2 decomposed into $T$ gates, showing the mechanism of error detection.  The red dashed lines indicate how any single $Z$ error will propagate to the readout qubit.  The measurement is in the $Z$ basis, and obtaining result $\ket{1}$ indicates an error was detected, so the qubits should be discarded.  As long as the ancilla qubits are initialized perfectly to $\ket{0}$ and the CNOT and $H$~gates have no errors, then only $Z$ errors in the $T$~gates matter, as $X$ errors cannot propagate to data qubits.  If the probability of a $Z$ error in each $T$~gate is i.i.d. $\texttt{Bernoulli}(p)$, then the success probability is $1-8p$ and the \emph{a posteriori} error probability is $28p^2$, to lowest order in $p$.  Modified from version published in Ref.~\cite{Jones2013_Toffoli}.}
  \label{error_detect_circuit}
\end{figure}

The D2 Toffoli circuit has a simple derivation, which I explain before adapting the circuit to the surface code.  It consists of two Toffoli$^{\star}$ gates acting on a target qubit which is in a bit-flip code~\cite{Shor1995,Nielsen2000}, as shown in Fig.~\ref{error_detect_circuit_simple}.  The gate with reversed triangles is the inverse operation (Toffoli$^{\star}$)$^{\dag}$.  Importantly, the controlled-$S$ and controlled-$S^{\dag}$ gates acting on the same qubits $x$ and $y$ are inverse operations, so they cancel.  A logically equivalent decomposition into $T$~gates is shown in Fig.~\ref{error_detect_circuit}; this circuit is convenient for analyzing how errors propagate.  The correspondence is achieved by placing $T$~gates on ancilla qubits with the aid of CNOT gates~\cite{Selinger2013_Tdepth}.  A single $Z$ error in any of the $T$~gates will necessarily propagate to the syndrome measurement for this bit-flip code, as indicated by the red dashed lines.  Upon such an event, all of the qubits are discarded.  Note that if a $T$~gate has an $X$ error, it does not propagate anywhere since it would commute with the CNOT gates.

\begin{figure}
  \centering
  \includegraphics[width=8cm]{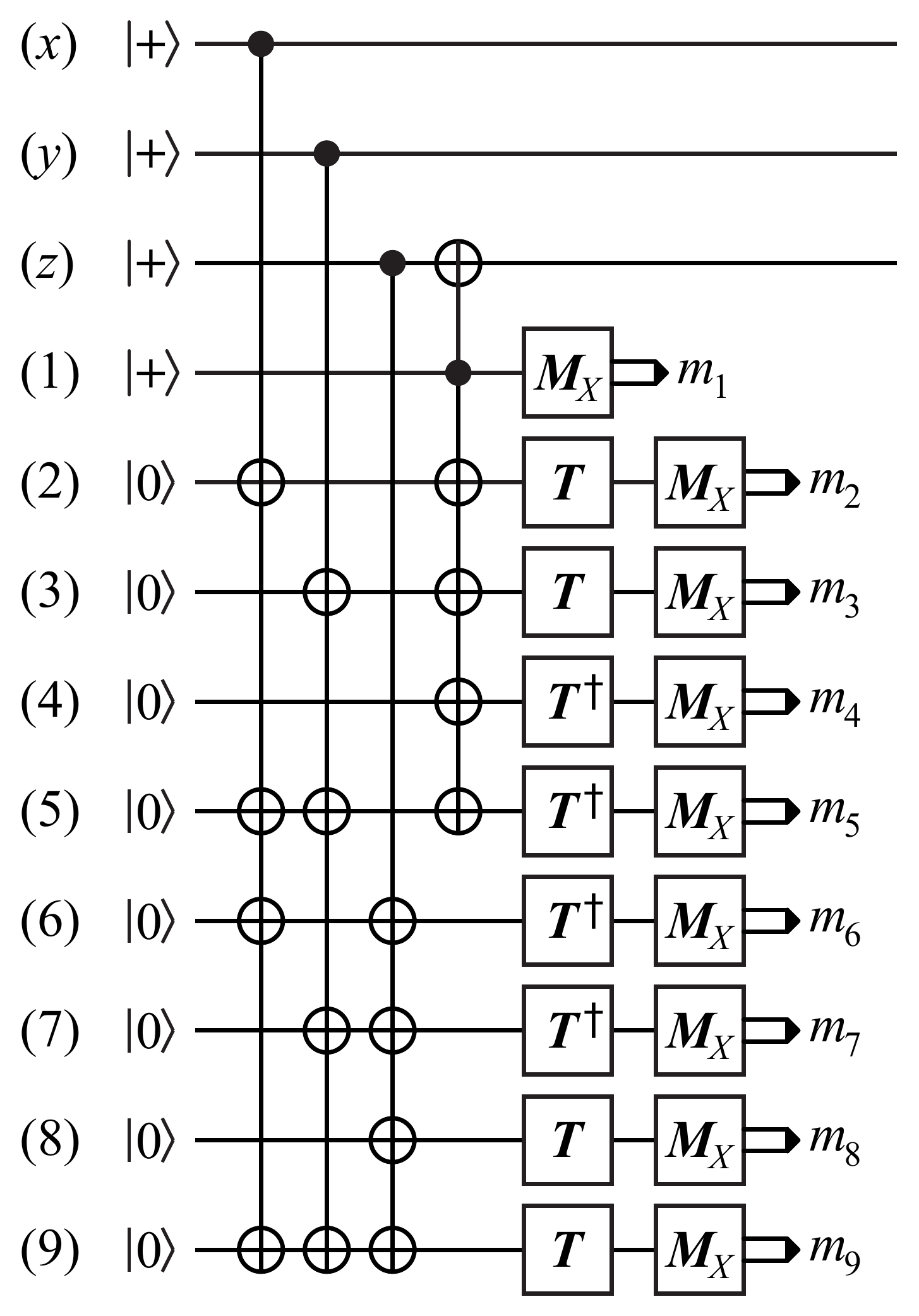}\\
  \caption[Circuit: D2 Toffoli prepared for surface code]{Circuit for D2 with preparations for surface code compilation.  The $T$~gates are placed on ancilla qubits and are followed immediately by $X$-basis measurement.  The circuit produces the QLUT for CCZ, though this can be converted to the Toffoli QLUT by applying a Hadamard gate to the target qubit.}
  \label{D2_prep_circuit}
\end{figure}

A modified but equivalent circuit diagram is shown in Fig.~\ref{D2_prep_circuit}, where many CNOTs have been replaced by $X$-basis measurement.  An error is detected if the parity of the measurements $(m_1,\ldots,m_9)$ is odd, and there are corrections to the Pauli frame for output qubits, conditioned on the measurement results.  By placing $T$~gates followed by measurement at the end of the circuit, this design is amenable to compilation in the surface code.

\begin{figure}
  \centering
  \includegraphics[width=9cm]{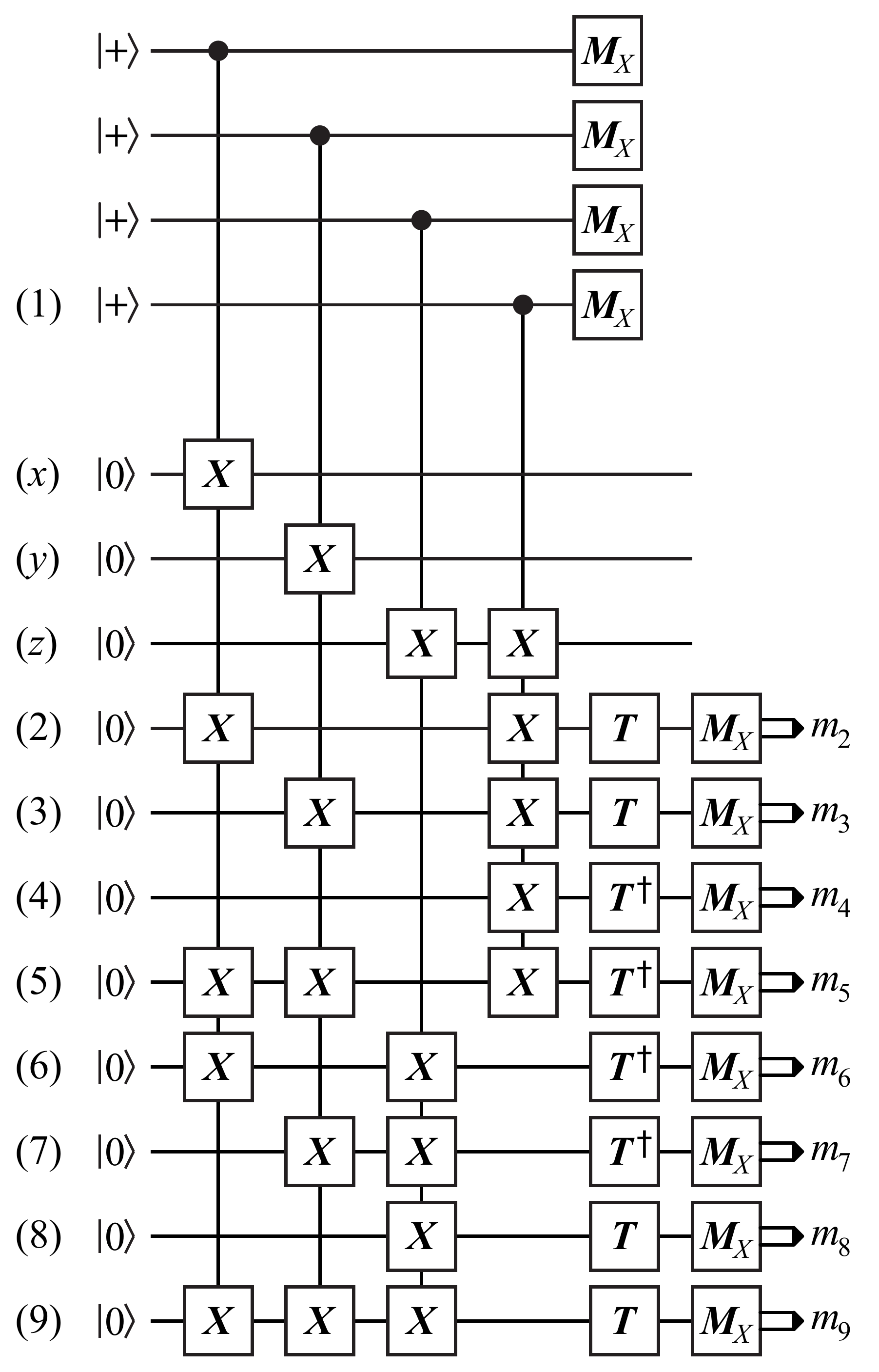}\\
  \caption[Circuit: D2 in stabilizer representation]{Circuit for D2 where all qubits are initialized to $\ket{0}$ and CNOT operations from Fig.~\ref{D2_prep_circuit} are replaced with multi-qubit $X$ stabilizers.}
  \label{D2_stabilizer_circuit}
\end{figure}

Special care was taken to optimize the D2 design in the surface code.  This required substantial rearrangement of qubits and stabilizer measurements, as I will now explain.  Like previous sections, the design here will produce CCZ instead of Toffoli.  The input states are all set to $\ket{+}$ instead of being arbitary; as explained later, this a QLUT for CCZ.  As shown in Fig.~\ref{D2_prep_circuit}, label the first three qubits from the top $(x)$, $(y)$, $(z)$, and the remaining qubits $(1)$,...,$(9)$, where numbers correspond to the measurement outputs.  Now, rewrite the CNOTs in terms of $X$-type stabilizer measurements.  For example, qubit (1) is a stabilizer measurement, which is a simple braiding operation and does not require a persistent logical qubit in the surface code.  Similarly, the other CNOT operations can be replaced with $X$-type stabilizer measurements on qubits that are initialized to $\ket{0}$, as shown in Fig.~\ref{D2_stabilizer_circuit}.  Changing CNOTs into stabilizer measurements actually simplifies the surface code compilation, because $X$-type stabilizers are simply dual braids (black pipes) that weave through the yellow primal braids.  The extra qubits in Fig.~\ref{D2_stabilizer_circuit} will not actually increase the volume of the design.  A brief note on diagrams: I will use controlled-$X$ operators for stabilizer measurements, while using the more familiar CNOT when it explicitly appears in the logic (for example, Fig.~\ref{Four_T_combined}(a)).  They are of course the same operation, and the distinction is made just to explain the methods developed here.

\begin{figure}
  \centering
  \includegraphics[width=\textwidth]{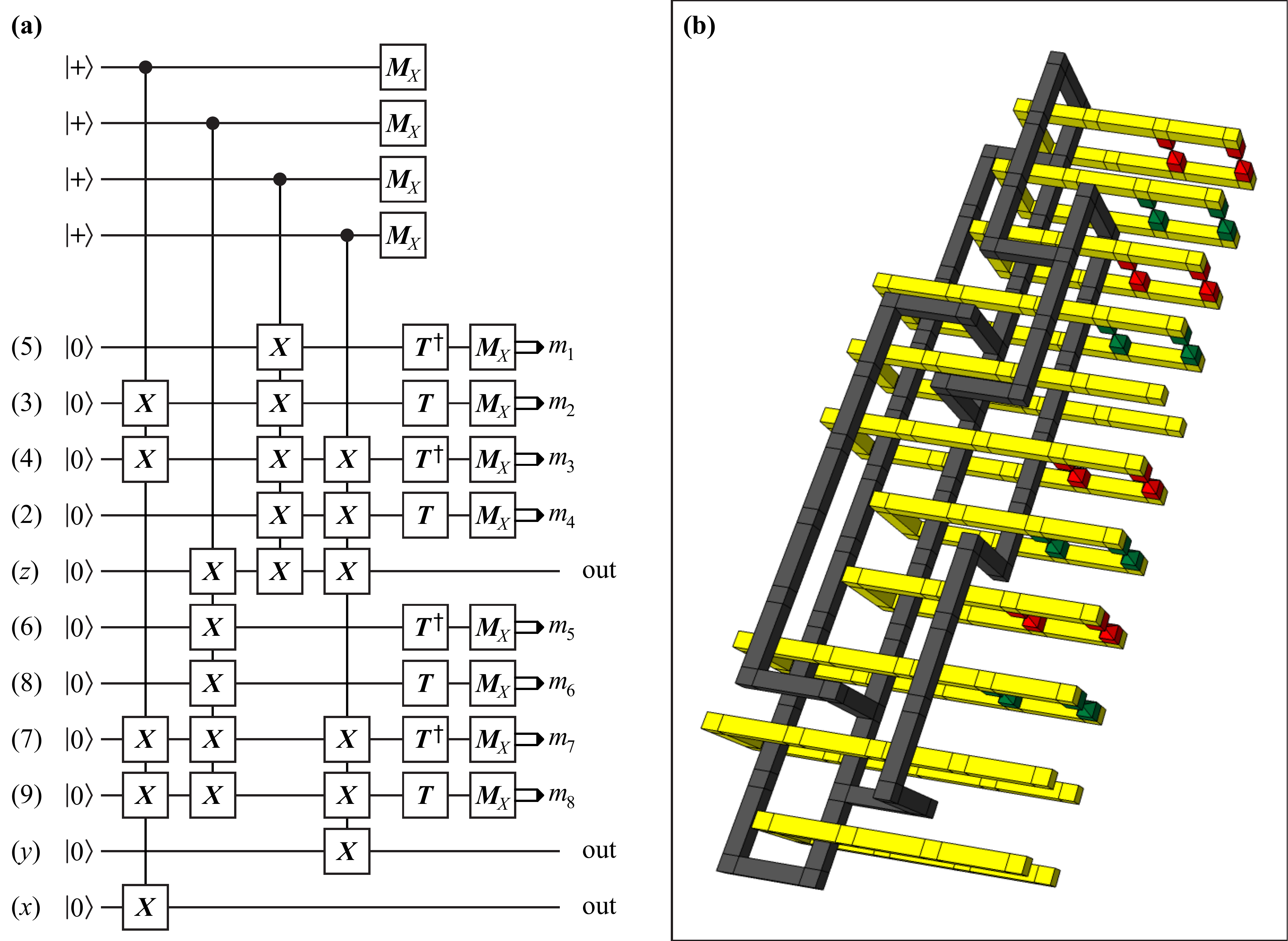}\\
  \caption[Circuit and Surface Code: Optimized D2 construction]{Circuit and corresponding surface code construction for D2.  (a)~Circuit that results from rearranging rows and modifying stabilizer measurements in Fig.~\ref{D2_stabilizer_circuit}.  (b)~Surface code design for D2.  The bounding box is $6 \times 12 \times 2 = 144$ plumbing pieces.}
  \label{D2_final}
\end{figure}

The circuit in Fig.~\ref{D2_stabilizer_circuit} is unchanged under two types of manipulation.  First, a single $X$-type stabilizer can be replaced with the product of itself and another stabilizer.  Because $X$ is Hermitian, the product operator has $X$ at qubits determined by the XOR of the original stabilizers.  Second, the order of qubits can be rearranged.  Denote the existing $X$-type stabilizers in Fig.~\ref{D2_stabilizer_circuit} as $S_1$,...,$S_4$.  Create a new set of stabilizers given by $(S_1 S_3 S_4, S_3, S_4, S_2 S_4)$ and reorder the rows according the labeling scheme in Fig.~\ref{D2_final}(a).  The row ordering was chosen to minimize surface code volume after trying many different combinations.  Under careful inspection, this circuit is equivalent to Fig.~\ref{D2_stabilizer_circuit}, with proper tracking of the stabilizer results.  A braiding pattern for D2 in the surface code is shown in Fig.~\ref{D2_final}(b), where horizontal yellow pipes correspond to qubits on the left, and $X$-type stabilizers are implemented with black pipes.  Since qubit (1) was converted to a dual braid, an error is detected if the total parity of the last $X$-type stabilizer and measurements $(m_2,\ldots,m_9)$ is odd.  The volume of D2 in Fig.~\ref{D2_final}(b) is $V_{\mathrm{D2}} = 144$ plumbing pieces.

\begin{figure}
  \centering
  \includegraphics[width=11cm]{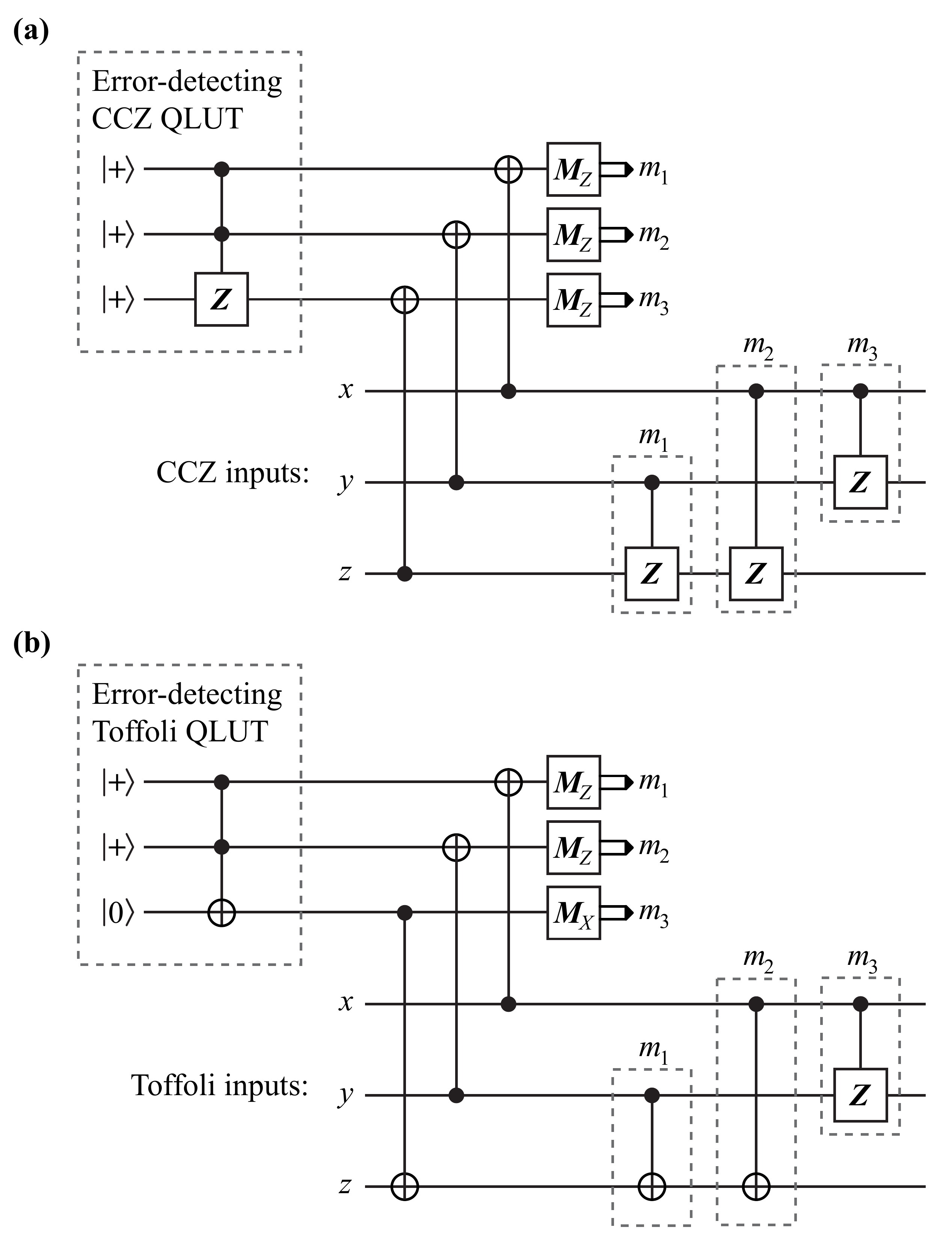}\\
  \caption[Circuit: CCZ and Toffoli gate using QLUT with teleportation]{Teleportation circuit for verified (a)~CCZ and (b)~Toffoli QLUTs.  The production of each QLUT is enclosed in the dashed box in upper left of each figure.  The QLUT could be formed using D2 or the C4C6 construction of Section~\ref{Sec_C4C6}.  QLUT production is probabilistic when error detection is used, so the QLUT couples to data only if no error is detected.  The subsequent CNOT gates and measurements teleport data qubits through the Toffoli gate encoded in the ancilla (\emph{cf.} p.~488 of Ref.~\cite{Nielsen2000}).  Clifford-group gates in dashed boxes are needed to correct errors from teleportation; they are conditioned on the measurements.  Corrections to the Pauli frame are also required but not shown.}
  \label{teleport_circuit}
\end{figure}

Because D2 only detects errors, it must be verified before interacting with data; otherwise, the data might be corrupted.  To avoid this scenario, one can produce a Toffoli QLUT, which is simply a Toffoli gate with $\ket{+}$ inputs to the controls and $\ket{0}$ input to the target~\cite{Shor1996,Preskill1998_FTQC,Nielsen2000}.  Since this chapter considers the CCZ gate, its QLUT is CCZ applied to three $\ket{+}$ qubits.  If the circuit fails because of a detected error, then the qubits are discarded, but no far-reaching damage occurs since this faulty circuit is not entangled to any data qubits.  Conditioned on the circuit succeeding, the QLUT enables teleportation of Toffoli (or CCZ) into data qubits, using only Clifford gates and measurement, as shown in Fig.~\ref{teleport_circuit}.  Using a representative value for $T$-gate error as $p = 10^{-8}$, the failure probability for creating the Toffoli ancilla with D2 construction is a modest $8 \times 10^{-8}$, which negligibly increases the number of times such preparation circuits must be repeated.

\section{Distance-Four Toffoli Gate (C4C6)}
\label{Sec_C4C6}
I present a second way to implement a verified Toffoli gate using an expandable code structure known as $C_4/C_6$~\cite{Knill2005}.  With respect to independent errors in $T$~gates, the distance of this design can be $2^t$ for any positive integer $t$.  The number of $T$~gates consumed is $8 \times 6^{t-1}$, though this may not be a meaningful measure.  In previous sections, I argued that $T$~gates were the dominant cost in quantum logic.  The results below increase the strength of error detection in the Toffoli so that, for sufficiently low values of gate error $p_g$, ``raw'' $T$~gates without distillation can replace distilled $T$~gates.  I focus on a distance-four construction that will be called ``C4C6'' (notation here is distinct from $C_4$/$C_6$, the underlying code family), but the technique can be generalized to detect more errors.

The $C_4$ and $C_6$ codes are four- and six-qubit CSS codes that each have distance two, meaning each can detect at least one error of arbitrary type~\cite{Knill2005}.  A $C_4$ code block encodes two logical qubits using four physical qubits at distance two.  A $C_6$ code block encodes two logical qubits using three pairs of qubits at distance $2d$, where $d$ is the distance for each of the three qubit pairs.  Beginning with $C_4$ at the lowest level, $C_6$ can be concatenated repeatedly, where each concatenation uses three independent pairs of qubits at a lower level.  With $t$ levels, meaning $t-1$ levels of $C_6$, the result is two logical qubits with distance $2^t$ using $4 \times 3^{t-1}$ physical qubits.  For more details on $C_4/C_6$, see the work by Knill~\cite{Knill2005}.  As a side note, the names for these codes has nothing to do with the $C_n$ set structure for unitary gates from the analysis by Gottesman and Chuang~\cite{Gottesman1999}.

\begin{figure}
  \centering
  \includegraphics[width=10cm]{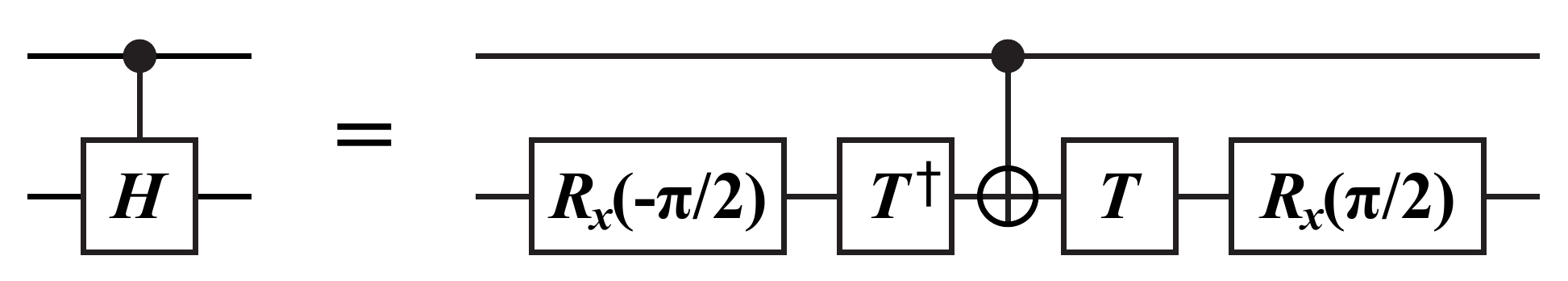}\\
  \caption[Circuit: Controlled-Hadamard from $T$ gates]{Circuit decomposition of Controlled Hadamard into $T$~gates and Clifford operations.}
  \label{Controlled_Hadamard_Circuit}
\end{figure}

The code $C_4/C_6$ has a transversal Hadamard operation at any number of levels of concatenation.  Applying transversal Hadamard at the physical level will implement Hadamard on both logical qubits, followed by a SWAP.  The strategy for C4C6 Toffoli will be to implement controlled-Hadamard, where the control is a bare qubit, on a $C_4/C_6$ block that encodes two logical qubits.  Controlled-Hadamard can be implemented with $T$~gates and Clifford gates, as shown in Fig.~\ref{Controlled_Hadamard_Circuit}.  Errors are detected with the code stabilizers, which is similar to the design of the distance-four, composite Toffoli in Ref.~\cite{Jones2013_Composite} and to techniques used in Section~\ref{Sec_Multilevel_Distillation}.  A Toffoli gate can be produced from two controlled-Hadamard gates and Clifford operations~\cite{Jones2013_Composite}, as explained below.

\subsection{Circuit Design for Controlled-Hadamard in $C_4/C_6$}
\begin{sidewaysfigure}
  \centering
  \includegraphics[width=\textwidth]{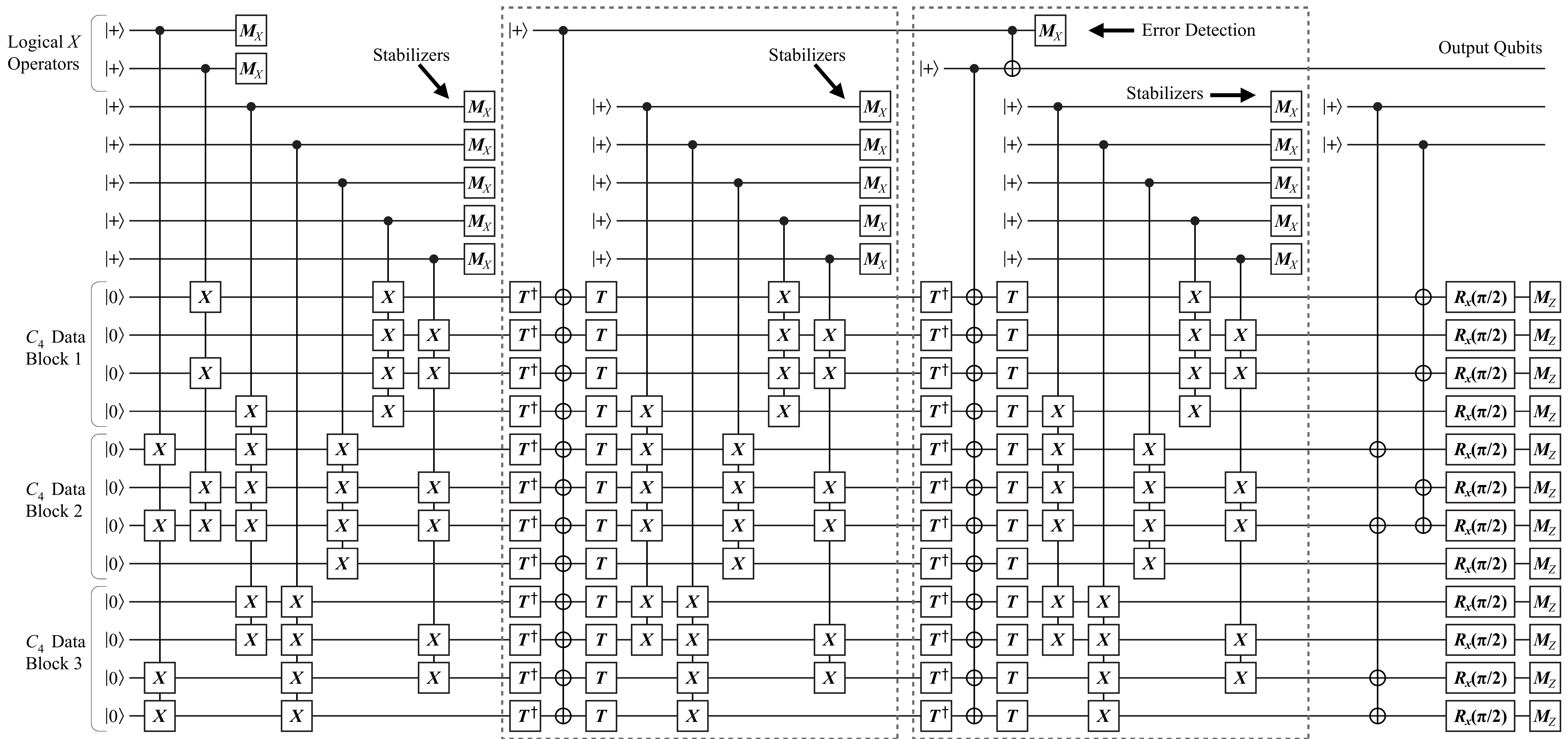}\\
  \caption[Circuit: QLUT for Toffoli from $C_4/C_6$-encoded Controlled-Hadamard]{Circuit to perform $C_4/C_6$-encoded controlled-Hadamard.  The control is a bare $\ket{+}$ qubit and the targets are two encoded $\ket{+}$ qubits.  The output state is a QLUT for a gate that differs from Toffoli by Clifford operations.  The first two CNOTs measure logical $X$ operators, initializing the encoded state in the $X$ basis.  Dashed boxes enclose repeated circuit elements consisting of stabilizer measurements and transversal controlled-Hadamard gates.  The Logical qubits are teleported out of the $C_4/C_6$ block.  The $R_x(-\pi/2)$ gates from Fig.~\ref{Controlled_Hadamard_Circuit} are absorbed into initialization.}
  \label{C4C6_Circuit}
\end{sidewaysfigure}

The core of the C4C6 construction implements controlled-Hadamard on a $C_4/C_6$ code block.  The operations are illustrated in the rather detailed circuit in Fig.~\ref{C4C6_Circuit}.  This circuit is designed with a surface code implementation in mind, which will come in Section~\ref{Sec_Compressed_Surface_Codes}.  As before, the $\ket{+}$-controlled $X$ operations denote stabilizers, whereas the CNOTs denote logic (in this case, controlled-Hadamard).  The circuit begins with 12 qubits initialized to $\ket{0}$.  The first two stabilizers are actually logical $X$ operators in the code, so their measurement in conjunction with measuring code stabilizers projects the 12 qubits into a logical $C_4/C_6$-encoded state in the $X$-basis.  The $T$-CNOT-$T^{\dag}$ (operator order, ``right comes first in time'') implements the core of the transversal controlled-Hadamard.  The $R_x(-\pi/2)$ gates are absorbed into initialization by flipping the sign of some $Z$ stabilizers.  At the end of the circuit, CNOTs followed by measurement teleport the encoded qubits to bare qubits, to facilitate the next step in processing.

The $C_4/C_6$ code has distance four with respect to $Z$ errors after the transversal $T$-CNOT-$T^{\dag}$.  However, if both $T$ and $T^{\dag}$ \emph{on the same qubit} have a $Z$ error, it will propagate to the bare $\ket{+}$ qubit, but the $Z$ errors cancel and are not detected by code stabilizers.  This is a weight-two error event, so it must be detected for C4C6 to have overall distance four.  By performing the $\ket{+}$-controlled Hadamard twice (see the two dashed boxes in Fig.~\ref{C4C6_Circuit}), error detection is implemented by comparing the phase on the two bare qubits with CNOT and $X$-basis measurement.  This a simple phase-flip code, and an error only passes the test if both bare qubits had $Z$ errors, which is now a weight-four error event.  Essentially the same technique was used to achieve high error distance for Hadamard-basis measurement in multilevel distillation in Section~\ref{Sec_Multilevel_Distillation}.

The analysis for output error probability in C4C6 resulting from $T$-gate errors is straightforward.  Within a $C_4$ block there are eight $T$/$T^{\dag}$ gates with independent error proability $p$, and each code block can always detect a single error.  The probability of a weight-two error appearing at the stabilizer measurement is $24p^2$.  There are $\binom{8}{2} = 28$ combinations of two errors, but four of these events are errors on the same qubit.  In the latter case, two errors on the same qubit cancel before stabilizer measurement, but the first error propagates to the bare qubit.  The $C_6$ code can detect an error in any $C_4$ code block, so the probability of a weight-four error not being detected is $\binom{3}{2}(24p^2)^2 = 1728p^4$.  Finally, there are 12 ways that a weight-two error can propagate to the bare qubit and not be detected by the code stabilizers, with total probability $12p^2$.  By repeating the transversal controlled-Hadamard, there is probability $144p^4$ for both bare qubits to be corrupted.  Since the $C_4/C_6$ error detection is performed twice with independent errors, the total error probability is $144p^4 + 2 \times 1728p^4 = 3600p^4$, to lowest non-vanishing order.  This approximation is valid when $p \ll 1$, and $p \le 10^{-2}$ works in practice.

\begin{figure}
  \centering
  \includegraphics[width=\textwidth]{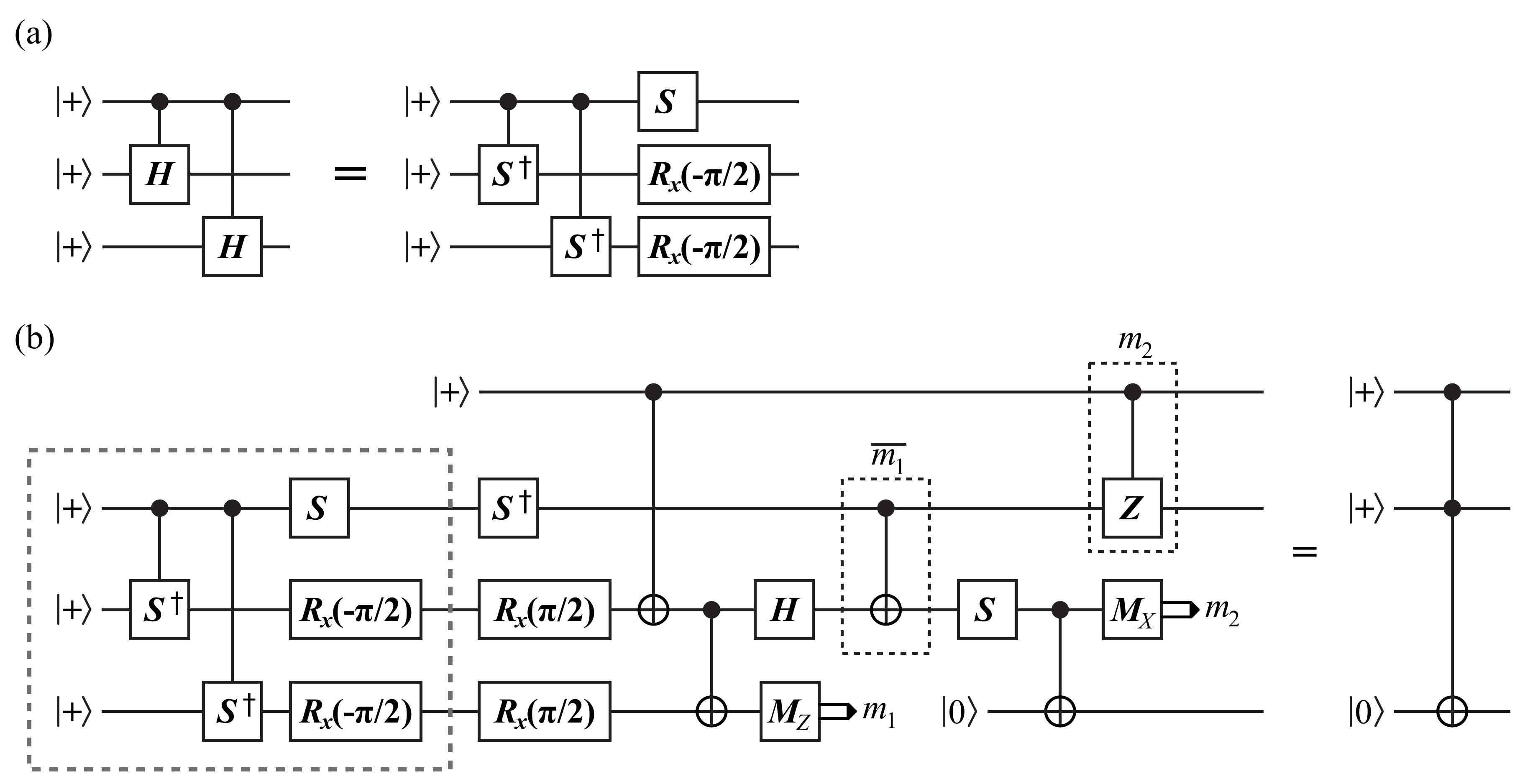}\\
  \caption[Circuit: Toffoli QLUT from Controlled-$H$]{Circuit diagram showing how to implement a Toffoli gate using the controlled-Hadamard operations produced by the C4C6 construction.  (a)~The output state of C4C6 is equivalent to controlled-$S^{\dag}$ gates and Clifford gates.  (b)~Toffoli QLUT is constructed with the output of C4C6 (dashed box) and Clifford gates with ancillas and measurement.  Two operations in dashed boxes are conditioned on measurement results, as indicated by $\overline{m_1}$ (gate is implemented if the measurement is $\ket{0}$) and $m_2$ (gate is implemented if the measurement result is $\ket{-}$).}
  \label{Toffoli_from_controlled_H}
\end{figure}

The output state of Fig.~\ref{C4C6_Circuit} can be used to make a Toffoli, with the aid of Clifford operations and measurement.  Figure~\ref{Toffoli_from_controlled_H} shows how to produce the Toffoli QLUT using the output of the C4C6 construction (controlled-Hadamard gates targeting with $\ket{+}$ qubits).  This QLUT can implement Toffoli using the circuit in Fig.~\ref{teleport_circuit}.

\subsection{Compressed Surface Codes and C4C6}
\label{Sec_Compressed_Surface_Codes}
Before proceeding with the surface code construction of C4C6, I introduce the notion of compressed surface codes.  $C_4/C_6$ has distance four with respect to all errors, which presents a novel opportunity.  In the surface code, defects are separated by distance $d$ to make logical errors highly improbable.  However, by concatenating surface code qubits with $C_4/C_6$, there is some redundancy in error correction.  Specifically, it is possible to relax the strength of error correction in the surface code if the errors will be caught by the next higher level of encoding.

For the compressed surface codes used to produce C4C6, the circumference of each defect will remain $4\lceil d/4 \rceil$, but the spacing between primal defects (yellow pipes) will be reduced to about $d/2$.  I will denote $d_1$ as the distance separating most operations in the surface code, except $d_2$ is the inter-defect spacing for qubits in the $C_4/C_6$ code block.  This modification makes logical $Z$ errors much more probable.  In particular, the logical errors are correlated, weight-two $Z$ errors on two qubits that are adjacent in the surface code.  These errors are still detectable because the $C_4/C_6$ code has distance four, as will be considered in Section~\ref{Sec_Compressed_Errors}.

Reducing the spacing between primal defects does introduce some complications.  Dual defects (black pipes) are now closer to the primal defects that they weave between, so care must be taken that there are no possible $X$ errors of weight less than $d$, which is a nontrivial path connecting primal defects.  By maintaining the circumference of the primal defects, dual errors which form a ring around primal defects have minimum distance $d$, so no new errors of this type are introduced.

\begin{figure}
  \centering
  \includegraphics[width=\textwidth]{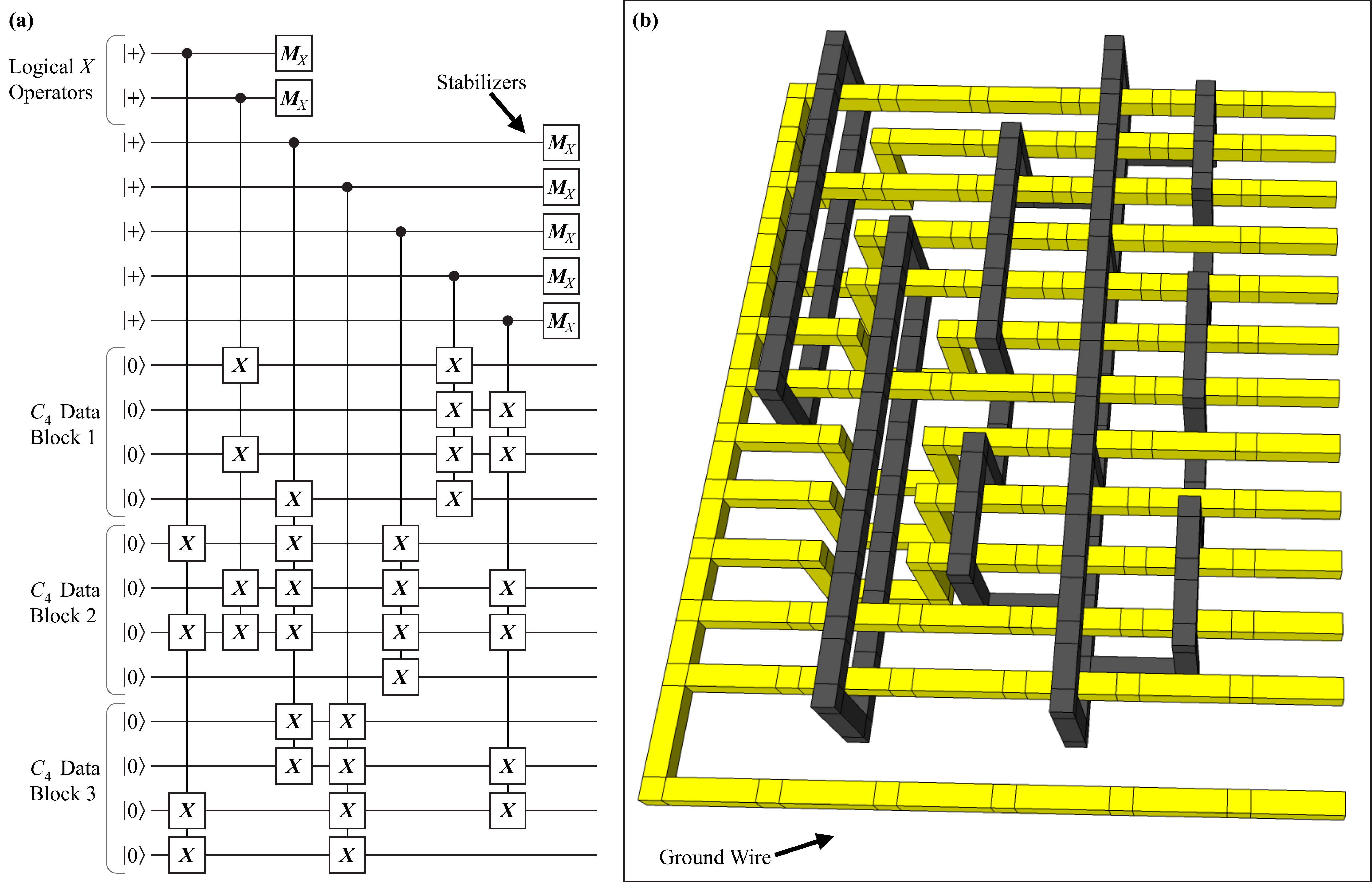}\\
  \caption[Circuit and Surface Code: Initialization of $C_4/C_6$-encoded $\ket{+}\ket{+}$]{Initialization of $C_4/C_6$ code block to logical $\ket{+}\ket{+}$.  (a)~Circuit diagram for initialization, which is the first component of Fig.~\ref{C4C6_Circuit}.  (b)~Compressed surface code braiding arrangement.  The first 12 primal defect rails correspond to the 12 $\ket{0}$ qubits in the circuit diagram.  The bottom defect rail is the ``ground wire,'' which forms the complementary defect to each of the other 12 primal rails.  The volume is $40.8 + 115.2r$ plumbing pieces, where $r < 1$ is the compression ratio.}
  \label{C4C6_init}
\end{figure}

An example of a compressed surface code program is shown in Fig.~\ref{C4C6_init}.  This program creates logical $\ket{+}\ket{+}$ encoded in the 12-qubit $C_4/C_6$ code, which is the first component of the C4C6 Toffoli construction.  The circuit diagram in Fig.~\ref{C4C6_init}(a) shows the 12 $\ket{0}$ qubits with two logical $X$ operators and the five $X$-basis code stabilizers being measured, just as in the beginning of Fig.~\ref{C4C6_Circuit}.  Fig.~\ref{C4C6_init}(b) depicts how this program is constructed in a compressed surface code.  There is direct correspondence between the top 12 horizontal, primal defect rails (yellow pipes) and the 12 $\ket{0}$ qubits on the left.  However, there is a new method for compressing space, which I call the ``ground wire.''  Instead of having 12 double-rail defects, the would-be lower rail for all primal defects has been combined into one rail at the bottom of Fig.~\ref{C4C6_init}(b).  This is topologically equivalent and saves space.  The name ground wire comes from the fact that this defect rail must eventually connect to each of the 12 defects at the end of the surface code program (when these logical qubits are measured), which is reminiscent of ground voltage in electrical circuits.  The volume of this compressed program is $V_{\mathrm{init}} = 6 \times (12(\frac{1+4r}{5}) + 1) \times 2 = 40.8 + 115.2r$ plumbing pieces.  The second dimension is the compressed one, and quantity $r$ is compression ratio: $d_2 = r \cdot d_1$.  Section~\ref{Sec_Compressed_Errors} will detail how $r$ is determined.

The program in Fig.~\ref{C4C6_init}(b) shows several novel features of compressed surface codes.  As advertised, the primal defects are spaced more closely, which reduces volume by about 35--40\%.  Also, the use of the ground wire frees space for primal defects to weave around dual rings, which means that any $X$-basis stabilizer can be measured in the depth of a single plumbing piece, as shown on left of the diagram.  The right implements more conventional braiding of dual defects around primal defects, because these particular stabilizer measurements have a compact implementation.

\begin{figure}
  \centering
  \includegraphics[width=\textwidth]{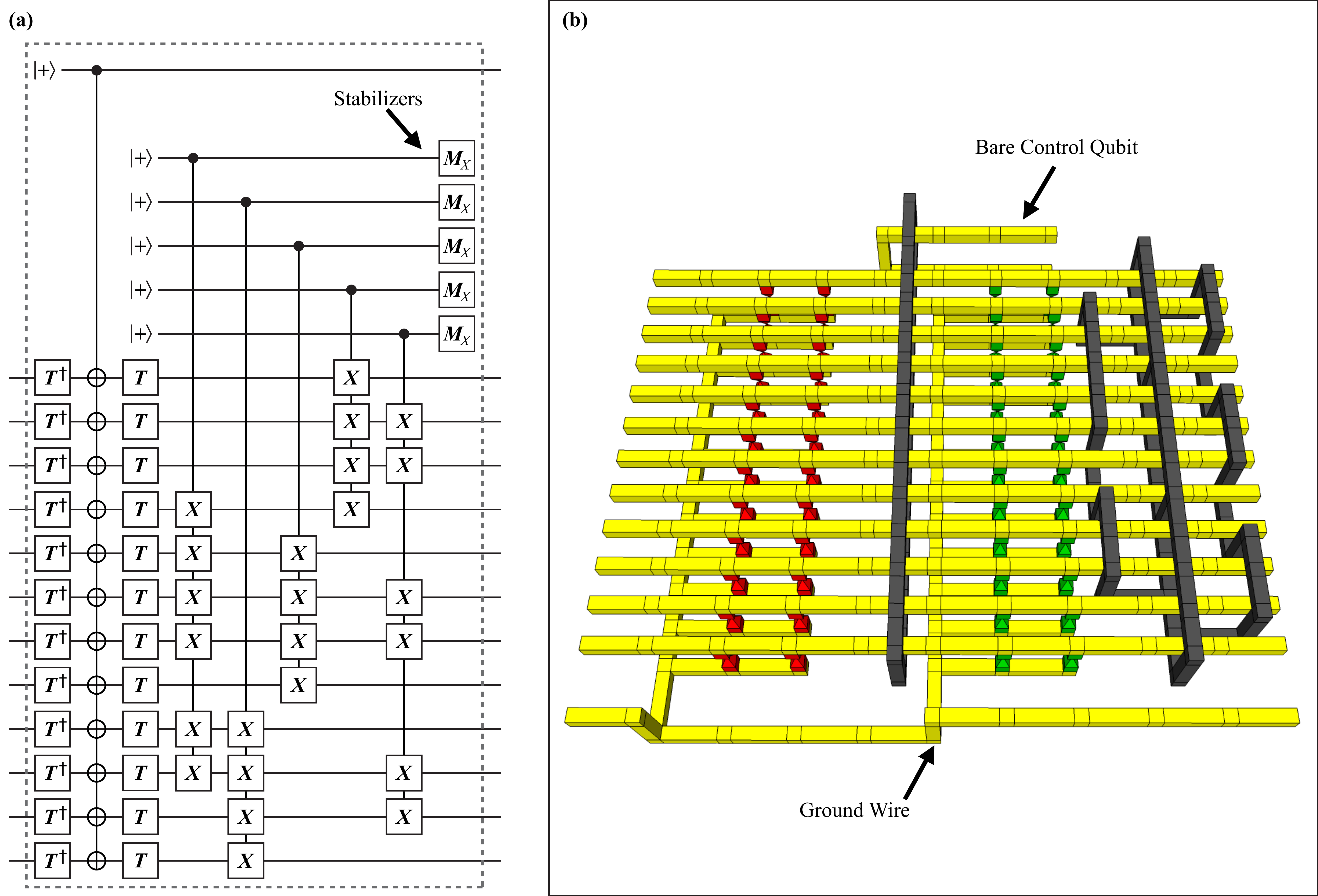}\\
  \caption[Circuit and Surface Code: Midsection of C4C6 construction]{Midsection of C4C6 construction where transversal $T$-CNOT-$T^{\dag}$ gates and error detection are implemented on the $C_4/C_6$ code block.  (a)~Circuit diagram for transversal operations and stabilizer measurements, which appears twice in the dashed boxes in Fig.~\ref{C4C6_Circuit}.  The $C_4/C_6$ stabilizers detect most of the errors.  The phase-flip code for the bare qubits detects the remaining errors of weight less than four, but this procedure is not shown because it is quite simple.  (b)~Compressed surface code braiding arrangement.  The top double-defect is the bare-qubit control for the transversal gates, and it has normal surface code spacing.  For simplicity, this top qubit does not connect to the ground wire.  The bottom defect rail is again the ground wire, which just passes through from the previous diagram in Fig.~\ref{C4C6_init}(b).  The volume is $79.2 + 172.8r$ plumbing pieces, where $r < 1$ is the compression ratio.}
  \label{C4C6_midsection}
\end{figure}

After initialization, the next step is to implement twice the transversal $T$-CNOT-$T^{\dag}$, followed by stabilizer measurements (shorthand: ``C4C6 midsection'').  Figure~\ref{C4C6_midsection} shows how to implement this sequence of operations in a compressed surface code.  The circuit in Fig.~\ref{C4C6_midsection}(a) appears twice in the entire circuit in Fig.~\ref{C4C6_Circuit}, so the surface code braiding in Fig.~\ref{C4C6_midsection}(b) is repeated as well.  The top of Fig.~\ref{C4C6_midsection}(b) shows the bare qubit which controls the transversal gates; this defect has normal surface code spacing because it is not embedded in a code.  The volume is $V_{\mathrm{mid}} = 9 \times (12(\frac{1+4r}{5}) + 2) \times 2 = 79.2 + 172.8r$ plumbing pieces for each instance of Fig.~\ref{C4C6_midsection}(b).

\begin{figure}
  \centering
  \includegraphics[width=\textwidth]{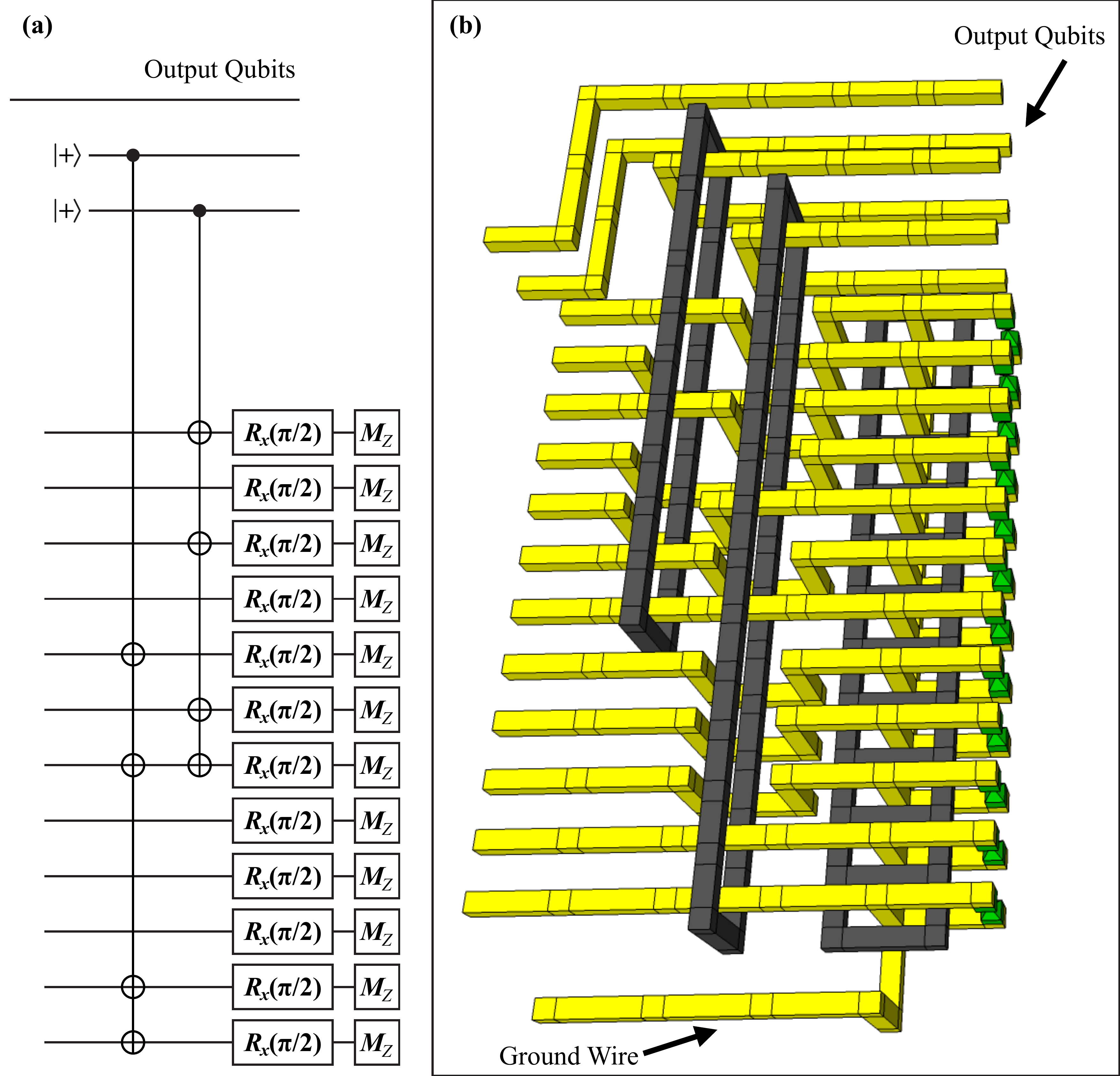}\\
  \caption[Circuit and Surface Code: Measurements for C4C6 construction]{Final teleportation and measure for C4C6 construction.  (a)~Final piece of Fig.~\ref{C4C6_Circuit}.  The CNOT gates will teleport logical qubits from the code block to bare qubits.  (b)~Compressed surface code implementation.  The final operations are $R_x(\pi/2)$ followed by $Z$-basis measurement.  The ground wire rejoins with the 12 defects in the $C_4/C_6$ block for measurement.  The top three double defects are output qubits, including one that was created in the preceding midsection step.  The volume is $64 + 96r$ plumbing pieces, where $r < 1$ is the compression ratio.}
  \label{C4C6_measurement}
\end{figure}

The final component of C4C6 is the teleportation and measurement step, as depicted in Fig.~\ref{C4C6_measurement}.  The CNOT gates in Fig.~\ref{C4C6_measurement}(a) are used to teleport logical qubits in the code block to bare qubits, which makes CCZ easier to implement.  The final step on the right performs the gate $R_x(\pi/2)$ using an ancilla followed by $Z$-basis measurement.  Because this is a Clifford operation, the ancilla can be produced temporally after the measurement by application of the Gottesman-Knill theorem~\cite{Nielsen2000}.  Figure~\ref{C4C6_measurement}(b) implements this sequence of operations in the compressed surface code.  Notably, the ground wire rejoins the other defects for the measurements.  The top of the diagram shows the three output qubits, including the bare qubit created in the previous ``midsection'' step.  The output qubits have normal surface code spacing because they are bare.  The volume is $V_{\mathrm{meas}} = 5 \times (12(\frac{1+4r}{5}) + 4) \times 2 = 64 + 96r$ plumbing pieces.

The compressed surface code construction of C4C6 is produced by combining all of the pieces in Figs.~\ref{C4C6_init}--\ref{C4C6_measurement}.  The total volume is approximately
\begin{eqnarray}
V_{\mathrm{total}}(r) & = & V_{\mathrm{init}} + 2 V_{\mathrm{mid}} + V_{\mathrm{meas}} \nonumber \\
& = & 263.2 + 556.8r
\end{eqnarray}
plumbing pieces.  The next section explains how to estimate error probability, which is used to determine $r$.

\subsection{Error Analysis in Compressed C4C6}
\label{Sec_Compressed_Errors}
New error events are introduced by compressing the spacing between primal defects.  Rigorous analysis of these errors has not yet been performed, because the analysis is complex.  Instead, I develop conservative bounds which should overestimate error.  I will refer frequently to Fig.~\ref{C4C6_midsection}(b), as this shows the core area of C4C6 where errors are a concern.  Compressed surface codes have two distance parameters: $d_1$, the longer distance which sets the pitch for all dual defects (black), as well as the temporal and vertical spacing for primal defects; and $d_2$, the compressed distance between primal defects.  When $d_2 = d_1$, the result is the typical defect spacing.  However, it is desirable to set $d_2 = r \cdot d_1$, where $r < 1$, to reduce volume.  Doing so makes some $Z$-type errors more likely than other logical errors in the surface code.

In particular, the errors made more likely by the compressed C4C6 construction are correlated weight-two $Z_i Z_{i+1}$ errors, where subscript refers to a qubit in the $C_4/C_6$ code block and $i \in [1,11]$.  Because of the linear arrangement, only nearest-neighbor correlated errors have increased likelihood.  Furthermore, separating the ground wire from the rest of the primal defects substantially simplifies analysis by making these weight-two errors the only ones requiring consideration.  If all of the errors considered here are weight-two, they cannot propagate to the bare qubit which controls the $T$-CNOT-$T^{\dag}$ operation.

Since $C_4/C_6$ is a distance-four code, it can always detect one of these weight-two errors.  The failure probability of C4C6 will depend on the number of ways two such weight-two errors can lead to undetected logical error; fortunately, this number is significantly less than all possible combinations of two correlated errors.  In particular, there must be errors in two different $C_4$ blocks, for which there are only 27 combinations.  Two errors in the same block either cancel (modulo the stabilizers) or lead to another weight-two error, which is detectable.  There are three nearest-neighbor error events per block, and $\binom{3}{2} = 3$ ways to pick two out of three blocks, leading to $3 \times 3^2 = 27$ combinations.

Examining Fig.~\ref{C4C6_midsection}(b) from the side, the total length over which such errors could occur is $13d_1$, which consists of $9d_1$ at the first level and $4d_1$ at the second level for $T$/$T^{\dag}$ gates.  The modified probability of error for the compressed surface code will be
\begin{equation}
P_{L,C}(p_g,d_1,d_2) = \frac{d_1}{3}\left(100 p_g\right)^{-(d_2+1)/2},
\label{Compressed_Surface_Code_Error}
\end{equation}
where dividing by 3 accounts for only $Z$ errors and subscript $C$ indicates that this function applies to compressed surface codes. The expression in Eqn.~(\ref{Compressed_Surface_Code_Error}) is a conservative bound, following the derivation in Ref.~\cite{Fowler2013_BlockCodes}.

A logical error could also occur via one weight-two correlated error, as just discussed, and two $T$-gate errors.  Both $T$-gate errors must occur before or after the CNOT, or else they will be detected by the phase-flip code on the bare qubits.  There are 288 possible combinations.  There are six ways to assign correlated error and $T$-gate errors to different blocks, three ways to have a correlated error in one block, and 12 different ways for $T$-gate errors to happen in the same block, with both occurring before or after the CNOT ($6 \times 3 \times 12 = 216$).  There are two ways for a correlated error to happen between two blocks, which must be matched with a $T$-gate error in each of those blocks, having six combinations each ($2 \times 6 \times 6 = 72$).  The total number of combinations is 288.

All of the errors can now be combined into one expression.  The volume of C4C6 was estimated in Section~\ref{Sec_Compressed_Surface_Codes} to be $V_{\mathrm{total}}(r) = 263.2 + 556.8r$.  The total error probability is
\begin{eqnarray}
P_{\mathrm{total}} & = & 2\left[27\left(13 P_{L,C}(p_g,d_1,d_2)\right)^2 + 288 {p_T}^2\left(13 P_{L,C}(p_g,d_1,d_2)\right)\right] \nonumber \\
& & + V_{\mathrm{total}}(r) P_L(p_g,d_1) + 3600 {p_T}^4,
\label{C4C6_total_error}
\end{eqnarray}
where $p_T$ is the probability of $T$-gate error.  The logical error per plumbing piece, $P_L(p_g,d)$, is defined in Eqn.~(\ref{Surface_Code_Error}), repeated here for convenience:
\begin{equation}
P_L(p_g,d) \approx d (100 p_g)^{(d+1)/2}.
\label{Surface_Code_Error_Repeat}
\end{equation}
The terms in Eqn.~(\ref{C4C6_total_error}) correspond to, in order: probability of two correlated errors, due to compression; probability of one correlated error and two $T$-gate errors; probability of undetected error of weight $d_1$ in the surface code; and probability of four $T$-gate errors.  The factor two at the beginning accounts for the midsection being implemented twice.  In the numerical search described below, $r \approx 0.6$ in efficient constructions of C4C6, so the volume becomes about 585 plumbing pieces.  I should also note that the volume has an overhead factor of $1/(1-p_{\mathrm{fail}})$, where $p_{\mathrm{fail}}$ is the probability of detecting an error, which causes C4C6 to fail.  Although many types of errors are listed above, the failure probability is dominated by the likelihood of just a single $T$-gate failure, so $p_{\mathrm{fail}} \approx 48 p_T$.

\section{Resource Analysis for Toffoli Gates}
\label{Sec_Toffoli_analysis}
\begin{figure}
  \centering
  \includegraphics[width=\textwidth]{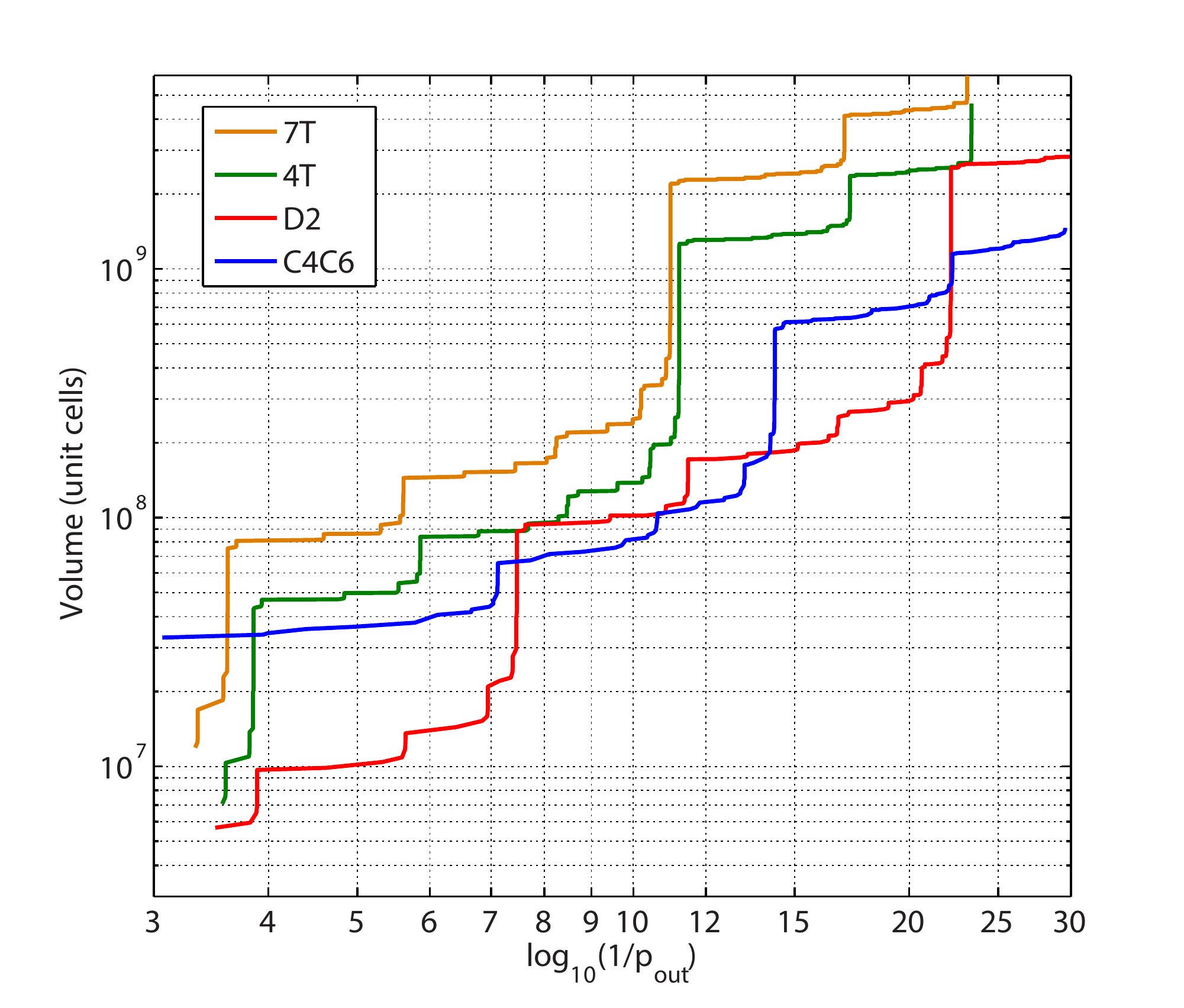}\\
  \caption[Resource costs for Toffoli gates in surface code]{Resource costs for Toffoli constructions with surface code error correction.  The error per physical gate is $p_g = 10^{-3}$, and each $T$~gate has error $p_{\mathrm{in}} = 10p_g = 10^{-2}$.  A unit cell in the surface code is the size of two stabilizer measurements, which encompasses two qubits (with an ancilla for measurement) and two stabilizer time steps.}
  \label{Toffoli_protocol_resources}
\end{figure}

Having constructed four different versions of the Toffoli gate, I will now compare their resource costs in the surface code, as a function of logical error probability.  The Toffoli gates and their properties are summarized in Table~\ref{Toffoli_summary}.  Using the analysis in preceding sections of this chapter and the resource analysis for $T$~gates in Chapter~\ref{Ch05}, I calculated the resource costs for each of the four Toffoli constructions.  As before, only efficient protocols are reported.  One protocol dominates another if the first has both lower volume and lower output error, and an efficient protocol is not dominated by any other.  The resource costs for $p_g = 10^{-3}$ are plotted in Fig.~\ref{Toffoli_protocol_resources}.

\begin{table}
  \centering
  \begin{tabular}{c c c m{8cm}}
    \toprule
    Label & Volume & $T$ Gates & Error Detection Properties \\
    \midrule
    7T & 154 & 7 & None \\
    4T & 126 & 4 & None \\
    D2 & 144 & 8 & Distance-two (detects one $T$-gate error) \\
    C4C6 & $263.2 + 556.8r$ & 48 & Distance-four (detects three $T$-gate errors); also detects some logical errors \\

    \bottomrule
  \end{tabular}
  \caption[Summary of Toffoli Gates]{Summary of Toffoli Gates}
  \label{Toffoli_summary}
\end{table}

Although C4C6 has impressive results, I hesitate to recommend its use at this time.  For one thing, C4C6 is proposed herein for the first time and has not yet passed through independent analysis.  Compressed surface codes play a role in its efficiency, and there may yet be an overlooked flaw there.  Moreover, when C4C6 beats D2, it outperforms by just a slim margin.  Therefore, I recommend the use of D2 until more careful analysis of compressed surface codes has been performed, like the work in Refs.~\cite{Wang2010,Fowler2010_Error,Fowler2012_Practical,Fowler2012_Timing}.  The design of D2 has been peer-reviewed in two independent publications from myself~\cite{Jones2013_Toffoli} and Eastin~\cite{Eastin2013_Toffoli}.  D2 is also simpler to examine and has a smaller volume, making it easier to reposition in a larger quantum program.  This prescription is conservative, because logic synthesis is a fairly new area of study, but it may change with further analysis of advanced Toffoli designs.

\begin{figure}
  \centering
  \includegraphics[width=\textwidth]{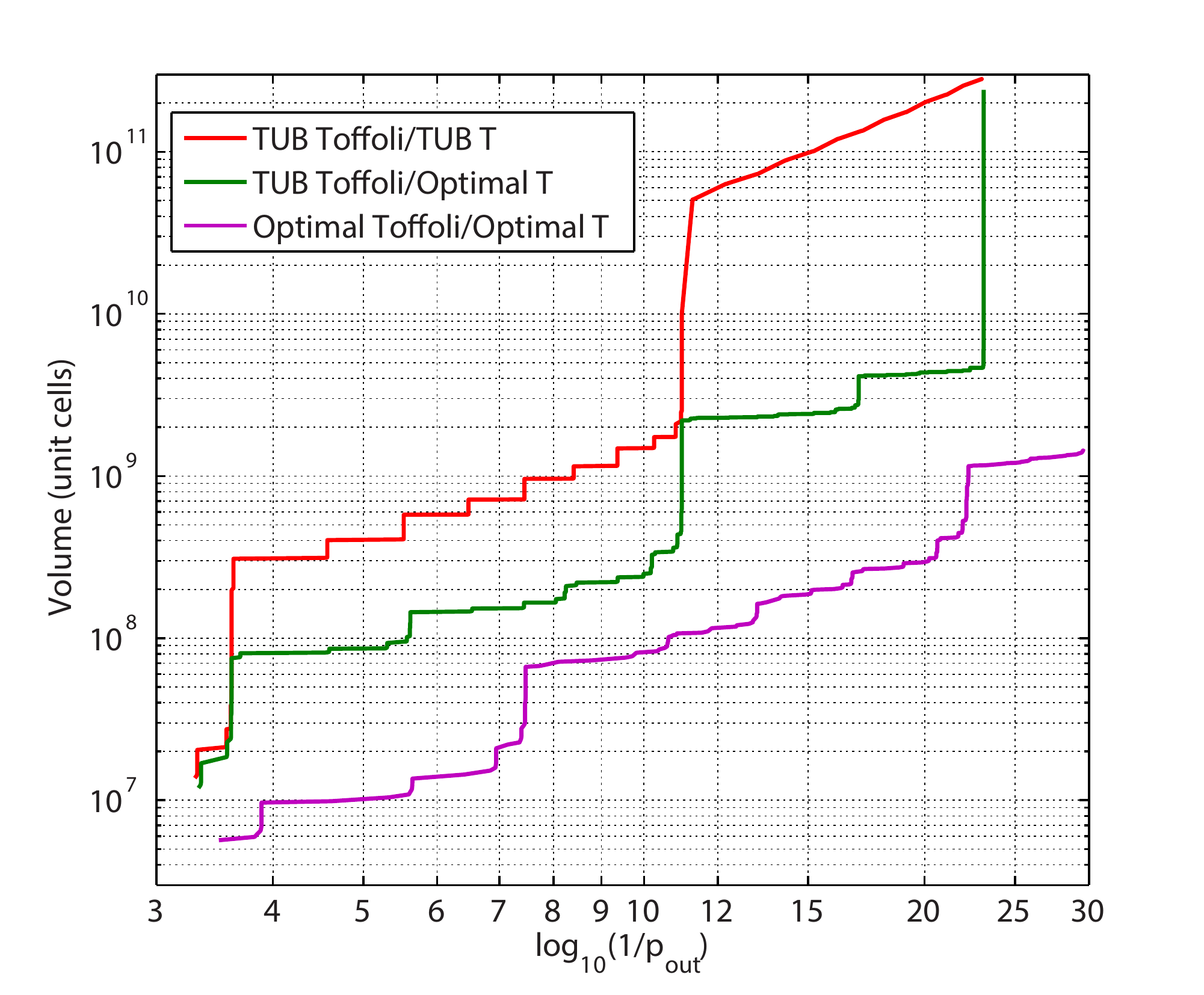}\\
  \caption[Toffoli resource costs for TUB and optimal designs]{Resource comparison for different levels of logic synthesis.  The TUB Toffoli is the 7T construction, and the TUB for $T$~gates is Bravyi-Kitaev distillation with all levels using the same code distance.}
  \label{Toffoli_TUB_comparison}
\end{figure}

As a concluding remark, the importance of logic synthesis becomes apparent when one calculates the resource costs for naive Toffoli constructions.  The 7T construction will serve as the Trivial Upper Bound (TUB) for Toffoli constructions.  Moreover, the Toffoli gates considered here depend on $T$~gates, so I will also consider what the costs will be if one uses the TUB for Bravyi-Kitaev distillation from Section~\ref{Sec_BK_resources}.  Figure~\ref{Toffoli_TUB_comparison} plots the resource costs for three different design approaches to the Toffoli gate, where again $p_g = 10^{-3}$.  The first (red) uses TUB for both $T$~gates and Toffoli.  The second (green) optimizes $T$~gates, but uses the naive 7T Toffoli.  The third (purple) uses optimized $T$~gates and the best Toffoli construction, which is the minimal resource cost over all methods plotted in Fig.~\ref{Toffoli_protocol_resources} for a given $p_{\mathrm{out}}$.  The improvement is significant for logical error rates relevant to quantum computing.  When $p_{\mathrm{out}} = 10^{-12}$, the optimized Toffoli (which is C4C6) beats the naive Toffoli by a factor of 20 using the same optimized $T$~gates.  The optimal design beats the TUB with both naive $T$ and Toffoli methods a factor of 500.  The simple resource analysis from Chapter~\ref{Ch02} and Ref.~\cite{Jones2012_PRX} used this TUB, showing that substantial reduction in resources can be achieved by applying logic synthesis to Toffoli gates.

\chapter{Approximating Phase Rotations}
\label{Ch07}
Earlier chapters established that some quantum gates are harder than others, when error correction is taken into account.  In the surface code, only a finite subset of gates are native to the code, making them easy to produce.  By contrast, a quantum algorithm could require any arbitrary unitary operation as a subroutine.  Such arbitrary gates must be approximated using combinations of easy and hard gates.  In this chapter, I will focus on phase rotations, an important class of hard operations that appears frequently in quantum algorithms.  Any arbitrary operation could be constructed from ``easy'' gates and phase rotations, but this is seldom required.  Each phase rotation could in principle be specified to infinite precision (such as an irrational angle), so one must approximate these gates, and it is crucial to understand how the resource costs increase as a function of approximation accuracy.

Phase rotations are any diagonal unitary operations, such as rotations around the $Z$ axis of the single-qubit Bloch sphere.  They appear ubiquitously in quantum simulation, as discussed in Chapter~\ref{Ch08}.  They also appear in the quantum Fourier transform and its approximate version~\cite{Barenco1996,Nielsen2000,Cleve2000,Weinstein2001}.  Consequently, any quantum algorithm that uses phase estimation requires phase rotations; examples include simulating energy eigenvalues in chemistry~\cite{Lloyd1996,Aspuru2005,Kassal2008,Kassal2011,Jones2012_NJP}, Shor's factoring algorithm~\cite{Shor1995,Isailovic2008,VanMeter2005,VanMeter2009,Jones2012_PRX}, and counting solutions in Grover's algorithm~\cite{Nielsen2000}.  Inspired by the importance of these gates, recent research has dramatically lowered their resource costs in comparison to the earliest proposals.  This chapter examines multiple schemes for producing phase rotations, but an exhaustive list is impractical in an active research field.  Several of the newer methods were devised using concepts from logic synthesis, which suggests the possibility of developing more quantum programs this way.

\begin{figure}
  \centering
  \includegraphics[width=10cm]{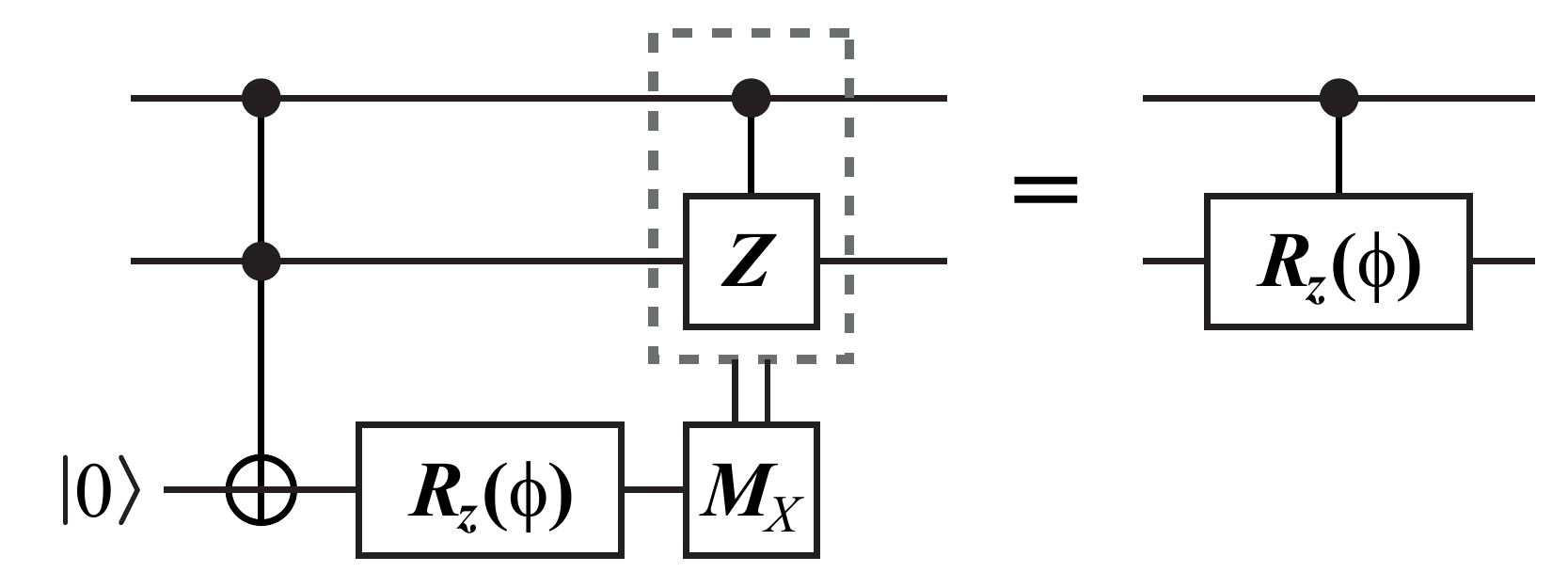}\\
  \caption[Circuit: Controlled rotation using Toffoli gate]{Circuit for implementing a controlled rotation using a Toffoli gate, a single-qubit rotation, and a $\ket{0}$ ancilla with quantum erasure.}
  \label{controlled_rotation}
\end{figure}

As a starting point, consider a single-qubit phase rotation $R_Z(\phi)$ defined as
\begin{equation}
R_Z(\phi) = \exp(i \phi(I-Z)/2) = \ket{0}\bra{0} + e^{i\phi}\ket{1}\bra{1}.
\label{single_qubit_rotation}
\end{equation}
Rotations on a single qubit are the simplest to examine, and they can be used to produce arbitrary rotations on larger Hilbert spaces.  For example, a controlled rotation on two qubits, defined as $CR_Z(\phi) = \ket{00}\bra{00} + \ket{01}\bra{01} + \ket{10}\bra{10} + e^{i\phi}\ket{11}\bra{11}$, can be implemented by the circuit in Fig.~\ref{controlled_rotation}.  This construction uses one Toffoli gate, one single-qubit phase rotation, and an ancilla erased by teleportation.

Because rotations are being approximated, I will quantify the distance between unitary gates using the distance proposed by Fowler in Ref.~\cite{Fowler2011_ArbitraryGates}:
\begin{equation}
\epsilon_F(A,B) = \sqrt{1-\frac{1}{2}\left|\mathrm{tr}(A^{\dag}B)\right|}.
\label{Fowler_distance}
\end{equation}
For example, $\epsilon_F(U,\tilde{U})$ is the distance between an ideal gate $U$ and some fault-tolerant sequence $\tilde{U}$ that approximates $U$.  Note that the denominator 2 applies to single-qubit gates and should be replaced with $d = \mathrm{tr}(I)$ for $d$-dimensional operators.  This quantity ranges $0 \le \epsilon_F \le 1$, achieving zero if and only if $U$ and $\tilde{U}$ are the same unitary, to within a meaningless complex factor.  Furthermore, it obeys the triangle inequality: $\epsilon_F(A,B) + \epsilon_F(B,C) \ge \epsilon_F(A,C)$~\cite{Fowler2011_ArbitraryGates}.  Achieving better accuracy (lower $\epsilon_F$) increases the resource costs of the approximation technique.  Moreover, these quantum programs are also susceptible to logical error.  After introducing several methods, I will compare their resource costs as functions of output error rate.

\section{Approximation Sequences}
\label{Sec_Approximation_Sequences}
As the name implies, approximation sequences are sequences of fundamental operations which approximate some arbitrary quantum gate. Fundamental here means that the operation has a fault-tolerant construction, and in practice the set of fundamental operations is finite, such as the Clifford group augmented with $T$~gates or Toffoli gates.  The task of enabling fault-tolerant, universal quantum computing is why so much attention was given in Chapters~\ref{Ch05} and~\ref{Ch06} to $T$ and Toffoli, operations which are costly to make fault-tolerant.

Determining an approximation sequence requires a classical algorithm.  Several have been proposed~\cite{Dawson2005,Fowler2011_ArbitraryGates,Duclos2012,Selinger2012,Kliuchnikov2012,Giles2013,Bocharov2013,Amy2013}, with varying tradeoffs in classical search time and quantum resources~\cite{Jones2012_NJP}.  This chapter will consider two general approaches: sequences that use $T$~gates, and sequences that use Toffoli gates.

As a historical aside, the Solovay-Kitaev algorithm~\cite{Nielsen2000,Dawson2005} was the first proposal for synthesizing arbitrary single-qubit gates from a finite set.  Although it has poly-logarithmic cost, the gate sequences are far from optimal.  Reference~\cite{Jones2012_NJP} shows that Solovay-Kitaev has cost in gates that is orders of magnitude greater than the competing methods discussed here.  Hence I will not analyze Solovay-Kitaev constructions in this chapter, because they are not competitive and should not be used.

\subsection{Approximation Sequences using $T$ Gates}
Several investigators have studied methods for approximating arbitrary gates (especially single-qubit gates) using the finite set of gates consisting of the Clifford group and $T = \exp(i \pi (I-Z)/8)$~gates.  As discussed in Chapter~\ref{Ch05}, $T$~gates are a popular choice for enabling universal quantum computing because they can be purified efficiently.

Fowler proposed a method to produce minimal-cost approximation sequences through an exhaustive search~\cite{Fowler2011_ArbitraryGates}.  The finite set of operations are the Clifford group and $T$~gates.  Because the search space is growing exponentially with sequence length, the classical algorithm has exponential time complexity.  Still, the method is interesting because it produces sequence having the minimum number of $T$~gates, and Chapter~\ref{Ch05} established that these operations are one of the main resource costs in fault-tolerant quantum computing.

Kliuchnikov, Maslov, and Mosca (KMM) have discovered a classical algorithm that achieves the performance of Fowler's algorithm for single-qubit phase rotations while having tractable computation time~\cite{Kliuchnikov2012}.  Their benchmarking results determine, through data fitting, that a gate with accuracy $\epsilon_F$ requires $10.7\log_{10} \epsilon_F - 23.0$ $T$~gates.  I will use this formula for resource estimates later.

\subsection{Approximation Sequences using $V$ Gates}
\begin{figure}
  \centering
  \includegraphics[width=8cm]{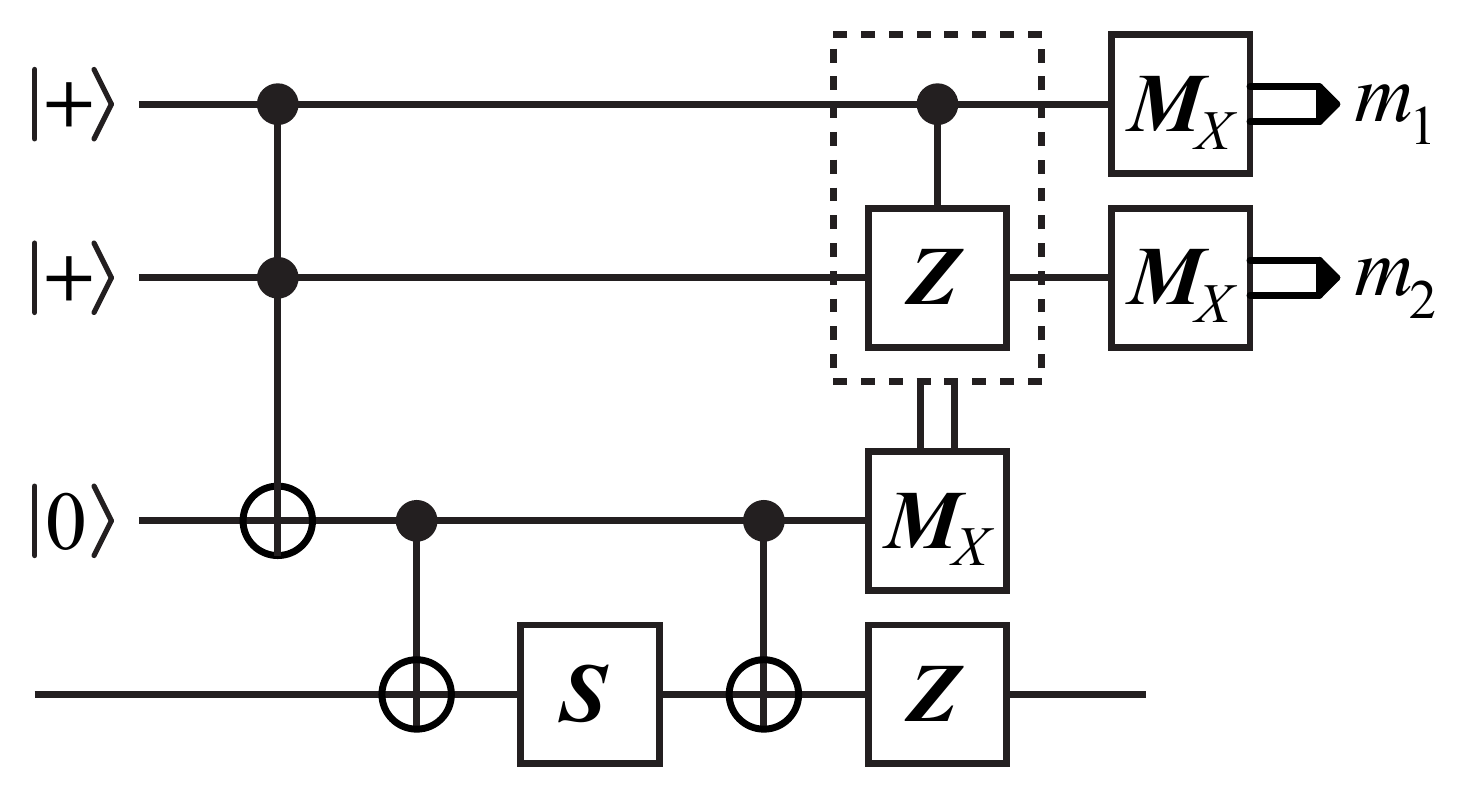}\\
  \caption[Circuit: $V_Z$ rotation from Toffoli]{Circuit diagram for implementing $V_Z$ using a Toffoli QLUT (see Chapter~\ref{Ch06}).  The circuit is probabilistic and implements $V_Z$ if both measurement results $m_1$ and $m_2$ are $\ket{+}$, which happens with 5/8 probability.  Otherwise, the circuit implements identity.  The $Z$ gate would be implicitly implemented using a Pauli frame (Section~\ref{Pauli_frames}).}
  \label{V_gate_circuit}
\end{figure}

A recent proposal for composing approximations from so-called $V$-basis rotations merits consideration~\cite{Bocharov2013}.  The $V$-basis rotations are defined as:
\begin{equation}
V_X = \frac{I + 2i X}{\sqrt{5}};
\end{equation}
\begin{equation}
V_Y = \frac{I + 2i Y}{\sqrt{5}};
\end{equation}
\begin{equation}
V_Z = \frac{I + 2i Z}{\sqrt{5}}.
\end{equation}
Several ways of producing a $V$-basis rotation have been proposed, but I offer another in Fig.~\ref{V_gate_circuit}.  This circuit produces $V_Z$ using one Toffoli gate, and it succeeds if both measurement results are $\ket{+}$, which happens with probability 5/8.  Rotations $V_X$ and $V_Y$ can be produced with the addition of Clifford gates; the inverse rotations only require substitution of $S^{\dag}$~gate for $S$.  Therefore, this construction has an average cost per $V$~gate of 1.6 Toffoli gates, which is the best I am aware of.  Alternative constructions are given in Ref.~\cite{Bocharov2013}.

Bocharov, Gurevich, and Svore (BGS) present a method which finds both optimal and near-optimal $V$-basis sequences, with tradeoffs in the classical complexity to determine said sequences~\cite{Bocharov2013}.  They suggest that the number of $V$~gates required should be $3\log_5 (1/\epsilon_F) = 4.29\log_{10}(1/\epsilon_F)$ in the ``optimal'' case; from their simulation results, I derive a fit of $4.14\log_{10}(1/\epsilon_F) - 0.07$ $V$~gates.  I will use the numerical results for a consistent comparison with KMM sequences.  Incorporating the factor of 1.6 Toffoli gates for each probabilistic $V$-gate construction, I estimate that each BGS construction requires $6.62\log_{10}(1/\epsilon_F) - 0.11$ Toffoli gates.

\section{Phase Kickback}
\label{Sec_Phase_Kickback}
Phase kickback~\cite{Kitaev1995,Cleve1998}, also known as the Kitaev-Shen-Vyalyi algorithm~\cite{Kitaev2002}, is an ancilla-based scheme that uses an addition circuit to impart a phase to a quantum register.  Phase kickback relies on a resource state $\ket{\gamma^{(k)}}$ which can be defined by the quantum Fourier transform (QFT)~\cite{Nielsen2000,Cleve2000,Weinstein2001}:
\begin{equation}
\label{QFT}
\ket{\gamma^{(k)}} = {U_{\mathrm{QFT}}}\ket{k} = \frac{1}{\sqrt{N}}\sum_{y=0}^{N-1}e^{i 2\pi ky/N}\ket{y}.
\end{equation}
Note that this is the sign convention of Ref.~\cite{Nielsen2000}, and opposite of that in Ref.~\cite{Jones2012_NJP}.  The register $\ket{k}$ contains $n$ qubits prepared in the binary representation of $k$, an odd integer.  The state $\ket{\gamma^{(k)}}$ is a uniform-weighted superposition state containing the ring of integers from $0$ to $N-1$, where $N = 2^n$, and each computational basis state has a relative phase proportional to the equivalent binary value of that basis state.  Section~\ref{Sec_FS_Distillation} provides a method to distill $\ket{\gamma^{(k)}}$ using addition circuits, which are composed of Toffoli gates.  This approach is an improvement over the phase-estimation procedure in Ref.~\cite{Kitaev2002}.

The register $\ket{\gamma^{(k)}}$ is a Fourier state, as introduced in Section~\ref{Sec_FS_Distillation}.  A notable property of Fourier states is that they are eigenstates of the modular addition operator $U_{\oplus u}\ket{x} = \ket{x+u \; (\mathrm{mod} \; N)}$:
\begin{equation}
U_{\oplus u}\ket{\gamma^{(k)}} = \frac{1}{\sqrt{N}}\sum_{y=0}^{N-1}e^{i 2\pi k(y-u)/N}\ket{y} = e^{-i 2\pi ku/N}\ket{\gamma^{(k)}},
\label{PK_definition}
\end{equation}
where $\oplus$ denotes addition modulo $N$ and $u$ is an integer.  Moreover, the eigenvalue of modular addition on $\ket{\gamma^{(k)}}$ is a phase factor proportional to the number $u$ added.  Note that the addition operation $U_{\oplus u}$ is readily implemented with a fault-tolerant quantum circuit~\cite{Vedral1996,Draper2000,VanMeter2005,Cuccaro2004,Draper2006}.  It is now clear how the method received its name: since $\ket{\gamma^{(k)}}$ is an eigenstate of addition, when an integer $u$ is added (using an addition circuit) to this register, a phase is ``kicked back.''

\subsection{Single-Qubit Rotation}
\label{Sec_PK_single}
\begin{figure}
  \centering
  \includegraphics[width=8cm]{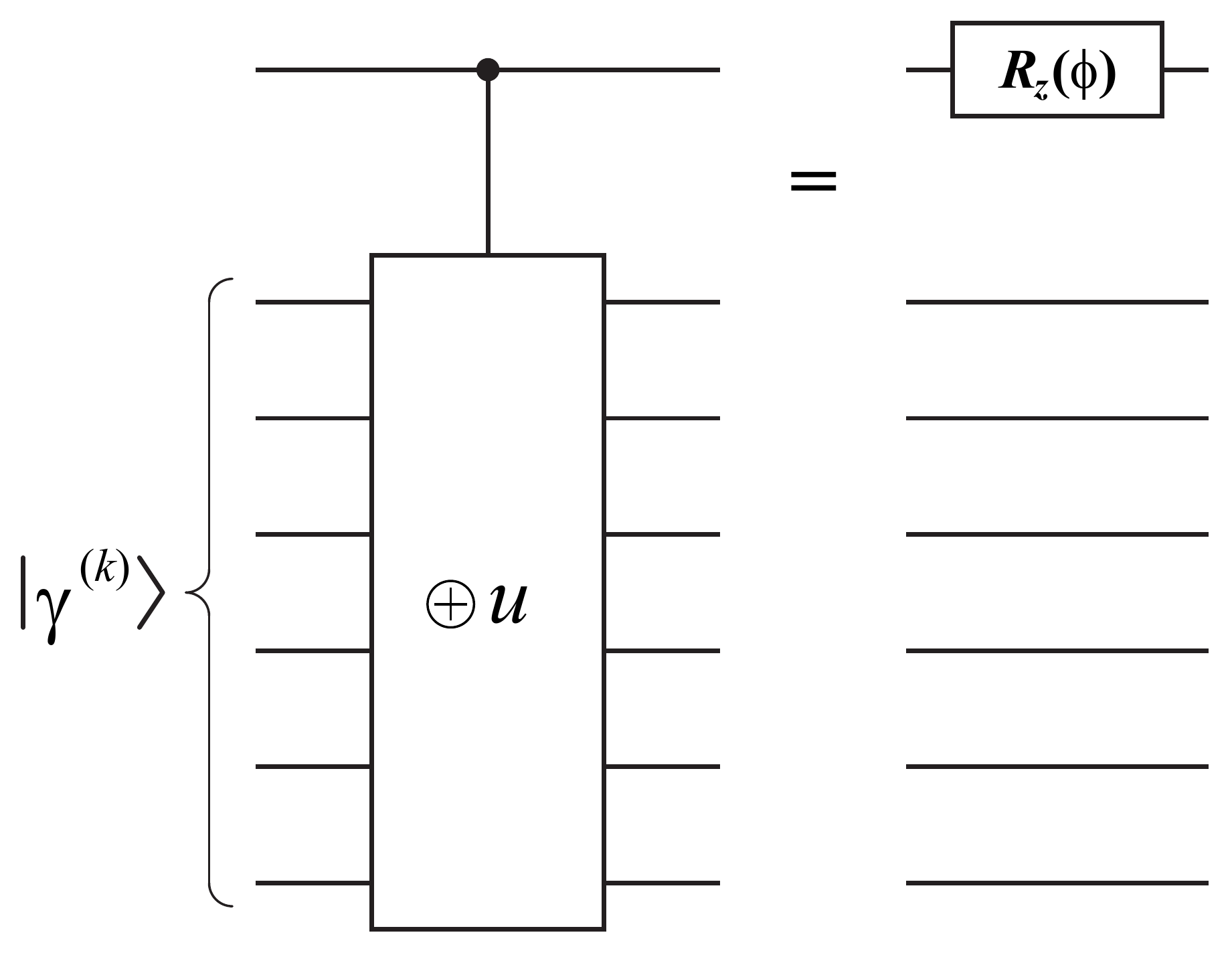}\\
  \caption[Circuit: Phase kickback for single-qubit phase rotation]{Circuit diagram showing how a rotation $R_Z(\phi)$ can be approximated with controlled addition on the Fourier state $\ket{\gamma^{(k)}}$.  The top qubit is the data qubit, while the bottom six (for example) make up the Fourier state.}
  \label{single_qubit_PK}
\end{figure}

Single-qubit phase rotations using phase kickback are constructed with a controlled addition circuit, as shown in Fig.~\ref{single_qubit_PK}.  Intuitively, a phase is kicked back to the control qubit if it is in the $\ket{1}$ state, which is equivalent to the phase rotation in Eqn.~(\ref{single_qubit_rotation}).  The accuracy of the phase gate and the quantum resources required depend on the number of bits in the ancilla state $\ket{\gamma^{(k)}}$.

To determine the value of $u$ in the addition circuit that produces a phase closest to $\phi$, one solves the modular equation
\begin{equation}
ku + \left\lfloor N \frac{\phi}{2\pi} \right\rceil \equiv 0 \; (\mathrm{mod} \; N),
\label{PK_determine_addend}
\end{equation}
which always has a solution since $k$ is odd and $N$ is a power of $2$ ($k$ and $N$ are coprime, so $k$ has a modular inverse).  The operation $\lfloor x \rceil$ denotes rounding any real $x$ to the nearest integer; any arbitrary rule for half-integer values suffices here.  By proper selection of $u$, one can approximate any phase rotation to within a precision of $|\Delta\phi| \le \frac{2\pi}{2^{n+1}}$ radians, where $|\Delta\phi| = \left[|\phi + \frac{2\pi}{N}ku| \; (\mathrm{mod} \; 2\pi)\right]$.  Note that $\frac{2\pi}{N}ku \approx -\phi$, explaining the plus sign to define $|\Delta\phi|$.  This is a consequence of the sign convention: $U_{\oplus u}\ket{\gamma^{(k)}} = e^{-i 2\pi ku/N}\ket{\gamma^{(k)}}$, and the phase $e^{i\phi}$ is intended.

After solving Eqn.~(\ref{PK_determine_addend}), the integer $u$ is added to $\ket{\gamma^{(k)}}$ using a quantum adder controlled by the qubit which is the target of the phase rotation.  There are various implementations of quantum adder circuits which have tradeoffs in performance between circuit depth, circuit size, and difficulty of implementation~\cite{Vedral1996,Draper2000,VanMeter2005,Cuccaro2004,Draper2006}.  Since $\ket{\gamma^{(k)}}$ is not altered by phase kickback, the number of such registers required for a quantum algorithm is equal to the maximum number of phase rotations which are computed in parallel at any point in the algorithm.

The Fourier state is reusable in phase kickback.  This is advantageous, since the register does not need to be reproduced for each phase rotation, but it also leads to a problem with correlated errors.  If there is an error in the addition circuit, it could corrupt the Fourier state, and this error corrupts all subsequent uses of this state.  These correlated errors might or might not be a problem, depending on the structure of the algorithm.  In particular, if there is no error suppression being invoked, then the correlated error is not a concern.  Any logical error leads to failure, so a set of \emph{logical} correlated errors is just as bad as any other.  However, if some verification protocol for the program segment is used, then Fourier states cannot spread outside the verified segment.  Instead, the Fourier state should be erased and a new one created.  In Section~\ref{Sec_Rotation_Resources}, I will show that the Fourier distillation procedure from Section~\ref{Sec_FS_Distillation} has cost similar to phase kickback, meaning that the overhead for producing these states is low.  As an example, one could use the Fourier state 10 times before erasing it, and the distillation overhead would be on order 10\%.

\subsection{Quantum-Variable Rotation (QVR)}
\label{Sec_QVR}
Phase kickback can be used for operations that are more complex than a single-qubit rotation.  Start with some arbitrary superposition $\ket{\psi} = \sum_x c_x \ket{x}$, where $c_x$ are complex probability amplitudes and each $\ket{x}$ is the binary representation of integer $x$ in qubits.  If the register $\ket{\psi}$ is added to some Fourier state $\ket{\gamma^{(k)}}$, the result is
\begin{eqnarray}
U_{\mathrm{add}}\ket{\psi}\ket{\gamma^{(k)}} &=& U_{\mathrm{add}}\left[\left(\sum_x c_x \ket{x}\right)\left(\frac{1}{\sqrt{N}}\sum_{y=0}^{N-1}e^{i 2\pi ky/N}\ket{y}\right)\right] \nonumber \\
&=& \frac{1}{\sqrt{N}}\sum_x \sum_{y=0}^{N-1} e^{i 2\pi ky/N} c_x  \ket{x} \ket{y+x} \nonumber \\
&=& \left(\sum_x e^{-i 2\pi kx/N} c_x  \ket{x}\right) \left(\frac{1}{\sqrt{N}} \sum_{z=0}^{N-1}  e^{i 2\pi kz/N}\ket{z}\right),
\end{eqnarray}
where in the last step the substitution $z = y+x$ was used to show that here a phase proportional to $x$ is kicked back to each element $\ket{x}$ of $\ket{\psi}$.  Since $\ket{\gamma^{(k)}}$ is returned unchanged, this a different form of phase kickback that effectively implements
\begin{equation}
\sum_x c_x\ket{x} \longrightarrow \sum_x e^{-i 2\pi \xi x} c_x \ket{x}.
\label{quantum_variable_rotation}
\end{equation}
The coefficient $\xi = k/N$, and I later show how to produce any $\xi$ to arbitrary precision.  The operation in Eqn.~(\ref{quantum_variable_rotation}) is a \emph{quantum-variable rotation} (QVR), since the phase rotation is encoded in a quantum state.  QVRs are useful in a variety of quantum algorithms, including quantum simulation~\cite{Kassal2008,Jones2012_NJP}, quantum Fourier transforms~\cite{Barenco1996,Nielsen2000,Cleve2000,Weinstein2001}, and the quantum linear-systems algorithm~\cite{Clader2013}.

\begin{figure}
  \centering
  \includegraphics[width=5cm]{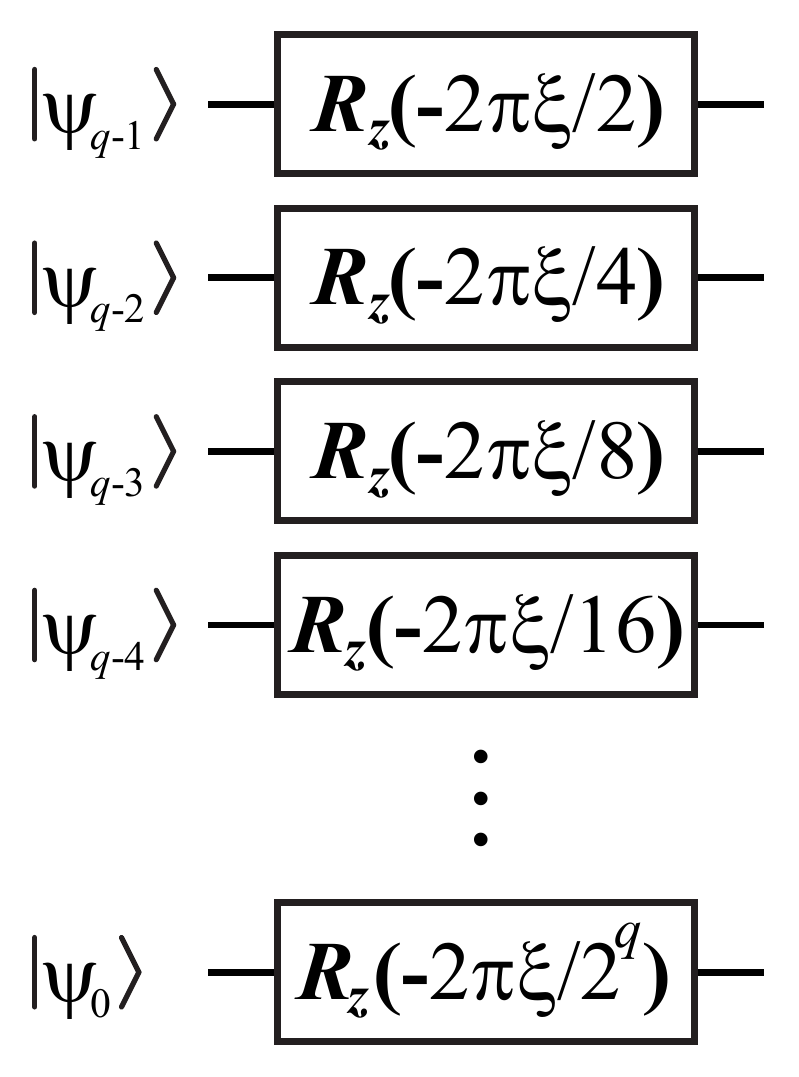}\\
  \caption[Circuit: Decomposing QVR into single-qubit rotations]{Circuit showing a quantum variable rotation decomposed into single-qubit rotations applied to each qubit in the $\ket{\psi}$ register consisting of $q$ qubits (see Eqn.~(\ref{quantum_variable_rotation})).  $\ket{\psi_{q-1}}$ refers to the most significant bit in the register $\ket{\psi}$, $\ket{\psi_0}$ refers to the least significant bit, \emph{etc}.  Modified from version published in Ref.~\cite{Jones2012_NJP}.}
  \label{QVR_bitwise}
\end{figure}

The QVR can be seen as parallel application of single-qubit rotations to each qubit in register $\ket{\psi}$, as shown in Fig.~\ref{QVR_bitwise}.  Instead of using phase kickback, each individual rotation could be created with the techniques in Section~\ref{Sec_Approximation_Sequences}.  Using this approach, QVR on state $\ket{\psi}$ having $t$ qubits requires $t$ separate bitwise rotations.  In order to control errors, each rotation requires accuracy $\epsilon_F \le p_{\mathrm{out}}/t$ to achieve final error probability $p_{\mathrm{out}}$ in the QVR, where I have used the fact that the distance measure in Eqn.~(\ref{Fowler_distance}) obeys the triangle inequality~\cite{Fowler2011_ArbitraryGates}.  The construction using phase kickback has fewer sources of error, making it more efficient.

\begin{figure}
  \centering
  \includegraphics[width=\textwidth]{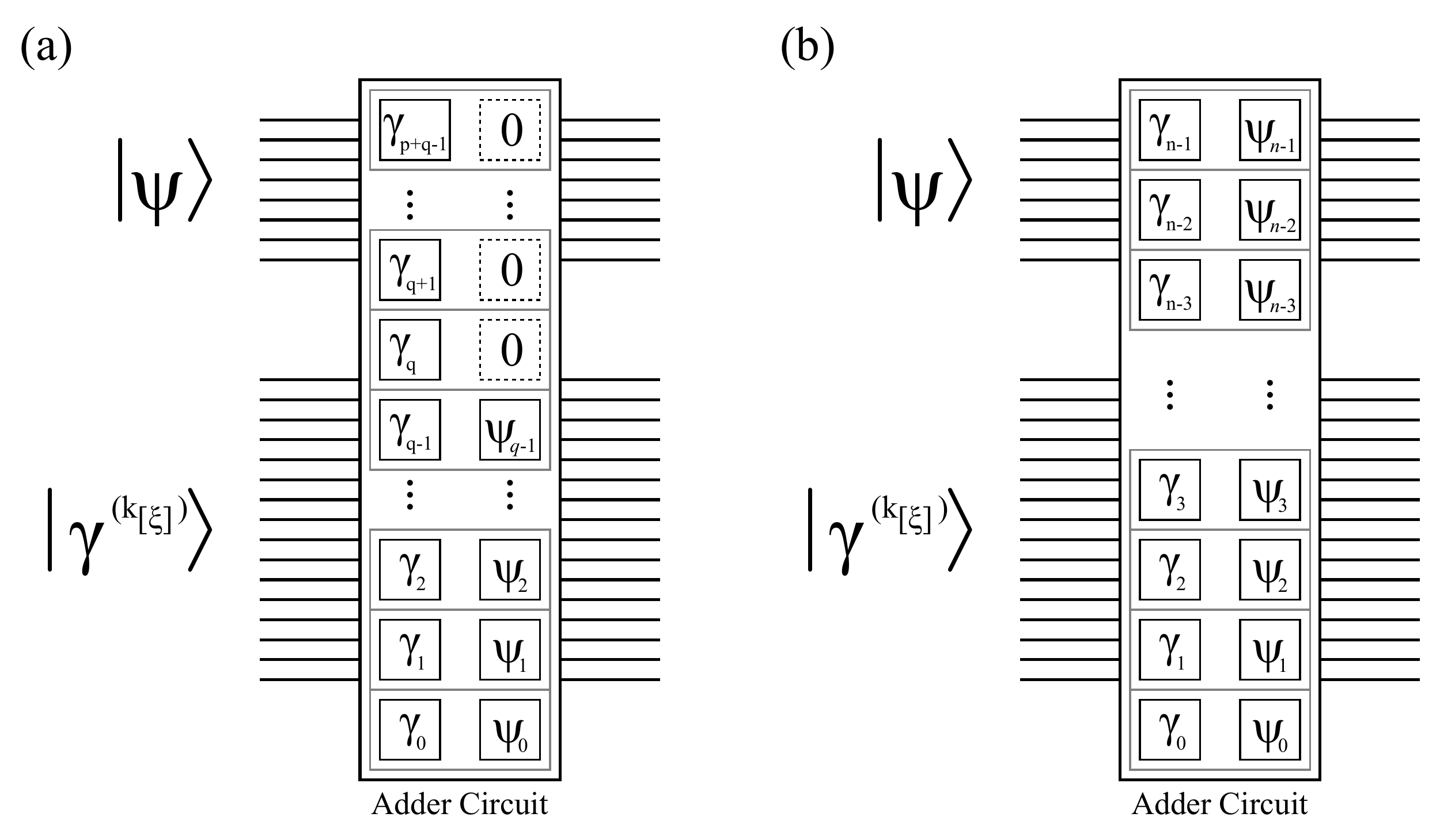}\\
  \caption[Circuit: Alignment of qubits in QVR adder]{Circuit for the addition operation in quantum variable rotation using phase kickback.  This circuit implements the operation in Eqn.~(\ref{quantum_variable_rotation}) with scaling factor $[\xi]$, which has been ``programmed'' into the phase kickback register $\ket{\gamma^{(k_{[\xi]})}}$.  This figure shows how the bits in the adder are aligned for different cases, using the method described in the text. (a)~The register $\ket{\psi}$ is shifted down $p$ bits since $p \ge 0$. $\psi_0$ is the least-significant bit in the $\ket{\psi}$ register, \emph{etc.}  The input qubits above $\ket{\psi}$ are logical zeros. (b)~The register $\ket{\psi}$ is shifted up $|p|$ bits since $p < 0$.  In this case, the $|p|$ most-significant bits of $\ket{\psi}$ are not used in the adder.  Modified from version published in Ref.~\cite{Jones2012_NJP}.}
  \label{PK_QVR}
\end{figure}

The phase-kickback procedure for QVR is as follows:
\begin{enumerate}[(1)]
\item Calculate a binary approximation to $\xi$, denoted $[\xi]$, to as many significant bits as desired.
\item Define some quantities that describe this quantum circuit:
    \begin{enumerate}[(a)]
    \item Let $m$ denote the number of significant bits in $[\xi]$, minus the number of trailing zeros.
    \item Define $w = \lfloor \log_2 [\xi] \rfloor$, or in other words, $w$ is the largest integer such that $2^w \le [\xi]$.
    \item Denote $p = (m-1) - w$, which is how many bits one must shift $[\xi]$ up to produce an odd integer (if $p < 0$, shift down).
    \item Let $q$ be the number of qubits in $\ket{\psi}$.
    \item Define odd integer $k_{[\xi]} = (2^p) [\xi]$.  Put another way, express $[\xi] = \frac{k_{[\xi]}}{2^p}$, where $k_{[\xi]}$ is an odd integer and $p$ is an integer.
    \end{enumerate}
\item Construct a Fourier state $\ket{\gamma^{(k_{[\xi]})}}$ of size $n = p + q$ qubits, using techniques in Section~\ref{Sec_FS_Distillation} or Ref.~\cite{Jones2012_NJP}.
\item Perform phase kickback with an addition circuit between registers $\ket{\psi}$ and $\ket{\gamma^{(k_{[\xi]})}}$ (in-place addition applied to $\ket{\gamma^{(k_{[\xi]})}}$), except this time the $\ket{\psi}$ register is shifted in one of two ways, as shown in Fig.~\ref{PK_QVR}.  If $p \ge 0$, then the $\ket{\psi}$ register is shifted down by $p$ qubits, and the $\ket{\psi}$ register is padded with $p$ logical zeros at the most-significant side of the adder input (Fig.~\ref{PK_QVR}(a)).  If $p < 0$, then $\ket{\psi}$ is shifted up by $|p|$ qubits, so that the $|p|$ most-significant bits of $\ket{\psi}$ are not used in the adder (Fig.~\ref{PK_QVR}(b)).  If $n \le 0$, then all rotations are identity and no QVR circuit is constructed.
\end{enumerate}
A simple proof confirms that this circuit implements the desired procedure:
\begin{equation}
\sum_x c_x\ket{x} \longrightarrow \sum_x e^{-i 2\pi (2^p k_{[\xi]}) x} c_x \ket{x} = \sum_x e^{-i 2\pi [\xi] x} c_x \ket{x}.
\end{equation}

QVR has interesting applications to other useful quantum circuits.  It can be used to make a fault-tolerant quantum Fourier transform (QFT).  One replaces each block of controlled rotations with a controlled-QVR, and the phase kickback register is $\ket{\gamma^{(1)}}$.  This approach uses substantially fewer of the expensive, non-Clifford gates than an equivalent circuit where each controlled rotation in the QFT is implemented using techniques in Section~\ref{Sec_Approximation_Sequences}.  The cost for phase-kickback QVR is $O(n^2)$ Toffoli gates (or $T$~gates), using $n-1$ adder circuits, \emph{vs.} $O(n^3)$ $T$~gates for approximation sequences, where each of the $O(n^2)$ controlled phase rotations requires $O(n)$ $T$~gates for sufficient accuracy.  The same approach can produce an approximate QFT~\cite{Barenco1996} by simply truncating the size of the $\ket{\gamma^{(1)}}$ register, in which case the Toffoli-gate or $T$-gate cost is $O(n \log n)$  \emph{vs.} $O(n \log^2 n)$.  Section~\ref{Sec_FS_Distillation} established that the costs of Toffoli gate and $T$~gate are comparable.

\section{Programmable Ancilla Rotations}
\label{Sec_PAR}
Another method for producing a phase rotation is the programmable ancilla rotation (PAR).  The name is derived from the fact that the desired rotation is ``programmed'' into ancillas that are produced before they are needed.  Shifting the computing effort to an earlier point in the program (assuming parallel computation) allows this method to achieve \emph{constant average depth} in the algorithm for any desired accuracy of rotation, which can be as small as 4 quantum gates.  The pre-calculated ancillas still require quantum programs of similar complexity to the preceding constructions for phase rotations, so this approach is best-suited to a quantum computer with many excess resources available for parallel computing.

\begin{figure}
  \centering
  \includegraphics[width=10cm]{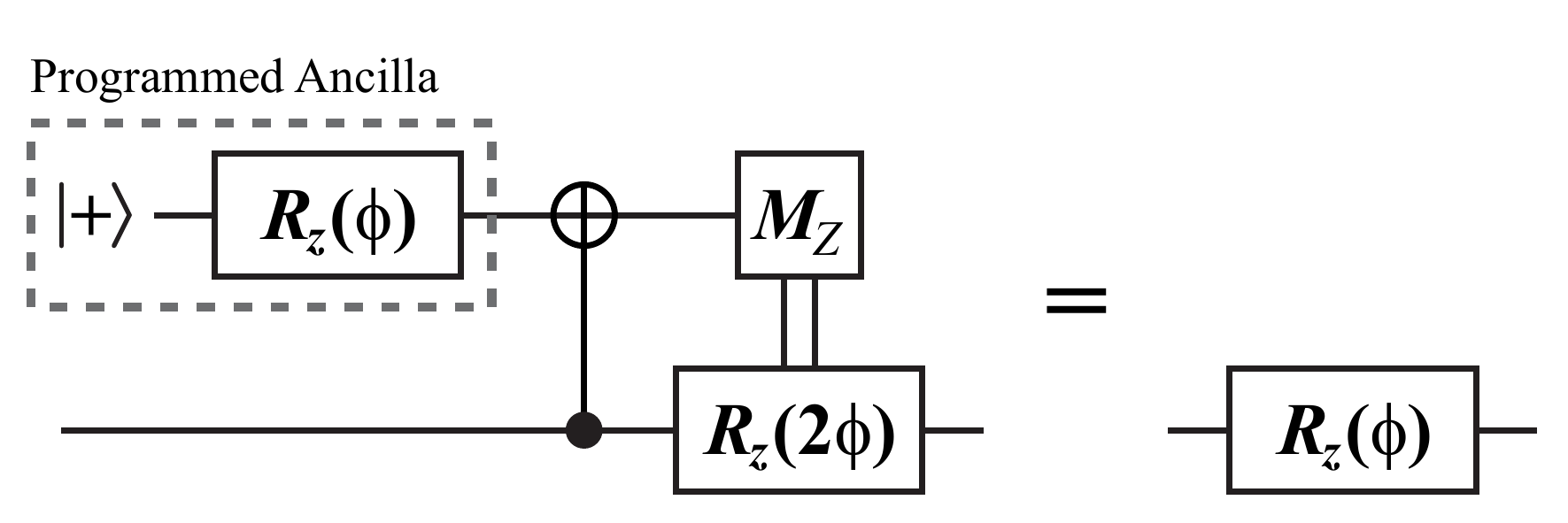}\\
  \caption[Circuit: Rotation using programmed ancilla]{Circuit diagram showing how a rotation $R_Z(\phi)$ can be implemented using a ``programmed'' ancilla $R_Z(\phi)\ket{+}$.  The correction $R_Z(2\phi)$ is implemented if the preceding measurement result is $\ket{1}$, \emph{etc}.}
  \label{ancilla_rotation}
\end{figure}

The PAR is based on a simple circuit which uses a single-qubit ancilla to make a phase rotation, which is a ``teleportation gate''~\cite{Gottesman1999,Zhou2000}, as shown in Fig.~\ref{ancilla_rotation}.  In this case, the operation $R_Z(\phi)$ is programmed into the ancilla $(1/\sqrt{2})(\ket{0} + e^{i\phi}\ket{1}) = R_Z(\phi)\ket{+}$.  However, the circuit in Fig.~\ref{ancilla_rotation} is probabilistic since there is a 50\% probability of enacting $R_Z(-\phi)$ instead of $R_Z(\phi)$; in such an event, one attempts the circuit again with angle $2 \phi$ (encoded in state $R_Z(2\phi)\ket{+}$), then $4 \phi$ (encoded in state $R_Z(4\phi)\ket{+}$) if the second attempt yields a negative-angle rotation, \emph{etc}.  This proceeds until the first observation of a positive-angle rotation after $m$ rounds of negative-angle rotations, in which case the result is a rotation $\phi_{\mathrm{total}} = 2^m \phi - \sum_{x=1}^{m-1}2^x\phi = \phi$.

\begin{figure}
  \centering
  \includegraphics[width=\textwidth]{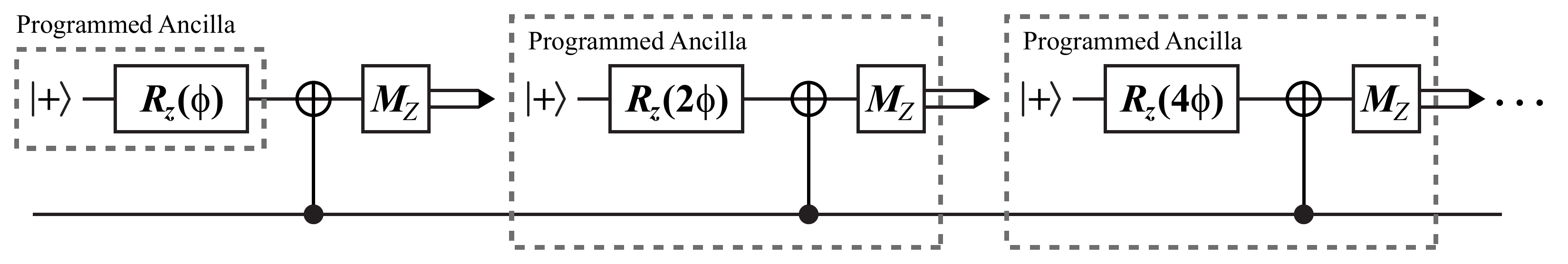}\\
  \caption[Circuit: PAR sequence]{Circuit diagram showing how a rotation $R_Z(\phi)$ can be implemented using the PAR sequence.  The correction operation from Fig.~\ref{ancilla_rotation} can be implemented with another programmed ancilla, and the procedure iterates until a measurement result of $\ket{0}$ is obtained.}
  \label{PAR_circuit}
\end{figure}

The complete circuit for the PAR is shown in Fig.~\ref{PAR_circuit}.  The programmed ancillas $R_Z(\phi)\ket{+}$, $R_Z(2\phi)\ket{+}$, \emph{etc}. are pre-computed using one of the phase rotation techniques in Section~\ref{Sec_Approximation_Sequences} or Section~\ref{Sec_Phase_Kickback}.  A very similar method was derived in Ref.~\cite{Isailovic2008}, but this section generalizes from $\phi = \frac{\pi}{2^k}$ to arbitrary rotation angles.  The iterated sequence of probabilistic rotations continues until the desired rotation is produced or the programmed ancillas are exhausted.  For practical reasons, one might choose to only produce a finite number of the PAR ancillas; if all ancilla-driven rotations fail, then a deterministic rotation using phase kickback or a gate approximation sequence is applied.  The probability of having to resort to this backstop is suppressed exponentially with the number of PAR ancillas computed in advance.

The expected number of rounds of the circuit in Fig.~\ref{PAR_circuit} before a successful rotation is simply given by $\sum_{m = 1}^{\infty} \frac{m}{2^m} = 2$.  Counting measurement as a gate, the number of gates per round is two, and the average number of gates per PAR is four.  With a finite number of pre-computed ancillas $M$, there is a probability $2^{-M}$ of having to implement the considerably more expensive (in circuit depth) deterministic rotation.  Nevertheless, if the computer supports the ability to calculate the programmed ancillas in advance, the PAR produces phase rotations that are orders of magnitude faster than other available methods, which also leads to faster execution of quantum algorithms~\cite{Jones2012_NJP}.

The phase kickback QVR can also be used to efficiently produce ancillas for PAR if the same rotation $R_Z(\phi)$ is required frequently.  Applying QVR with the appropriate $\xi$ to $\ket{+}\ket{+}\ket{+}...$ will produce the set of ancillas needed for PAR, but requiring only one addition circuit for the entire set instead of a phase kickback addition or approximation sequence for each ancilla qubit, which can be seen by comparing Fig.~\ref{QVR_bitwise} with the pre-computed ancillas in Fig.~\ref{PAR_circuit}.  Creating the necessary Fourier state $\ket{\gamma^{(k_{[\xi]})}}$ is more expensive than a single phase rotation, so there is a net gain only if a certain rotation $R_Z(\phi)$ is used multiple times.

Another application of combining QVR and PAR is to use the $n$-qubit Fourier state $\ket{\gamma^{(1)}}$ to approximate a phase rotation $R_Z(\phi)$ with a modified PAR technique.  Calculate $a = \left\lfloor 2^n \frac{\phi}{2\pi} \right\rceil$ as a temporary variable.  For each qubit in $\ket{\gamma^{(1)}}$, beginning with the least-significant, apply this procedure iteratively:
\begin{itemize}
\item If $a$ is odd, apply the ancilla rotation in Fig.~\ref{ancilla_rotation} using the least-significant remaining qubit from $\ket{\gamma^{(1)}}$.  If the measurement result is $\ket{0}$ (positive-angle rotation), set \hbox{$b \leftarrow (a+1)/2$}; if result is $\ket{1}$, set \hbox{$b \leftarrow (a-1)/2$}.
\item If $a$ is even, discard the least-significant qubit and set $b \leftarrow a/2$.
\item Set $a \leftarrow b$ and iterate from the beginning.
\end{itemize}
Since the least-significant bit of $\ket{\gamma^{(1)}}$ is either consumed or discarded in each iteration, the size of this register is shrinking.  After iterating through all the qubits, the Fourier state has been used to implement phase rotation $R_Z(\left\lfloor 2^n \frac{\phi}{2\pi} \right\rceil 2\pi/2^n)$, which is the closest $n$-bit-discretized rotation to $R_Z(\phi)$.  This is very similar to the technique in Ref.~\cite{Isailovic2008}, except the Fourier state is produced more efficiently through distillation rather than approximation sequences.

\section{Resource Analysis for Phase Rotations}
\label{Sec_Rotation_Resources}
This section estimates the resource costs for the phase rotations considered above.  The list of methods considered is not exhaustive, so the estimates here are not the final word in fault-tolerant phase rotations.  Still, the examples chosen show a variety of techniques in fault-tolerant quantum computing, so the resource costs should be representative of the intrinsic costs for approximating rotation gates by any method.  All of these techniques depend on the results from Chapters~\ref{Ch05} and~\ref{Ch06}.

The distance metric in Eqn.~(\ref{Fowler_distance}) is not a probability of error.  To determine the probability of an error introduced by approximating $U$ with $\tilde{U}$, one would use
\begin{equation}
\max_{\ket{\psi}}\left|\bra{\psi}U^{\dag}\tilde{U}\ket{\psi}\right| = {\epsilon_F}^2.
\end{equation}
Probability of error is the combination of approximation errors and logical failure:
\begin{equation}
p_{\mathrm{out}} = {\epsilon_F}^2 + p_L.
\end{equation}
When many approximate rotation gates are used in an algorithm, these probabilities sum together in a manner that implicity assumes that approximation errors sum incoherently.  This assumption simplifies the analysis, and it will hold in practice if there is sufficient variety in the rotation paths through Hilbert space (\emph{i.e.}, the sequence of gates appears random).  On the other hand, it is trivial to construct examples where this assumption breaks down, so the circuit designer must be aware of the dangers.

The best results for approximating single-qubit rotations with sequences of $T$~gates and Clifford gates each require about 3--4 $T$~gates per bit of precision~\cite{Fowler2011_ArbitraryGates,Kliuchnikov2012}.  The KMM method requires at most 2 ancilla qubits, as opposed to $(2n-1)$ for phase kickback, but the total resource costs from non-Clifford gates will dominate in most cases.  Reference~\cite{Kliuchnikov2012} estimates the number of $T$~gates is
\begin{equation}
C_{\mathrm{KMM}}^{(T)} = 10.7\log_{10}(1/\epsilon_F) - 23.0.
\end{equation}
Similarly, the BGS approach~\cite{Bocharov2013} produces approximation sequences using $V$~gates and Clifford operations, where each $V$~gate can be produced probabilistically using a Toffoli gate.  In this case, the cost in Toffoli gates is
\begin{equation}
C_{\mathrm{BGS}}^{(\mathrm{Tof})} = 6.62\log_{10}(1/\epsilon_F) - 0.11,
\end{equation}
including the probabilistic nature of the $V$-gate construction in Fig.~\ref{V_gate_circuit}.  The coefficients of $\log_{10}(1/\epsilon_F)$ are similar, so incorporating the previously calculated costs for $T$ and Toffoli gates is necessary to accurately compare the costs of these methods.

The accuracy of phase kickback depends on the number of bits in the Fourier state and adder.  A rotation using $p$ bits of precision has accuracy
\begin{equation}
\epsilon_F = \sqrt{1-\frac{1}{2}\left|1+\exp(i\pi/2^p)\right|} \approx \frac{1}{\sqrt{8}}\left(\frac{\pi}{2^p}\right),
\end{equation}
where approximation is correct to at least four significant figures for $p \ge 6$.  A phase kickback rotation accurate to $p$ bits has the succinct error expression $\log_2(1/\epsilon_F) \approx p - 0.15$, so the number of bits in the Fourier state is very nearly the number of bits of accuracy.  For explicit comparison with the preceding methods, the cost in Toffoli gates is
\begin{equation}
C_{\mathrm{PK}}^{(\mathrm{Tof})} = 3.32 \log_{10}(1/\epsilon_F) - 0.50.
\end{equation}

Using phase kickback to approximate an arbitrary phase rotation with bounded error probability $p_{\mathrm{out}} < 2^{-2(n-1.15)}$ requires an $n$-qubit, $\ket{\gamma^{(1)}}$ Fourier state, including the residual error in the Fourier state from distillation.  As shown in Section~\ref{Sec_FS_Distillation}, preparing such a state requires $O(n \log n)$ Toffoli gates, but this initialization need only be performed once.  Each phase-kickback rotation uses an addition circuit, which requires at most $2(n-2)$ Toffoli gates~\cite{Cuccaro2004}.  However, for this special case, one of the addends is a known value (the quantity $u$ from Eqn.~(\ref{PK_determine_addend})).  Using this fact, one can simplify the adder and ``short-circuit'' half of the Toffoli gates, replacing them with Clifford gates.  A single-qubit phase rotation with a precision of $\pi/2^{n-1}$ radians, which is $(n-1)$ bits of precision, requires just $(n-2)$ Toffoli gates, $n$ qubits for the Fourier states, and $(n-1)$ ancilla qubits for the internal carry operations of the adder.  Forming controlled-rotation gates is simple as well.  Each additional control input to the multi-qubit gate requires one more Toffoli gate and one more ancilla qubit (see Fig.~\ref{controlled_rotation}).

\begin{figure}
  \centering
  \includegraphics[width=\textwidth]{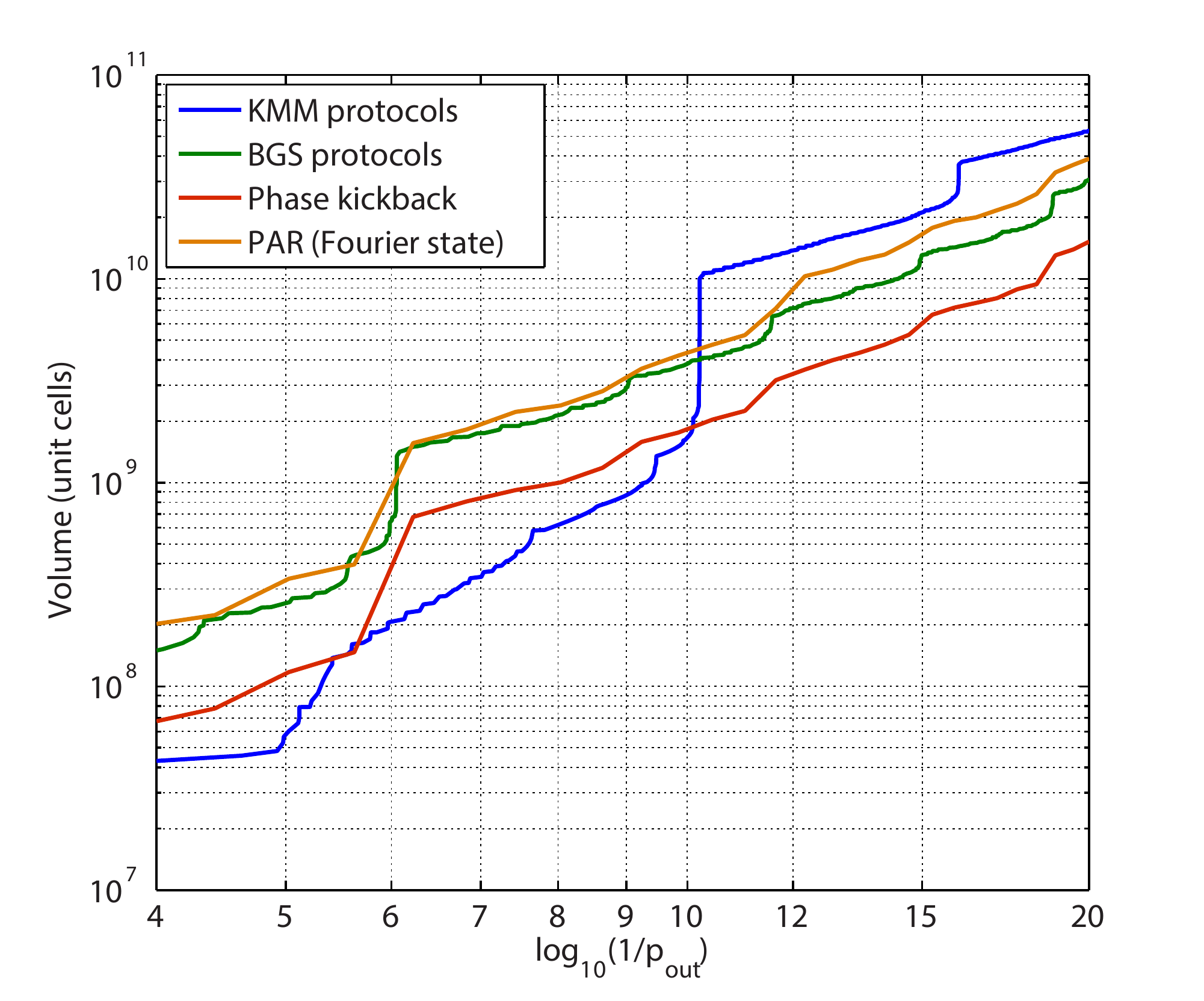}\\
  \caption[Resource costs for fault-tolerant phase rotations]{Resource costs for fault-tolerant phase rotations.  The error rate at the hardware level is $p_g = 10^{-3}$.  The KMM~\cite{Kliuchnikov2012} and BGS~\cite{Bocharov2013} protocols are named for the authors which propose them.  The sharp jump in KMM resources around $p_{\mathrm{out}} = 10^{-10}$ is due to a sharp jump in the cost of $T$~gates around $p_T = 10^{-12}$.  By contrast, the other three protocols use Toffoli gates.  Phase kickback uses the CDKM adder~\cite{Cuccaro2004} with short-circuit, and the cost of distillation is presumed to be negligible if the Fourier state is used sufficiently often.  The PAR consumes a Fourier state for each rotation, so its volume cost is dominated by Fourier-state distillation.}
  \label{Rotation_resources}
\end{figure}

The resource costs for a fault-tolerant, single-qubit rotation are plotted in Fig.~\ref{Rotation_resources}.  In addition to the approximation error $\epsilon_F$, each rotation has positive error probability associated with logical error in the surface code.  As in previous chapters, only efficient protocols (\emph{i.e.} not dominated by other protocols) are plotted.  Each of the methods from this chapter are considered: KMM~\cite{Kliuchnikov2012} and BGS~\cite{Bocharov2013} approximation sequences; phase kickback~\cite{Cleve1998,Kitaev2002,Jones2012_NJP}; and programmable ancilla rotations~\cite{Isailovic2008,Jones2012_NJP}.  The KMM protocol uses $T$~gates and other Clifford-group gates.  The BGS protocol uses $V$~gates that are produced using the Toffoli-based construction in Fig.~\ref{V_gate_circuit}.  Phase kickback uses a Fourier state $\ket{\gamma^{(1)}}$ and a CDKM adder, both of which are analzyed in Section~\ref{Sec_FS_Distillation}.  The PAR consumes a Fourier state $\ket{\gamma^{(1)}}$ using the alternative construction at the end of Section~\ref{Sec_PAR}, as opposed to programmed states.

Examining Fig.~\ref{Rotation_resources}, the resource costs are very similar for all of the methods, despite the methods themselves being significantly different.  This might suggest some fundamental lower bound to the cost of approximating phase rotations in the surface code, similar to the theoretical results of Ref.~\cite{Harrow2002}, but the evidence here is far from conclusive.  Additionally, I do not use a Trivial Upper Bound for the resource estimates in this chapter, because it is not obvious what it would be.  All that can be said is that these resource costs are a sampling of the least costly methods available, implemented in the best error correction scheme available under present assumptions.  In terms of policy, Fig.~\ref{Rotation_resources} suggests that one should use KMM protocols for rotations with error probability above $p_{\mathrm{out}} = 10^{-10}$, and phase kickback for rotations with error probability below $p_{\mathrm{out}} = 10^{-10}$.





\chapter{Quantum Simulation Algorithms}
\label{Ch08}
Quantum simulation inspired Richard Feynman to propose the idea of a quantum computer in 1982~\cite{Feynman1982}.  His reasoning was that while it is difficult for a classical computer perform calculations in the exponentially expanding Hilbert space of a simulation problem, quantum systems evolve in this manner naturally.  Therefore, it should be possible for one controllable quantum system to reproduce the physics of another quantum system.  Despite this early start, algorithms for simulation using quantum computers were not developed until several years later~\cite{Lloyd1996,Zalka1998}.  Aspuru-Guzik \emph{et al.} demonstrated that these simulation algorithms could be applied to quantum chemistry, which showed how quantum computers might be useful for everyday problems in science and engineering~\cite{Aspuru2005}.  Quantum chemistry simulations are considered throughout this chapter because they exhibit most of the features of quantum simulation. However, many other proposals exist, such as spin lattice models~\cite{Lidar1997}, lattice gas automata~\cite{Boghosian1998} and lattice gauge theories~\cite{Byrnes2006}, or quantum chaos theories~\cite{Levi2003}.  As with preceding chapters, I focus here on fault-tolerant circuit-model computation, but there are alternative proposals for special-purpose simulators as well~\cite{Buluta2009,Kassal2011,Barreiro2011,Simon2011,Ma2011}.

Quantum simulation is now one of the most prominent and intensely researched quantum algorithms.  In fact, there is not one algorithm but instead a broad family of algorithms.  At the heart of each is the time-dependent Schr\"{o}dinger's equation.  In essence, each simulation algorithm propagates the state vector in time with dynamics governed by the Hamiltonian.  Just as Feynman envisioned, an artificial quantum system (in the quantum computer) evolves in simulated time as an approximation to the natural evolution of a quantum system in real time.  Because the quantum computer must represent states in its own form of encoding, such as qubits, simulation algorithms introduce approximations that must be understood.  Moreover, simulation algorithms can calculate many different properties of a quantum system, including energy eigenvalues, chemical reaction rates, and polarizability~\cite{Kassal2008,Kassal2011}.  I will consider the problem of using simulation to determine an energy eigenvalue for a chosen Hamiltonian.  Energy eigenvalues are calculated using the phase estimation algorithm, which is an important application of the quantum Fourier transform.  A reader interested in calculating other quantities is referred to several review papers~\cite{Kassal2008,Brown2010,Kassal2011}.

Quantum chemistry simulation may be one of the most useful applications of quantum computers.  Quantum chemistry and band structure calculations account for up to 30\% of the computation time used at supercomputer centers~\cite{NERSC2010}, and \emph{ab initio} chemistry is one of the two physics-simulation applications which dominate the use of supercomputing resources (the other being fusion-energy research).  The most-employed techniques include density functional theory and polynomially-tractable approximate quantum chemistry methods~\cite{HeadGordon2008}.  Despite the success of these methods, for example, in simulating the dynamics of a small protein from first principles~\cite{Ufimtsev2011} or in predicting novel materials~\cite{Sokolov2011}, they are still approximate, and much work is carried out in developing more accurate methods.  Quantum simulators offer a fresh approach to quantum chemistry~\cite{Kassal2011}, as they are predicted to allow for the exact simulation (within a selected basis) of a chemical system in polynomial time. A quantum computer of a sufficient size, say 128 logical quantum bits~\cite{Aspuru2005,Kassal2008}, would already outperform the best classical computers for \emph{exact} chemical simulation.  This would open the door to high-quality \emph{ab initio} data for parameterizing force fields for molecular dynamics~\cite{Huang2012} or understanding complex chemical mechanisms such as soot formation~\cite{Wheeler2007}, where a number of different chemical species must be compared.  The difficulty of solving these problems with conventional methods suggests that computational chemistry would be one of the first novel applications of universal quantum computers.

This section provides an overview of some common forms of quantum simulation, particularly for quantum chemistry.  Simulation algorithms depend heavily on phase rotations, so the methods in this chapter demonstrate the importance of the logic constructions developed in Chapters~\ref{Ch05}--\ref{Ch07}.  In addition to building on results from prior chapters, several improvements to simulation algorithms are considered here.

\section{Schr\"{o}dinger Equation in a Quantum Computer}
It is straightforward to show that a universal quantum computer can simulate other quantum systems.  Quantum computers themselves are quantum systems that (ideally) evolve according to the time-dependent Schr\"{o}dinger equation:
\begin{equation}
i \hbar \frac{\partial}{\partial t} \ket{\psi} = \hat{H} \ket{\psi}.
\end{equation}
Given some initial state $\ket{\psi(0)}$, the propagator for time evolution is $\ket{\psi(t)} = U(t)\ket{\psi(0)}$ given by
\begin{equation}
U(t) = \mathcal{T}\exp\left(\frac{-i}{\hbar}\int_0^t \hat{H}(\tau) d\tau\right),
\label{Eqn_propagator}
\end{equation}
where $\mathcal{T}$ is the time-ordering operator that applies if commutator $\left[\hat{H}(t_i),\hat{H}(t_j)\right]$ is not zero for all $t_i \neq t_j$.  If one has a universal quantum computer of arbitrary size, then in principle one can represent $\ket{\psi(0)}$ and apply any operation $U(t)$, reproducing the evolution of any quantum system.

The way in which the simulated system is encoded into qubits is a defining aspect of the simulation algorithm.  Encoding dictates both the memory requirements (number of qubits) and computation efficiency (number of gates).  This chapter considers two forms of encoding for chemistry, second-quantized and first-quantized.  There are, of course, tradeoffs between these approaches, where second-quantized tends to require fewer qubits while first-quantized has lower asymptotic execution time~\cite{Kassal2008,Jones2012_NJP}.  In most cases, encoding is approximate, and higher accuracy comes at the cost of more qubits and gates.

The Schr\"{o}dinger equation determines the dynamics of a quantum system, so the quantum computer must reproduce the propagator in Eqn.~(\ref{Eqn_propagator}) for the chosen encoding.  In most cases, the propagator cannot be implemented exactly, so it must be approximated.  A common approach is to separate the Hamiltonian into terms which can be implemented efficiently and form an approximation using a Trotter-Suzuki sequence~\cite{Suzuki1976,Suzuki1992,Wiebe2010,Kassal2011}.  For example, if $\hat{H} = A + B$, where $e^{-iAt}$ and $e^{-iBt}$ can be implemented efficiently, then full propagator can be approximated as
\begin{equation}
e^{-i\hat{H}t} \approx \left[e^{-iAt/n}e^{-iBt/n}\right]^n.
\end{equation}
Where the approximation becomes exact in the limit $n \rightarrow \infty$, with leading error term $O(t^2/n)$.  However, only finite $n$ can be used because of practical resource constraints, so one might instead use a second-order Trotter-Suzuki sequence
\begin{equation}
e^{-i\hat{H}t} \approx \left[e^{-iAt/(2n)}e^{-iBt/n}e^{-iAt/(2n)}\right]^n,
\end{equation}
which has leading error term $O(t^3/n^2)$.  Wiebe~\emph{et al.} analyzed this problem and found that the optimal number of exponential operations is approximately linear in $t$, the length of simulated time~\cite{Wiebe2010}.

\begin{figure}
  \centering
  \includegraphics[width=\textwidth]{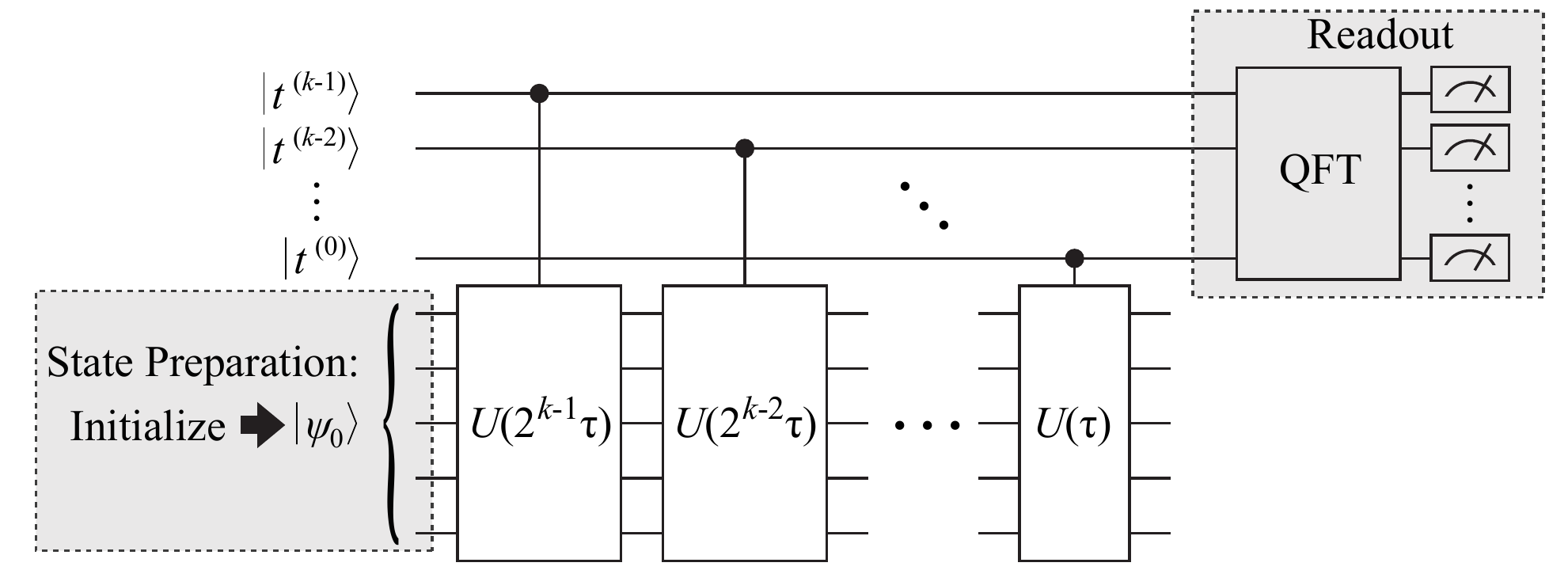}\\
  \caption[Circuit: Energy eigenvalue estimation]{Circuit for energy eigenvalue estimation~\cite{Lloyd1996,Aspuru2005}.  After preparing an initial state $\ket{\psi_0}$, the system is evolved in simulated time by solving the time-dependent Schr\"{o}dinger equation.  Note the system propagators $U(2^n \delta t)$ are controlled by qubits $\{\ket{t^{(n)}}\}$, where $n$ is the bit index in a register counting simulated time steps of length $\delta t$.  A quantum Fourier transform (QFT) on the time register provides an estimate of an energy eigenvalue.  Originally published in Ref.~\cite{Jones2012_NJP}.  {\copyright}2012 IOP Publishing Ltd.}
  \label{Sim_circuit}
\end{figure}

A digital quantum simulation algorithm consists of three primary steps (Fig.~\ref{Sim_circuit}): state preparation, simulated time evolution, and measurement readout.  This chapter focuses on the second step, evolving the system in simulated time, because this represents the core of the algorithm.  Simulation of time evolution on a quantum computer is a sequence of quantum gates which closely approximates the evolution propagator in Eqn.~(\ref{Eqn_propagator}) for the Hamiltonian $\hat{H}$ being simulated.  In the case of a time-independent Hamiltonian, $U(\delta t) = \exp\left(-\frac{i}{\hbar} \hat{H} \delta t\right)$, as in Fig.~\ref{Sim_circuit}.  The increment $\delta t$ is a single time step of simulation, and a simulation algorithm often requires many time steps, depending on the desired result (\emph{e.g.} energy eigenvalue).  State preparation and measurement readout are necessary steps which are not discussed here, but details can be found in references~\cite{Abrams1997,Zalka1998,Abrams1999,Nielsen2000,Grover2002,Mohseni2006,Ward2009}.

The circuit in Fig.~\ref{Sim_circuit} implements energy eigenvalue estimation~\cite{Lloyd1996,Aspuru2005}, which is a special case of phase estimation~\cite{Cleve1998,Abrams1999,Cleve2000,Parker2000,Nielsen2000}.  Applying the quantum Fourier transform (QFT) to the simulated time vector transforms the simulated system into the frequency domain, where frequency of complex-phase evolution is directly proportional to the energy eigenvalue associated with each eigenstate: $\exp\left(-\frac{i}{\hbar} \hat{H} \delta t\right)\ket{E_j} = \exp\left(-\frac{i}{\hbar} E_j \delta t\right)\ket{E_j}$.  By measuring an energy eigenvalue in the top register of Fig.~\ref{Sim_circuit} (upper right), the remaining register is projected into the associated eigenstate~\cite{Abrams1999}.  It is worth noting that the full QFT is shown here for clarity, but in practice one would use the iterative phase estimation algorithm, which produces the same result at lower resource cost~\cite{Parker2000}.

\section{Second-Quantized Encoding}
\label{Sec_Second_Quant}
Chemistry simulation in the second-quantized form expresses the electronic Hamiltonian $\hat{H}$ in terms of the creation operators ${a_p}^{\dag}$ and the wavefunction in terms of fermionic (or bosonic) modes $\ket{p}  \equiv {a_p}^{\dag} \ket{0}$ (\emph{i.e.}, occupation number representation).  The single-electron molecular orbital picture has provided a practical method for approximating an $N$-electron wavefunction.  Using second-quantized algorithms, basis sets in computational chemistry can be imported directly into quantum computational algorithms.  For this reason, both theoretical~\cite{Aspuru2005,Wang2008,Whitfield2011} and experimental~\cite{Lanyon2010,Du2010} investigations in second-quantized simulation have been performed.

\begin{figure}
    \centering
    \includegraphics[width=\textwidth]{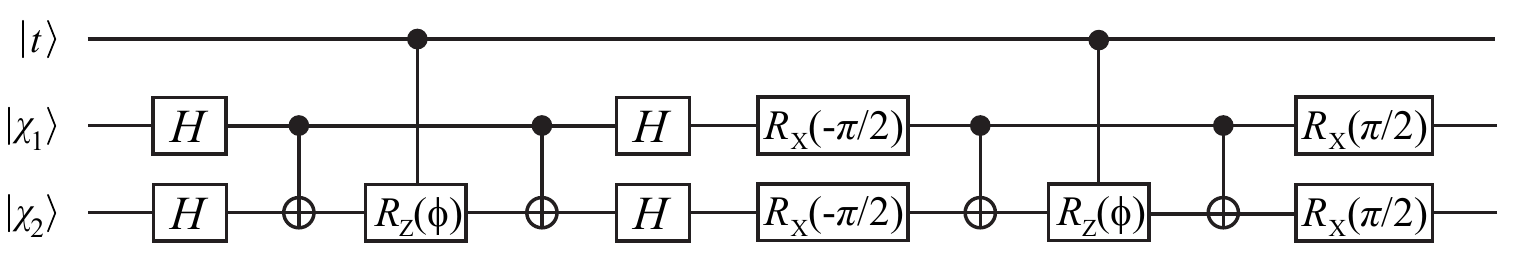}\\
    \caption[Circuit: Excitation operator in second-quantized form]{Excitation operator $e^{-ih_{12}({a_1}^{\dag} a_{2}  + {a_2}^{\dag} a_{1})\delta t}$ encoded into a quantum circuit~\cite{Whitfield2011}.  Above, $\phi = h_{12} \delta t$.  The gate $R_X(\pi/2) = HSH$ is in the Clifford group.  In this example, the control qubit $\ket{t}$ is used for phase estimation, and the qubits $\ket{\chi_1}$ and $\ket{\chi_2}$ are basis functions (\emph{e.g.} molecular orbitals).  The controlled phase rotations $CR_Z(\phi)$ must be approximated using circuits of available fault-tolerant gates, as in Chapter~\ref{Ch07}.  Modified from version published in Ref.~\cite{Jones2012_NJP}.}
    \label{excitation_operator}
\end{figure}

Following the standard construction (see \emph{e.g.} Ref.~\cite{Kassal2011}), an arbitrary molecular Hamiltonian in second-quantized form can be expressed as
\begin{equation}
\label{Eqn_second_hamiltonian}
\hat{H} = \sum_{p,q}h_{pq}{a_p}^{\dag}a_q + \frac{1}{2}\sum_{p,q,r,s}h_{pqrs}{a_p}^{\dag}{a_q}^{\dag}a_r a_s ,
\end{equation}
where $h_{pq} =\langle p | (\hat{T} + \hat{V}_N) | q \rangle$ are one-electron integrals ($\hat{T}$ is the kinetic energy operator, and $\hat{V}_N$ is the nuclear potential) and $h_{pqrs} = \langle pq | \hat{V}_e | rs \rangle$ represent the Coulomb potential interactions between electrons.  All of the terms $h_{pq}$'s and $h_{pqrs}$'s are pre-computed numerically with classical computers, and the values are then used in the quantum computer to simulate the Hamiltonian evolution through the operators:
\begin{equation}
U_{pq} = e^{-ih_{pq}({a_p}^{\dag}a_q+{a_q}^{\dag}a_p) \delta t};
\end{equation}
and
\begin{equation}
\label{Eqn_two_body_propagator}
U_{pqrs} = e^{-ih_{pqrs}({a_p}^{\dag}{a_q}^{\dag}a_r a_s + {a_s}^{\dag}{a_r}^{\dag}a_q a_p)\delta t}.
\end{equation}
These operators are constructed with a Jordan-Wigner transform and an arbitrary controlled phase rotation $CR_Z(\phi)$~\cite{Whitfield2011}, as shown in Fig.~\ref{excitation_operator}.  The controlled rotations can be constructed using methods in Chapter~\ref{Ch07}.  The Jordan-Wigner transform requires Hadamard, $S$, and CNOT gates, which are often readily available in fault-tolerant settings (see Section~\ref{quantum_programs_section}).  Section~\ref{Sec_teleport_JW} shows how to implement the Jordan-Wigner transform efficiently.

\subsection{Finite Precision in Precomputed Integrals}
\label{finite_integrals}
The execution time of a second-quantized simulation algorithm is proportional to the number of integral terms $h_{pq}$ and $h_{pqrs}$, as indicated by Eqns.~(\ref{Eqn_second_hamiltonian}--\ref{Eqn_two_body_propagator}).  In general, the number of terms is $O(M^4)$ for $M$ single-particle orbitals.  However, these integral terms vary substantially in magnitude, so it is possible to reduce simulation computation time by omitting the integral terms that are negligibly small, while introducing only small error.  In this way, the effort for evaluating these integrals often scales somewhere between $O(M^2)$ and $O(M^3)$ with modern implementations~\cite{Helgaker2000}, because typically many integral terms fall below a chosen threshold and can be dropped from the simulation.  Consequently, the execution time of second-quantized simulation is determined by the number of pre-computed integrals of the form $h_{pq}$ and $h_{pqrs}$ of sufficiently large magnitude, as well as the efficiency of producing the corresponding arbitrary phase rotations in the quantum computer, such as $CR_Z(h_{pq} \delta t)$ in the gate sequence for $e^{-ih_{pq}(a_p^{\dag}a_q+a_q^{\dag}a_p) \delta t}$~\cite{Whitfield2011}.

\begin{figure}
    \centering
    \includegraphics[width=\textwidth]{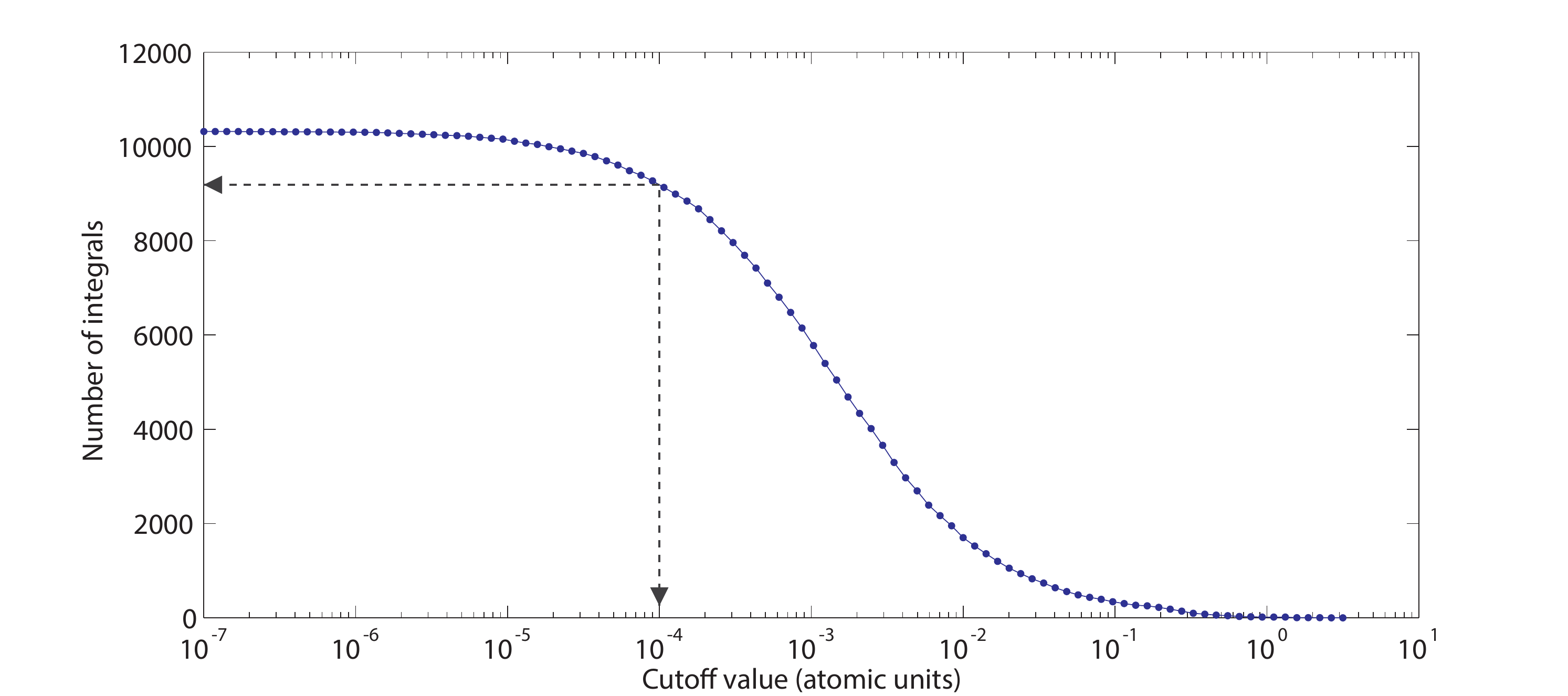}\\
    \caption[Plot of number of integral terms above threshold magnitude]{The number of integral terms implemented in a second-quantized simulation of LiH using a TZVP basis, as a function of cutoff threshold.  Only integral terms with absolute value above the threshold are implemented in circuits, and the rest are neglected.  As shown in the figure, a cutoff of $10^{-4}$ would require the algorithm to implement just over 9000 integral terms.  Originally published in Ref.~\cite{Jones2012_NJP}.  {\copyright}2012 IOP Publishing Ltd.}
    \label{Integral_cutoffs}
\end{figure}

To illustrate how many integral terms are present in a typical chemical problem, Ref.~\cite{Jones2012_NJP} calculated the integrals for a second-quantized simulation of lithium hydride, LiH.  The integrals are calculated in the minimal basis and in a triple-zeta basis, using the GAMESS quantum chemistry package~\cite{Schmidt1993,Gordon2005}, at a bond distance of 1.63 \AA, with an integral term cutoff of $10^{-10}$ in atomic units.  Reference~\cite{Jones2012_NJP} then computed the number of integrals above cutoff using the STO-3G basis~\cite{Hehre1969} containing 12 spin-orbitals (6 spatial orbitals) and the TZVP basis~\cite{Dunning1971} containing 40 spin orbitals (20 spatial orbitals).  The cumulative number of integral terms as a function of cutoff in TZVP basis is plotted in figure~\ref{Integral_cutoffs}.  With the STO-3G basis, there were 231 non-zero molecular integrals, but only 99 of them were greater than 10$^{-10}$ atomic units in magnitude.   This is an order of magnitude below what is expected from $O(M^4)$ scaling.  Considering the larger, more accurate basis set (TZVP), there were 22155 non-zero integrals, but only 10315 were greater than the cutoff $10^{-10}$.  Figure~\ref{Integral_cutoffs} shows that a higher cutoff, such as $10^{-4}$, can further reduce the number of integrals in TZVP basis implemented in the simulation.  As a result, the effective number of integral terms the quantum computer must implement as phase rotations is nearly two orders of magnitude less than the asymptotic analysis would suggest, an example of the over-estimation of the resource costs that can occur when using asymptotic estimates.  This technique becomes particularly relevant in large molecules since distant particles interact weakly, and in such an event, many of the associated integral terms may be negligibly small.  Raising the cutoff threshold impacts the accuracy of the simulation, so one must attempt to balance the resource costs of simulation with the usefulness of the result.

\subsection{Jordan-Wigner Transform using Teleportation}
\label{Sec_teleport_JW}
\begin{figure}
    \centering
    \includegraphics[width=\textwidth]{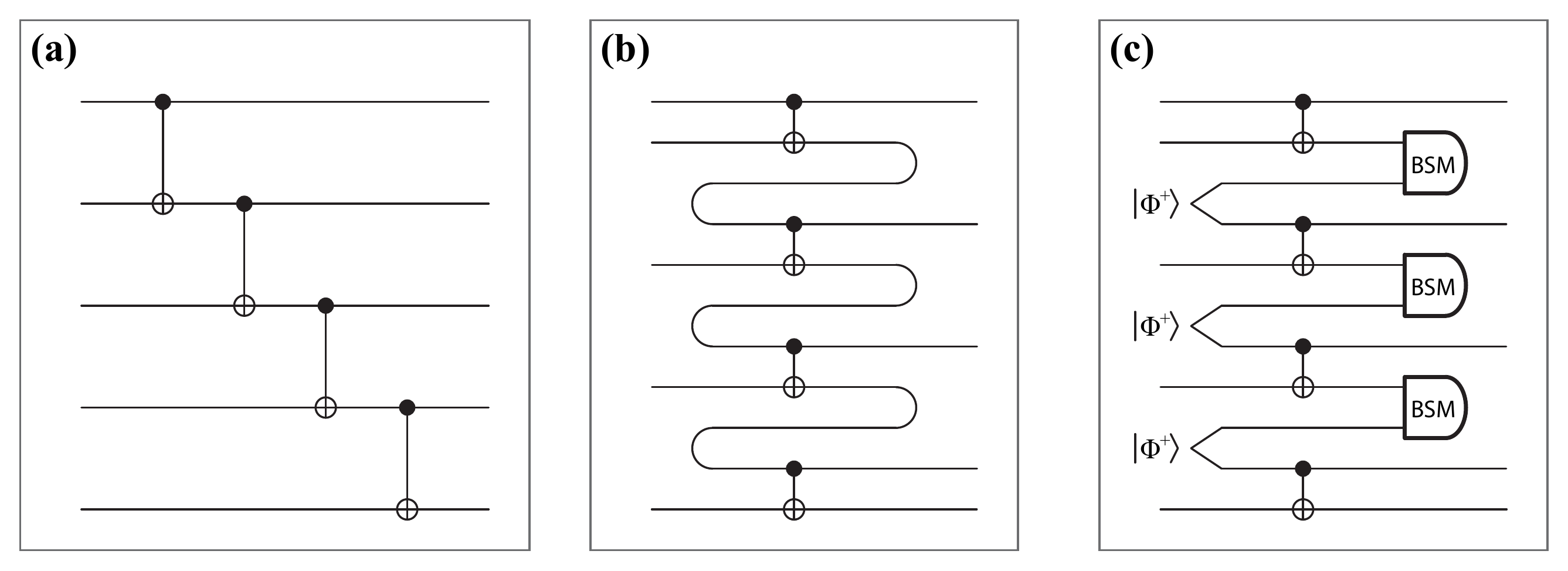}\\
    \caption[Circuit: Jordan-Wigner transform in constant time]{Rearrangement of the CNOT ladder common in Jordan-Wigner transforms using teleportation.  (a)~The original CNOT ladder requires an execution time that grows with the extent of the simulation in qubits.  (b)~A conceptual diagram of what teleportation accomplishes.  The qubits ``move'' backwards in time.  (c)~A valid quantum circuit that uses teleportation~\cite{Bennett1996,Gottesman1999,Zhou2000} to move qubits in a manner which allows parallel computation of the CNOTs.  The symbol ``BSM'' is the Bell state measurement which teleports the qubits; the result of this measurement indicates the Pauli errors which are tracked by the Pauli frame~\cite{Jones2012_PRX}.  The Bell state $\ket{\Phi^{+}} = \frac{1}{\sqrt{2}}\left(\ket{00} + \ket{11}\right)$ can be prepared from two $\ket{0}$ ancillas using one Hadamard gate and one CNOT gate.  Similarly, the BSM can be implemented using one Hadamard, one CNOT, and measurement of the two qubits in the computational basis.  Originally published in Ref.~\cite{Jones2012_NJP}.  {\copyright}2012 IOP Publishing Ltd.}
    \label{JWTeleport}
\end{figure}

The second-quantized algorithm uses Jordan-Wigner transforms to implement operators such as $e^{-ih_{pq}({a_p}^{\dag}a_q+{a_q}^{\dag}a_p) \delta t}$, and this section shows how to perform such transforms in constant time.  As elaborated in Ref.~\cite{Whitfield2011}, the circuits for Jordan-Wigner transforms often consist of ladders of CNOT gates, such as the one in Fig.~\ref{JWTeleport}(a).  In a simulation with $M$ basis states, these ladders can extend across the entire register of qubits corresponding to these basis states, which leads to the $O(M^5)$ asymptotic runtime quoted in Ref.~\cite{Kassal2011} when there are at most $O(M^4)$ integral terms.

The CNOT ladder is a sparse network of Clifford gates, so I show how it may be implemented in constant time using teleportation~\cite{Bennett1996,Gottesman1999,Zhou2000}.  Figure~\ref{JWTeleport}(b) gives an intuitive picture for what will be accomplished.  If the path of the qubits could be rearranged to somehow propagate backwards in time, the CNOT gates could be implemented simultaneously.  Qubits cannot move backwards in time \emph{per se}, but they can be moved arbitrarily using teleportation; notice how the conceptual (but unphysical) circuit in Fig.~\ref{JWTeleport}(b) is realized by a physical circuit in Fig.~\ref{JWTeleport}(c).  Ancilla Bell states $\ket{\Phi^{+}} = \frac{1}{\sqrt{2}}\left(\ket{00} + \ket{11}\right)$ are used to teleport qubits in this rearranged CNOT ladder.  Teleportation introduces a random Pauli error on the teleported qubit, but it is possible to track these errors and their propagation through CNOT gates using Pauli frames (see Section~\ref{Pauli_frames}).  With this modification, it is possible to implement the Jordan-Wigner transform in constant time, which removes one of the bottlenecks to high-speed second-quantized simulation.  This method could be adapted to implement other Clifford-group circuits in constant time, at the expense of requiring enough ancilla Bell states.

\section{First-Quantized Encoding}
\label{first_quantized}
The first-quantized simulation algorithm is in some ways more complex than the second-quantized algorithm, but for problems in chemistry larger than a handful of particles, it is computationally faster~\cite{Kassal2008}.  A first-quantized simulation is essentially a finite-difference method for solving the Schr\"{o}dinger equation.  Configuration space is discretized into a Cartesian grid, and each particle (\emph{e.g.} electron) has a wavefunction expressed in a quantum register that which encodes a probability amplitude at each coordinate on the grid.  For example, form a position-basis representation for a single electron on a $2^{p} \times 2^{p} \times 2^{p}$ grid, which requires only $3p$ qubits.  Explicitly, the electronic wavefunction is represented as
\begin{equation}
\label{first_quant_wavefunction}
\ket{\psi_e} = \sum_{x,y,z = 0}^{2^p - 1} c(x,y,z)\ket{x}\ket{y}\ket{z} = \sum_{\bf r} c({\bf r})\ket{\bf r},
\end{equation}
where $c(x,y,z)$ is the complex probability amplitude for the electron to occupy the volume element centered at the position ${\bf r} \equiv (x,y,z)$.  The rightmost part of Eqn.~(\ref{first_quant_wavefunction}) is shorthand that will be used throughout this section.  The spin degree of freedom can be incorporated by including an extra qubit, and to describe a many-electron state, the wavefunction has to be properly anti-symmetrized~\cite{Abrams1997,Ward2009}.

Methods to simulate the evolution of a time-independent molecular Hamiltonian $\hat{H}$ for problems in quantum chemistry were developed in Refs.~\cite{Zalka1998,Kassal2008}.  The complete Hamiltonian in first-quantized form can be expressed as the sum of the kinetic ($\hat{T}$) and potential ($\hat{V}$) operators
\begin{equation}\label{Hamiltonian}
\hat{H} = \hat{T} + \hat{V} =  - \sum_j \frac{\hbar^2 {\nabla_j}^2}{2m_j}  + \frac{1}{2}\sum_{j \ne k} \frac{q_j q_k }{4\pi \epsilon _0 r_{jk}},
\end{equation}
where the indices $j$ and $k$ run over all particles (electrons and nuclei) of any given molecule. Here $r_{jk}  \equiv \left| {\bf r}_j  - {\bf r}_k \right|$ is the distance between particles $j$ and $k$, which carry charges $q_j$ and $q_k$ respectively.

Let us outline how first-quantized simulation works before delving into details.  The core of the algorithm is evolving the Hamiltonian in simulated time, achieved by applying the propagator $U(t) = \exp(-i\hat{H}t)$ (setting $\hbar = 1$ and assuming $\hat{H}$ is time-independent), which solves the time-dependent Schr\"{o}dinger equation~\cite{Lloyd1996}. This process is readily achieved using the split operator approximation, a form of Trotter-Suzuki decomposition~\cite{Suzuki1992,Brown2010,Wiebe2010,Kassal2011}, where the kinetic and potential energy operators are simulated in alternating steps as
\begin{equation}
U(t) = e^{-i\hat{H}t} \approx \left[e^{-i \hat{T} \delta t/2}e^{-i \hat{V} \delta t}e^{-i \hat{T} \delta t/2}\right]^\frac{t}{\delta t}.
\end{equation}
The exponent $t/\delta t$ is the number of times the circuit corresponding to the expression in brackets is implemented, so it is always an integer.  The operators $e^{-i \hat{V} \delta t}$ and $e^{-i \hat{T} \delta t}$ are diagonal in the position and momentum bases, respectively.  One can switch the encoded configuration space representation between these two bases by applying the quantum Fourier transform to each spatial dimension of the wavefunction (\emph{cf.} Eqn.~(\ref{first_quant_wavefunction})), which can be efficiently implemented in a quantum computer~\cite{Weinstein2001}.  Reference~\cite{Kassal2008} shows how to construct quantum circuits for operators $e^{-i \hat{V} \delta t}$ and $e^{-i \hat{T} \delta t}$, as explained in greater detail below.

To make the first-quantized algorithm fault-tolerant, its constituent operations must be decomposed into circuits of fault-tolerant gates such as those in Section~\ref{quantum_programs_section}.  Consider the potential energy propagator $e^{-i \hat{V} \delta t}$ as an example. Given a $b$-particle wavefunction in the position basis as
\begin{equation}
\ket{\psi_{1,2,...,b}} = \sum_{{\bf r}_1,{\bf r}_2,...,{\bf r}_b} c({\bf r}_1,{\bf r}_2,...,{\bf r}_b)\ket{{\bf r}_1{\bf r}_2...{\bf r}_b},
\end{equation}
where $c(\cdot)$ is the complex amplitude as a function of position in configuration space and subscripts correspond to particles in the system, one calculates the phase evolution of the potential operator $e^{-i\hat{V}\delta t}$ in three steps, as follows:
\begin{eqnarray}
& \sum_{{\bf r}_1,...,{\bf r}_b} & c({\bf r}_1,...,{\bf r}_b)\ket{{\bf r}_1...{\bf r}_b}\ket{000...} \nonumber \\
& \longrightarrow & \sum_{{\bf r}_1,...,{\bf r}_b} c({\bf r}_1,...,{\bf r}_b)\ket{{\bf r}_1...{\bf r}_b}\ket{V({\bf r}_1,...,{\bf r}_b)} \label{compute_V} \\
& \longrightarrow & \sum_{{\bf r}_1,...,{\bf r}_b} e^{-iV({\bf r}_1,...,{\bf r}_b)\delta t} c({\bf r}_1,...,{\bf r}_b)\ket{{\bf r}_1...{\bf r}_b}\ket{V({\bf r}_1,...,{\bf r}_b)} \label{first_quant_phase} \\
& \longrightarrow & \sum_{{\bf r}_1,...,{\bf r}_b} e^{-iV({\bf r}_1,...,{\bf r}_b)\delta t} c({\bf r}_1,...,{\bf r}_b)\ket{{\bf r}_1...{\bf r}_b}\ket{000...}. \label{uncompute_V}
\end{eqnarray}
First, Eqn.~(\ref{compute_V}) calculates the potential energy $V({\bf r}_1,...,{\bf r}_b)$ as a function of position coordinates~\cite{Kassal2008} (note that $\hat{V}$ is diagonal in this basis) and stores the result in a quantum register $\ket{V({\bf r}_1,{\bf r}_2,...,{\bf r}_b)}$ to some finite precision.  Section~\ref{Sec_First_Quant_Operators} describes how to implement this operator using fault-tolerant gates for molecular Hamiltonians.  Second, Eqn.~(\ref{first_quant_phase}) uses the $\ket{V({\bf r}_1,{\bf r}_2,...,{\bf r}_b)}$ register in a quantum variable rotation (QVR from Section~\ref{Sec_QVR}) that imparts a phase to each grid point of the wavefunction in position basis proportional to the potential energy at those coordinates.  Finally, the quantum circuit from the first step is reversed in Eqn.~(\ref{uncompute_V}) to reset the $\ket{V({\bf r}_1,{\bf r}_2,...,{\bf r}_b)}$ register to $\ket{000...}$, also known as ``uncomputation''~\cite{Nielsen2000}.  The sequence of these three steps is equivalent to the operation $e^{-i\hat{V}\delta t}\ket{\psi}$.

The kinetic energy propagator $e^{-i \hat{T} \delta t}$ is calculated similarly in three steps, with the second also being a QVR.  This operator is diagonal in momentum basis, which can be reached by transforming the representation of the system wavefunction from position basis $\{x,y,z\}$ to momentum basis $\{k_x,k_y,k_z\}$ by applying a QFT along each spatial dimension of the encoding in Eqn.~(\ref{first_quant_wavefunction}).  This form permits efficient calculation of the kinetic energy operator~\cite{Kassal2008}, which is described in Section~\ref{Sec_First_Quant_Operators}.

\subsection{Constructions for First-Quantized Operators}
\label{Sec_First_Quant_Operators}
First-quantized molecular simulation represents the simulated system wavefunction on a Cartesian grid, and the Hamiltonian is calculated with digital arithmetic acting on this coordinate space.  Similar methods were discussed in the supplementary material for Ref.~\cite{Kassal2008}, but I provide analysis incorporating the QVR operation introduced in Section~\ref{Sec_QVR}.  The potential energy operator is diagonal in position basis, and it is the sum of Coulomb interactions between electrons and nuclei in the system: $\hat{V} = \frac{1}{2} \sum_{j \ne k} \hat{V}_{jk}$, where
\begin{equation}
\hat{V}_{jk} = \frac{q_j q_k}{4 \pi \varepsilon_0} \left(\frac{1}{|{\bf r_j} - {\bf r_k}|} \right)
\label{potential_term}
\end{equation}
and $q_j$ is the charge of particle $j$.  The prefactor on the RHS of Eqn.~(\ref{potential_term}) is a constant for any given pair of particles, so this scaling factor can be encoded into the QVR operation.  What remains is to calculate $\frac{1}{|{\bf r_j} - {\bf r_k}|}$ over the position-encoded wavefunction.  Each position register can be decomposed in Cartesian components $\ket{{\bf r}} = \ket{x}\ket{y}\ket{z}$.  For a given pair of particles, one calculates
\begin{equation}
\ket{{r_{jk}}^2} = \ket{\left(x_j - x_k\right)^2 + \left(y_j - y_k\right)^2 + \left(z_j - z_k\right)^2}.
\end{equation}
The required multiplication operations can be implemented using quantum adder circuits~\cite{Cuccaro2004,VanMeter2005,Draper2006}.  Next the quantity $\ket{\frac{1}{r_{jk}}}$ is calculated using the Newton-Raphson method with the iterative equation
\begin{equation}
a_{n+1} = \frac{1}{2} a_n \left(3 - {a_n}^2 {r_{jk}}^2 \right).
\label{Newton_Raphson}
\end{equation}
With suitably chosen initial value $a_0$, Eqn.~(\ref{Newton_Raphson}) converges within 5 iterations at 32-bit arithmetic, and typically less precision is required for simulation.  The register $\ket{\frac{1}{r_{jk}}}$ is used in a QVR with scaling factor $\xi = \frac{q_j q_k \delta t}{8 {\pi}^2 \varepsilon_0 \hbar}$ from above, where $\delta t$ is the time-step of this simulated evolution and an additional factor $1/{2 \pi}$ comes from Eqn.~(\ref{quantum_variable_rotation}).  Note that each component of $\ket{\frac{1}{r_{jk}}}$ is entangled to a position-basis component of the system wavefunction, so the QVR effectively kicks back a phase to the wavefunction.  Each of the steps prior to the QVR is uncomputed, and the net effect of this sequence of operations is to implement the potential energy propagator $e^{-i \hbar^{-1} \hat{V}_{jk} \delta t}$, as in Eqns.~(\ref{compute_V}--\ref{uncompute_V}).

The kinetic energy operator is calculated using a similar approach as the potential energy.  The kinetic energy is the sum of individual kinetic energy operators on each particle: $\hat{T} = \sum_j \hat{T}_j$, where
\begin{equation}
\hat{T}_j = \frac{{\hat{p}_j}^2}{2 m_j} = \frac{{\hbar^2 |{\bf k}_j|^2}}{2 m_j}.
\end{equation}
The quantity $m_j$ is the mass and ${\bf k}_j = {\bf p}_j/\hbar$ is the non-relativistic wavevector corresponding to particle $j$.  By performing a quantum Fourier transform along each spatial dimension of the wavefunction, the system representation is transformed from position basis to momentum basis: $\{x,y,z\} \rightarrow \{k_x,k_y,k_z\}$.  This form permits immediate calculation of magnitude squared of the wavevector:
\begin{equation}
\ket{|{\bf k}|^2} = \ket{{k_x}^2 + {k_y}^2 + {k_z}^2}.
\end{equation}
The $\ket{|{\bf k}|^2}$ register is used in a QVR with scaling factor $\xi = \frac{\hbar \delta t}{4 \pi m_j}$.  Afterwards, the intermediate registers used in the calculation of $\ket{|{\bf k}|^2}$ are uncomputed, and the end result is the operator $e^{-i \hbar^{-1} \hat{T}_{j} \delta t}$.

Each application of QVR requires a resource-costly Fourier state, but three properties of the first-quantized simulation algorithm make this approach efficient.  First, there are only a polynomial number of such operations: for $b$ particles, there are $b$ QVRs in the kinetic energy operator and $\frac{1}{2}b(b-1)$ QVRs in the potential operator.  Second, many of these QVRs have the same scaling factor $\xi$, so a phase kickback register can be reused many times without modification.  For example, the scaling factor in the kinetic energy operator is the same for all electrons and nuclei with the same mass.  Third, each $\ket{\gamma^{(k_{[\xi]})}}$ Fourier state can be calculated independently of other operations in the algorithm, so the impact of this process on circuit depth is minimal.

\subsection{Parallel Computation in Potential Energy Operator}
The majority of the circuit effort in first-quantized simulation is devoted to calculating the potential energy~\cite{Kassal2008}.  I consider here a technique to substantially reduce the time needed to calculate the potential energy operator $\hat{V}$, which is simply the sum of the Coulomb interactions $\hat{V}_{jk} = \frac{q_j q_k}{4\pi \epsilon_0 r_{jk}}$ between all pairwise combinations of the electrons and nuclei.  Note that this operator is a function of the positions ${\bf r}_i$ of the system particles only, so it is diagonal in the position basis $\ket{{\bf r}_1 {\bf r}_2 ...{\bf r}_b }$.  This fact means that all terms $\hat{V}_{jk}$ commute with each other, so they may be calculated in any order.  Moreover, there are many sets of the $\hat{V}_{jk}$ operators that are disjoint, which means that each particle in the system is acted on by just one operator in the set.  Using this observation, for example, one may calculate the Coulomb interaction $\hat{V}_{12}$ between particles 1 and 2 at the same time as $\hat{V}_{34}$ between particles 3 and 4, and so on.  In general, for a system of $b$ particles, there are $\frac{1}{2}b(b-1)$ pairwise interactions, and the algorithm can perform $\lfloor \frac{b}{2} \rfloor$ pairs in parallel, which means that a potential energy operator with $O(b^2)$ terms can be calculated in $O(b)$ time.  This parallelism can increase the speed of simulation significantly since evaluation of the potential energy dominates resource costs~\cite{Jones2012_PRX}.

The potential operator calculation can be further parallelized to achieve $O(\log b)$ or $O(1)$ (constant) circuit depth.  Exploiting the fact that all $\hat{V}_{jk}$ are diagonal in position basis (and hence commute), use transversal CNOT gates to copy the data in position-basis particle wavefunction onto multiple empty quantum registers.  For a single particle, this process is
\begin{flalign}
&\left(\sum_{x,y,z = 0}^{2^p - 1} c(x,y,z)\ket{x}\ket{y}\ket{z}\right)\ket{000...}\ket{000...} ... \nonumber \\
&\rightarrow \sum_{x,y,z = 0}^{2^p - 1} c(x,y,z)\left(\ket{x}\ket{y}\ket{z}\right) \left(\ket{x}\ket{y}\ket{z}\right) \left(\ket{x}\ket{y}\ket{z}\right) ...
\end{flalign}
For $b$ particles, the copy operation is performed $b-2$ times (for $b-1$ total copies), which can be fanned out using a binary tree with depth $\lceil\log_2(b-1)\rceil$; constant depth can be achieved in some quantum computer architectures which support one-control/many-target CNOTs, such as the surface code~\cite{Fowler2009_PRA,Jones2012_PRX} or in general architectures using a teleportation circuit similar to those described in Section~\ref{Sec_teleport_JW}.  This approach is similar to that employed in Ref.~\cite{Cleve2000} to produce a parallel circuit for the QFT.  The system wavefunction is now expanded to the state
\begin{equation}
\ket{\psi_{\mathrm{expand}}} = \sum_{{\bf r}_1,...,{\bf r}_b} c({\bf r}_1,...,{\bf r}_b)\left(\ket{{\bf r}_1}\right)^{\otimes (b-1)}...\left(\ket{{\bf r}_b}\right)^{\otimes (b-1)},
\end{equation}
which requires $O(b^2)$ memory space.  Note that this process is not cloning---the position-basis particle registers are still entangled to one another.  With multiple accessible copies of each particle's position-basis information, the particles are matched in all $b(b-1)$ possible pairings, and the potential energy operator applied to each pairing in parallel, which can be accomplished in constant time, but still requires $O(b^2)$ circuit effort.  After each of the potential energy operators $\hat{V}_{jk}$ kicks back a phase, the excess copies of each particle wavefunction are uncomputed by reversing the tree of CNOTs above.  The preceding example demonstrates that it is possible to calculate $\hat{V}$ in time which is sub-linear in the number of particles, even if each $\hat{V}_{jk}$ is treated as a black box operator.  In practice, more efficient circuits can be produced by generating the internal ``workspace'' registers of $\hat{V}$ in parallel, rather than making copies of the input registers $\sum_{{\bf r}_1,...,{\bf r}_b} c({\bf r}_1,...,{\bf r}_b)\ket{{\bf r}_1...{\bf r}_b}$ (see Section~\ref{Sec_First_Quant_Operators}).


\chapter{Strategies for Quantum Logic Synthesis}
\label{Ch09}
This chapter serves as a high-level summary of the methods for synthesizing quantum logic.  In particular, this chapter identifies patterns in the methods from preceding chapters.  These patterns may point to new methods for reducing resource costs, or they may suggest limitations on what is possible.  Either way, the patterns serve as guidelines to applying logic synthesis with current methods.  Each section of this chapter considers a strategy to synthesize resource-efficient programs, except the last, which makes some general observations and speculates on the limits of logic synthesis.

\section{Checkpoint States and Validation before Teleportation}
The resource overhead for error correction can be greatly reduced by using post-selected computation.  In this paradigm, intermediate states of the algorithm are verified before proceeding.  I will call these ``checkpoint states.''  The only requirement of a checkpoint state is that it must be sufficiently well described that a validation procedure is possible.  A checkpoint state is useful in practice when the validation procedure requires fewer resources than a naive fault-tolerant program which creates the same state.

Encoding processes into quantum look-up tables (QLUTs) is an effective use of checkpoint states.  The maxim ``validation before teleportation'' is directly related to QLUTs.  By design, each QLUT is independent of the rest of the computation.  If a faulty QLUT is identified, it may be discarded with no far-reaching consequences, and the only penalty is that the QLUT preparation must be repeated.  If the probability of the QLUT being rejected is modestly low (say $1\%$ or less), then the overhead of occasionally rejecting QLUTs is negligible.  At the same time, a well-designed validation protocol can produce a high-fidelity QLUT using fewer resources than the logically equivalent but naive procedure.  When the QLUT preparation succeeds, the associated gate is teleported into the computation.

\section{Validation for Asymmetric Error Models}
Asymmetric error models present an opportunity to reduce resource costs.  Simply put, one should allocate error correction only where it is needed.  Several examples of this heuristic appeared in Chapters~\ref{Ch05} and~\ref{Ch06}.  The reason for using magic-state distillation is that it completes universality, filling in the gaps of an error correction scheme like the surface code.  When universality is achieved through distillation, most of the underlying gates used by the protocol are protected by the surface code.  The distillation code is chosen in such a way that best addresses $T$-gate errors, and often these codes are not especially good for other purposes.

Toffoli gates can also use codes suited to a very specific error model.  The D2 construction (Section~\ref{Sec_D2}) fundamentally relies on a phase-flip code to catch an error in one of the $T$~gates.  The CNOT gates are arranged in a way that ensures such an error is caught, using the assumption that these gates are made sufficiently reliable by the surface code.  The C4C6 construction expands on this approach to catch as many as three errors in $T$~gates, but C4C6 introduces a new trick.  Because the validation in C4C6 can catch logical $Z$ errors, the demands on the surface code can be relaxed by placing primal defects closer together (Section~\ref{Sec_C4C6}).  The reduced spacing of defects saves resources, but it depends on the overlapping nature of the surface code and the $C_4/C_6$ code, which work together instead of independently.  Because logical $Z$ errors are only detected (not corrected), the state must be discarded upon such an event, but this post-selection was already incorporated for $T$-gate errors.

The distillation of Fourier states in Section~\ref{Sec_FS_Distillation} is another area where error detection is customized.  The distillation protocol acts in the basis of Fourier states, and a quantum Fourier transform (QFT) is required to measure in this basis.  The trick here is that the protocol only needs to detect the Fourier state $\ket{\gamma^{(0)}} = \ket{+}\ket{+}\ldots$, which happens to be a state that can be detected with $X$-basis measurements native to the surface code (and many other codes).  If any other result is obtained, the protocol has certainly failed, so the relatively costly QFT is avoided.

\section{Hierarchical Logic Synthesis}
\label{Sec_Hierarchical}
Hierarchical logic synthesis is a design principle where error suppression is achieved in multiple stages.  The quantum program is separated into subprograms, each having economical validation.  These subprograms are further divided into verifiable components, and so on.  By distributing error correction across multiple levels, there is no wasted effort.  A counter-example is the naive design principle where one distills very high fidelity $T$~gates, then produces programs from them without any subsequent validation.  This was the approach taken in Chapter~\ref{Ch02} and Ref.~\cite{Jones2012_PRX}.  This was also the Trivial Upper Bound for Toffoli gates in Section~\ref{Sec_Toffoli_analysis}, and the optimized logic that divided error correction between $T$~gates and Toffoli (C4C6 in this case) reduced cost by a factor of 20 for output error $10^{-12}$ (see Fig.~\ref{Toffoli_TUB_comparison}).

Another example of hierarchical logic is Fourier-state distillation from Section~\ref{Sec_FS_Distillation}, which is tasked with producing the state $\ket{\gamma^{(1)}}$.  The Fourier state is separable, meaning it could be prepared by independent phase rotations from gate approximation sequences (Chapter~\ref{Ch07}); this could serve as the Trivial Upper Bound on resources.  Instead, Fourier-state distillation distills noisy Fourier states using adders (a form of phase kickback QVR, Chapter~\ref{Ch07}), which are made of optimized Toffoli gates (Chapter~\ref{Ch06}), which also use distilled $T$~gates (Chapter~\ref{Ch05}) for high-fidelity quantum programs.  Efficient validation is available for each process, so there is less burden on surface code error correction.  Importantly, the best overall design uses validation techniques in all stages, as opposed to, say, producing high-fidelity $T$~gates and naively building Toffoli~gates and adders from them.  If drawn out, the flow of logic components is a tree structure, where $T$~gates flow into Toffoli gates, which flow into adders; the adders themselves vary in size and error probability with the round of distillation.

The general strategy for hierarchical logic synthesis begins by finding checkpoint states which are easy to verify.  As shown in Chapters~\ref{Ch05} and~\ref{Ch06}, $\ket{A} = (1/\sqrt{2})(\ket{0} + e^{i\pi/4}\ket{1})$ (QLUT for $T$~gate), the QLUT for the Toffoli gate, and Fourier states make good checkpoints.  The next step is to synthesize a quantum program in a tree fashion, where each level is a checkpoint state.  As many states from level $n-1$ feed into level $n$, the error probability is approximately the sum of errors for all input states.  The error at level $n$ is then suppressed using the appropriate validation procedure.  In this manner, the error probability in any component of the program can actually be much higher than the naive approach that suppresses errors at the lowest level (think high-fidelity $T$~gates, as in Chapter~\ref{Ch02}) and does not verify intermediate states.  Less error correction means lower resource costs, which is why hierarchical designs emerged as the efficient solution in several places throughout this thesis.  I believe this hierarchical approach will play an important role in the future of logic synthesis.

\section{General Observations and Speculation}
Multiple examples of checkpoint states were identified in Chapters~\ref{Ch05}, \ref{Ch06}, and~\ref{Ch07}.  These include: $\ket{A}$, the QLUT for $T$~gates (Sections~\ref{Sec_MS_Distillation} and~\ref{Sec_Multilevel_Distillation}); Fourier states (Section~\ref{Sec_FS_Distillation}); the QLUT for controlled-controlled-$Z$ and Toffoli gates (Chapter~\ref{Ch06}); and programmable ancilla rotations (Section~\ref{Sec_PAR}).  All of these checkpoint states have notable common features.  Each is a QLUT that replaces an otherwise costly quantum program.  For each, the corresponding program is either diagonal in the computational basis or simple to diagonalize, such as diagonalizing Toffoli with Hadamard gates.  Finally, the eigenvalues of the associated program can be compactly described, either because the program acts on one to three qubits or the eigenvalues follow a simple pattern, such as the eigenvalues of phase kickback QVR (associated with a Fourier state) being roots of unity and regularly spaced around the unit circle.

The observed patterns are directly related to a QLUT being useful in practice.  A checkpoint state having a simple description is necessary for a simple validation procedure to exist.  This allows us to speculate, is it always advisable to transform a program into a teleportation procedure using a checkpoint state?  Probably not.  Technically, one can validate any state by, for example, finding a unitary gate for which this state is an eigenvector with eigenvalue distinct from all other eigenvectors, then performing phase estimation; and one can encode any program into a QLUT, as shown in Chapter~\ref{Ch04}.  However, an important practical criterion for checkpoint states and specifically QLUTs is that their use saves resources.  With current understanding, the use of checkpoint states only yields a net reduction in resources for special cases like those considered above.

To see why it important to understand the limits on checkpoint states, consider an important unsolved case.  Simulation algorithms often repeat the same $U(t) = e^{-i \hat{H} t}$ process, where $\hat{H}$ is some encoded Hamiltonian, many times as a subroutine in phase estimation (see Chapter~\ref{Ch08}).  Having a QLUT for $U(t)$ could in theory save substantial resources.  However, the point of the simulation algorithm is to diagonalize $U$, because for general Hamiltonians this is a hard problem.  A paradox seems to exist: designing a QLUT with validation requires diagonalizing $U$, but diagonalizing $U$ obviates the need for the QLUT.  Without a simple description for the eigenvalues/vectors of $U$, there is no currently known method to produce the associated QLUT-driven program efficiently.  Finding a more general procedure for synthesizing QLUT-driven programs, especially those associated with processes like $U(t)$ that in general do not have compactly described eigenvalues and eigenvectors, would be a valuable tool in logic synthesis. 


\chapter{Discussion}
\label{Ch10}
Quantum logic synthesis is maturing as a research field.  The earliest incarnations were concerned with counting the number of logical qubits or gates used by an algorithm~\cite{Vedral1996,Zalka1998,Nielsen2000,Beauregard2003,Takahashi2006}.  After the results of magic-state distillation percolated through the community, it was appreciated that some gates, like the $T$~gate, were much more costly than others~\cite{Knill2005,Isailovic2008,VanMeter2009,Jones2012_PRX,Fowler2012_Architecture}.  The maxim that emerged was ``minimize number of $T$~gates.'' (Arguably, this is still conventional wisdom at the time of writing).  However, algorithms rarely need $T$~gates explicitly.  The most frequent use of $T$~gates is to make Toffoli gates, and Chapter~\ref{Ch06} shows that there are better ways to make Toffoli gates than naively producing high-fidelity $T$~gates, where the latter was the basis of the Trivial Upper Bound for a Toffoli gate.  The evolving role of logic synthesis is to find the true costs of fault-tolerant quantum computing, and minimize them however possible.

The importance of logic synthesis was demonstrated here multiple times, with the most dramatic results given in Chapters~\ref{Ch05} and~\ref{Ch06}.  Scaling surface code distance to a level appropriate to each round of Bravyi-Kitaev distillation reduces volume by about a factor of 25, for $T$-gate error of $10^{-12}$ (see Fig.~\ref{BK_resource_scaling}).  For Toffoli gates with error $10^{-12}$, selecting the optimal C4C6 design over the 7T design reduces volume by a factor of 20, when both have access to optimized $T$~gates (Fig.~\ref{Toffoli_TUB_comparison}).  Without optimized $T$~gates, the disparity is a factor of 500.  Considering not-optimized logic was used to produce the resource estimates in Chapter~\ref{Ch02} and Ref.~\cite{Jones2012_PRX} (as well as others in the literature), there is substantial room to lower the requirements for quantum hardware.  The intimidating predictions of ``100 million qubits,'' given just a year ago~\cite{Jones2012_PRX,Fowler2012_Architecture}, are not the final word in fault-tolerant quantum logic.

There are limitations in the analysis I provide, though this represents an opportunity for future work.  The only form of magic-state distillation analyzed fully in the surface code was Bravy-Kitaev 15-to-1 distillation for $T$~gates.  I give two justifications for this.  First, as discussed in Section~\ref{Sec_Multilevel_Distillation}, codes which distill multiple output states lead to correlated errors, which are problematic.  The 15-to-1 protocol is the only one I considered that does not have this complication.  Second, a list of protocols was analyzed under the auspices of multilevel distillation.  Therein, I did find protocols that outperform Bravyi-Kitaev \emph{as measured by counting input states to output states} by about a factor of two or three around error $10^{-12}$ (Fig.~\ref{Protocols_plot}).  This counting of states is easy to do, but it is not a complete representation of resource costs.  I avoided distillation protocols that distill multiple qubits with correlated errors because they are complicated to construct in the surface code.  However, these same protocols might be more efficient than Bravyi-Kitaev, as shown in Ref.~\cite{Fowler2013_BlockCodes}, depending on how much overhead is required for ``routing'' qubits around.  Given the importance of magic-state distillation in general, this matter deserves further analysis.

Chapter~\ref{Ch07} concludes with an interesting result: a handful of methods for approximating phase rotations, developed in different ways by different authors, have similar performance.  I noted before that this could suggest some fundamental lower bound on resource costs, but this is still speculative.  The findings do show that no single method is always the best, and I selected what I think are the most promising techniques developed so far.  For even such a narrowly defined problem, opportunities remain for optimization and custom tailoring to yet-to-be-built machine architectures.  I find this exciting, and I look forward to seeing how the field evolves.

The next frontier of logic synthesis, I believe, is hierarchical quantum logic.  The argument to support this claim is inductive.  When a program is divided into subprograms, the resource cost is minimized when those subprograms have efficient verification.  If one zooms into each subprogram, then the best strategy is to verify each of its subroutines, and so on.  Fourier-state distillation, proposed in Section~\ref{Sec_FS_Distillation} and discussed further in Section~\ref{Sec_Hierarchical}, provides an excellent example.  The optimal design, which comes naturally from a broad search over possible protocols, divides the problem into multiple levels of verification, from $T$~gates to Toffoli gates to adders of varying size and reliability.  This hierarchical design simply emerges as the one that minimizes resource costs, but it also highlights what I think is a valuable principle for logic synthesis in the future.

Finally, the tone throughout much of this thesis has been one of ``survival of the fittest,'' where different logic constructions go head-to-head, vying for the title of Most Efficient.  I believe this was appropriate, but not for the apparent reason of finding the Most Efficient.  The true value of this work is the methodology, which has two components: (1)~knowing how to design logic, both for correctness and with good heuristics; and (2)~knowing how to evaluate the resource cost of logic.  Although I am satisfied with my analysis and, in some cases, proud of the techniques I develop, I sincerely hope that the logic constructions considered here are made obsolete as soon as possible.  To do so would require novel advances in logic synthesis, which is what this has all been about.





    \bibliographystyle{unsrt}

\onlinesignature

\end{document}